\documentclass{elsart}

\usepackage{epsfig}
\usepackage{amsmath,amssymb,graphicx,tabularx}
\usepackage{epstopdf}

\newcounter{saveeqn}

\def\lambdabar{\protect\@lambdabar}
\def\@lambdabar{%
\relax
\bgroup
\def\@tempa{\hbox{\raise.73\ht0
\hbox to0pt{\kern.25\wd0\vrule width.5\wd0
height.1pt depth.1pt\hss}\box0}}%
\mathchoice{\setbox0\hbox{$\displaystyle\lambda$}\@tempa}%
{\setbox0\hbox{$\textstyle\lambda$}\@tempa}%
{\setbox0\hbox{$\scriptstyle\lambda$}\@tempa}%
{\setbox0\hbox{$\scriptscriptstyle\lambda$}\@tempa}%
\egroup
}
\def\chem#1#2{$\rm{}^{#1}\kern-0.8pt#2$}
\def\reac#1#2#3#4#5#6{$\rm\,{}^{#1}\kern-0.8pt{#2}\,({#3}\,,{#4})\,
{}^{#5}\kern-0.8pt{#6}\,$}
\def\gsimeq{\,\,\raise0.14em\hbox{$>$}\kern-0.76em\lower0.28em\hbox  
{$\sim$}\,\,}  
\def\lsimeq{\,\,\raise0.14em\hbox{$<$}\kern-0.76em\lower0.28em\hbox  
{$\sim$}\,\,}  
\def\beqy{\begin{eqnarray}}
\def\eeqy{\end{eqnarray}}
\def\bmlet{\begin{mathletters}}
\def\emlet{\end{mathletters}}

\begin{document}
 \begin{frontmatter}  
 
\title{The r-process of stellar nucleosynthesis: Astrophysics and nuclear
physics achievements and mysteries}
\author{M. Arnould, S. Goriely, and K. Takahashi}
\address{Institut d'Astronomie et d'Astrophysique, Universit\'e Libre de Bruxelles,\\
 CP226, B-1050 Brussels, Belgium}
  
\begin{abstract}

 The r-process, or the rapid neutron-capture process, of stellar nucleosynthesis is called for 
to explain the  production of the stable  (and some long-lived radioactive) neutron-rich nuclides 
heavier than iron that are observed in stars of various metallicities, as well as in the solar
system.

    A very large amount of nuclear information is necessary in order to
    model the r-process. This concerns the static characteristics of a large
    variety of light to heavy nuclei between the valley of stability
    and the vicinity of the neutron-drip line, as well as their beta-decay
    branches or their reactivity. Fission probabilities of very neutron-rich
    actinides have also to be known in order to determine the most massive nuclei
    that have a chance to be involved in the r-process. Even the properties of
    asymmetric nuclear matter may enter the problem. The enormously challenging
    experimental and theoretical task imposed by all these requirements is
    reviewed, and the state-of-the-art development in the field is presented.

    Nuclear-physics-based and astrophysics-free r-process models of
    different levels of sophistication have been constructed over the years.
    We review their merits and their shortcomings. The ultimate goal of r-process studies is
    clearly to identify realistic sites for the development of the r-process.
    Here too, the challenge is enormous, and the solution still eludes us.
    For long, the  core collapse supernova of massive stars
    has been envisioned as the privileged r-process location. We present a brief summary
    of the one- or multidimensional spherical or non-spherical
    explosion simulations available to-date.  Their predictions are confronted
    with the requirements imposed to obtain an r-process. The possibility of r-nuclide 
    synthesis during the decompression of the matter of neutron stars following their 
    merging  is also discussed.

    Given the uncertainties remaining on the astrophysical r-process site
    and on the involved nuclear physics, any confrontation between predicted
    r-process yields and observed abundances is clearly risky. A comparison
    dealing with observed r-nuclide abundances in very metal-poor stars and in
    the solar system is attempted on grounds of r-process models based on
    parametrised astrophysics conditions. The virtues of the r-process product
    actinides for dating old stars or the solar system are also critically reviewed.
 
\end{abstract}

\end{frontmatter}
\vskip0.5truecm

\section{Introduction}
\label{intro}
 
A myriad of observations provide a picture of the composition of the various
constituents of the Universe that is getting quickly more and more complete, and
concomitantly more and more complex. Despite this spectacular progress, the solar system
(hereafter SoS) continues to provide a body of abundance data whose quantity, quality and
coherence remain unmatched. This concerns especially the heavy elements (defined here as
those with atomic numbers in excess of the value $Z = 26$ corresponding to iron), and 
in particular  their isotopic compositions, which are the prime fingerprints of astrophysical
nuclear processes. Except in a few instances, these isotopic patterns indeed remain
 out of reach even of the most-advanced stellar spectroscopic techniques available 
today. No wonder then that, from the early days of its development, the theory of 
nucleosynthesis has been deeply rooted in the SoS composition, especially in the 
heavy element domain.

Since \cite{BBFH57}, it has proved operationally most rewarding to introduce three categories
of heavy nuclides referred to as s-, p-, and r-nuclides. This splitting is not a mere game. It
corresponds instead to the `topology' of the chart of the nuclides, which exhibits three
categories of stable heavy nuclides: those located at the bottom of the valley of nuclear
stability, called the s-nuclides, and those situated on the neutron-deficient or neutron-rich
side of the valley, named the p- or r-nuclides, respectively. Three different mechanisms are
called for to account for the production of these three types of stable nuclides. They are
naturally referred to as the s-, r-, and p-processes. An extensive survey of the p-process has
been published recently \cite{arnould03}, and an overview of the s-process in low- and
intermediate mass stars has been prepared by \cite{busso99}. 

The main aim of this review is to expand on the limited surveys of the r-process of 
\cite{meyer94,AK99} both in its astrophysics and nuclear physics components. It is of course 
impossible to do justice to the very abundant literature on the subject. It may not be 
desirable either to try to come close to exhaustiveness in this matter, as it
 would bring more confusion than otherwise.  Also note that no work published or made 
available after 30 September 2006 is referred to in this review.

\section{Observed abundances of the r-nuclides}
\label{observations}

\subsection{The bulk solar system composition}
\label{obs_solar}

Much effort has been devoted over the years to the derivation of a meaningful set of
elemental abundances representative of the composition of the bulk material from which the
SoS formed some 4.6 Gy ago. This bulk material is made of a well-mixed blend of many
nucleosynthesis agents that have contributed to its composition over the approximate 10 Gy
that have elapsed between the formations of the Galaxy and of the SoS. The latest
detailed analysis of the SoS is found in \cite{lodders03}. As in previous compilations,
the selected abundances are largely based on the analysis of a special class of rare
meteorites, the CI1 carbonaceous chondrites, which are considered as the  least-altered
samples of primitive solar matter available at present. Materials from other origins may
exhibit substantial deviations from the CI1 in their elemental compositions. This results from
the physio-chemical processes that may have operated at different levels in different 
phases of the solar system material.

Solar spectroscopic data for some elements up to Fe have been reanalysed in the framework of
 time-dependent three-dimensional hydrodynamical atmosphere models, which has led to spectacular 
revisions of the solar photospheric abundances of some major elements lighter than Ne 
\cite{asplund05}.  In general, the solar abundances now come in quite good agreement with the 
 CI1 data for a large variety of elements. 
Some notable differences result from depletion in the Sun (Li) or in meteorites (H,
C, N, O). The newly-determined solar Na abundance does not quite overlap with the meteoritic
value. A marginal agreement (within quite large uncertainties) is found for Cl and Au, while
the results for Ga, Rb, Ag, In and W are discordant. At least some of these discrepancies may
be attributed to spectroscopic or atomic data problems. The abundances of the noble gases Ar,
Kr and Xe, as well as of some specific elements like Hg, have still to rely on theoretical considerations.  

The isotopic composition of the elements in the SoS is mostly based on the terrestrial data, except for H and the noble gases \cite{lodders03}, where some adjustments are also applied for
Sr, Nd, Hf, Os, and Pb. The practice of using terrestrial isotopic data
is justified by the fact that, in contrast to the elemental abundances, the isotopic patterns
are not affected to any significant level by geological processes. Only some minor
mass-dependent fractionation may operate. A notable exception to the high bulk isotopic
homogeneity comes from the decay of relatively short-lived radio-nuclides that existed in the
early  SoS and decayed in early-formed solids in the solar nebula. Also
interplanetary dust particles contain isotopic signatures apparently caused by chemical
processes. Additional isotopic `anomalies' are observed in some meteoritic inclusions or
grains. Isotopic anomalies in the SoS are discussed further in Sect.~\ref{anomalies}.   
 
The SoS nuclidic abundance distribution exhibits a high `iron peak'
centred around \chem{56}{Fe} followed by a broad peak in the  mass number $A \approx 80-90$ 
region, whereas double peaks show up at $A = 130\sim 138$ and $195\sim 208$. These peaks are 
superimposed on a curve decreasing rapidly with increasing $A$. It has
been realised very early that these peaks provide a clear demonstration that a
tight correlation exists between SoS abundances and nuclear neutron shell closures. 

\begin{figure}
\center{\includegraphics[width=0.9\textwidth,height=0.60\textwidth]{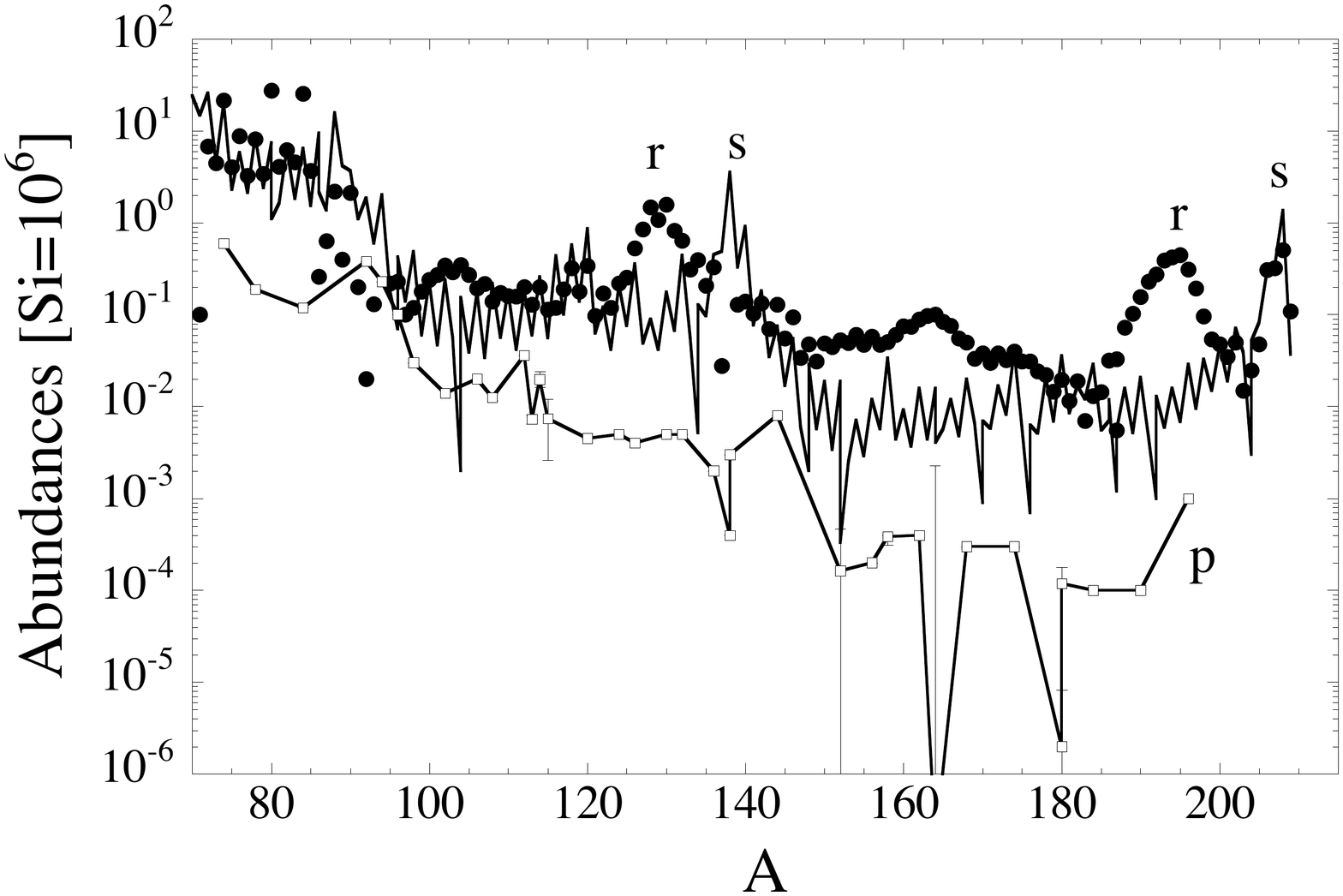}}
\vskip-0.7cm
\caption{Decomposition of the solar abundances of heavy nuclides into s-process ({\it solid
line}), r-process ({\it dots}) and p-process ({\it squares}) contributions. The
uncertainties on the abundances of some p-nuclides that come from a possible s-process contamination
 are represented by vertical bars (from \cite{arnould03}). See Figs.~\ref{fig_ssol_elem} -
 \ref{fig_rsol_isot} for the uncertainties on the s- and r-nuclide data}
\label{fig_solar}
\end{figure}  

\begin{figure}
\center{\includegraphics[width=0.9\textwidth,height=0.60\textwidth]{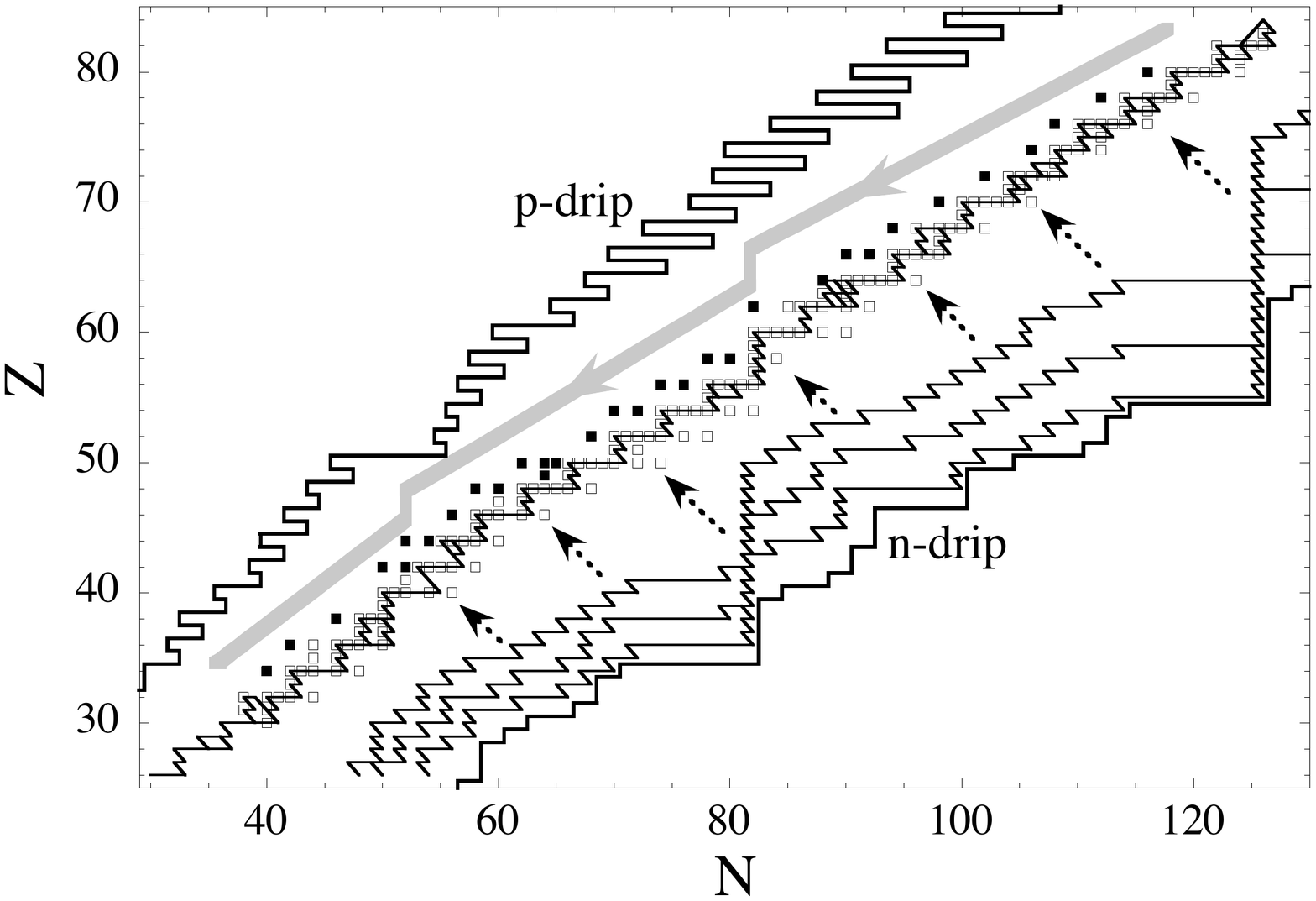}}
\vskip-0.9cm
\caption{Location in the $(N,Z)$-plane of the stable isotopes of the elements between Fe and
Bi. The p-isotopes are represented by black squares, while both the s- and r-isotopes 
(including the `sr'-isotopes of mixed origins, along with a few  `sp'-isotopes)
are identified with open squares (see Figs.~\ref{fig_ssol_elem} - \ref{fig_rsol_isot} for 
details).  The p-nuclides are the progeny of
unstable neutron-deficient isobars located on the down-streaming p-process flow (thick
greyish line with arrows; for more details on the p-process flow, see \cite{arnould03}).
  The r-process contribution to the r-only and `sr'-nuclides is provided by
the decay (represented by arrows) of the neutron-rich nuclides located on the
up-streaming r-process flow (three such flows are represented by solid zigzag lines) 
 associated with some r-process models (for more details, see Sects.~\ref{high_t_param} and 
 \ref{r_dccsn}; r-process flows of a quite different nature are possible, however, as 
discussed in Sects.~\ref{HIDER} and \ref{compact_general}). The
up-streaming s-process flow (thin black line) is confined at the bottom of the valley of
nuclear stability and brings the s-process contribution to the s-only and sr-nuclides. The
p- and n-drip lines represent the approximate
  locations of zero proton and neutron separation energies (see Sect.~\ref{nuc_static})}
\label{fig_srp}
\end{figure}  

\subsection{The s-, r- and p-nuclides in the solar system: generalities}
\label{obs_srp}
 
It is very useful to split the abundance distribution of the
nuclides heavier than iron into three separate distributions giving the image of the SoS
content of the p-, s- and r-nuclides. A rough representation of this splitting is displayed
in Fig.~\ref{fig_solar}. In its details, the procedure of decomposition is not as obvious as
it might be thought from the very definition of the different types of nuclides, and is to
some extent dependent on the models for the synthesis of the heavy nuclides. These models
predict in particular that the stable nuclides located on the neutron-rich/neutron-deficient
 side of
the valley of nuclear stability are produced, to a first good approximation, 
by the r-/p-process only.
 Figure~\ref{fig_srp} provides a schematic view of the flows resulting
from the action of the nuclear transmutations making up the p- and r-processes. 
The details of the flow patterns depend on the astrophysical models and
 on the adopted nuclear physics, as discussed in Sects.~\ref{high_t_param},
\ref{HIDER}, \ref{r_dccsn} and \ref{compact_general} for the r-process, and as reviewed by 
\cite{arnould03} for the p-process. In all cases it remains true, however, that highly 
neutron-rich/deficient and  
$\beta$-unstable nuclides are involved in the r-/p-process and cascade to the stable
neutron-rich/deficient nuclides when the nuclear transformations end for one reason or 
another.   These stable nuclides are naturally called `r-only' and `p-only' nuclides, and 
their abundances are deduced directly from the SoS abundances (Sect.~\ref{obs_solar}).  The
situation is more intricate for the nuclides situated at the bottom of the valley of nuclear
stability. Some of them are produced solely by the s-process, the typical flow of which runs
very close to the bottom of nuclear stability, as illustrated in Fig.~\ref{fig_srp}.
They are referred to as `s-only' nuclides, and are encountered only when a stable 
r-isobar exists, which `shields' the s-isobar from the r-process.

 As a result, only even-$Z$ heavy elements possess an s-only isotope. In general, a 
 phenomenological model of the s-process is used to fit at best the abundances of all 
the s-only nuclides (Sect.~\ref{sr-splitting}). Once the
parameters of this model have been selected in such a way, it is used to predict the s-process
contributions to the other s-nuclides. The subtraction of these s-process  contributions from
the observed solar abundances leaves for each isotope a residual abundance that represents the
contribution to it of the r-process (if neutron-rich) or  p-process (if neutron-deficient).
These nuclides of mixed origins are called `sr' or `sp' nuclides. 
 
Figure~\ref{fig_solar} shows that about half of the heavy nuclei in the solar material come
from the s-process, and the other half  from the r-process, whereas the p-process is
responsible for the production of about 0.01 to 0.001 of the abundances of the s-  and
r-isobars. It also appears that some elements have their abundances
dominated by an s- or r-nuclide. They are naturally referred to as s- or r-elements.
Clearly, p-elements do not exist. If this statement remains valid in other locations than the 
SoS, stellar spectroscopy can provide information on the s- or r-  (but not the p-) abundances 
outside of the  SoS. Even if the dominance of the s- or r-processes on a given element remains
 true in all astrophysical locations, a wealth of observations demonstrate significant
departures from the  SoS s- or
r-element abundances. Such departures exist in the SoS itself in the form of
`isotopic anomalies' (Sect.~\ref{anomalies}), or in stars with different ages, galactic
locations, or evolutionary stages (Sect.~\ref{galaxy}). The SoS abundances and their s-, r-
 and p-process contributions  do not have any `universal' character.
       
From the above short description of the splitting procedure between s-, r- and p-nuclides, it
is easily understood that uncertainties affect the relative s- and r-(p-) process
contributions to the SoS abundances of the sr(p)-nuclides. Even so, they are quite
systematically swept under the rug, as exemplified by the
recent work of \cite{simmerer05}. This question of the uncertainties clearly deserves a careful study,
especially in view of the sometimes very detailed and far-reaching considerations that have
the s-r  SoS splitting as an essential starting point.

\subsection{A detailed analysis of the s- and r-process contributions to the solar system s-
 and r-nuclides}
\label{sr-splitting}
 
We limit ourselves here to a discussion of the abundance distribution of the s-
and r-nuclides, the p-nuclides being dealt with in substantial detail by \cite{arnould03}. We
closely follow the analysis of \cite{goriely99}, which is the only attempt to evaluate on a
quantitative basis the uncertainties in the derived  SoS s- and r-abundances. These
uncertainties stem from various sources, including the measured abundances themselves, as
well as from the s-process models used in the splitting procedure. 
 
As recalled in Sect.~\ref{obs_srp}, the  SoS r-nuclide abundance distribution is
obtained by subtracting from the observed  SoS abundances
those predicted to originate from the s-process. These predictions are classically
based on a parametric model, referred to as the canonical exponential model initially
developed by \cite{clayton61}, and which has received some refinements over the years (e.g.
\cite{kappeler89}).   This model assumes that  stellar material composed only of iron nuclei
is subjected to neutron densities and  temperatures that remain constant over the whole
period of the neutron irradiation.  In addition, the SoS s-abundance pattern is
viewed as originating from a superposition of two exponential distributions of the
time-integrated neutron exposure,
$\tau_{\rm n}=\int_0^t N_{\rm n} v_T {\rm d}t$ (where $v_T$ is the most probable relative 
neutron-nucleus velocity
at temperature $T$). These distributions are traditionally held responsible for the so-called
weak ($70 \lsimeq A \lsimeq 90$) and main ($A \gsimeq 90$) components of the s-process. A third
 exponential
distribution is sometimes added in order to account for the $204 < A \le 209$ s-nuclides.
Through an adequate fitting of the parameters of the three $\tau$-distributions, the
superposition of the two or three resulting abundance components reproduces quite
successfully the abundance distribution of the s-only nuclides in the SoS, from which
it is concluded that the s-contribution to the sr-nuclides can be predicted reliably. It has
to be stressed that this result is rooted only in the nuclear properties of the species
involved in the s-process, and does not rely at all on specific astrophysics scenarios.  Many
s-process calculations have been performed in the framework of models for stars of various
masses and initial compositions (e.g. \cite{arnould97} - \cite{goriely05}). 
Some model calculations along the line have been used to obtain the contributions of the s- and 
r-processes to the SoS abundances   
(\cite{arlandini99,simmerer05}). This procedure is currently not advisable.
Large uncertainties remain in the s-abundances predicted from model stars. 
In addition, the SoS s-nuclide abundances result from a long evolution of the galactic composition
that cannot be mimicked reliably enough.

Despite the success of the canonical model in fitting the solar s-nuclide distribution,
some of its basic assumptions deserve questioning. This concerns in particular a 
presumed exponential form for the distribution of the neutron exposures $\tau$,
 which has been introduced by \cite{clayton61} in view of their mathematical ease in abundance
 calculations. In addition, the canonical model makes it difficult in the s-nuclide abundance 
predictions to evaluate uncertainties of nuclear or observational nature. As a result,
the concomitant uncertainties in the solar r-abundances are traditionally not evaluated.
The shortcomings of the canonical model are cured to a large extent by the so-called
multi-event s-process model (MES) \cite{goriely99}. In view of the importance to evaluate
the uncertainties affecting the solar distribution of the abundances of the r-nuclides, we
review the MES in some detail. A similar multi-event model has also been developed for the
r-process (MER), and is presented in Sect.~\ref{MER}.  

The MES relies on a superposition of a given number of canonical events, each of them being
 defined  by a neutron irradiation on the \chem{56}{Fe} seed nuclei during a time 
$t_{\rm irr}$ at a constant temperature $T$ and a constant neutron density $N_{\rm n}$. In 
contrast to the canonical model,  no hypothesis is made concerning any particular 
distribution of the neutron exposures. Only a set of canonical events that are
considered as astrophysically plausible is selected a priori.  We adopt here about 500
 s-process canonical events  covering ranges of astrophysical conditions that are
identified as relevant by the canonical model, that is, $1.5 \times 10^8 \le T \le 4 \times
10^8$ K, $7.5 \le \log  N_{\rm n} [{\rm cm^{-3}}] \le 10$, and 40 chosen $t_{\rm irr}$-values,
corresponding to evenly distributed values of $n_{\rm cap}$ in the $5 \le n_{\rm cap} \le 150$ range,
where
 
\begin{equation}
n_{\rm cap} = \sum_{Z,A}A~N_{Z,A}(t=t_{\rm irr}) - \sum_{Z,A}A~N_{Z,A}(t=0) 
\label{eq_ncap}
\end{equation}

 is the number of neutrons captured per seed nucleus (\chem{56}{Fe}) on the timescale
 $t_{\rm irr}$, the summation extending over all the nuclides involved in the s-process. For
 each of the selected canonical events, the abundances $N_{Z,A}$ are obtained by solving a 
reaction network including 640 nuclear species between Cr and Po. Based on these calculated abundances,
 an iterative inversion procedure described in \cite{goriely97} (see also Sect.~\ref{MER}) 
allows us to identify a combination of events from the considered set that provides the best
 fit to the solar abundances of a selected ensemble of nuclides.  This set includes 35 
nuclides comprising the s-only nuclides, complemented with $^{86}{\rm Kr}$ and $^{96}{\rm Zr}$ 
(largely produced by the s-process in the canonical model), $^{152}{\rm Gd}$ and 
$^{164}{\rm Er}$ 
(unable in  the p-process and able in the s-process  to be produced in solar abundances
\cite{arnould03}), and $^{208}{\rm Pb}$ (possibly produced by the strong s-process component
in the canonical model).  

\begin{figure}
\center{\includegraphics[scale=0.45]{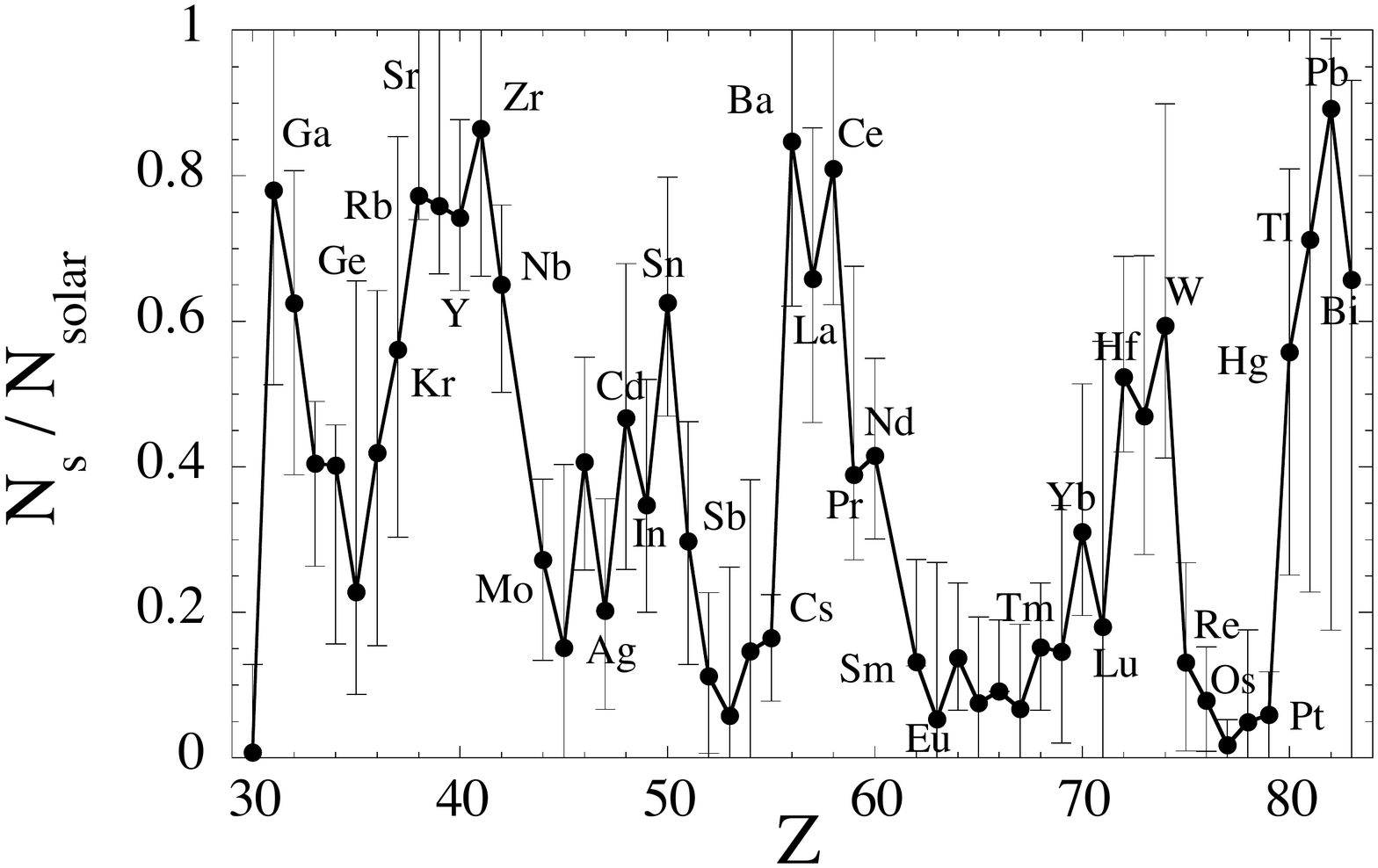}}
\vskip-1.3cm
\caption{MES predictions of the s-process contribution to the SoS abundances 
$N_{\rm solar}$ \cite{palme93} of the elements with $Z \geq 30$. Uncertainties are 
represented by vertical bars (from the calculations of \cite{goriely99}) }
\label{fig_ssol_elem} 
\end{figure}

\begin{figure}
\center{\includegraphics[scale=0.45]{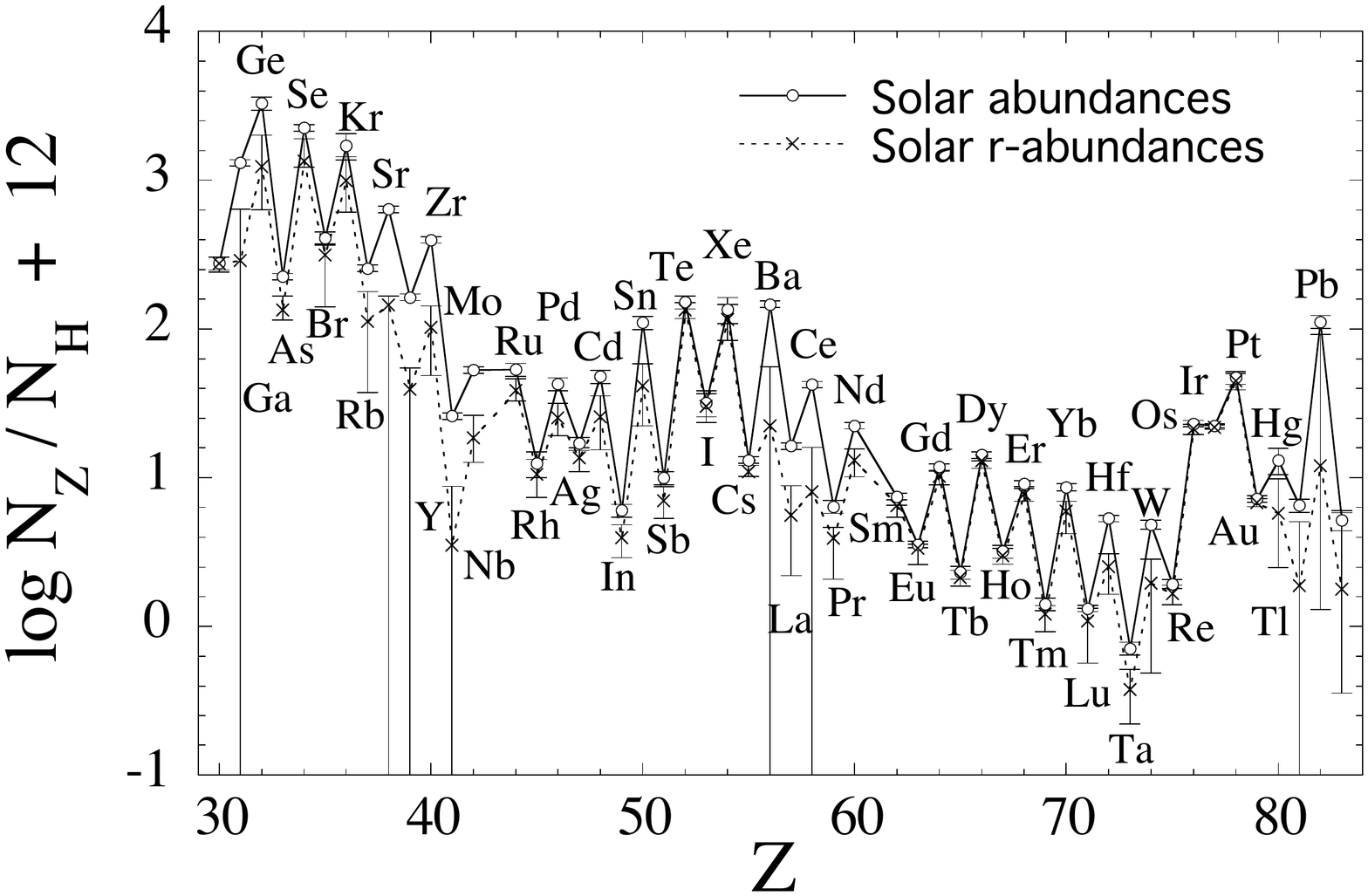}}
\vskip-1.3cm
\caption{SoS r-residuals and their uncertainties for the $Z \geq 30$ elements based on 
the s-abundances of Fig.~\ref{fig_ssol_elem}. The abundances $N_{\rm Z}$ and $N_{\rm H}$
 refer to element Z and to H.  The ordinate used here is about 1.55 dex larger than 
the [Si=10$^6$] unit used elsewhere}
\label{fig_rsol_elem} 
\end{figure}

\begin{figure}
\center{\includegraphics[scale=0.45]{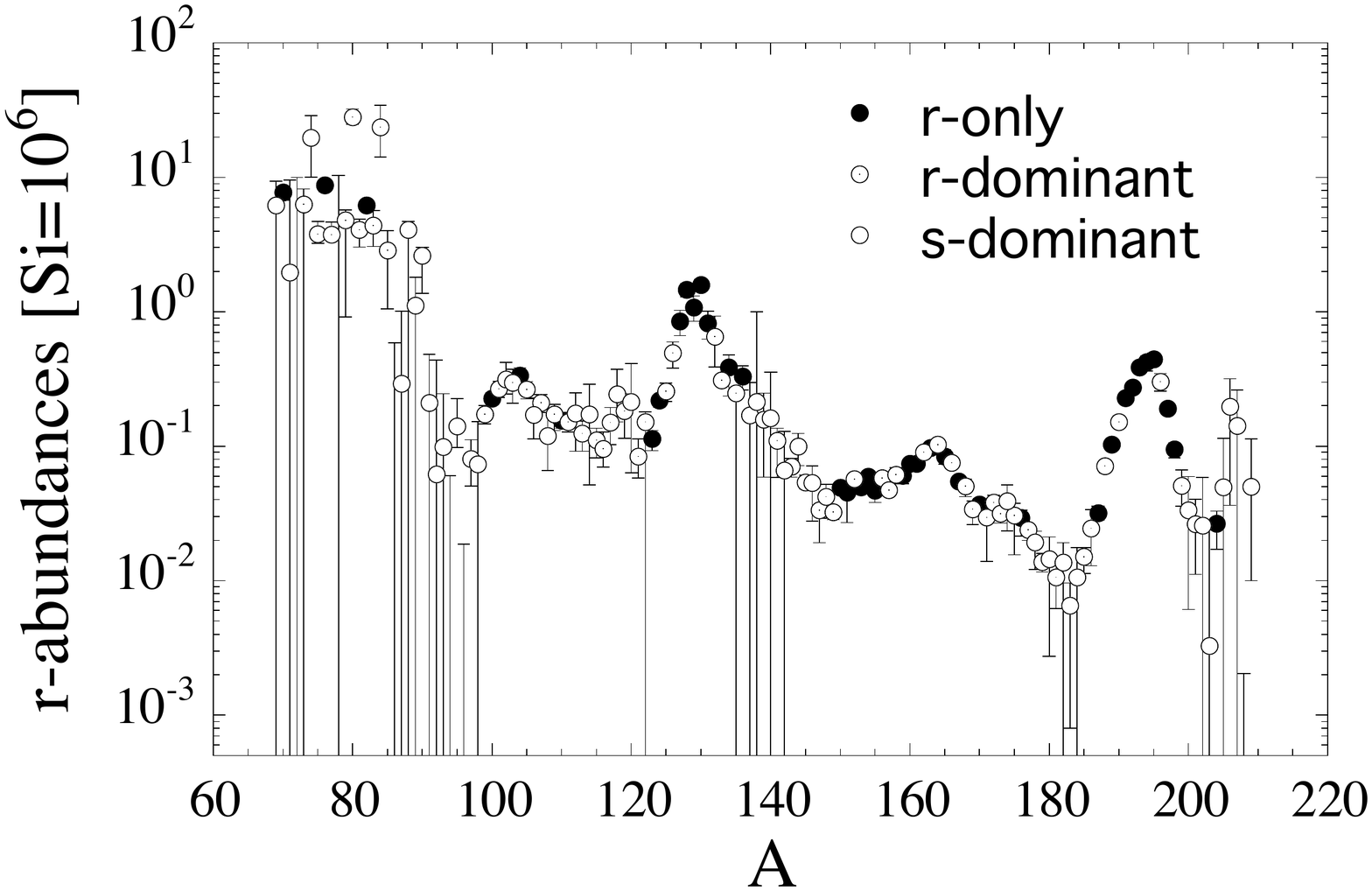}}
\vskip-1.0cm
\caption{SoS  isotopic r-residuals corresponding to the elemental abundances of 
Fig.~\ref{fig_rsol_elem}. Different symbols identify different relative levels of r-process
 contribution. The s-dominant nuclides are defined here as 
those predicted by MES to have more than 50\% of
 their abundances  produced by the s-process. The s-process contribution varies between 10 
and 50\% in the case of the r-dominant species, and does not exceed 10\% for the r-only nuclides}
\label{fig_rsol_isot} 
\end{figure}
 
On grounds of the solar abundances of \cite{palme93} (which are in close
agreement with those of \cite{anders89}), it has been demonstrated in \cite{goriely99} that
the derived MES distribution of neutron irradiation agrees qualitatively with the exponential
distributions assumed in the canonical model, even though some deviations are noticed with
respect to the canonical weak and strong components.\footnote{A MES calculation with the
revised solar abundances \cite{lodders03,asplund05} has not been done, but is expected not to
give significantly different results from those reported here}  
The MES provides an excellent fit to the
abundances of the 35 nuclides included in the considered set of species, and 
 in fact performs to a quite-similar overall quality as that of 
the exponential canonical model predictions of \cite{palme93}. 
Even a better fit than in the canonical framework is obtained
for the s-only nuclides (see \cite{goriely99} for details).  The MES model is therefore 
expected to provide a decomposition of the solar abundances into their s- and r-components
 that is likely to be more reliable than the one derived from the canonical approach 
for the absence of the fundamental assumption of exponential distributions of neutron exposures.

 
Compared with the canonical approach, the MES model has the major advantage of allowing a
systematic study of the various uncertainties affecting the abundances derived from
the parametric s-process model, and consequently the residual r-nuclide abundances. The
 uncertainties in these residuals have been evaluated in detail by \cite{goriely99} from due
 consideration
of the uncertainties in (i) the observed SoS abundances as given by \cite{palme93}
(see footnote$^1$), (ii) the experimental and theoretical radiative neutron-capture rates
involved in the s-process network, and  in (iii) the relevant $\beta$-decay and
electron-capture rates. Total uncertainties resulting from a combination of (i) to (iii) have
finally been evaluated. The results of such a study for the elements with $Z \geq 30$ are 
displayed in Figs.~\ref{fig_ssol_elem}  and \ref{fig_rsol_elem}. The corresponding SoS isotopic
 r-residuals and their uncertainties are shown in Fig.~ \ref{fig_rsol_isot} and listed in
 Table~\ref{tab_r}.  Different situations can be identified concerning the uncertainties 
affecting the r-residuals. Many sr-nuclides are predicted to have a small s-process component only.
The r-process contribution to these species, referred to as r-dominant, is clearly quite 
insensitive to the s-process uncertainties. The situation is just the opposite in the case of
 s-dominant nuclides.

\renewcommand{\arraystretch}{0.80}
\renewcommand{\baselinestretch}{0.5}
\begin{table*}
\label{tab_r}
\begin{center}
\caption{Standard, minimum and maximum r-process contributions to the
SoS abundances from \cite{palme93}  in the Si =  $10^6$ scale (see also Fig.~\ref{fig_rsol_isot})}
\begin{tabular}{ccccc|ccccc}
\hline
Z & A & standard & min & max & Z & A & standard & min & max\\
\hline
31	 & 	69	 & 	6.1800	 & 	0.0000	 & 	9.3700	 & 	58	 & 	140	 & 	0.1610	 & 	0.0000	 & 	0.3570	\\
30	 & 	70	 & 	7.7400	 & 	6.8000	 & 	8.5500	 & 	59	 & 	141	 & 	0.1100	 & 	0.0545	 & 	0.1360	\\
31	 & 	71	 & 	1.9600	 & 	0.0000	 & 	9.6100	 & 	58	 & 	142	 & 	0.0660	 & 	0.0000	 & 	0.1310	\\
32	 & 	72	 & 	0.0000	 & 	0.0000	 & 	9.9300	 & 	60	 & 	143	 & 	0.0706	 & 	0.0526	 & 	0.0811	\\
32	 & 	73	 & 	6.3100	 & 	0.0000	 & 	8.1900	 & 	60	 & 	144	 & 	0.0998	 & 	0.0582	 & 	0.1240	\\
32	 & 	74	 & 	19.700	 & 	9.9400	 & 	28.900	 & 	60	 & 	145	 & 	0.0540	 & 	0.0456	 & 	0.0611	\\
33	 & 	75	 & 	3.7800	 & 	3.2400	 & 	4.6800	 & 	60	 & 	146	 & 	0.0533	 & 	0.0145	 & 	0.0711	\\
32	 & 	76	 & 	8.7800	 & 	7.8400	 & 	9.6800	 & 	62	 & 	147	 & 	0.0334	 & 	0.0156	 & 	0.0347	\\
34	 & 	77	 & 	3.7600	 & 	3.4800	 & 	4.6500	 & 	60	 & 	148	 & 	0.0421	 & 	0.0221	 & 	0.0522	\\
34	 & 	78	 & 	0.0000	 & 	0.0000	 & 	10.300	 & 	62	 & 	149	 & 	0.0323	 & 	0.0278	 & 	0.0328	\\
35	 & 	79	 & 	4.8100	 & 	0.9180	 & 	5.7100	 & 	60	 & 	150	 & 	0.0490	 & 	0.0459	 & 	0.0515	\\
34	 & 	80	 & 	28.100	 & 	24.800	 & 	32.200	 & 	63	 & 	151	 & 	0.0452	 & 	0.0267	 & 	0.0482	\\
35	 & 	81	 & 	4.0700	 & 	3.0400	 & 	4.8700	 & 	62	 & 	152	 & 	0.0571	 & 	0.0498	 & 	0.0622	\\
34	 & 	82	 & 	6.2000	 & 	5.8300	 & 	6.5100	 & 	63	 & 	153	 & 	0.0495	 & 	0.0460	 & 	0.0526	\\
36	 & 	83	 & 	4.3800	 & 	3.0500	 & 	5.6800	 & 	62	 & 	154	 & 	0.0595	 & 	0.0505	 & 	0.0609	\\
36	 & 	84	 & 	23.600	 & 	14.200	 & 	34.500	 & 	64	 & 	155	 & 	0.0468	 & 	0.0364	 & 	0.0500	\\
37	 & 	85	 & 	2.8700	 & 	1.0500	 & 	4.0100	 & 	64	 & 	156	 & 	0.0579	 & 	0.0501	 & 	0.0634	\\
36	 & 	86	 & 	0.0000	 & 	0.0000	 & 	0.5870	 & 	64	 & 	157	 & 	0.0471	 & 	0.0429	 & 	0.0508	\\
37	 & 	87	 & 	0.2920	 & 	0.0000	 & 	1.0100	 & 	64	 & 	158	 & 	0.0614	 & 	0.0497	 & 	0.0694	\\
38	 & 	88	 & 	4.0900	 & 	0.0000	 & 	4.7500	 & 	65	 & 	159	 & 	0.0601	 & 	0.0517	 & 	0.0672	\\
39	 & 	89	 & 	1.1100	 & 	0.0000	 & 	1.8100	 & 	64	 & 	160	 & 	0.0741	 & 	0.0655	 & 	0.0787	\\
40	 & 	90	 & 	2.6100	 & 	1.2600	 & 	3.0100	 & 	66	 & 	161	 & 	0.0741	 & 	0.0684	 & 	0.0745	\\
40	 & 	91	 & 	0.2100	 & 	0.0000	 & 	0.4840	 & 	66	 & 	162	 & 	0.0900	 & 	0.0795	 & 	0.0917	\\
40	 & 	92	 & 	0.0620	 & 	0.0000	 & 	0.4370	 & 	66	 & 	163	 & 	0.0972	 & 	0.0890	 & 	0.0980	\\
41	 & 	93	 & 	0.0987	 & 	0.0000	 & 	0.2700	 & 	66	 & 	164	 & 	0.1030	 & 	0.0827	 & 	0.1040	\\
40	 & 	94	 & 	0.0000	 & 	0.0000	 & 	0.0602	 & 	67	 & 	165	 & 	0.0839	 & 	0.0728	 & 	0.0941	\\
42	 & 	95	 & 	0.1400	 & 	0.0976	 & 	0.2260	 & 	68	 & 	166	 & 	0.0753	 & 	0.0691	 & 	0.0833	\\
40	 & 	96	 & 	0.0000	 & 	0.0000	 & 	0.0250	 & 	68	 & 	167	 & 	0.0546	 & 	0.0495	 & 	0.0586	\\
42	 & 	97	 & 	0.0808	 & 	0.0496	 & 	0.1120	 & 	68	 & 	168	 & 	0.0506	 & 	0.0420	 & 	0.0570	\\
42	 & 	98	 & 	0.0739	 & 	0.0000	 & 	0.1530	 & 	69	 & 	169	 & 	0.0340	 & 	0.0250	 & 	0.0391	\\
44	 & 	99	 & 	0.1730	 & 	0.1460	 & 	0.2000	 & 	68	 & 	170	 & 	0.0369	 & 	0.0283	 & 	0.0407	\\
42	 & 	100	 & 	0.2260	 & 	0.2100	 & 	0.2500	 & 	70	 & 	171	 & 	0.0297	 & 	0.0107	 & 	0.0326	\\
44	 & 	101	 & 	0.2670	 & 	0.2300	 & 	0.3050	 & 	70	 & 	172	 & 	0.0381	 & 	0.0323	 & 	0.0432	\\
44	 & 	102	 & 	0.3150	 & 	0.2440	 & 	0.4260	 & 	70	 & 	173	 & 	0.0316	 & 	0.0266	 & 	0.0353	\\
45	 & 	103	 & 	0.2970	 & 	0.2090	 & 	0.3750	 & 	70	 & 	174	 & 	0.0391	 & 	0.0229	 & 	0.0515	\\
44	 & 	104	 & 	0.3370	 & 	0.2980	 & 	0.3830	 & 	71	 & 	175	 & 	0.0305	 & 	0.0156	 & 	0.0374	\\
46	 & 	105	 & 	0.2660	 & 	0.2240	 & 	0.3030	 & 	70	 & 	176	 & 	0.0292	 & 	0.0177	 & 	0.0334	\\
46	 & 	106	 & 	0.1710	 & 	0.1130	 & 	0.2280	 & 	72	 & 	177	 & 	0.0238	 & 	0.0186	 & 	0.0263	\\
47	 & 	107	 & 	0.2110	 & 	0.1780	 & 	0.2440	 & 	72	 & 	178	 & 	0.0192	 & 	0.0100	 & 	0.0236	\\
46	 & 	108	 & 	0.1190	 & 	0.0660	 & 	0.1930	 & 	72	 & 	179	 & 	0.0138	 & 	0.0109	 & 	0.0160	\\
47	 & 	109	 & 	0.1720	 & 	0.1310	 & 	0.2070	 & 	72	 & 	180	 & 	0.0145	 & 	0.0000	 & 	0.0214	\\
46	 & 	110	 & 	0.1560	 & 	0.1360	 & 	0.1740	 & 	73	 & 	181	 & 	0.0106	 & 	0.0042	 & 	0.0144	\\
48	 & 	111	 & 	0.1520	 & 	0.1270	 & 	0.1790	 & 	74	 & 	182	 & 	0.0136	 & 	0.0000	 & 	0.0215	\\
48	 & 	112	 & 	0.1760	 & 	0.0921	 & 	0.2500	 & 	74	 & 	183	 & 	0.0065	 & 	0.0000	 & 	0.0100	\\
48	 & 	113	 & 	0.1240	 & 	0.0916	 & 	0.1550	 & 	74	 & 	184	 & 	0.0106	 & 	0.0000	 & 	0.0179	\\
48	 & 	114	 & 	0.1720	 & 	0.0515	 & 	0.2910	 & 	75	 & 	185	 & 	0.0151	 & 	0.0110	 & 	0.0176	\\
49	 & 	115	 & 	0.1110	 & 	0.0816	 & 	0.1360	 & 	74	 & 	186	 & 	0.0245	 & 	0.0073	 & 	0.0337	\\
48	 & 	116	 & 	0.0955	 & 	0.0697	 & 	0.1270	 & 	75	 & 	187	 & 	0.0318	 & 	0.0270	 & 	0.0359	\\
50	 & 	117	 & 	0.1500	 & 	0.1030	 & 	0.1930	 & 	76	 & 	188	 & 	0.0708	 & 	0.0633	 & 	0.0781	\\
50	 & 	118	 & 	0.2440	 & 	0.1510	 & 	0.3750	 & 	76	 & 	189	 & 	0.1030	 & 	0.0961	 & 	0.1090	\\
50	 & 	119	 & 	0.1840	 & 	0.1150	 & 	0.2470	 & 	76	 & 	190	 & 	0.1520	 & 	0.1370	 & 	0.1680	\\
50	 & 	120	 & 	0.2140	 & 	0.0634	 & 	0.4120	 & 	77	 & 	191	 & 	0.2290	 & 	0.2210	 & 	0.2370	\\
51	 & 	121	 & 	0.0836	 & 	0.0578	 & 	0.1130	 & 	76	 & 	192	 & 	0.2730	 & 	0.2520	 & 	0.2890	\\
50	 & 	122	 & 	0.1520	 & 	0.0000	 & 	0.1800	 & 	77	 & 	193	 & 	0.3880	 & 	0.3740	 & 	0.4020	\\
51	 & 	123	 & 	0.1130	 & 	0.0925	 & 	0.1310	 & 	78	 & 	194	 & 	0.4210	 & 	0.3620	 & 	0.4700	\\
50	 & 	124	 & 	0.2200	 & 	0.1950	 & 	0.2420	 & 	78	 & 	195	 & 	0.4450	 & 	0.3940	 & 	0.4930	\\
52	 & 	125	 & 	0.2560	 & 	0.2170	 & 	0.2950	 & 	78	 & 	196	 & 	0.3020	 & 	0.2560	 & 	0.3470	\\
52	 & 	126	 & 	0.4920	 & 	0.3810	 & 	0.6010	 & 	79	 & 	197	 & 	0.1910	 & 	0.1790	 & 	0.2040	\\
53	 & 	127	 & 	0.8480	 & 	0.6630	 & 	1.0300	 & 	78	 & 	198	 & 	0.0950	 & 	0.0805	 & 	0.1050	\\
52	 & 	128	 & 	1.4700	 & 	1.2900	 & 	1.6200	 & 	80	 & 	199	 & 	0.0507	 & 	0.0357	 & 	0.0682	\\
54	 & 	129	 & 	1.0800	 & 	0.8510	 & 	1.3100	 & 	80	 & 	200	 & 	0.0334	 & 	0.0061	 & 	0.0640	\\
52	 & 	130	 & 	1.5800	 & 	1.4200	 & 	1.7400	 & 	80	 & 	201	 & 	0.0265	 & 	0.0111	 & 	0.0426	\\
54	 & 	131	 & 	0.8220	 & 	0.6270	 & 	1.0100	 & 	80	 & 	202	 & 	0.0257	 & 	0.0000	 & 	0.0677	\\
54	 & 	132	 & 	0.6530	 & 	0.3890	 & 	0.9260	 & 	81	 & 	203	 & 	0.0033	 & 	0.0000	 & 	0.0271	\\
55	 & 	133	 & 	0.3090	 & 	0.2830	 & 	0.3410	 & 	80	 & 	204	 & 	0.0266	 & 	0.0171	 & 	0.0330	\\
54	 & 	134	 & 	0.3850	 & 	0.2310	 & 	0.4770	 & 	81	 & 	205	 & 	0.0497	 & 	0.0000	 & 	0.1150	\\
56	 & 	135	 & 	0.2480	 & 	0.0000	 & 	0.2720	 & 	82	 & 	206	 & 	0.1970	 & 	0.0364	 & 	0.3790	\\
54	 & 	136	 & 	0.3300	 & 	0.2600	 & 	0.3960	 & 	82	 & 	207	 & 	0.1420	 & 	0.0000	 & 	0.4330	\\
56	 & 	137	 & 	0.1700	 & 	0.0000	 & 	0.2960	 & 	82	 & 	208	 & 	0.0003	 & 	0.0000	 & 	1.7800	\\
56	 & 	138	 & 	0.2140	 & 	0.0000	 & 	1.0000	 & 	83	 & 	209	 & 	0.0501	 & 	0.0100	 & 	0.1640	\\
57	 & 	139	 & 	0.1570	 & 	0.0183	 & 	0.2480	 & 				 &       &          &          &         \\
\hline
\end{tabular}
\end{center}
\end{table*}
 
Some r-process residuals are seen to suffer from remarkably large uncertainties, which quite
clearly cannot be ignored when discussing the r-process and the virtues of one or another
model for this process. This concerns in particular the elements  Rb, Sr, Y,
Zr, Ba, La, Ce and Pb. Some of them, and in particular Ba or La, are often used as tracers of the
levels of s- or r-processing during the galactic history (see Sect.~\ref{galaxy}). Lead has also a special status in the studies of the s-process (e.g. \cite{goriely05} for references), as well as of the r-process (see Sect.~\ref{actinides}). It could well be of pure s-nature if a strong s-process component can indeed develop in some stars, but a pure r-process origin cannot be excluded. These uncertainties largely blur any picture one might try to draw from spectroscopic observations and from simplistic theoretical considerations. 

\subsection{Isotopic anomalies in the solar composition}
\label{anomalies}

The bulk SoS composition has been of focal interest since the very beginning of
the development of the theory of nucleosynthesis. Further astrophysical interest
and excitement have developed with the discovery of the fact that a minute fraction of the SoS
material has an isotopic composition deviating from  that of the bulk. Such `isotopic
anomalies' are observed in quite a large suite of  elements ranging from C to Nd (including
the rare gases), and are now known to be carried by high-temperature inclusions of primitive 
meteorites, as well as by various types of meteoritic grains. The inclusions are formed from
SoS material out of equilibrium  with the rest of the solar nebula. The grains are
considered to be of circumstellar origin, and to have survived the process of incorporation into
the SoS.

These anomalies contradict the canonical model of an homogeneous and gaseous protosolar
nebula, and provide new clues to many astrophysical  problems, like the physics and chemistry
of interstellar dust grains, the  formation and growth of grains in the vicinity of objects
with active  nucleosynthesis, the circumstances under which stars (and in particular 
SoS-type structures) can form, as well as the early history of the Sun (in the
so-called `T-Tauri' phase) and of the SoS solid bodies. Last but not least, they
raise the question of their nucleosynthesis  origin and offer the exciting perspective of
complementing the  spectroscopic data for chemically peculiar stars in the confrontation 
between abundance observations and nucleosynthesis models for a very limited number of
stellar sources, even possibly a single one. This situation is in marked contrast with the
one encountered when trying  to understand the bulk SoS composition, which results
from the  mixture of a large variety of nucleosynthesis events, and consequently 
requires the modelling of the chemical evolution of the Galaxy.

\begin{figure}
\center{\includegraphics[scale=0.5]{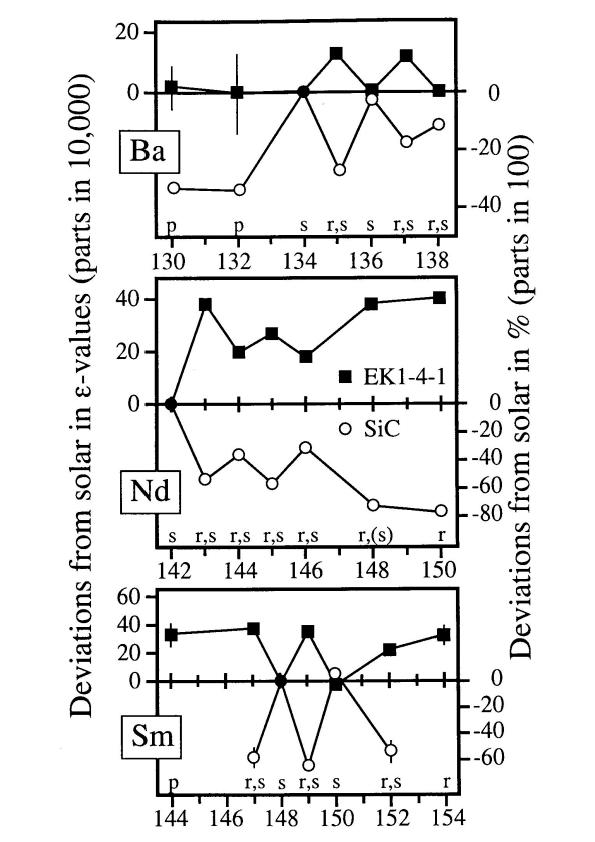}}
\caption{Isotopic compositions of Ba, Nd, and Sm in the FUN inclusion EK1-4-1 (black squares)
and in presolar SiC grains of the so-called `mainstream' type, which constitutes more
than 90\% of the identified presolar SiC grains (open dots). The
$\varepsilon$ values displayed on the left scale refer to EK1-4-1 and are defined as $\varepsilon_i =10^4 \times [(N_i/N_{\rm norm})_{\rm sample}/(N_i/N_{\rm norm})_{\rm standard} - 1]$, where
$N_i$ is the abundance of {\it i} in the considered  sample (inclusion or presolar grain)
or in the (`solar') material adopted as the isotropically normal 
(subscript: standard).
The isotopes used for normalisation (\chem{134}{Ba}, \chem{142}{Nd} and \chem{148}{Sm}) are
made only or predominantly by the s-process and have an abundance $N_{\rm norm}$ (the
EK1-4-1 values are additionally normalised in order to suppress any anomaly for
\chem{138}{Ba} and \chem{150}{Sm}, and to obtain  the equal excesses for \chem{143}{Nd} and
\chem{148}{Nd}). Positive(negative) $\varepsilon$ values thus represent excesses(deficits)
with respect to the s-isotopes used for normalisation.  The values on the right scale refer to the
SiC grains and are expressed in units of $100 \varepsilon$, indicating the large
difference in the levels of anomalies between the presolar grains and the SoS
inclusions (from \cite{begemann93})}
\label{comp_grain_inclusion}
\end{figure}

Among the identified anomalies, several concern the p-, s- and r-nuclides. They provide the
clear demonstration that the products of the three nucleosynthesis processes responsible for
their production have not been perfectly mixed in the forming SoS. Broadly speaking,
they can be divided into three categories. The first one involves anomalies attributable to
the decay of manufactured radio-nuclides, the lifetimes of which may be long enough ($\tau
\gsimeq 10^5$ y) for having been in live form in the early SoS before their eventual
in-situ decay in meteoritic solids. The second category relates largely to presolar
grains found in meteorites (e.g. \cite{zinner03} for a review). The third category involves
anomalies discovered in bulk meteoritic samples or in specific meteoritic inclusions,
particularly of the Ca-Al-rich (CAI) type. A rare class of CAIs dubbed FUN (for
Fractionation and Unknown Nuclear processes) carry an especially remarkable suite of isotopic
anomalies. In contrast to the presolar grains, the CAIs are considered to have formed in the
SoS itself, even if some aspects of their origins remain puzzling. This is particularly
the case for their FUN members. In fact, the CAIs are presumably the first solids to have
formed in view of their high condensation temperatures. Generally speaking, the levels of
the anomalies observed in presolar grains are much higher than in the material of SoS origin. 

Except in some cases to be mentioned below, the heavy
element isotopic anomalies observed up to now in presolar grains are in general characterised
by a deficit of p- and r-isotopes relative to the s-isotopes. As illustrated in
Fig.~\ref{comp_grain_inclusion}, this is in striking contrast with the patterns observed in the
bulk meteoritic material or in inclusions.
 
We concentrate in the following on anomalies made by an excess of r-nuclides with respect to
the s-nuclides. We do not discuss the various anomalous abundances of neutron-rich isotopes 
of elements of the Fe peak or lighter ones which have  been identified in FUN and non-FUN CAI
inclusions or in hibonite bearing inclusions (e.g. \cite{meyer05}), as they are
generally not attributed to the r-process, but rather to a quasi-statistical equilibrium
established in neutron-rich explosive environments.  The
anomalies involving s- or p-nuclides are not reviewed either. The interested reader is
referred instead to \cite{meyer05} or to \cite{arnould03} for a discussion of the s- and
p-anomalies. 

\begin{figure}
\centerline{\epsfig{figure=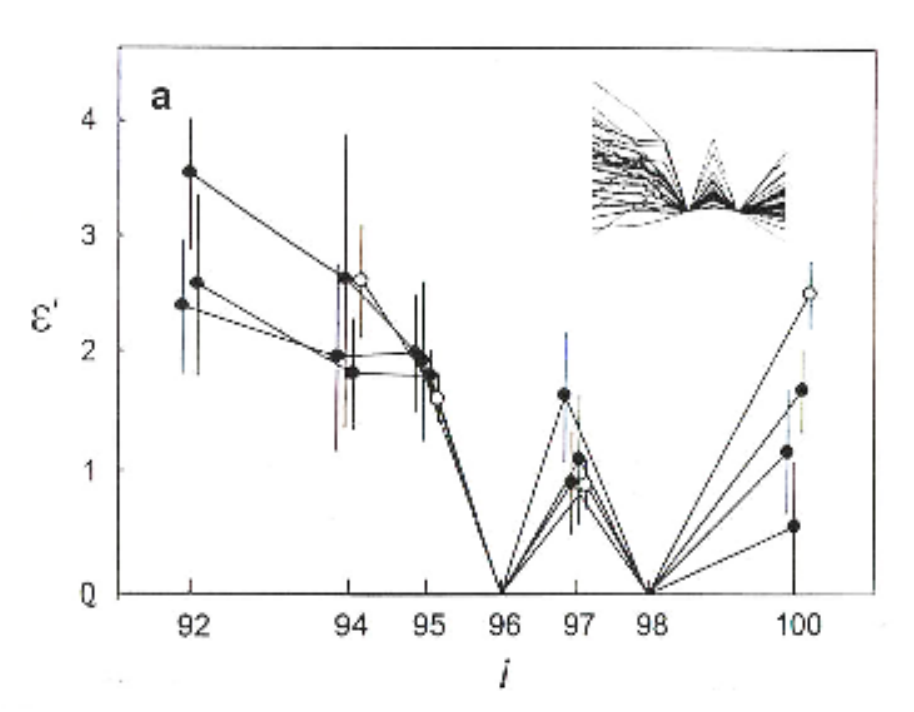,height=9cm,width=6cm,angle=270}}
\caption[]{Anomalous Mo isotopic patterns observed in the bulk of the carbonaceous chondrite
Allende, as well as of various meteorites of the iron, mesosiderite and pallasite types (top
right insert, where the uppermost lines correspond  to Allende). The $\varepsilon$ scale is
defined as in Fig.~\ref{comp_grain_inclusion}, and the normalising isotope is \chem{96}{Mo}
 (from \cite{dauphas02}). The data obtained by \cite{yin02} for bulk samples of the Murchison and
Allende chondrites largely agree with those displayed here}
\label{figMO1}
\end{figure}

\begin{figure}
\centerline{\epsfig{figure=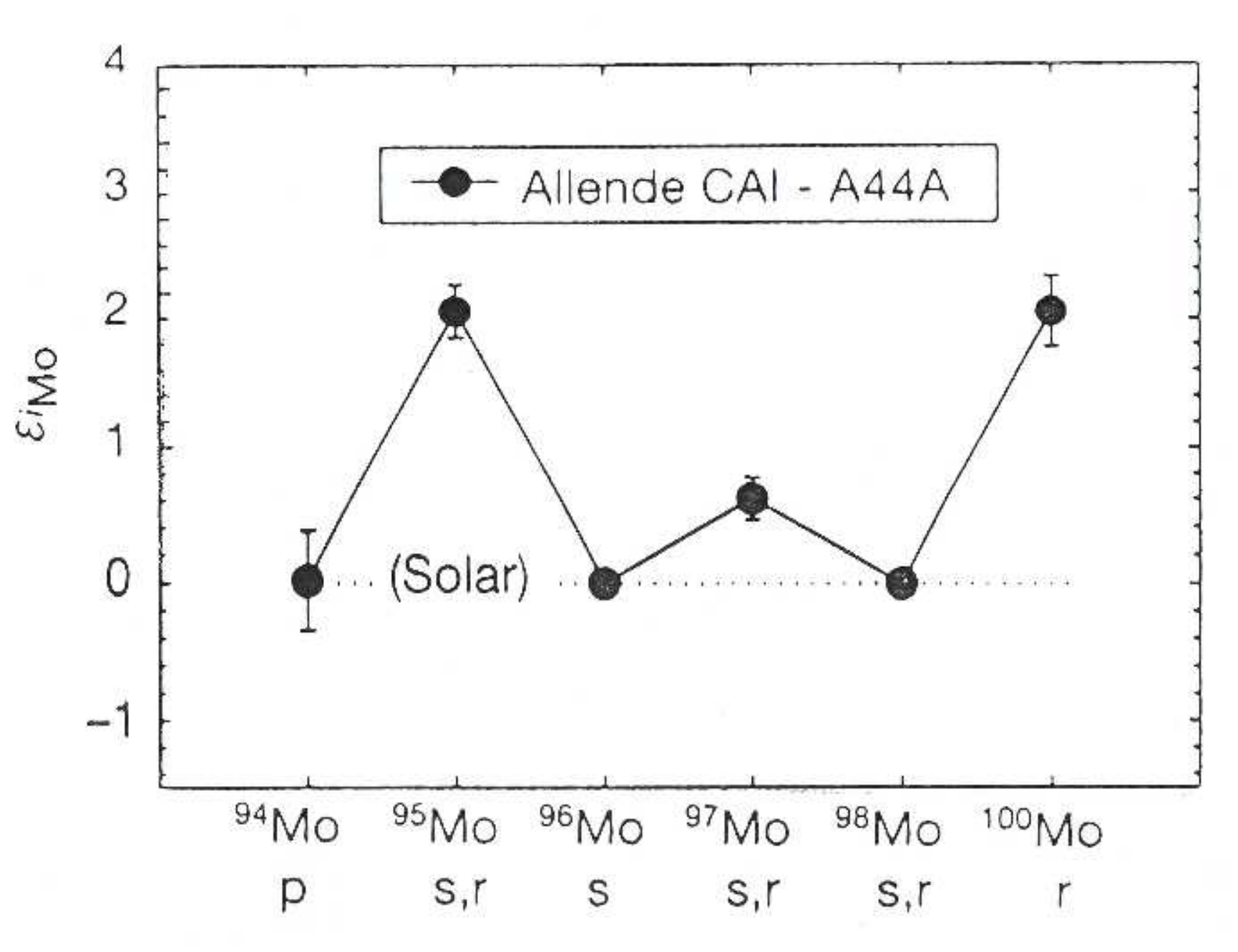,height=6cm,width=9cm}}
\caption[]{Anomalous Mo in the Allende CAI inclusion A44A (from \cite{yin02})}
\label{figCAI3}
\end{figure}
 
\begin{figure}
\center{\includegraphics[scale=0.3]{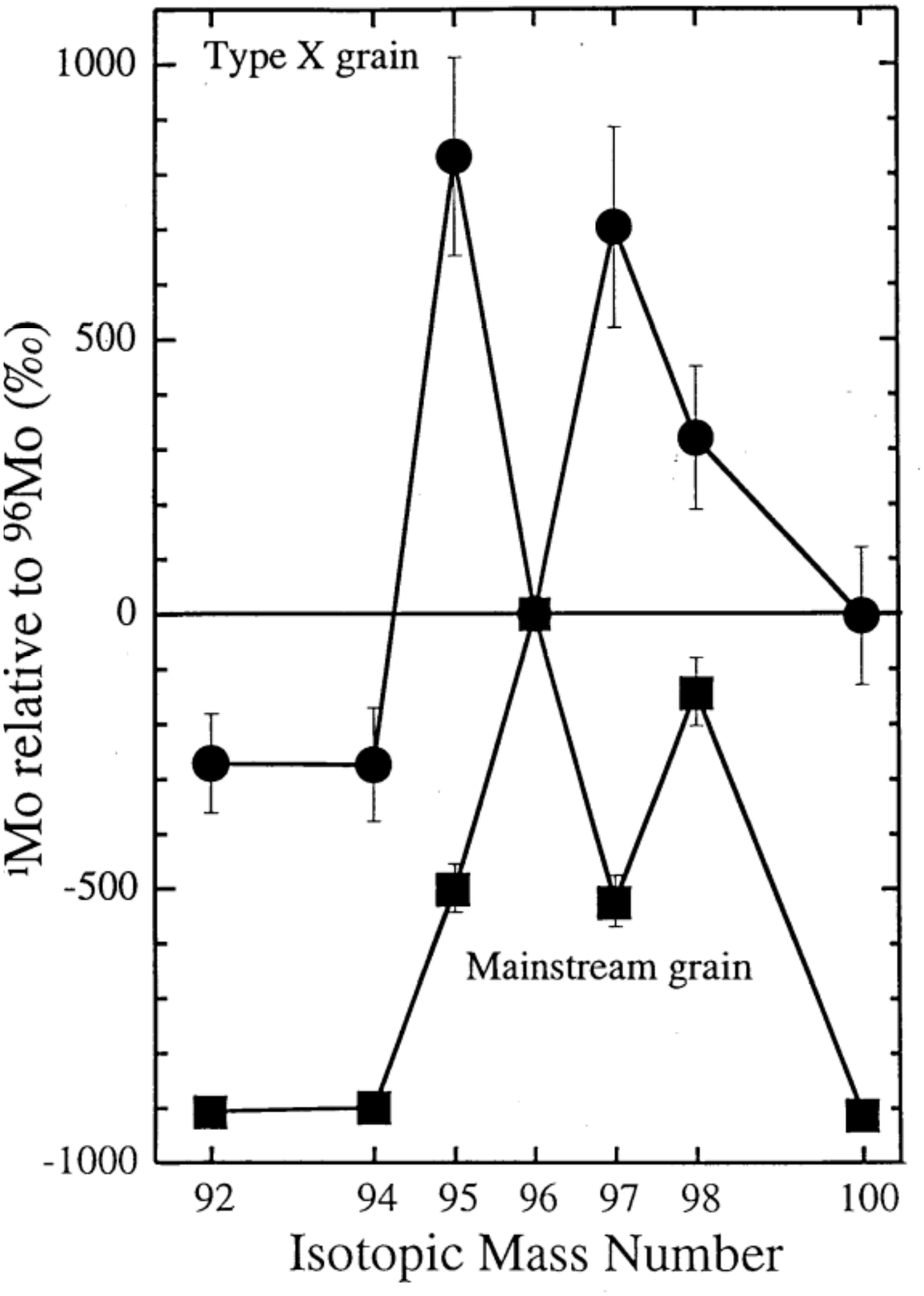}}
\caption{Mo isotopic compositions measured in a Type X grain, showing excesses of the
sr-nuclides \chem{95}{Mo} and \chem{97}{Mo}. For comparison, the data for a mainstream SiC
grain are also displayed, which show an excess of the s-isotope \chem{96}{Mo}, to which all
data are normalised. This pattern is reminiscent of the SiC s-nuclide excesses for
Ba, Nd and Sm shown in Fig.~\ref{comp_grain_inclusion}. The level of the anomalies is
expressed in units of $\delta = 0.1 \epsilon$, with $\epsilon$ defined in
Fig.~\ref{comp_grain_inclusion} (from \cite{pellin99})}
\label{figMOX}
\end{figure}

\subsubsection{The Mo anomalies}
\label{anomaly_Mo}

The Mo isotopic composition has raised much excitement recently. This element
exhibits various anomalous patterns in bulk meteoritic material of the
chondritic or differentiated types, as well as in CAIs
\cite{dauphas02,yin02}. As displayed in Figs.~\ref{figMO1} and \ref{figCAI3}, enhancements  of
the abundances of the sr-isotopes \chem{95}{Mo}, \chem{97}{Mo} 
and of the r-only isotope \chem{100}{Mo} are observed.
 As in Fig.~\ref{comp_grain_inclusion}, this is just the opposite of the s-isotope excess observed
in presolar SiC grains. Accompanying excesses of the p-isotopes \chem{92}{Mo} and
\chem{94}{Mo} are also found in bulk material, but not in the A44A CAI inclusion. From these
anomalous patterns, it is concluded that the p- and r-nuclide anomalies are largely correlated,
even though some decoupling is not excluded. In contrast, resolvable excesses of only
\chem{92}{Mo} and of \chem{95}{Mo} are found by \cite{chen04} in their analysis of bulk iron
and carbonaceous meteorites, as well as of some Allende CAIs, leading 
\cite{chen04} to conclude that
the p- and r-processes are essentially decoupled. This disagreement in the precise
characterisation of the anomalous Mo patterns is ascribed by \cite{chen04} to possible
technical issues.  Another problem concerns the positive \chem{97}{Mo} anomaly. It naturally
raises the prospect of a contribution to this nuclide of the in-situ decay of live
 \chem{97}{Tc} ($t_{1/2} = 2.6 \times 10^6$ y) in the early SoS. This interpretation
is not favoured, however, by \cite{dauphas02}. Other observations \cite{yin02} leave the door
open to a \chem{97}{Tc} decay origin, but do not demonstrate it. The absence of any Mo isotopic anomaly is
also claimed by \cite{chen04} in some iron meteorites, in pallasites and in ordinary
chondrites. Additionally, the Mo isotopic composition has been analysed in presolar grains
of the so-called X-type
\cite{pellin99,pellin00}, which are generally considered to be supernova condensates. As
shown in Fig.~\ref{figMOX}, excesses in the sr-isotopes  \chem{95}{Mo} and \chem{97}{Mo},
 as well as in \chem{98}{Mo} are found in this
case.  A resolvable anomaly at the r-nuclide \chem{100}{Mo} is also found in some of these
 grains. Additional excesses in the sr-nuclides \chem{88}{Sr} and \chem{138}{Ba} are 
identified in two X-grains \cite{pellin00}.
Several SiC grains of Type A+B have also been studied \cite{savina03}. One of them shows a
pattern similar to the one of Fig.~\ref{figMOX}. Finally, let us note that the unusual Mo
isotopic pattern in X-grains is associated with large enhancements of the
sr-nuclides \chem{88}{Sr} and \chem{96}{Zr} \cite{pellin00}.

\subsubsection{The Xe-HL and Te-H anomalies}
\label{anomaly_Xe}

\begin{figure}
\centerline{\epsfig{figure=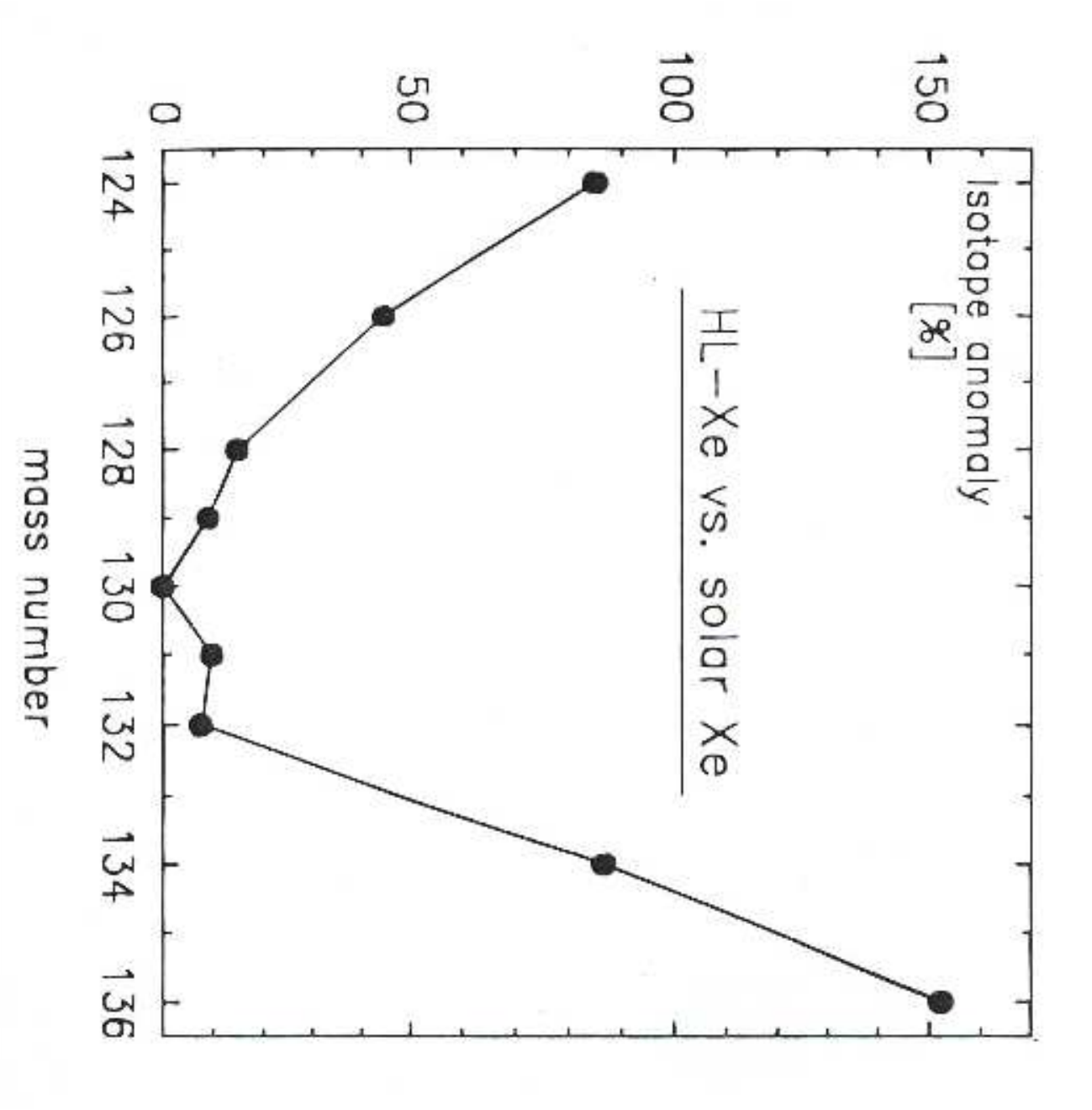,height=8.5cm,width=8.0cm,angle=90}}
\caption{Typical isotopic pattern of the Xe-HL component carried by circumstellar
diamonds (from \cite{huss95}). The over-abundances (in \%) are relative to the solar Xe
\cite{pepin95}. The two lightest and two heaviest isotopes are of pure p- and
r-origin, respectively. Normalisation is to the s-only isotope \chem{130}{Xe}.}
\label{figXeHL}
\end{figure}

Among the discovered anomalies, one of the most puzzling ones concerns the so-called
Xe-HL, which is characterised by excesses of the light (L) isotopes \chem{124}{Xe} and
\chem{126}{Xe} and to a less extent of \chem{128}{Xe}, correlated with enhancements 
 of  the r-process heavy (H) isotopes \chem{134}{Xe} and \chem{136}{Xe}. These remarkable
anomalies are carried by still-enigmatic diamond grains generally presumed to be of supernova
origin. A typical Xe-HL isotopic pattern is displayed in Fig.~\ref{figXeHL}. 

The case has become even more exciting with the discovery of a Te component, referred to
as Te-H, made solely of the r-isotopes \chem{128}{Te} and \chem{130}{Te} accompanying Xe-HL
in presolar diamonds \cite{richter98}.

\subsubsection{The Ba, Nd and Sm anomalies}
\label{anomaly_Ba}

Isotopic anomalies in Ba, with its seven stable isotopes produced to various
levels by the p-, s-, and r-processes, have been searched for with special vigour. Excesses of the
sr-isotopes \chem{135}{Ba} and \chem{137}{Ba} with respect to the s-isotopes \chem{134}{Ba}
and \chem{136}{Ba} have been observed in whole rocks of the Allende and Orgueil carbonaceous
chondrites, as well as in Allende CAIs, and especially in the FUN CAI EK1-4-1, which exhibits
the largest observed excesses as shown in Fig.~\ref{comp_grain_inclusion} (\cite{hidaka03}, and references
therein). These anomalies point again to a differentiated blend of the p-, s- and r-nuclides
in the SoS. A complication arises at mass 135 as a result of the possible
contribution to \chem{135}{Ba} of the decay of the now-extinct radionuclide
\chem{135}{Cs} (half-life $t_{1/2} = 2.3$ My).
 In view of the lack of carbonaceous chondritic phases
with high Cs/Ba ratios, a clear correlation between the Cs abundance and the \chem{135}{Ba}
excess that would result from the decay of \chem{135}{Cs} has not been
demonstrated yet, even if some hint has been identified \cite{hidaka03}. Regarding presolar
grains, evidence for an excess of the sr-nuclide \chem{138}{Ba} in two X-grains has been
presented by \cite{pellin00}, accompanied with deficits and/or excesses of \chem{135}{Ba} and
\chem{137}{Ba}. The situation in this respect thus remains quite
confusing.
 The anomalous Ba
abundance pattern observed in EK1-4-1 is complemented with excesses of the sr-or r-only
isotopes of Nd and Sm, as illustrated in Fig.~\ref{comp_grain_inclusion}.

As a very brief summary of Sect.~\ref{anomalies}, one can state that various blends of p, s-,
and r-nuclides that differ more or less markedly from the bulk SoS mixture depicted
in Sect.~\ref{obs_srp} are identified in a variety of meteorites at various scales, including
bulk samples, refractory inclusions or grains interpreted from their many highly anomalous
isotopic signatures as grains of  circumstellar origins. This is generally interpreted in
terms of the decoupling between the three mechanisms producing these nuclides. One of the
surprises of main relevance to this review is that those grains that are generally
interpreted in terms of supernova condensates do not carry the unambiguous signature of the
r-process that would be expected if indeed supernovae are the privileged r-process providers
(see Sect.~\ref{r_dccsn}).

\subsection{Evolution of the r-nuclide content of the Galaxy}
\label{galaxy}

The disentangling of the SoS s- and r-process components (Sect.~\ref{obs_srp}) is
complemented nowadays by a substantial observational work that aims at tracing the 
contribution of these two processes to the composition of the Galaxy throughout its history. 

Before going into some detail, recall that spectroscopic studies provide at
best elemental abundances for the species heavier than Fe, the isotopic wavelength offsets
being in general too small to be distinguishable. The only known exceptions in the $Z > 30$ region
 concern Ba, Eu, Hg, Pt, Tl and, tentatively, Pb (see Sect.~\ref{galaxy_isotopes}). Therefore,
 the contributions of the s- and r-processes to the composition of a given star are customarily
 evaluated from elements whose SoS abundances are predicted to be dominated by one or the other of
 these neutron capture processes. The traditional probe of the s-process has long been Ba with 70 
to 100 \% contribution from this process to the SoS (see Fig.~\ref{fig_ssol_elem}). To avoid the 
difficulty  of Ba with several stable isotopes which have different s- and r-contributions and whose 
precise abundance determinations raise specific problems \cite{simmerer05}, the 
essentially mono-isotopic La has
been used instead in recent works. It is classically considered that about 75\% of the SoS
La originates from the s-process (e.g. \cite{burris00}). A more careful examination of the
situation leads to values ranging all the way from about 45 to 100\% (Table~\ref{tab_r}),
which indicates that a roughly equal contribution of the s- and r-processes cannot be
excluded. This may be an embarrassment when interpreting the observations of both Ba and La.
Their s- and r-process contributions may be different from star to star and/or
with metallicity.\footnote{Here and in the following, the
metallicity is defined in the standard stellar spectroscopic notation by the
ratio $[\rm Fe/\rm H] = \log_{10}N({\rm Fe})/N({\rm H})|_\ast - 
log_{10}N({\rm Fe})/N({\rm H})|_\odot$, where $N$(X) is the abundance by number of 
 element X, the indices $\ast$ and $\odot$ referring to a given star and to the Sun.
 The metallicity is related to, but must not be identified to,
  the metal content generally expressed in terms of the mass 
fraction $Z$ of all the elements heavier than H and He}
In particular, Ba might be of pure r-process venue in certain cases.
A substantial r-process contribution to La cannot be excluded either because of
 the uncertainties mentioned above concerning the SoS. The r-process is classically traced by Eu,
 which is estimated to be 80 to 100\% of r-process origin (Table~\ref{tab_r}).

Ba, La and Eu all probe the production of the heavy neutron-capture elements. It is widely
considered that the $A \lsimeq 100$ s-nuclides are produced in different environments ($M
\gsimeq$ 10 $M_\odot$ stars) than the heavier ones (Asymptotic Giant Branch stars).
Recently, it has been repeatedly claimed that a similar situation prevails for the $A \lsimeq
130$ and heavier r-nuclides (see Sect~\ref{galaxy_universality}). It is thus of interest to 
compare the trends
of the abundances of the light and heavy neutron-capture elements, especially at low
metallicity. In this respect, many observations are available in particular for Sr, Y and Zr.
As derived from Table~\ref{tab_r}, the r-process contributions to Sr, Y and Zr in the SoS 
amount
to about 0 - 20\%, 0 - 40\% and 10 - 40\%, respectively. These large uncertainties may
again endanger the disentangling of the s- and r-process contributions to the observed
abundances. It is even more so as there is no reason for the contributions of these two 
processes to the production of a given element to be the same as in the SoS at all times and at all locations in the 
Galaxy. An additional potential difficulty might come from the fact that some fraction at 
least of elements like Sr, Y or Zr could be produced by a nucleosynthesis mechanism, referred
 to as the $\alpha$-process (Sect.~\ref{DYR}), involving charged-particle captures.

In the analysis of the neutron-capture elements, it is of interest to adopt a classification 
introduced by
\cite{christlieb04} for a very rare class of objects known as
`r-process-enhanced metal-poor' stars.\footnote{The term `metal-poor' does not
necessarily refer to the overall metal content in case the CNO elements have large
over-abundances}. 
Following \cite{christlieb04,barklem05}, r-only stars (sometimes referred to 
as `pure' r-process stars)
characterised by [Ba/Eu] $< 0$ are classified into so-called r-II stars if [Eu/Fe] $ > +1.0$,
and r-I stars if $+0.3 \leq$ [Eu/Fe] $\leq +1.0$. A list of these stars and relevant
references can be found in \cite{christlieb04,barklem05}. The r-II stars appear to be extreme
counterparts of, and much less common than, the r-I objects. In addition, some of the `pure'
r-process stars show no r-process enhancement ([Eu/Fe] $ <0.3$), or  even have an r-process
deficiency ([Eu/Fe] $< -0.2$) \cite{ishimaru04}. A class of C-rich low-metallicity ([Fe/H]
 around -2.55 with a scatter of about 0.26 dex) `r+s stars' has also been identified
 \cite{jonsell06}. They are characterised by an enhancement of both Ba ([Ba/Fe] $>$ 1) and Eu 
([Eu/Fe] $>$ 1.0), and additionally by [Ba/Eu] $>0.0$. These constraints are not precisely 
defined, however, and the classification relies on the SoS separation between s- and r-process 
contributions to the heavy elements, which may not be strictly valid for the considered 
low-metallicity stars.

\begin{figure}
\center{\includegraphics[scale=0.5,angle=0]{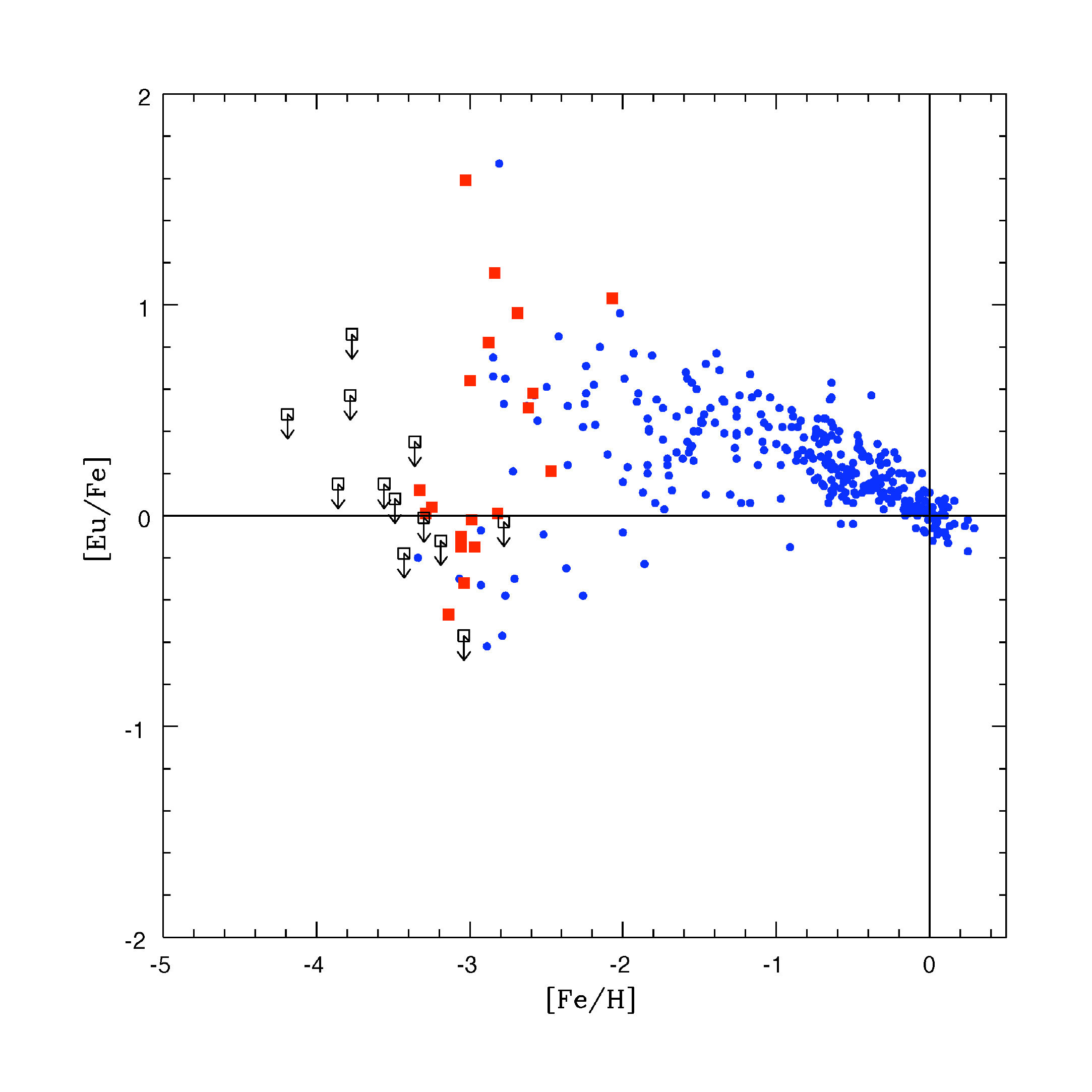}}
\vskip-0.5cm
\caption{Ratio [Eu/Fe] versus [Fe/H] for a large stellar sample. The data are from various
 sources, the details of which can be found in \cite{cescutti06}}
\label{galaxy_eu}
\end{figure}

\begin{figure}
\center{\includegraphics[scale=0.5,angle=90]{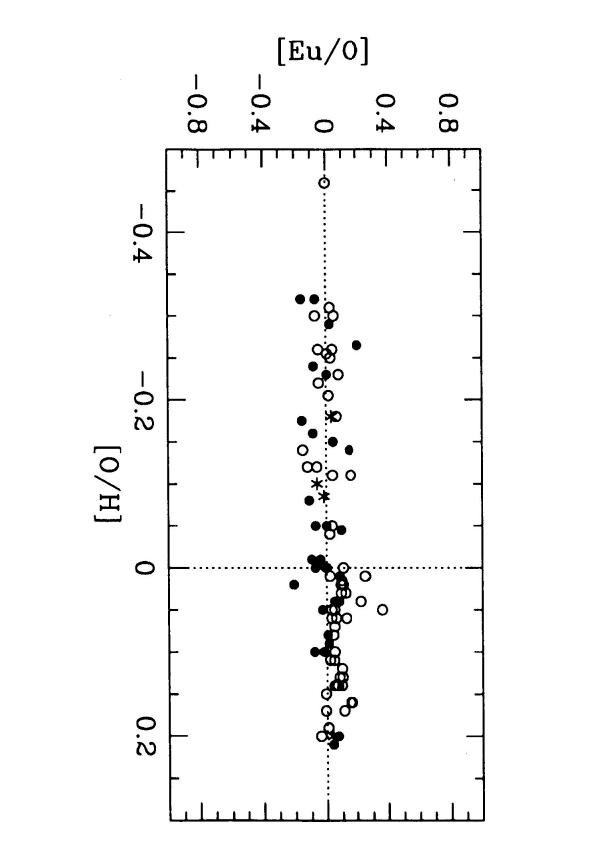}}
\vskip-1.5cm
\caption{Values of [Eu/H] versus [O/H] for stars in the thick- (solid circles) and thin- 
(open circles) discs (from \cite{bensby05})}
\label{galaxy_eu_o}
\end{figure}

\subsubsection{How did the neutron-capture element abundances evolve over the galactic
history?}
\label{galaxy_s_r_evolution}

The large variety of possible [Eu/Fe] ratios that has inspired the above-mentioned
classification is demonstrated in Fig.~\ref{galaxy_eu}, which features in particular a large 
scatter in Eu/Fe at low [Fe/H] for field halo and disc stars. The scatter decreases 
substantially
with increasing [Fe/H], as demonstrated by the analysis of mostly disc stars. This situation
suggests composition inhomogeneities in the early Galaxy resulting from the unmixed output
from a more or less limited amount of individual events that do not all yield similar Eu/Fe
abundance ratios. Another noticeable feature is the tendency for [Eu/Fe] to decrease with
increasing metallicity for [Fe/H] $\gsimeq -1$. This tendency is nicely confirmed by the 
recent study
of a large sample of thick and thin disc stars with -1 $\lsimeq$ [Fe/H] $\lsimeq$ 0.4
\cite{bensby05}. In contrast, Eu is found to follow the O abundances very well in the
above-mentioned [Fe/H] range, as demonstrated in Fig.~\ref{galaxy_eu_o}. These trends probably
reflect a common origin of O and Eu, possibly in Type II supernovae or at least in their 
massive star progenitors, and the increase of the
Fe content of the Galaxy when Type Ia supernovae start contributing to the galactic Fe. 

\begin{figure}
\center{\includegraphics[scale=0.8,angle=0]{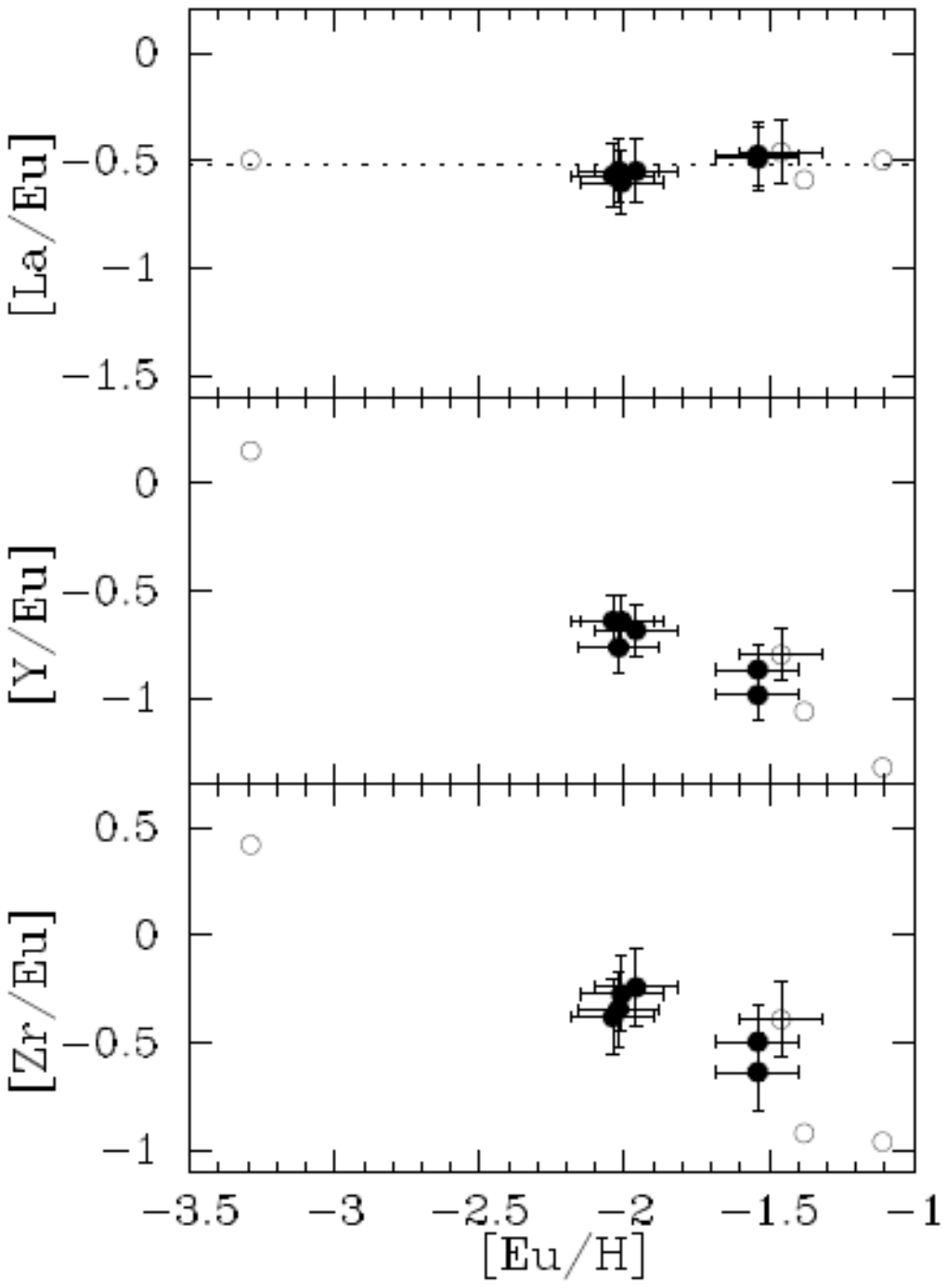}}
\vskip-8.5cm
\caption{Values of [La/Eu], [Y/Eu] and [Zr/Eu] versus [Eu/H] for stars in the globular
 cluster M15 (filled circles). The dotted line in the top panel indicates the SoS value 
proposed by \cite{simmerer05} (from \cite{otsuki06}. For comparison,  data for some field 
halo stars are also displayed (open circles)} 
\label{glob}
\end{figure}

\begin{figure}
\center{\includegraphics[scale=0.5]{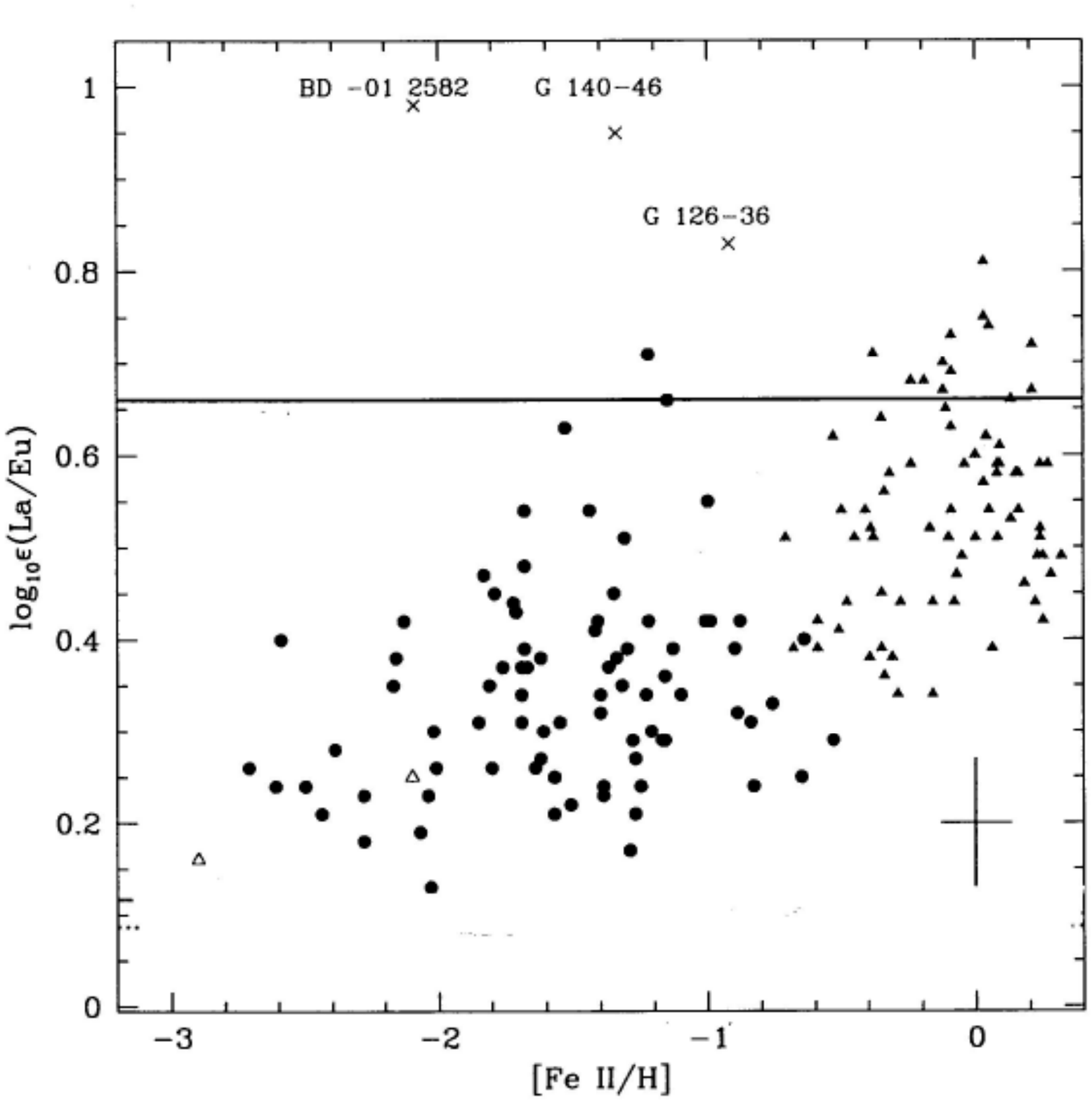}}
\vskip -2.2cm
\caption{Values of $\log_{10}\epsilon(\rm La/\rm Eu) = \log_{10}(N_{\rm La}/N_{\rm H}) -
\log_{10}(N_{\rm Eu}/N_{\rm H})$ versus [Fe/H] from various studies (different symbols). 
A typical error is shown as a dagger. The three labelled points
correspond to metal-poor La-rich stars. The solid line is the total SoS La/Eu ratio (from 
\cite{simmerer05}, where the references to the observations are provided)}
\label{galaxy_la_eu}
\end{figure}

The data of Fig.~\ref{galaxy_eu} are also classically used to support the idea that the
r-process has contributed very early to the heavy element content of the
Galaxy (e.g. \cite{truran02}, and references therein).  However, the observed scatter
clearly introduces some confusion when one tries to establish a more detailed trend of the Eu
enrichment of the Galaxy with metallicity. The question of the metallicity lag between 
the onsets of the r-
and s-processes can also be tackled through an examination of the time variation of the La/Eu
abundance. Figure~\ref{glob} exhibit a remarkable constancy of [La/Eu] versus the Eu abundance
 for stars in the globular cluster M15. This can be interpreted in terms of a purely and common
 r-process origin of La and Eu \cite{otsuki06}. On the other hand,  Fig.~\ref{galaxy_la_eu} 
shows the results of the analysis of a quite large sample of giant and dwarf stars in
 the $-3 <$ [Fe/H] $< 0.3$  range \cite{simmerer05}. From this, \cite{simmerer05} concludes 
that there is
no unambiguous [Fe/H] value at which the s-process signature becomes identifiable, considering
in particular the large scatter in the La/Eu ratio, even near solar metallicities. The
possible non-negligible contribution of the r-process to La might in fact blur the picture
further, this effect being always forgotten in the published discussions.

\begin{figure}
\center{\includegraphics[scale=0.5,angle=90]{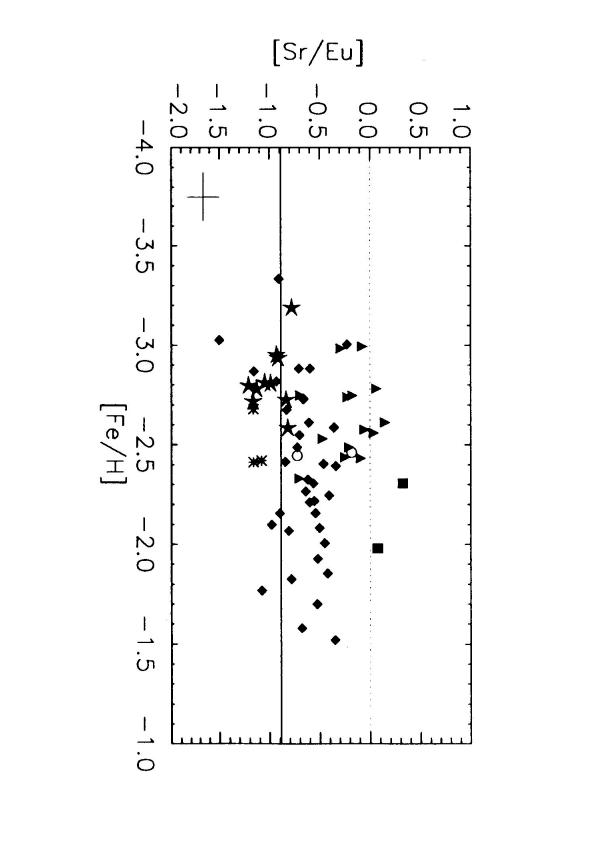}}
\vskip -1.5cm
\caption{Values of [Sr/Eu] versus [Fe/H] for an ensemble of r-I (diamonds) and
r-II (stars) r-process-rich stars, as well
as for pure r-only stars without excess of r-nuclides (triangles), for r+s stars
(asterisks) and two other s-process-rich stars (squares). These various stellar types are
defined in the text. The average relative error is shown at the bottom
left. The horizontal line represents an estimate of the pure r-process SoS value (from
\cite{barklem05})}
\label{galaxy_sr_eu}
\end{figure}

\begin{figure}
\center{\includegraphics[scale=0.5,angle=180]{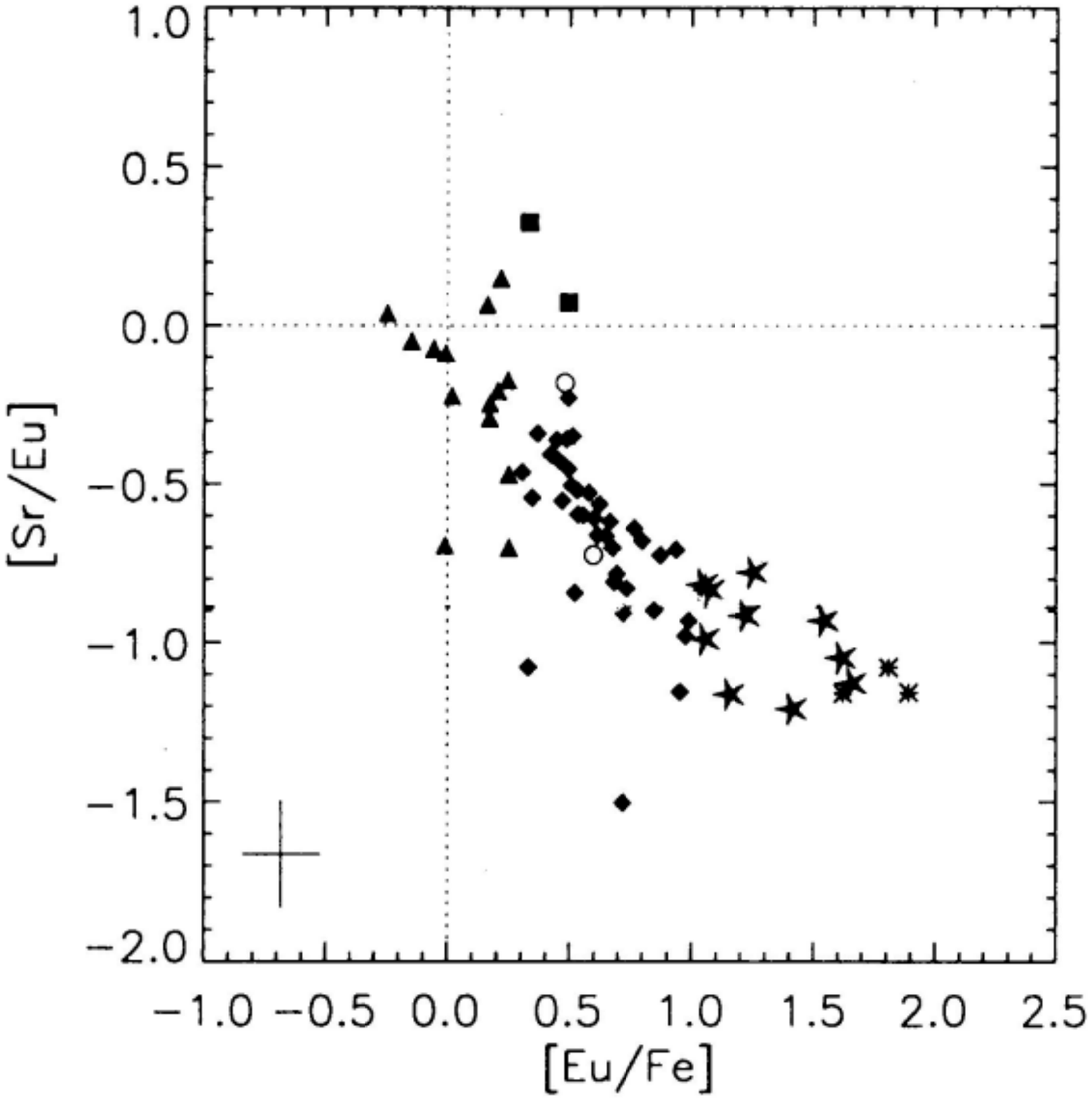}}
\vskip -2.1cm
\caption{Same as Fig.~\ref{galaxy_sr_eu}, but for [Sr/Eu] versus [Eu/Fe] (from
\cite{barklem05})}
\label{galaxy_sr_eu_1}
\end{figure}

As stressed above, the information provided by the heavy neutron-capture elements
(like La and Eu) might be usefully complemented with data on the lighter elements (like Sr, Y
or Zr). Abundance trends of Zr or Y with respect to Eu are seen in Fig.~\ref{glob} to depart
 from  that of La/Eu
 for stars in the globular cluster M15. Substantial differences between the  La/Eu and Sr/Eu 
patterns versus metallicity are also seen for field stars from a comparison of 
Figs.~\ref{galaxy_la_eu}  and \ref{galaxy_sr_eu}. Disregarding s-process-rich stars or not, it
 appears that Sr has a quite different nucleosynthesis history than the heavier La, with no 
clear identifiable trend of [Sr/Eu]
with [Fe/H], in contrast to [La/Eu]. This situation can be put in agreement
with the traditional views about the s-process which predict Sr to be produced in more massive
stars, i.e. earlier in the galactic history, than La. Figure \ref{galaxy_sr_eu_1} also
demonstrates that Sr appears to have had a different enrichment history than Eu. More
precisely, [Sr/Eu] is especially high in r-only stars in which Eu is not enhanced
([Eu/Fe] $ <0.3$) or in the two s-process-rich stars already displayed in
Fig.~\ref{galaxy_sr_eu}, this ratio decreasing all the way from r-I to r-II stars. The r+s
stars of Fig.~\ref{galaxy_sr_eu} are also characterised by low [Sr/Eu] ratios. This pattern
of [Sr/Eu] versus [Eu/Fe] can be made compatible with the idea that Sr has been produced by
the s-process in stars which did not produce much Eu.\footnote{Note that this interpretation
is not in contradiction to the fact that most of the stars exhibiting the highest [Sr/Eu]
ratio are defined as r-only stars. This classification is indeed based on the
fact that the heavy neutron-capture element Ba has an abundance such that [Ba/Eu] $< 0$, and
makes no reference to the possible enrichment of light neutron-capture elements such as Sr}
This s-process origin of Sr is compatible with the high [Sr/Eu] ratios observed in the two 
s-process-rich stars displayed in Figs.~\ref{galaxy_sr_eu} and \ref{galaxy_sr_eu_1}. Still, 
it remains to be demonstrated that very-low metallicity stars can produce enough Sr  to
 account for the observations. This might require special mixing processes as those called 
for in Asymptotic Giant Branch stars (e.g. \cite{goriely05}). On the other hand, it cannot 
be excluded that some fraction of the observed Sr could have been produced by the r-process 
along with Eu in quantities that could have been different from star to star and from the
 SoS mix. 

\subsubsection{Can the available isotopic data tell us something about the prevalence of
the s- or of the r-process at early galactic times?}
\label{galaxy_isotopes}

\begin{figure}
\center{\includegraphics[scale=0.33,angle=0]{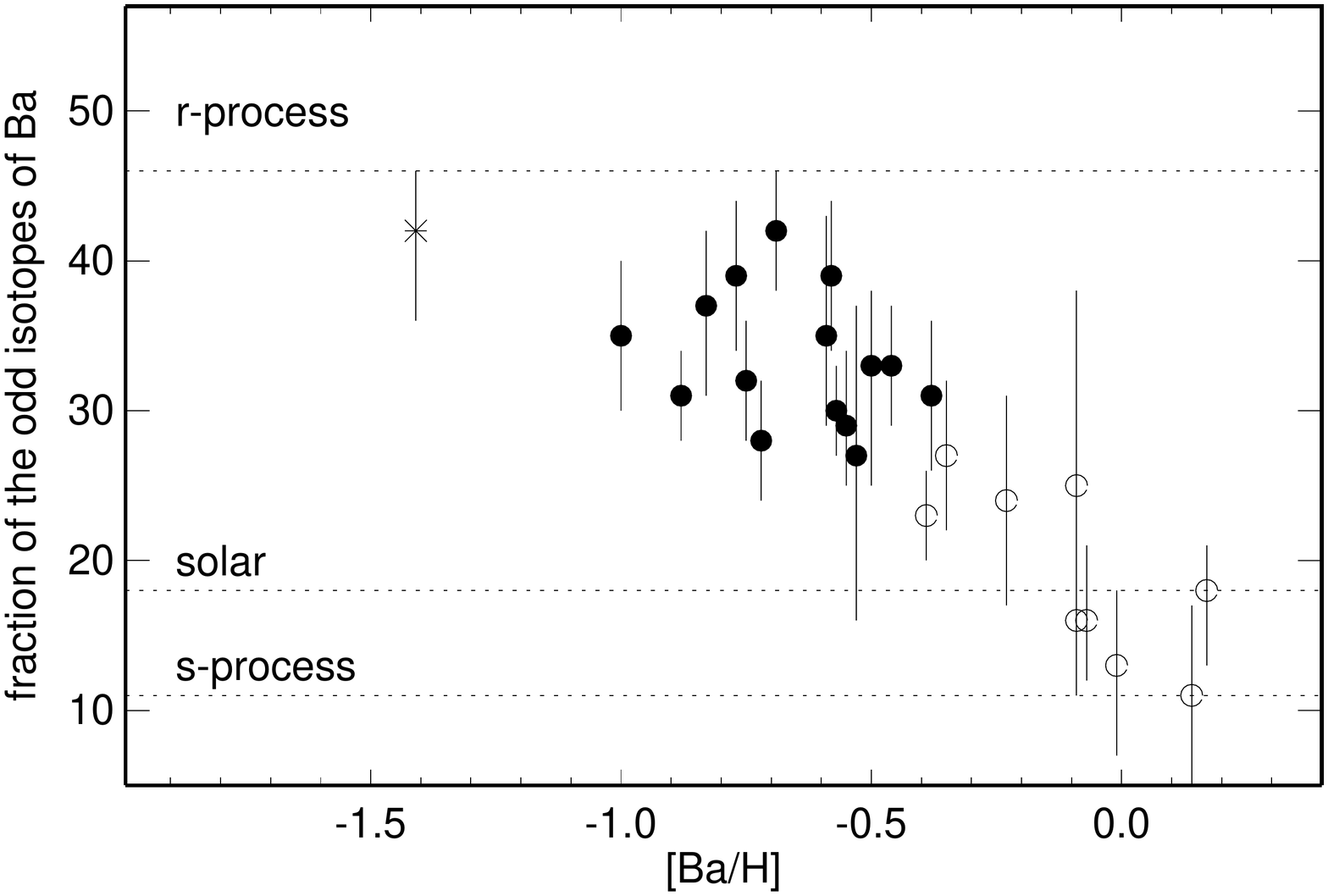}}
\vskip-0.3cm
\caption{Values of $f_{\rm odd}$  (in \%) versus [Ba/H] for a sample of thick disc (solid circles)
 and thin disc (open circles) stars with $-1.35 \leq {\rm [Fe/H]} \leq 0.25$ 
(from \cite{mashonkina06})}
\label{fig_ba_odd}
\end{figure}

The considerations above rely on elemental abundance data. A most useful information
regarding the relative evolution of the s- and r-process efficiencies in the Galaxy would be
provided by the knowledge of the isotopic composition of the neutron-capture elements. Such
data are unfortunately very scarce, and sometimes still debated.  Isotopic information 
concerning Ba (\cite{mashonkina06}, and
references therein) and Eu \cite{sneden02,aoki03} have been discussed in terms of s- and 
r-contributions. The hyperfine splitting of Ba spectral lines has been used by \cite{magain95}
 to analyse its isotopic composition in the metal-poor subgiant HD 140283, and in particular to
determine the fractional abundance $f_{\rm odd} = [N(^{135}{\rm Ba}) + 
N(^{137}{\rm Ba})]/ N({\rm Ba})$ of the odd Ba isotopes. This ratio has indeed been 
considered as a measure of the relative contributions of the r- and s-processes to Ba 
 for the canonical model analyses (e.g. \cite{arlandini99}) allow little room for 
 r-process contributions to the even-mass Ba isotopes. 
 The SoS $f_{\rm odd}$ is about 0.18.  The r-process fraction to Ba is given by 
$r/(r+s) = [f_{\rm odd} - f^{\rm s}_{\rm odd}] / [f^{\rm r}_{\rm odd} - 
f^{\rm s}_{\rm odd}].$ 
Here, the fractions $f^{\rm s}_{\rm odd}$ and  $f^{\rm r}_{\rm odd}$ are of the odd-mass Ba 
isotopes in the cases of {\it pure} s- and {\it pure} r-processes, respectively, which for the
 time being cannot be  evaluated in any other way than by analysing the SoS Ba isotopic 
compositions. 
Table~\ref{tab_r}, in reference to \cite{palme93}, gives 
$0.06 \lsimeq f^{\rm s}_{\rm odd} \lsimeq  0.23$ with
 the `standard' value being $f^{\rm s}_{\rm odd} \approx 0.10$. On the other hand,
 any value of $f^{\rm r}_{\rm odd}$ ($0 \lsimeq f^{\rm r}_{\rm odd} \lsimeq 1$) is permissible
 in consideration of the largely uncorrelated uncertainties, whilst its standard value is 0.66. 
 On top of the uncertainties in $f^s_{\rm odd}$ and $f^r_{\rm odd}$, the derivation of
$f_{\rm odd}$ from observation is also hampered with substantial difficulties, as
illustrated by the diverging conclusions concerning HD 140283 by \cite{magain95} who
concludes that Ba in this star is of typical SoS s-r mix, and by \cite{lambert02} who instead
claims that it is of pure r-process origin. A solar mixture of Ba isotopes is also
preferred by \cite{mashonkina99} for a sample of metal-poor main sequence or close to main
sequence cool stars.  Note that all these statements are based on roughly the same
$f_{\rm odd}$ values. The Ba isotopic composition has been analysed recently in 25 cool thick
 and thin disc dwarf stars with $-1.35 \leq {\rm [Fe/H]} \leq 0.25$ \cite{mashonkina06}. The 
derived  $f_{\rm odd}$ values are displayed in Fig.~\ref{fig_ba_odd}. They are seen to be 
smaller in the thin disc than in the thick disc, whose age is estimated to be comparable to
 the one of the halo. 
Note that the $f^s_{\rm odd}$- and $f^r_{\rm odd}$-values (0.10 and 0.46) indicated 
in Fig.~\ref{fig_ba_odd}
are those derived from the predictions of certain AGB models, and thus may not be reliable
to be applied in the problem at hand because of the large intrinsic uncertainties of the 
models themselves. In particular, to combine (\cite{mashonkina06}) the metallicity-dependent
 AGB s-process  models along the chemical evolution history of the Galaxy with the SoS 
isotopic composition is quite a dubious practice because it 
is then effectively dictating the possible metallicity dependence of the r-process.
Nonetheless, with these or our `standard' $f^s_{\rm odd}$- and $f^r_{\rm odd}$-values taken
 for granted  as a measure, the observed $f_{\rm odd}$ trend seemingly suggests that the r-process 
contribution to Ba decreased during the galactic evolution. However, before deriving firm 
conclusions, the uncertainties in the $f_{\rm odd}$ values certainly do not have to be swept 
under the rug.

The isotopic compositions obtained by \cite{sneden02,aoki03} for some r-only stars of
the r-I and r-II types are all close to \chem{151}{Eu}/\chem{153}{Eu} $\approx 0.5$. This is
not in contradiction with the SoS r-process abundance ratio  which lies in the 0.6 - 0.9 range,
following Table~\ref{tab_r}, but does not exclude a contribution from the s-process.

\subsubsection{How do the patterns of abundances of r-nuclides in metal-poor stars compare
with the SoS counterpart?}
\label{galaxy_universality}

\begin{figure}
\center{\includegraphics[scale=0.5,angle=90]{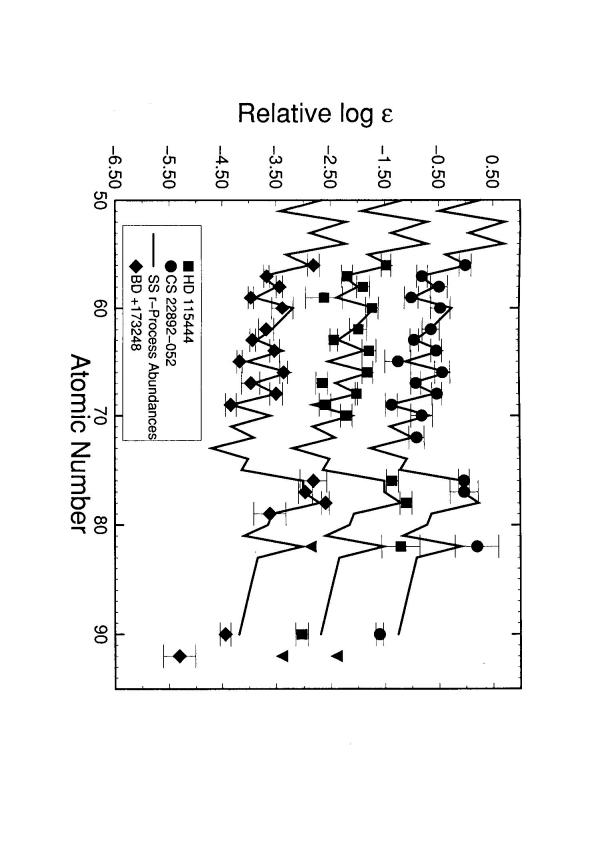}}
\vskip -.9cm
\caption{Heavy-element abundance patterns for one r-II (CS 22892-052) and two r-I (HD 115444
and BD +173248) stars, the [Eu/Fe] ratios of which are included in
Fig.~\ref{galaxy_eu}. The solid lines represent a scaled SoS r-nuclide distribution. Inverted
triangles indicate upper limits (from \cite{truran02})}
\label{fig_galaxy_specific1}
\end{figure}
 
\begin{figure}
\center{\includegraphics[scale=0.5,angle=89.4]{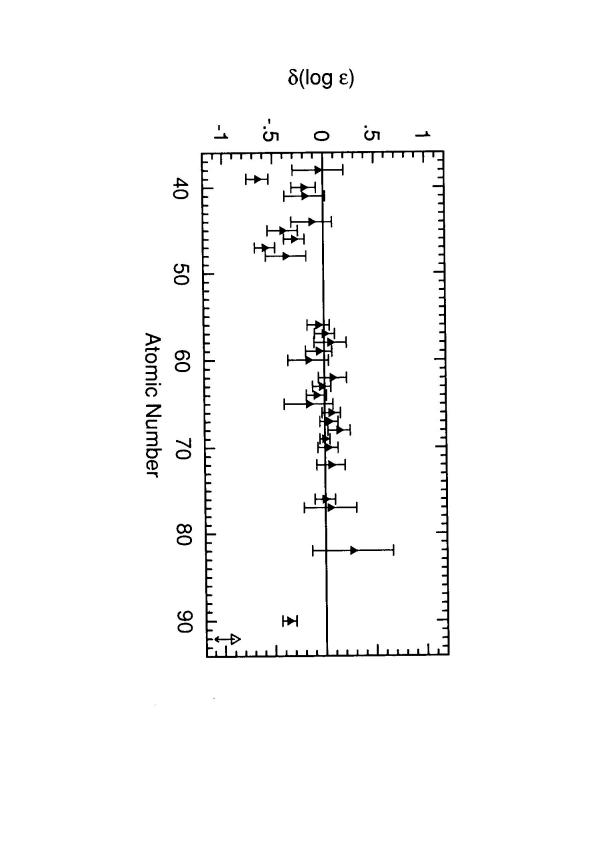}}
\vskip -2.3cm
\caption{Differences in log$_{10}\epsilon$ (as defined in Fig.~\ref{galaxy_la_eu}) between
the heavy-element abundances observed in the r-II star CS 22892-052 and a scaled SoS
r-nuclide distribution (from \cite{truran02})}
\label{fig_galaxy_specific2}
\end{figure}

\begin{figure}
\center{\includegraphics[scale=0.57,angle=90]{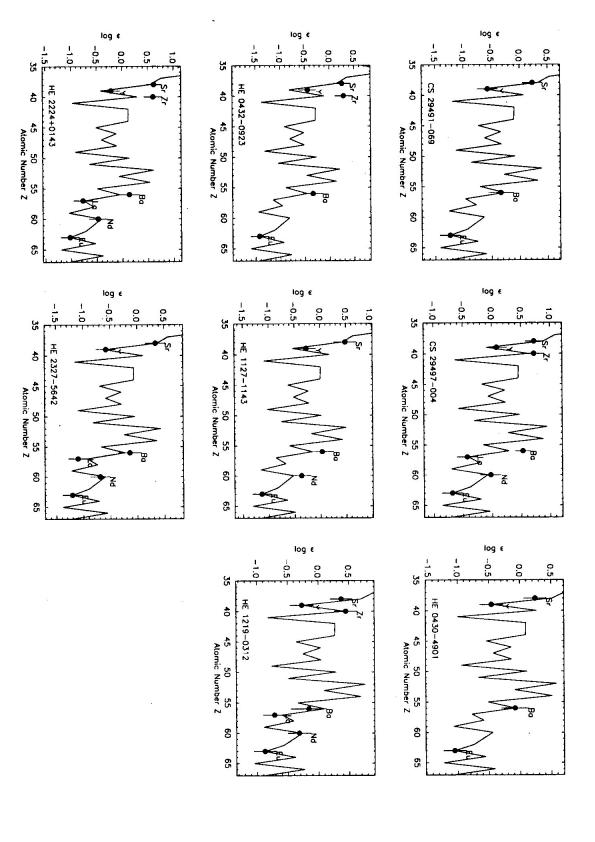}}
\caption{Similar to fig.~\ref{fig_galaxy_specific1} for an ensemble of r-II stars. The solid
line represent a scaled SoS r-nuclide distribution normalised to the Eu abundances derived
from the observations (from \cite{barklem05})}
\label{fig_galaxy_specific3}
\end{figure}

Excitement has been raising in recent years following the mounting evidence that the patterns
of abundances of heavy neutron-capture elements in the range from around Ce ($Z = 58$) to
Os ($Z = 76$) observed in r-process-rich metal-poor stars are remarkably similar to the SoS one. 
Such a convergence appears in Figs.~\ref{fig_galaxy_specific1} -
 \ref{fig_galaxy_specific3} for a sample of r-I and r-II stars.\footnote{Note that Ba
 ($Z = 56$) is often included in the convergence range as well. We
avoid this practice here. The observational situation is somewhat confusing indeed.
While \cite{barklem05} find that the [Ba/Eu] scatter is small among pure r-process halo stars,
which seemingly implies that Ba is co-produced with Eu at a close-to-constant value by the
r-process in the early Galaxy, \cite{truran02} conclude instead that Ba/Eu shows substantial
scatter. This may point to uncertainties in Ba abundance determinations, as already mentioned.
On the other hand, the level of s-process contamination to Ba in metal-poor stars is
difficult to ascertain} 
This has led in the literature to the recurrent claim that the r-process is `universal'. 

\begin{figure}
\center{\includegraphics[scale=0.53,angle=0]{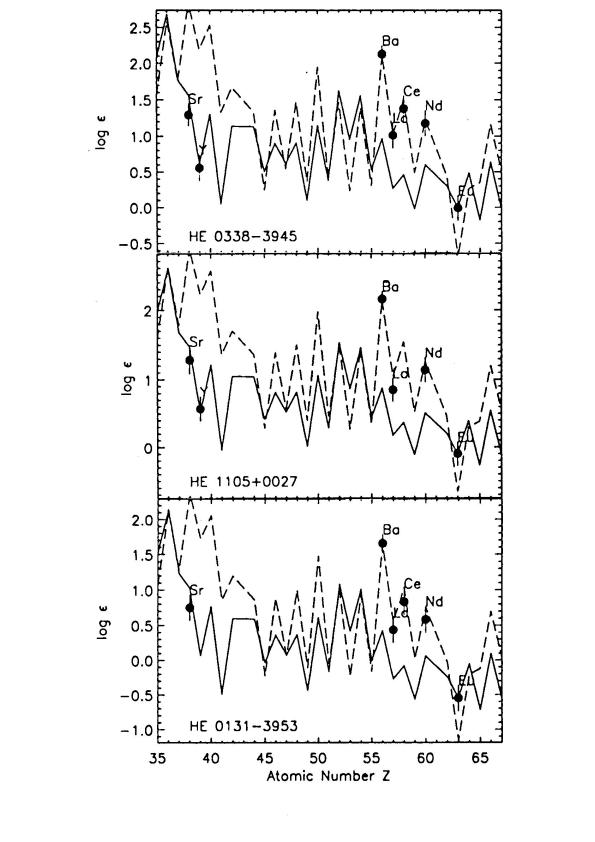}}
\vskip-0.8cm
\caption{Abundance patterns of some neutron-capture elements in three strongly Eu enhanced
r+s stars. The solid line represents a scaled SoS r-process abundance
distribution normalised to Eu, while the dashed one shows the corresponding s-nuclide
distribution normalised to Ba (from \cite{barklem05})} 
\label{fig_galaxy_specific4}
\end{figure}

Some words of caution are in order here regarding this claim. First, as illustrated in 
Fig.~\ref{fig_galaxy_specific4}, the situation in r+s stars is different from the one 
encountered in r-I and r-II stars. The La, Ce and Nd abundances fit a scaled SoS s-process 
abundance curve which does not account for Eu. Second, one may wonder about the real relevance
 of scaled SoS abundance distributions in the case of low-metallicity stars, for which the 
s-process abundance pattern may be quite different from the solar one (e.g. \cite{goriely05}).
 In the third place, an  early critical examination of the claimed universality of the 
r-process \cite{goriely97a} has demonstrated that the abundance convergence in the 
$58 \lsimeq Z \lsimeq 76$ range can be the natural signature of nuclear properties, and does
 not tell much about the astrophysics of the r-process (see Sect.~\ref{MER_universality}). 

Outside the $58 \lsimeq Z \lsimeq 76$ range,  reservations  have been repeatedly expressed 
on the universality of the r-process  (\cite{goriely97a,goriely99a,goriely01a}). These
 reservations have received mounting support from observation. This concerns in particular 
the Pb-peak elements and the actinides \cite{plez04,yushchenko05}. This non-universality 
has far-reaching consequences, in particular in attempts of building galactic chronologies 
based on the actinides content of very metal-poor stars (see also Sect.~\ref{actinides} and 
Sect.~\ref{chronometry_lowz}).

 The similarity observed in the $Z = 58$ to 76 abundance patterns extends in some cases to
the lighter neutron-capture elements. As shown in Fig.~\ref{fig_galaxy_specific2}
 (see also Fig.~\ref{fig_galaxy_specific3}),
 this is especially the case for Sr and Y in some r-II
stars, the Sr/Eu ratio being quite close to solar. This also holds for
the three r+s stars displayed in Fig.~\ref{fig_galaxy_specific4}. In fact, the Sr/Eu ratio
deviates more and more markedly from its SoS value with increasing [Eu/Fe], as seen in
Fig.~\ref{galaxy_sr_eu_1}. Other deviations of light neutron-capture elements with respect to
a scaled SoS r-pattern are observed, as shown in Fig.~\ref{fig_galaxy_specific2}. The 
observation of these deviations have led to many speculations concerning the decoupling of
 the r-process production of the $Z \lsimeq 58$ and $58 \lsimeq Z \lsimeq 76$ nuclides.
 The existence of two so-called r-process components have been hypothesised on such grounds. 
We do not want to dwell on an exercise of identifying the number of these components, 
which may not be as productive as it might look at first sight. Just a word of caution is 
in order here: the observed neutron-capture element data are classically discussed with 
reference to the SoS s- or r-process distribution curves. It is certainly informative to find 
that, as reviewed above, some relative abundances appear to fit closely the SoS patterns. 
However, one must also keep in mind that the s-process, as well as various aspects of
 the r-process, have no fundamental reason to have close similarities in the SoS and in
 individual stars, particularly very metal-poor ones. It is worth stressing this as
it is not often duly appreciated in the literature.

In conclusion, metal-poor r-process-rich stars exhibit a pattern of r-nuclide abundances in
the approximate $58 \lsimeq Z \lsimeq 76$ that is remarkably similar to a scaled SoS
r-nuclide abundance pattern. The situation is not as clear for lighter neutron-capture
elements whose abundances may deviate from the SoS distributions, as exemplified by the
Sr/Eu ratio.

\subsection{Actinides in the Solar System, in the Local Interstellar Medium and in stars}
\label{actinides}

Actinides have a very special status in the theory of nucleosynthesis, as they are the only 
ones of clear and unique r-process origin. In addition, their radioactivity, and in some cases
 their lifetimes commensurable with the presumed age of the Galaxy, makes them potentially 
suited for chronological considerations.

For long, the abundances of the actinides have been known only in the SoS
essentially through meteoritic analyses \cite{lodders03}. Since the much-celebrated piece
 of work of 
\cite{fowler60}, the SoS \chem{232}{Th}, \chem{235}{U} and \chem{238}{U} have been widely
 used in attempts for estimating the age of the Galaxy. Some details on this chronological
 technique can be found in Sect.~\ref{chronometry_solar}.

\begin{figure}
\center{\includegraphics[scale=0.5,angle=90]{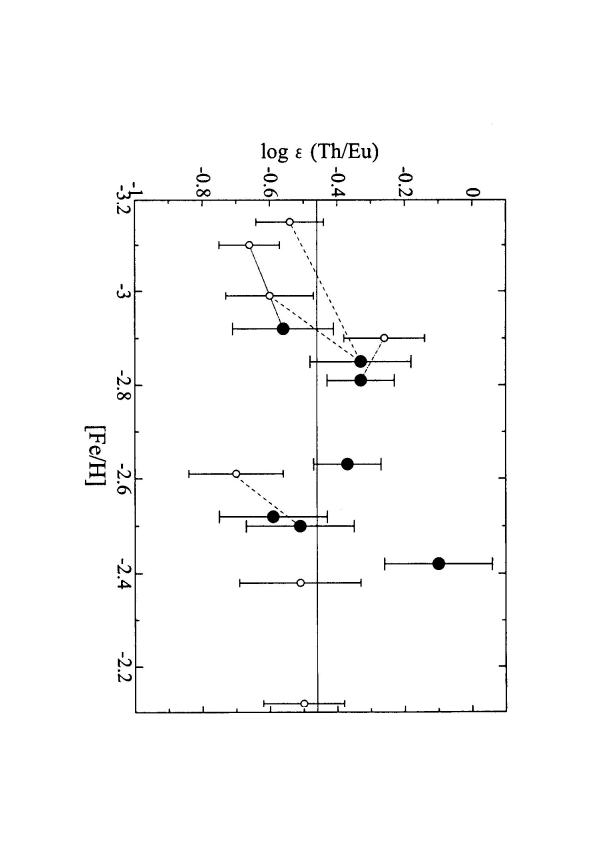}}
\vskip -1.2cm
\caption{Values of Th/Eu (in the $\epsilon$ scale defined in Fig.~\ref{galaxy_la_eu}) versus
 [Fe/H] obtained from different 
observations (from  \cite{honda04}). The closed circles are determined by \cite{honda04},
whereas the open circles are earlier observations with sometimes different [Fe/H] 
assignments. The solid line correspond to the SoS value
}
\label{th_eu}
\end{figure}

The astrophysical importance of Th and U has been enhanced further with the first observation
 of Th in stars with close-to-solar composition \cite{butcher87}, and later in metal-poor 
stars \cite{francois93}. Observations of this kind have been actively pursued, so that Th 
has by now been measured in 14 stars with [Fe/H] in the approximate range from -3.2 to -2.2, 
as seen in Fig.~\ref{th_eu}. The Th data have been used by several authors in combination 
with the corresponding Eu abundances to evaluate the ages of the concerned stars. In this 
respect,  it would clearly be desirable to rely instead on the Th/U abundance ratio in stars,
 as it is done in the SoS case . Unfortunately, detection of U is difficult because of the 
weakness of its spectral lines, combined with its present low abundance in the studied stars.
 In spite of these difficulties, U has been successfully measured in  the [Fe/H] = -2.9 giant 
CS 31082-001 (\cite{plez04}, and references therein). Only upper limits have been obtained 
by \cite{honda04} for the 7 stars in which they have derived Th abundances. The reliability 
of the age determination of specific stars based on the use of the Th/Eu and Th/U derived 
from observations is discussed in Sect.~\ref{chronometry_lowz}.

It has to be noted that \chem{232}{Th}, \chem{235}{U} and \chem{238}{U} all decay to Pb. 
The Pb abundance has been measured in CS 31082-001 \cite{plez04}. From this observation, it 
is concluded \cite{plez04} that more than 50\% of the total Pb in this star are the actinide 
progeny. This does not provide any strong constraint on the fraction of the Pb in CS 31082-001
 that is a direct (instead of an actinides  decay) product of the r-process. Lead in very
 metal-poor stars can indeed originate from the s-process as well \cite{vaneck03}. It has
 also to be remarked that the SoS r-process Pb is highly uncertain, the fractional 
contribution of this process derived from Table~\ref{tab_r}  lying between 1 and 80\%! 
 In such conditions, Pb data in low-metallicity stars or in the SoS can hardly provide 
useful information on the r-process.

Finally, let us recall the attempts to measure the  \chem{244}{Pu} content in the local
 interstellar medium (ISM), which may have some interesting astrophysics implications. At
present, this can be done through the analysis of dust grains of identified interstellar
origin recovered in deep-sea sediments (e.g. \cite{paul01}). In a near future, the
determination of elemental and isotopic composition of the ISM grains will be a major
goal of research with their recovery to Earth by the Stardust mission \cite{browlee96}.

\subsection{The r-nuclide content of Galactic Cosmic Rays}
\label{GCR}

The measured abundances in the Galactic Cosmic Rays (GCRs) of all the elements in the 
approximate $30 \lsimeq Z \lsimeq 60$ range of atomic numbers are roughly consistent with a solar 
composition at the source of the GCRs (the very nature of which remains quite mysterious)
 once corrections for atomic selection effects have been duly taken into account. In 
particular, no clear trend is identified in this range of atomic numbers  for a specific enhancement
or deficiency of either r- or s-nuclides.
The elements with $Z \gsimeq 40$ might be overabundant {\it relative to} Fe. This conclusion
 is very sensitive, however, to the modelling of the GCR propagation conditions  (see e.g. 
\cite{meyer97}).

The situation appears to be different in the $Z \gsimeq 60$ range, where observations suggest
 that the abundances in the `Pt group' (Pt, Ir and Os) are in excess to those around Pb
 (`the Pb group') relative to the SoS composition (e.g. \cite{westphal98}), this conclusion 
 being relatively insensitive to the propagation conditions. Various interpretations of this
relative excess in the Pt group have been proposed. One of them \cite{meyer97} relates to 
the high volatility of Pb and of other `Pb group' elements, which contrasts with the refractory 
character of the `Pt group' elements. In these views, the observations could well be consistent 
with a mixture of the SoS type. Another interpretation (e.g. \cite{westphal98}) calls for 
an r-process enrichment of the material to be accelerated to GCR energies. A note of caution 
is in order at this point. As in the analysis of stellar spectroscopic data, the discussion
 concerning the Pt group over the Pb group abundance ratio relies heavily on the SoS splitting
 between s- and r-nuclides. As already stressed before, these two contributions to Pb are 
very uncertain (these uncertainties are much lower in the Pt group case, the r-process 
contribution to which varying from about 85 to 100\%; see Table~\ref{tab_r}). Recall that 
GCRs are made of much younger material than the SoS (i.e., $20 \sim 30$ My old).
There is no proof at 
this time that the r-process(es) that has(have) contributed to this recent sample of 
galactic material is(are) similar to the one(s) contained in the SoS material.

The identification of actinides in the GCRs has been made possible quite recently by the use of
the Trek detector \cite{westphal98}. An accurate measurement of their abundances relative 
to each other and to the Pt group  is within the reach of the
planned Extremely Heavy Cosmic Ray Composition Observer (ECCO) \cite{westphal00}. Such 
data would in particular help discriminating between various GCR sources that have been
 proposed, including fresh supernova ejecta, supper-bubble material, or old galactic material.  

It is generally taken for granted today that supernova explosions are the most probable 
GCR energy source. It is believed that individual supernova remnants may be responsible 
for the acceleration of external, swept-up interstellar matter, with at most a very tiny 
contribution of internal, nucleosynthetically processed material \cite{meyer99,ellison99}.
 Observation of GCR actinides  could confirm  this scenario. Massive star supernova
 explosions 
 are not random in the Galaxy, however, and concentrate strongly in OB associations. In fact,
fireworks of sequential explosions of tens of massive stars lasting for periods as short as a
 few million years could create `multiple supernova remnants'. These can grow into so-called 
`super-bubbles' made of hot tenuous plasma most commonly observed from their
x-ray emission in our and nearby galaxies (e.g. \cite{spitzer90}). Super-bubbles might well be
privileged galactic locations for the acceleration of matter to GCR energies 
\cite{higdon98,parizot00,parizot01,bykov01}. Just as in the case of isolated remnants, each 
super-bubble remnant accelerates external, swept-up super-bubble material. This material,
though predominantly ordinary ISM evaporated from nearby clouds, is significantly
 contaminated by the recent ejecta of previous local supernovae.  In addition, turbulent
 acceleration should take place steadily throughout the super-bubble gas.  So, more fresh 
supernova ejecta (say with typical ages shorter than about 30 My) may be expected in GCRs 
from super-bubbles than from isolated supernovae.
This results from the study of both the super-bubble dynamics and from considerations about 
the synthesis of the light elements Li, Be and B in the early galaxy (e.g.
 \cite{parizot00,parizot01}). This increased fraction of fresh ejecta also nicely accounts 
for the GCR \chem{22}{Ne} anomaly \cite{binns05}. As noted by \cite{westphal00}, GCRs 
originating from supper-bubbles would likely be young enough for
containing a significant amount of \chem{244}{Pu} and \chem{247}{Cm}, the lifetimes of which
 are commensurable with those of the super-bubbles. Concomitantly, the presence of \chem{247}{Cm}
 in the GCRs and their implied young age would be the indication that Th, U, and Pu have
 abundance ratios close to their r-process production ratios. In such conditions, first-hand
 constraints on the actinides production by the r-process could be gained in addition to 
quality information on the origin and age of the GCRs.
 
 As in the stellar case, information on the isotopic composition of the GCRs would be of
 prime interest in helping to evaluate the fractional contribution of freshly synthesised 
r-process material to this quite recent sample of galactic material. In fact, GCR composition
 measurements with isotopic resolution up to at least $Z \approx 40$ are within the reach of
 present detector technology, as exemplified  by the  R-process Isotope Observer RIO project 
currently under study \cite{weaver05}

\section{The nuclear physics for the r-process}                                                  
\label{nucphys_general}

As shall be concluded from Sects.~\ref{high_t_param}, \ref{HIDER}, \ref{evol_massive}, 
and \ref{compact_general}, the site(s) of
the r-process is (are) not identified yet,  all the proposed scenarios facing serious
problems. It is easily conceivable that the nuclear physics that enters the r-process
 modelling depends to a more or less large extent on the astrophysics conditions. From the 
variety of available r-process studies, it  appears reasonable to say that an incredibly huge 
body of nuclear data are potentially needed for the purpose of r-process nucleosynthesis
 predictions. This includes the static properties (like masses, matter and charge 
distributions, single particle spectra, or pairing characteristics) of thousands of nuclides
 from hydrogen to the super-heavy region  located between the  valley of $\beta$-stability and
the neutron-drip line. Their  decay characteristics 
($\beta$-decay, $\alpha$-decay, various kinds of $\beta$-delayed processes,
 spontaneous or induced fission), and reactivity (nucleon or
 $\alpha$-particle captures, photo-reactions, or neutrino captures) may be needed as well. 

A major effort has been devoted in recent years to the measurement of nuclear data of 
relevance to the r-process. Still, a large body of information in quest remains,  and will remain in 
a foreseeable future, out of reach of experimental capabilities. This is of course the direct 
consequence of the huge number of nuclear species that may be involved in one r-process or 
another, along with the fact that nuclei very far from the valley of stability are likely to
 enter the process. Theory has thus mandatory to complement the laboratory measurements.  

In order to meet at best the demanding r-process nuclear-physics needs, the nuclear models
of choice have to satisfy to the largest possible extent two basic requirements: they have to be 
as  {\it microscopic} and {\it universal} as possible. The microscopic nature of the underlying
 models is essential as a large amount, if not all, of data need to be extrapolated far away
 from experimentally known regions. In these situations, two
characteristics of the nuclear theories have to be considered. The first one is the accuracy
of a model. In most nuclear applications, this criterion has been the main, if not the
unique, one for selecting a model. The second one is the reliability of the
predictions. A physically-sound model that is as close as
possible to a microscopic description of the nuclear systems is expected to provide the
best possible reliability of extrapolations.
Of course, the accuracy of such microscopic models in reproducing experimental
data may be poorer than the one obtained from more phenomenological models in which
enough free parameters can guarantee a satisfactory reproduction of the data at the
expense of the quality of the input physics, and consequently of the reliability. The
coherence (or `universality') of these microscopic models (through e.g. the use of the
same basic nuclear inputs, like the effective nuclear forces) is also required as
different ingredients have to be prepared in order to evaluate each nuclear-reaction rate.
Failure to meet this requirement could lead to inaccurate rate evaluations. Much
progress has been made recently in the development of models that are microscopic and
 universal to 
the largest possible extent, although much remains to be worked out.

\subsection{Nuclear ground state properties}
\label{nuc_static}

Impressive progress has recently been made in the measurement of the masses of unstable  
nuclei (\cite{lunney03} for a review). The advance results mainly from the use of 
Penning-traps \cite{sch01} or Schottky spectrometers \cite{nov02}.  The 2003 Atomic Mass 
Evaluation \cite{awt03} contains 2228  measured masses, i.e 263 more than the one in 1995  
\cite{aw95}. More accurate mass determinations are also available for about 130 nuclei.  The new 
data concern only 47 neutron-rich nuclides, almost none of them being involved in the main 
nuclear r-process flows predicted by most models.  Under such circumstances, theoretical
 predictions are called for, not only to provide masses, separation energies or reaction 
Q-values, but also to predict  the ground state properties entering the calculation of the 
reaction and decay rates, such as deformations, density distributions, single-particle level 
schemes, and pairing gaps.

Attempts to estimate nuclear masses go back over seventy years to the
liquid drop semi-empirical mass formula  \cite{we35}. Improvements to this intuitive
 model have been brought little by little, leading to the construction of macroscopic-microscopic 
formulae, where microscopic ('shell'-) corrections to the liquid-drop (or the more generalised `droplet')
 part are introduced in a phenomenological way (for a review, see \cite{lunney03}). In this framework, the
 macroscopic and microscopic contributions are treated independently, 
 and are connected only through a parameter fit to experimental masses.
 Further  developments have been made in relation to the macroscopic properties of 
infinite and semi-infinite nuclear matter and the finite-range character of the nuclear forces. 

Until recently the atomic masses have been calculated on the basis of one extension or
another of the droplet approximation, the most sophisticated version of which being the FRDM 
model \cite{frdm}. Despite the success of this formula in fitting experimental data (the 2149
 $Z\ge8$ measured masses \cite{awt03} are reproduced with an rms error of 0.656 MeV), it
 suffers from some shortcomings, such as the incoherent link between the macroscopic part 
and the microscopic corrections, the instability of the mass predictions to different 
parameter sets, or the instability of the shell corrections.  
These aspects are worrisome when extrapolations are required as in 
astrophysics applications, and especially in the r-process modelling.
In fact, the important question of the reliability does not only concern 
masses,  but more generally  the predictions of experimentally unknown ground- and excited-state
properties. As already stressed above, the predictive power of a model in these respects is 
expected to improve with the increase of its  microscopic character. This point of view has 
driven the efforts to construct mass models relying on a  global mean field approach.  

The Extended Thomas-Fermi plus Strutinsky Integral (ETFSI) model is a first step in this 
direction. It is a high-speed approximation to the Skyrme-Hartree-Fock (HF) method, with 
pairing handled in the BCS approximation. ETFSI  has led to the construction of a complete
 mass table \cite{abo95}, and has shown the way to a full HF calculation based on a Skyrme
 force fitted to essentially all the mass data. It has been demonstrated by \cite{sg00} that 
this approach is not only feasible, but that it can compete with the most accurate
 droplet-like formulae. The adopted Skyrme force has its conventional form including nine
 free parameters (e.g.  \cite{Vautherin72}), and is complemented with a $\delta$-function
 pairing force 

\begin{equation}
v_{\rm pair}(\mbox{\boldmath$r$}_{ij})=
V_{\pi q}~\left[1-\eta
\left(\frac{\rho}{\rho_0}\right)^\alpha\right]~
\delta(\mbox{\boldmath$r$}_{ij}) \quad 
\label{eq2}
\end{equation}

acting between like nucleons. In this expression,   $\rho$ is the density distribution,
 $\rho_0$ being its saturation value. The strength parameter $V_{\pi q}$ is allowed to
 be different for neutrons and protons, and also to be stronger for an odd  number of 
nucleons ($V_{{\pi q}}^-$) than
for an even one ($V_{{\pi q}}^+$). A Coulomb  energy and  a phenomenological Wigner term 
of the form

 \begin{equation}
\label{eq3}
E_{\rm W} =  V_{\rm W}\exp\Bigg\{-\lambda\Bigg(\frac{N-Z}{A}\Bigg)^2\Bigg\}
+V_{\rm W}^{\prime}|N-Z|\exp\Bigg\{-\Bigg(\frac{A}{A_0}\Bigg)^2\Bigg\}  
\end{equation}

are added.

The first fully-microscopic (Skyrme-based HF) mass-table ever constructed is referred to 
 as HFBCS-1 \cite{sg00}. It makes use of the BCS approximation for pairing, and involves all
 the nuclei with $Z, N \ge 8$ and $Z \le 120$ lying between the drip lines. 
In order to improve
the description of  highly neutron-rich nuclei that could be of special relevance to the 
r-process, the HF + BCS approach has subsequently been replaced by a full HF-Bogoliubov (HFB) 
calculation \cite{sam01}, referred to as HFB-1. The HFB model has the important advantage of
being able to treat the mean and pairing fields self-consistently and on the same footing.  
In contrast with the BCS method, it also allows any pair of nucleons to scatter and does not 
create any spurious gas outside the nucleus, a necessary condition to safely 
describe exotic nuclei.  

The HF+BCS and HFB-1 models achieve comparable fits (typically with a rms deviation of about
 0.75~MeV) to some 1888 masses of  the $N, Z \ge$ 8 nuclei reported in the 
 compilation in 1995 \cite{aw95}. The HF+BCS model can in fact be shown to be a very good approximation
 to the HFB model provided that both models are fitted to experimental masses. The unmeasured
 masses indeed never differ by more than 2 MeV below $Z\le 110$. The reliability of the 
HF predictions far away from the experimentally known region, and in particular towards the 
neutron-drip line, is improved when 
the Bogoliubov treatment of the pairing correlations is adopted. 

The data made available in 2001 \cite{aw01}, including 382 new measured masses since the 1995
compilation \cite{aw95}, out of which only 45 concern  neutron-rich nuclei, have in fact 
revealed
significant limitations in  both the HFBCS-1 and HFB-1 models. This deficiency has been cured
in the subsequent HFB-2 mass model \cite{sg02} through a modification of the  prescription for
 the cutoff of the spectrum of single-particle
states over which the pairing force acts. The rms error with respect to the measured
masses of all the 2149 nuclei included in  the latest 2003 atomic mass evaluation
\cite{awt03} with $Z,N \geq 8$ is $0.659$~MeV \cite{sg02}.  Despite the success of the HFB-2
 predictions,  it has been considered desirable to carry out further studies based on 
modifications of  the Skyrme force and of some aspects of the computational techniques
 \cite{sam03,gor03,sam04,gor04}. The aim of these complementary calculations was primarily to try
 improving the quality of the mass fits. It was also to examine the level of convergence
 of the predictions toward the neutron-drip line of mass models that give essentially the 
same rms deviations to measured masses. Even if this rms constraint is met, this convergence 
far from stability is not guaranteed indeed. The analysis of the convergence provides a possible way to 
estimate the reliability of HFB mass evaluations in regions where no experimental guide is 
available to-day. Additionally, HFB calculations are underlying the predictions of other 
nuclear quantities of interest for the r-process modelling, like  the
nuclear matter equation of state \cite{gor04}, fission barriers (Sect.~\ref{fission}),
 $\beta$-decay  strength
functions (Sect.~\ref{beta_sub2}), giant dipole resonances (Sect.~\ref{th_strength}), nuclear
level densities (Sect.~\ref{th_leveldens}) and neutron optical
potential  of highly unstable nuclei (Sect.~\ref{th_pot}). It may well be that different models
 that are equivalent from the standpoint of masses may  give different results for these other
properties. It is therefore of particular interest to develop different HFB mass models with 
the quest for a universal framework on which all the different nuclear properties can
be based.

\begin{figure}
\center{\includegraphics[scale=0.5]{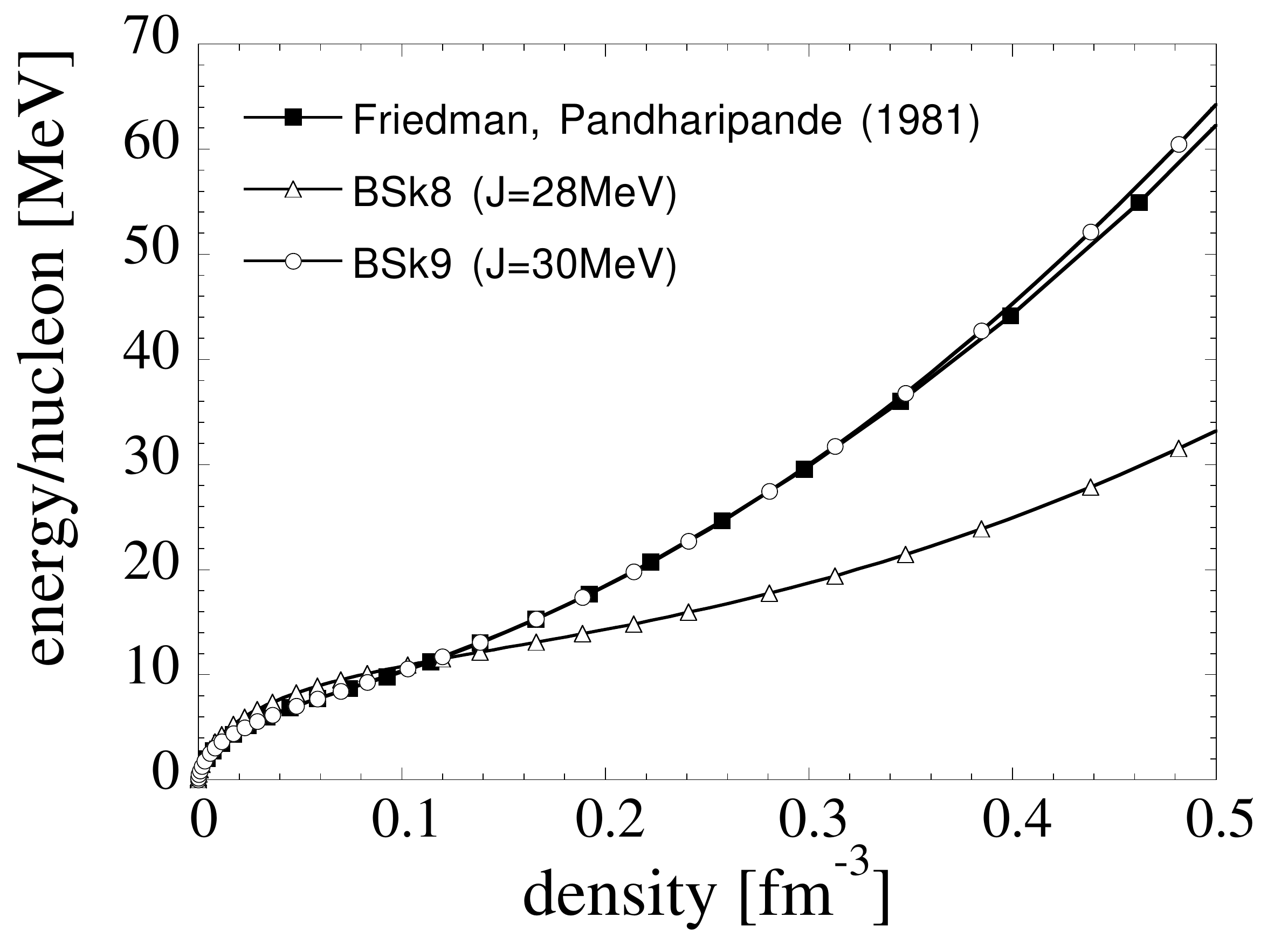}}
\caption{Energy per nucleon versus neutron matter density as predicted from the HFB 
calculations with the forces BSk8 and BSk9, and from the variational calculations of
 \cite{fp81} (closed squares)}
\label{fig_nm}
\end{figure}

For the reasons mentioned above, seven additional new mass-tables, referred to as
 HFB-3 $\sim$ HFB-9, based on the Skyrme forces BSk3 to BSk9, have been constructed,
 some of them differing in the form of their parametrisation.  In particular, \\
(1) HFB-3,5,7 \cite{sam03,gor03} are characterised by a density dependence of the pairing 
force inferred from the pairing gap in infinite nuclear matter at
different densities \cite{gar99} calculated with a `bare' or  `realistic' nucleon-nucleon
interaction (corresponding to
$\eta=0.45$ and $\alpha=0.47$ in Eq.~\ref{eq2});\\
(2) HFB-4,5 (HFB-6,7,8,9) are based on a low isoscalar effective mass $M^*_s=0.92$
 ($M^*_s=0.8$) \cite{gor03} derived from microscopic (Extended  Br\"uckner-Hartree-Fock)
 nuclear matter calculations \cite{zuo99};\\
(3)  HFB-8 and HFB-9 allow  the
particle number symmetry to be restored by applying the projection-after-variation technique
 to the HFB wave function \cite{sam04}, and\\
(4) BSk9 used in HFB-9 has been tailored in order to reproduce at best the results of 
microscopic descriptions of neutron matter. As already recognised by \cite{rayet82}, this 
requirement on the Skyrme forces is essential in attempts to predict the properties of highly 
neutron-rich nuclei all the way to the neutron-drip line, as it is the case in r-process 
simulations (see especially Sect.~\ref{compact_general}). The early HFBCS-1 mass model leads 
to a neutron matter that is so soft that it collapses at supra-nuclear densities. The situation 
improves with the adoption in the BSk1 - 8 forces of the value $J = 28$~MeV for the nuclear-matter
 symmetry coefficient, this value being the lowest acceptable one to avoid the collapse.
 Still,  the neutron matter remains a little softer 
than predicted by realistic neutron matter calculations, as illustrated in  Fig. \ref{fig_nm} 
 for the case of BSk8. The situation is essentially unchanged for BSk1- 7. Fig.~\ref{fig_nm} 
shows that the problem is cured with BSk9  by forcing $J$ to increase from the value 
$J=28$~MeV to $J = 30$~MeV. Future accurate measurements of the neutron-skin thickness of 
finite nuclei would help better constraining $J$ \cite{gor04}.

\begin{figure}
\center{\includegraphics[scale=0.5]{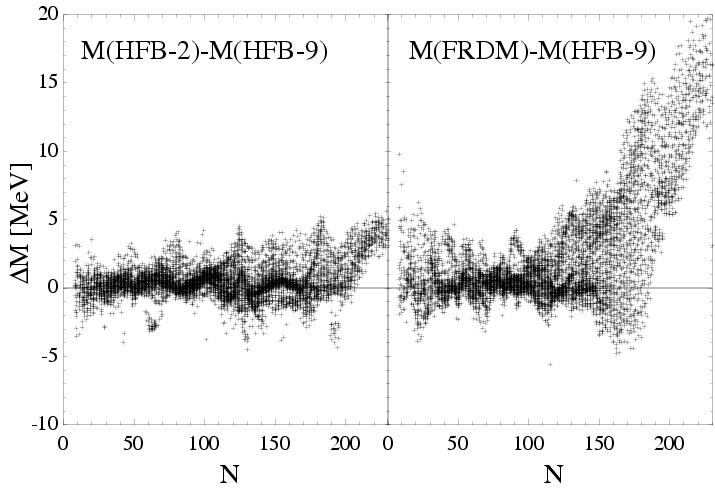}}
\caption{Comparison 
of the mass predictions by HFB-2, HFB-9 and by the microscopic-macroscopic FRDM
for nuclei with $8\le Z \le 110$ lying between the proton- and neutron-drip lines}
\label{fig_extra}
\end{figure}

\begin{figure}
\center{\includegraphics[scale=0.5]{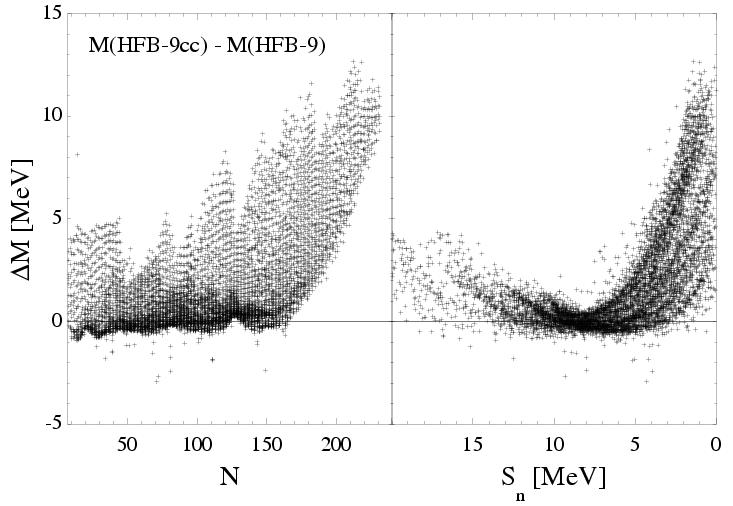}}
\caption{ Differences between the HFB-9cc and HFB-9 masses  (left
panel) and neutron separation energies $S_{\rm n}$ (right panel) for all nuclei with $8\le Z
\le 110$ lying between the proton and neutron-drip lines}
\label{fig_cc}
\end{figure}

The HFB-1- 8 mass tables reproduce the 2149 experimental masses \cite{awt03} with  an rms 
deviation as low as about $0.65$~MeV. Even if this value increases slightly up to 0.73~MeV 
for HFB-9, which is the price to pay in order for BSk9 to feature a better reproduction of 
the neutron matter properties, one may state that HFB-1 to HFB-9 are very close in their mass 
predictions. This is illustrated in Fig.~\ref{fig_extra}  for the HFB-2 and HFB-9 masses. 
Although these two models rely on significantly different Skyrme forces, deviations no
 larger than 5~MeV are obtained for all nuclei with $Z \le 110$. In contrast, larger deviations
are seen in Fig.~\ref{fig_extra} between the predictions from HFB-9 and the 
macroscopic-microscopic droplet-model FRDM. This is especially the case for the heaviest
 nuclei. For lighter ones, the mass differences remain below 5~MeV, but significant 
differences are observed locally in the shell and deformation effects. In this respect, it is
 especially noticeable that
the HFB calculations predict a weaker (though not totally vanishing) neutron-shell closure
close to the neutron-drip line than FRDM (e.g \cite{sam03,gor03}). 

The HFB rms charge radii and radial  density distributions are also in excellent agreement 
with the experimental data \cite{sam04}. More specifically,
the rms deviation between the HFB-9 and experimental rms charge radii for the
782 nuclei with  $Z, N \ge 8$ listed in the  2004 compilation \cite{ang04} amounts to
only 0.027~fm. The ability of the Skyrme forces to reproduce excited-state properties has
 also been tested. In particular, the giant dipole resonance properties 
obtained within the HFB plus Quasi-particle Random Phase Approximation (QRPA) framework with 
the BSk2-7 forces have been found to agree satisfactorily with the experiments
\cite{khan04}. 

Although complete HFB mass-tables are available now, one still has to aim at further 
improvements that could have an impact on mass-extrapolations towards the neutron-drip line. 
 In particular, all the HFB mass fits show a strong pairing effect that is most probably one 
of the manifestations of the neglect of extra correlations in the calculation of the
total binding energy.  In view of the good mass fits that are obtained, it is likely that the 
adjustments of the Skyrme force parameters have further masked some of the neglected physics. 
 In particular, corrections for vibrational zero-point motions should have to be included
 explicitly.
 A first step in this direction has been achieved \cite{bender05}, but the current computing
 capabilities prevent the application of this approach to global mass fits.  The interplay 
between the Coulomb and strong interactions should also be scrutinised. It could lead to an
 enhancement of the Coulomb energy at the nuclear surface \cite{bu99} that might explain the 
Nolen-Schiffer anomaly, i.e the systematic reduction in the estimated binding energy 
differences between mirror nuclei with respect to experiment. This Coulomb correlation effect
 could in fact significantly affect the nuclear mass predictions close to the neutron-drip 
lines. To analyse its impact,  BSk9 has been refitted by excluding the contribution from the 
Coulomb exchange energy as the Coulomb correlation energy seemingly cancels the Coulomb
 exchange energy to a good approximation \cite{bu99}. The force derived in such a way leads 
to so-called HFB-9cc masses that differ from the 2149 experimental ones by an rms value of 
0.73~MeV, which is identical to the HFB-9 result.  In spite of this, Fig.~\ref{fig_cc} shows
 that the HFB-9cc predictions  differ from those of HFB-9 by more than 10~MeV close to the 
neutron-drip line. This quite large effect clearly needs to be studied further. 

\subsection{Beta-decay properties of neutron-rich nuclei}   
\label {beta}
  
The role that $\beta^-$ decays of very neutron-rich nuclei play in the r-process is twofold. 
Most important, they allow high-$Z$ nuclei to be produced from lighter `seed nuclei'  during 
timescales over which the r-process could operate. In addition, $\beta^-$ decays influence to
 some extent the relative abundances of product r-nuclides, at least `locally' in the chart of
 nuclides. As most of the very neutron-rich nuclei which are supposedly involved in the
 r-process have yet to be discovered in the laboratory,  predictions of $\beta^-$ decay 
properties are unavoidable in the modelling of the r-process. These predictions, and more 
generally weak interaction processes, are discussed in e.g.  \cite{borzov05,langanke03} 
(see also the textbook  \cite{grotz90}).
 
Most reviews of the r-process seem to conspicuously avoid discussions of the physics of 
 $\beta$-decays of heavy nuclei. Given this, we consider that it is of some interest to 
summarise some of the basics of the topic and to clarify the state-of-the-art techniques
 of the $\beta$-decay predictions. 

\subsubsection{Half-lives as critical data}
\label{beta_sub1}

\begin{figure}
\center{\includegraphics[width=1.0\textwidth,height=0.8\textwidth]{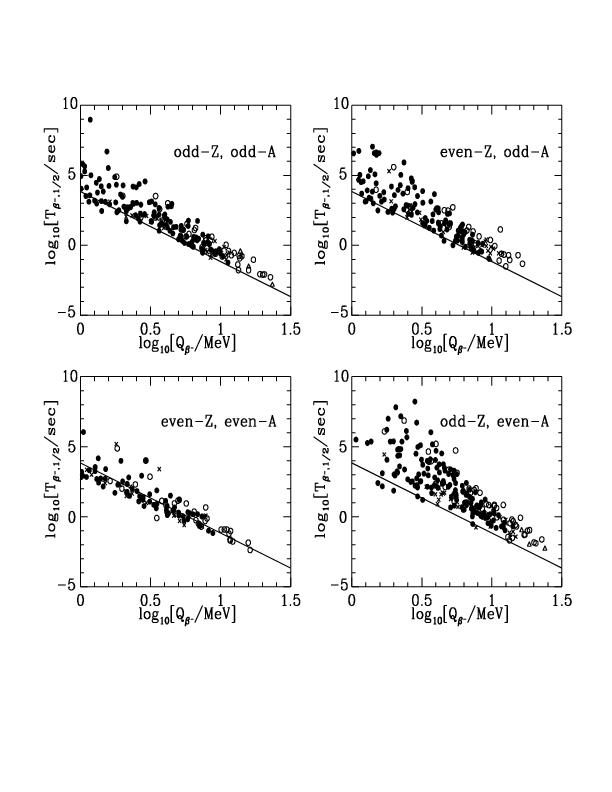}}
\vskip-2.4cm
\caption{Experimentally known $\beta^-$ decay half-lives, $T_{\beta^-,\frac{1}{2}}$
versus $\beta^-$-decay $Q$-values.  The open  and closed dots  are  for  `light'
($Z < 26$) and  `heavy' ($Z \ge 26$) nuclei with experimentally known $Q_{\beta^-}$ values 
\cite{audi03a,awt03}.
These symbols are replaced by triangles  and crosses if the  $Q_{\beta^-}$ values are derived 
 from mass systematics. Each panel corresponds to a specific  even-oddness of the numbers 
of nucleons in the parent nuclei in order to illustrate the role of the nuclear pairing. 
The solid lines corresponding to the decay rates $\lambda_{\beta^-} \equiv $
 ln~2$ /T_{\beta^-,\frac{1}{2}}\ = 10^{-4} \times [Q_{\beta^-}$ in MeV]$^5$ s$^{-1}$ are added 
as a guide to the eye. It is generally expected that heavy nuclei with $Q_{\beta^-} \sim 
10-20$ MeV are involved in  the r-process}
\label{fig:beta_tq}
\end{figure}

Figure~\ref{fig:beta_tq} displays the  known  $\beta^-$-decay half-lives versus 
 $\beta^-$-decay $Q_{\beta^-}$ values separately for the classes of even-even, odd-odd and
 odd-even nuclei.\footnote{$Q_{\beta^-}$ is conventionally defined as the  difference
 between the ground-state atomic masses of the the parent $(Z,A)$ and  daughter $(Z+1,A)$
 nuclei, no matter how  efficiently $\beta$-transitions  take place between  the
 ground states. It thus represents  the maximum nuclear energy release available for electron
 kinetic energy if generally tiny corrections for the atomic binding energy difference, recoil
  energy, and neutrino mass are neglected}
The decrease of the half-lives with increasing $Q_{\beta^-}$ comes 
primarily from the larger phase volumes available for the emitted leptons (e$^-$ and 
$\bar{\nu}_{\rm e}$). The significant scatter of the data  points for  low $Q_{\beta^-}$ values 
is blamed on various (such as the spin-parity) selection rules for the $\beta$-transitions, 
the impact of which is increasing with the decreasing number of suitable final 
states. The spread is reduced for higher $Q_{\beta^-}$ cases as the number of such states becomes 
larger. On the other hand, the different trends seen in the four panels of  
Fig.~\ref{fig:beta_tq}  are ascribed to the effect of nuclear  pairing on the $Q_{\beta^-}$ 
values, combined with the dependence on the `pairing gaps' of the number of low-lying states of the 
daughter nucleus that can be fed by $\beta$-decay.
 %

%

Experimental data are largely lacking for heavy nuclei in  the approximate $Q_{\beta^-} 
\gsimeq 10$ MeV  range. This is unfortunate as these decays are likely of  prime concern in 
the r-process. The good news is that,  for the reasons sketched above, the half-lives are 
close to decrease monotonically, and even to converge to a common value (within perhaps a 
factor of even less than about five) in the range of the highest  $Q_{\beta^-}$ values of 
relevance to the r-process. This convergence even holds when comparing `light' and 'heavy'
 nuclei, which may be surprising, given that
 (i) heavier nuclei have more neutrons to decay, and (ii) 
 the probability of finding an electron to be emitted near the nucleus is very much enhanced in
 heavy nuclei because of the large Coulomb attraction. This relative hindrance in heavy nuclei
 is explained in terms of the Gamow-Teller giant resonance (GTGR) and of the associated sum 
rules \cite{ikeda63}. Before discussing this hindrance in some detail, let us recall that the 
 $\beta$-decay rate $\lambda_{\beta} (\equiv$ ln~2$/T_{\beta,\frac{1}{2}}$) is expressed as a
 sum over the energetically possible final states  

\begin{equation}
 \lambda_{\beta}\ =\sum_{f(\Omega)} \frac{G_{\Omega}^2}{2 \pi^3}
\vert\langle f|~\Omega~|i \rangle\vert^2 f_{\Omega}(E_i-E_f)  \approx\ \sum_{f(\Omega)}
\int_{-Q_{\beta}}^0
\frac{G_{\Omega}^2}{2 \pi^3} S_{\Omega} f_{\Omega}(-E) {\rm d} E,
\label{eq:beta_rate}
\end{equation}

where  $\Omega$ stands for a $\beta$-decay operator, $G_{\Omega}^2\vert\langle 
f|\Omega|i \rangle\vert^2$  is the square (or in some cases a cross product) of either the
 Vector or/and Axial-vector coupling constant(s) times the corresponding nuclear matrix 
element between  the initial $|i \rangle$ and final $|f \rangle$ states
 (e.g.\cite{konopinski66,konopinski66a}), whereas  $f_\Omega$ is the `integrated Fermi
 function'  measuring the lepton  phase volume for a given nuclear energy release $E_i - E_f$. 
This sum may
 be approximated by an integral through the introduction of the  $\beta$-strength function 
$S_\Omega$. This  approximation is especially appropriate when  $Q_\beta$ is high.

The leading terms in Eq.~\ref{eq:beta_rate} are the `allowed' transitions of the
 Fermi ($G_{\rm V}^2$) and Gamow-Teller ($G_{\rm A}^2$) types, for which $\Omega = 
\sum_k \tau_k^\mp$ and $\Omega = \sum_k \tau_k^\mp {\bar {\bf \sigma}}_k$, 
respectively, where  $\tau_k^\mp$ (for $\beta^{\mp}$) and  ${\bar {\bf \sigma}}_k$ 
are the isospin-laddering and spin operators acting on the $k$-th nucleon. 
The selection rules on the nuclear spin $J$ and parity $\pi$ changes
 between  the parent and daughter nuclei are $\Delta J =0$ and $\Delta \pi =$ no
 for the Fermi transitions, and  $\Delta J =0$ (only if $J \neq 0$), $\pm 1$  and 
$\Delta \pi =$ no  for the Gamow-Teller (GT) transitions.
For the allowed $\beta^-$-transitions (and essentially for all the cases of relevance here), 
 $f_\Omega(|E|)$ can be replaced by an integral over  the electron
energy (including the electron rest-mass $m_{\rm e}c^2$) 

\begin{equation}
          f_{0}(|E|)\ =\ \int_1^{W_0} p W (W_0 - W)^2 F(Z,W) {\rm d}W,
\label{eq:beta_f0}
\end{equation}

where  $W_0 = |E|/m_{\rm e}c^2 +1$, $p = [W^2-1]^{1/2}$, and the so-called 
Fermi function $F(Z,W)$ is the ratio of the probability of finding an electron at the
nucleus to that in the absence of the Coulomb attraction
 (e.g. \cite{konopinski66} - \cite{blatt52}). This integral can be calculated quite accurately
 with the inclusion of the finite-size 
nuclear corrections and screening effects ({\rm e.g.} \cite{gove71}). However, the use 
in Eq.~\ref{eq:beta_f0} of an analytic form of $F(Z,W)$ that is valid for the pure Coulomb 
field of a point-charged nucleus (e.g. \cite{blatt52}) more than suffices  when dealing with
 $\beta^-$-decays of heavy neutron-rich nuclei. For high-energy transitions of special interest
 here, $f_0$ increases as $\sim |E|^{4 \sim 5}$.
 
 Note that it is customary to express the $\beta$-decay nuclear matrix  elements squared in 
terms of  the product denoted  $f_0t \equiv f_0 T_{\beta,\frac{1}{2}}$ (with $T_{\beta,\frac{1}{2}}$ in s). 
The log$_{10}  f_0t$-values of allowed  transitions have typical values of $\sim 5$, but may
 range all the way  from $\sim 4$ to $\sim 9$. Some transitions characterised by log$_{10}
 f_0t \sim 3$ are called `super-allowed', and are observed  in light nuclei under specific
 circumstances (see Sect.~\ref{beta_sub2}). An additional contribution to the $\beta$-decay
 rates that is significant in a number of cases comes from `first-forbidden' transitions. 
They obey the selection rules $|\Delta J| \le  2$ and $\Delta\pi =$ yes, and their probability
 depends on six  major matrix elements (see e.g. \cite{konopinski66a}). Their log$_{10} f_0t$ 
values are typically of the order of 5 if $|\Delta J| \le  1$ (`non-unique' 
transitions).\footnote{In the case of `unique' first-forbidden transitions 
($|\Delta J| =  2$), 
 $f_0$ has to be replaced by the function $f_1$ obtained in most cases by multiplying the 
integrand in Eq.~\ref{eq:beta_f0} by a factor proportional to $(W_0 - W)^2$. Typically 
log$_{10} f_1t \sim 8 - 10$}
 Some $\beta$-decays also take place through transitions of higher order forbiddeness. 
They can  safely be neglected in r-process studies. 
  
\subsubsection{General structure of the $\beta$-strength distributions} 
\label{beta_sub2}

In order to understand $\beta$-decays in general, and those of nuclei  far off the line  of 
stability in particular, it is often much more profitable to work with  $\beta$-strength 
functions $S_\Omega$ (Eq.~\ref{eq:beta_rate}) rather than with matrix elements of individual 
 transitions.  Sect.\ref{beta_sub3} briefly reviews some models developed in order to evaluate
 $S_\Omega$.

It can be demonstrated \cite{yamada65,takahashi69} that the general behaviour of $S_\Omega$ is 
largely determined by the commutator $[H,\Omega]$, where $H = H_{\rm N} + H_{\rm C}$ is the 
total (nuclear $H_{\rm N}$  + Coulomb $H_{\rm C}$) Hamiltonian\footnote{If $\Omega$ is a 
vector as in the GT case, $\Omega$ in the following has to be replaced by each of its three
 components}. As $\beta$-decay transforms a neutron into a proton or vice versa and creates
 an electron or a positron,  $\Omega$ does not  commute with $H_{\rm C}$. The long-range 
character of the Coulomb interaction dictates that $[H_{\rm C},\Omega] \approx 
\Delta_{\rm C} \Omega$, where  the Coulomb displacement energy $\Delta_{\rm C} \approx
\pm [ 1.44~Z/(r_0~A^{\frac{1}{3}}/1.2)-0.7825]$ MeV  for $\beta^{\mp}$-decays if the nucleus 
is a uniformly-charged  sphere with radius $r_0A^{\frac{1}{3}}$ fm. If $\Omega$ commutes with 
 $H_{\rm N}$, the  $\beta$-strength is concentrated at energy $E_i + \Delta_{\rm C}$, $E_i$ 
being  the energy of the parent state. This is exactly the case of the Fermi strength 
distribution  because the nuclear force is essentially charge-independent. It exhibits a 
narrow resonance at the `Isobaric Analog State' $IAS|i\rangle$ \cite{alford61}. 
Only when the parent and daughter nuclei are the Isobaric Analog of each other (or more 
generally, when $IAS|i\rangle$ falls in 
the $Q_\beta$ window), can one observe a strong (`super-allowed') Fermi transition, as in some 
light nuclei.
The extremely small Fermi $\beta$-decay matrix-elements observed in heavy nuclei
demonstrate the smallness of the isospin impurity.

In analogy to the isospin-multiplet structure for the Fermi transitions,  the `persistence' of 
the Wigner supermultiplet structure for the GT transition has been postulated by 
\cite{ikeda63}.
The resonance is expected to be much broader than in the IAS case as a consequence of the 
spin-dependent parts  of $H_{\rm N}$ with which the GT operator does not  commute. 
The confirmation of that theoretical conjecture has been provided more than a decade later by
 the  experimental discovery of the GTGR based on the finding that  $(p,n)$ reaction cross 
sections at forward angles are proportional to a linear combination of the GT and Fermi 
strengths (\cite{bainum80,goodman80}; see also \cite{doering75}).
 
Much of the GT strength is exhausted by the GTGR located only about $\Delta_{\rm N} \approx 
6.7 - 30 (N-Z)/A$ MeV above the IAS (e.g. \cite{gaarde81}). Consequently, the $f_0t$ values 
of the GT $\beta^-$ transitions in heavy nuclei are much larger than  those of the 
super-allowed neutron and tritium decays for they occur in the tail region of  the strength
 distribution.\footnote{Some super-allowed GT transitions have recently been found to occur 
in very light neutron-halo nuclei near the neutron-drip line \cite{borge91}. They may not be 
related to the halo structure, however (\cite{hamamoto94} and references therein).}
 
The $(p,n)$ experiments have revealed that the observed GT strengths  are systematically 
 much lower than the expected values of the sum rule often referred to as the Ikeda sum rule 
(e.g.  \cite{gaarde80,gaarde83}). This has inflamed a long-standing holy battle over the 
possible causes of this so-called GT quenching. Recent $(p,n)$ experiments 
\cite{wakasa97,yako05} lend support to an early assessment \cite{bertch82} that, by and large,
 the quenching finds its origin in configuration mixing.
 
\subsubsection{$\beta$-decay models}
 \label{beta_sub3}

Different approaches have been proposed to understand (and wishfully predict)  $\beta$-decays 
of heavy nuclei. Two of them in the extreme can be clearly identified: a macroscopic model 
referred to as the Gross Theory on one hand,  and on the other the (large-scale) shell model,
 which is fully microscopic. Those in between are global  approaches of various 
kinds, the microscopic character of which is more or less pronounced.  In the first instance,
 they differ  in  their ways of describing the initial ground state and final excited states.

{\it Macroscopic Approach:}\ 
 The Gross Theory of $\beta$-decay \cite{yamada65,takahashi69,koyama70,takahashi71} aims at
 describing the general behaviour of  the $\beta$-strength distributions in a statistical
manner. With an assumed large number of final states $|f\rangle$, $S_\Omega$ for allowed and
 first-forbidden transitions are constructed by folding `one-particle strength functions' via 
a very simple pairing scheme taking into account the corresponding sum rules and even-odd 
effects. This is in fact the first attempt to predict $\beta$-decay half-lives of nuclei far
 off the line of stability that has taken into account the then-conjectured existence of the 
GTGR \cite{takahashi73}. 

\begin{figure}
\center{\includegraphics[width=1.04\textwidth,height=0.92\textwidth]{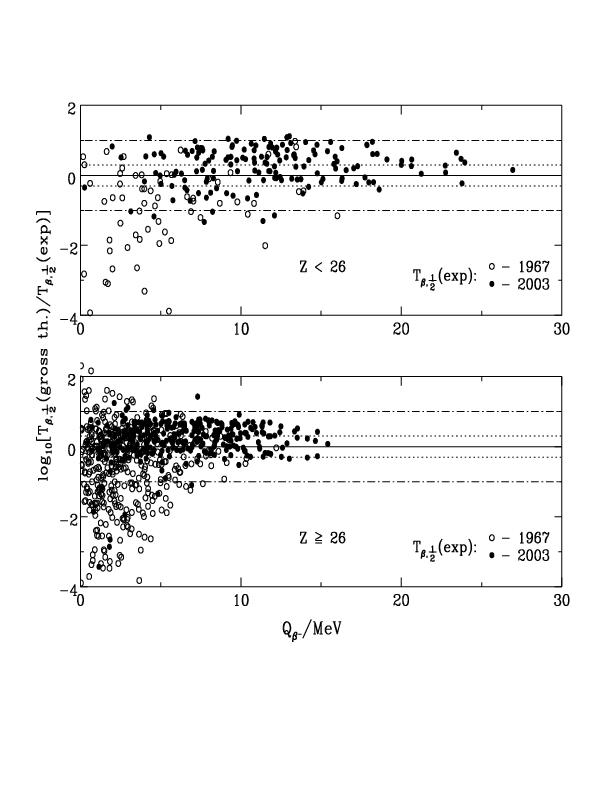}}
\vskip-2.5cm
\caption{Ratios of $\beta^-$ decay half-lives calculated according to the original Gross 
Theory to the experimental data \cite{audi03a}. The upper (lower) panel refer to `light' 
(`heavy') parent nuclei with atomic numbers of $Z <\ 26\ (Z \ge 26)$, respectively. Open dots refer
 to those nuclei which  were included in the 1967 compilation \cite{lederer67}, and which
 served as the main basis for the early theoretical analysis. The black dots
 indicate subsequent
 experimental  data. The experimental $Q_{\beta^-}$ values  (including those from systematics)
 \cite{awt03} are used as input rather than those of \cite{myers66} used originally. 
As in \cite{takahashi73}, however, the modified-Lorentz form for the `one-particle strength
 distribution' with $\sigma_{\rm N} = 12$ MeV is adopted. This value was derived in 
\cite{takahashi71} from a fit to 32 half-lives only
(and prior to the discovery of the GTGR!). Dotted and dash-dotted lines delimit the mismatch
 factor of two and ten, respectively. The theoretical results for heavy nuclei with high 
$Q_{\beta^-}$ values can be well reproduced by the analytic fit formulae given in 
\cite{kodama75}}
\label{fig:beta_gross}
\end{figure}  

The original Gross Theory has been improved in several ways, especially by 
\cite{tachibana90,nakata97}. Use is made in particular of some empirical and theoretical
 considerations in order to modify the one-particle strength functions and the pairing 
scheme. A version that is referred to as `GT2' (the 2nd-generation Gross Theory) is used 
extensively in Sects.~\ref{high_t_param} - \ref{r_dccsn} (see also Fig.~\ref{fig:beta_n82}).

With regard to the `predictive power',  it may be of some interest to see how the original
 version of the Gross Theory \cite{takahashi73} fares with the experimental half-lives 
gathered over the years. In fact, evidence has accumulated that newly-measured 
$\beta^-$ decay half-lives of neutron-rich nuclei are often shorter than the predicted
 values. It has been argued \cite{takahashi88} that, in some cases, this tendency comes 
simply from the use of the relatively low $Q_{\beta^-}$\ values predicted by the adopted
 liquid-drop-type mass formula of \cite{myers66}. Considering this possibility,  the original
 calculations are repeated  with the use of the empirical $Q_{\beta^-}$\ values \cite{awt03} 
instead. This is justifiable as  the overall $\beta$-strength distributions do not depend on 
the $Q_{\beta}$ values. On the other hand,  we leave unchanged the value of the sole 
adjustable (constant)  parameter, $\sigma_{{\rm N}}$, of the theory which is the partial
  width of the  one-particle strength functions caused by the  non-vanishing
 $[H_{{\rm N}},\Omega]$ commutator.  Figure~\ref{fig:beta_gross} compares the
 $T_{\beta^-,\frac{1}{2}}$ values calculated in such a way with the experimental data
 \cite{audi03a}. It demonstrates that the original Gross Theory modified as described above
 has a very satisfactory predictive power, particularly when high $Q_{\beta^-}$ transitions 
are involved, as it is the case for the decay of very neutron-rich nuclei. We conclude that 
the simple Gross Theory remarkably succeeds  in capturing the essence of $\beta$-strength tail
distributions throughout the chart of nuclides at the expense of the adjustment of just one 
overall parameter.

{\it Global (semi-)microscopic approaches:}\ 
 From the  viewpoint of  microscopic physics, the most efficient way  to get the essential
 features of the GT strength distribution, and of the GTGR in particular, is to embed at the 
onset of the modelling an effective nucleon-nucleon interaction, notably of the spin-isospin
 ${\bar{\sigma}} 
\cdot {\bar{\sigma}} \tau \cdot \tau$  type. This so-called Gamow-Teller force allows 
 particle-hole excitations of the charge-exchange collective mode. Given the residual
 interaction, the final GT states can be constructed in an approximate way from the model
 ground-state. Essentially all the calculations of relevance for the r-process adopt the 
`random phase  approximation (RPA)', which provides the simplest description of the excited
 states of a nucleus which allows the ground state not to have a purely independent particle 
character, but may instead contain correlations \cite{eisenberg72}. A casual inspection of 
the literature reveals a slight problem, however:  there appear RPA, QRPA, CQRPA, RQRPA, DRPA,
 SRPA, ERPA,  ESRPA, you name it, SCQRPA, FR-QRPA, SRQRPA, PQRPA, ... Although only a few of
 them are of concern here, they still have to be combined with as  many  acronyms standing for
 the various nucleon-nucleon interactions in use and for the models adopted to describe the 
 ground-state nucleus, a few  of which  make Sect.~\ref{nuc_static} lively. In this respect,
 a tabular comparison in \cite{borzov05} of the basic ingredients  of various models that have
 been employed so far in the predictions of $\beta$-decay half-lives is quite instructive.

Early in the game, a commuter method has been introduced \cite{fujita65} in order to minimise 
 cancellation errors in the evaluation of the $\beta$-decay matrix elements (i.e. the GT 
strengths in the tail). These errors might relate to the near-exhaustion of  of the sum rule 
by the GTGR .
The close relations between this technique and the RPA in general, and the Tamm-Dancoff 
 Approximation (TDA) in particular, have been discussed by \cite{morita71,ejiri71}. The TDA 
describes the ground state as single-particle states successively filled up to the Fermi level
 (so to say, a closed Hartree-Fock state). It thus denies  `ground-state correlations.'
 Accordingly, an asymmetry is introduced in the treatment of the ground and excited 
1p(article)1h(ole) states. The first large-scale microscopic model fit (and subsequent 
predictions) of $\beta^-$ decay half-lives \cite{klapdor84} is based on the TDA with an 
empirical potential for the ground state, and a schematic GT force whose strength is an 
adjustable parameter. 

More recently, the RPA approach, in which the excited states can be constructed by creating 
 or  destroying a particle-hole pair in the ground state, have been used for systematic 
predictions of $\beta$-decay half-lives of heavy nuclei. In particular,  large-scale 
computations of $\beta$-decay half-lives have been made by somehow introducing the pairing 
interaction, a most important residual force in open-shell nuclei. As an example,  such a 
calculation complemented with the evaluation of some other properties such as masses  has 
been made within the FRDM framework (Sect.~\ref{nuc_static}) making use of an empirical 
potential  and a BCS-type pairing scheme for the ground state \cite{moeller97}. Meanwhile,
 the TDA \cite{klapdor84} has been extended into a RPA approximation adopting the BCS pairing
 in the ground state, and formally including both particle-hole and  particle-particle (or
 the `$T=0$ pairing')  interactions required for treating the  pairing in the ground and 
excited states in a symmetry non-violating manner \cite{staudt90}. Such a method is generally
 and quite naturally referred to as  the `quasi-particle RPA (QRPA).'

In recent years, most attempts to make the QRPA models more sophisticated are heading 
primarily for a self-consistent description at least of the ground state, but ideally of the 
residual interaction as well. From the large volume of the literature on the subject (e.g. 
\cite{terasaki05} and references therein), we just pick up three models that explicitly deal
 with the $\beta^-$ decay half-lives of neutron-rich nuclei with a stated interest in the
 r-process. The first one \cite{borzov03,borzov05} adopts a density functional (DF) to 
describe the ground state self-consistently, while the QRPA-like framework of the `Theory of
 Finite Fermi System (FFS)' (or `Migdal Theory' \cite{migdal65}) is exploited to take 
advantage of a well-determined phenomenological effective nucleon-nucleon interaction.
 In the particle-hole channel, it consists of a local interaction whose strength is given by
 the Landau-Midgal parameter $g'_0$, augmented by the one-pion and rho-meson exchange 
re-normalised by the nuclear medium.
This approach provides an appropriate balance of the repulsive ($g'_0$) and attractive 
($f_\pi$) terms, as well as of the local and finite-range components, which appears to be of 
prime importance in $\beta$-decay studies. The continuum states are properly included in the
 $\beta$-strength calculations, leading to the so-called `continuum  QRPA (CQRPA)' approach.
 The formalism has been developed not only for the GT, but also for the first-forbidden
  transitions. 
(See \cite{krumlinde84} for an early work on the RPA description of the first-forbidden
 decay.) An earlier version of the CQRPA that uses the Skyrme-ETFSI approximation 
(Sect.~\ref{nuc_static}) had been applied to the GT $\beta$-decay half-lives of as many as 
800  near-spherical nuclei \cite{borzov00}.

A fully consistent HFB/QRPA model has been developed with the use of a Skyrme interaction
 both for the ground state and for the effective interaction \cite{engel99}.  HFB 
calculations with  `relativistic mean field (RMF)' phenomenological effective interactions 
have also been performed (RQRPA, \cite{niksic05}). Both methods have so far treated the
 GT-decay half-lives of a limited number of even-even  nuclei only.   

{\it Large-Scale Shell-Model Approach:}\ 
Through the use of a realistic (two-body) effective interaction, the standard nuclear
 shell-model tactics would  eventually, after much configuration mixing, be  capable of
 successfully describing the $\beta$-decay matrix elements or strength distributions along 
with other nuclear properties, such as the  excitation spectra. This exercise is severely 
hampered, however, by the difficulty of assuring a large-enough model configuration space,
 and is next to impossible to perform for heavy nuclei. In practice, therefore, some drastic
 truncation of the model space is often required. Just as an example, let us compare the 
descriptions of the two mid-shell nuclei $^{28}$Si and $^{56}$Ni in the full major ($sd$- and
 $fp$-) shell space. The two systems are viewed as being made of  $^{16}$O + 6 protons + 6 
neutrons and $^{40}$Ca + 8 protons + 8 neutrons, respectively. When the so-called $m$-scheme
 is adopted, the total number of uncoupled Slater determinants to be considered amounts to
 about 94,000 for \chem{28}{Si} after spin and isospin projections  ($j_z=0, T=0$), while this
 number is increased to nearly one billion for $^{56}$Ni.  With the advent of increased 
computational capabilities, nevertheless, much progress is being made with regard to 
`large-scale' shell-model calculations, as reviewed by \cite{caurier05}.

We note here that the consideration of the basis dimension of one billion does not at all
 imply that as much resolved eigenvectors have to be obtained, as made clear in particular 
by the shell model Lanczos iteration method. In this approach, the eigenvectors are 
successively generated so as to be orthogonal to the previous ones (e.g.
 \cite{caurier05}-\cite{caurier99}). As  shown explicitly for GT $\beta$-strength functions 
(e.g. \cite{caurier05,mathews83,takahashi86}), their essential characteristics are revealed 
by rapidly converging iterations. The relation between the dimensions of the adopted basis
 and the actual computational times is discussed by  \cite{caurier05} (Fig.~9 of this 
reference).

\subsubsection{Comparison between some model predictions}
\label{beta_sub4}

\begin{figure}
\center{\includegraphics[width=1.04\textwidth,height=0.87\textwidth]{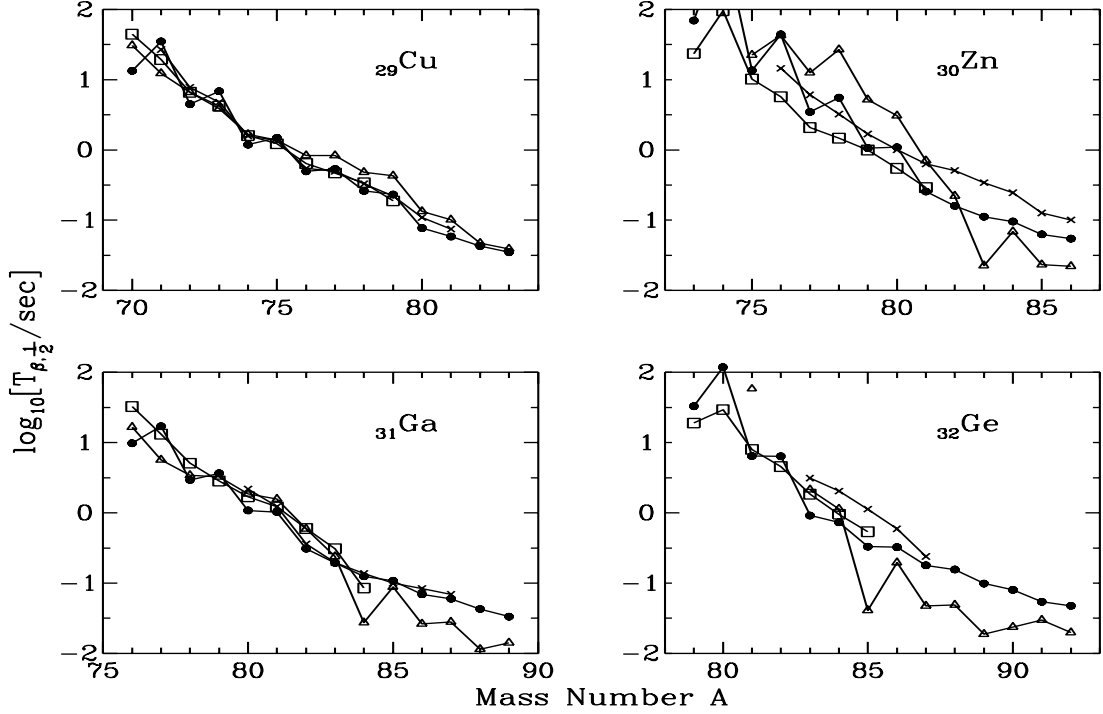}}
\vskip-2.4cm
\caption{Comparison of theoretical $\beta^-$ decay half-lives  of Cu - Ge isotopes 
and the  experimental data \cite{audi03a} (open squares). The triangles are from the FRDM + 
RPA model \cite{moeller97}, with some values obviously out of the scale. The closed circles
correspond to the original  Gross Theory \cite{takahashi73} but with the same $Q_{\beta^-}$
 as used in \cite{moeller97}. The crosses for  some isotopes are the CQRPA results 
\cite{borzov05a}. The latter two models include the contributions from both the GT and
 first-forbidden transitions} 
\label{fig:beta_cuge}
\end{figure}  

\begin{figure}
\center{\includegraphics[width=1.0\textwidth,height=1.2\textwidth]{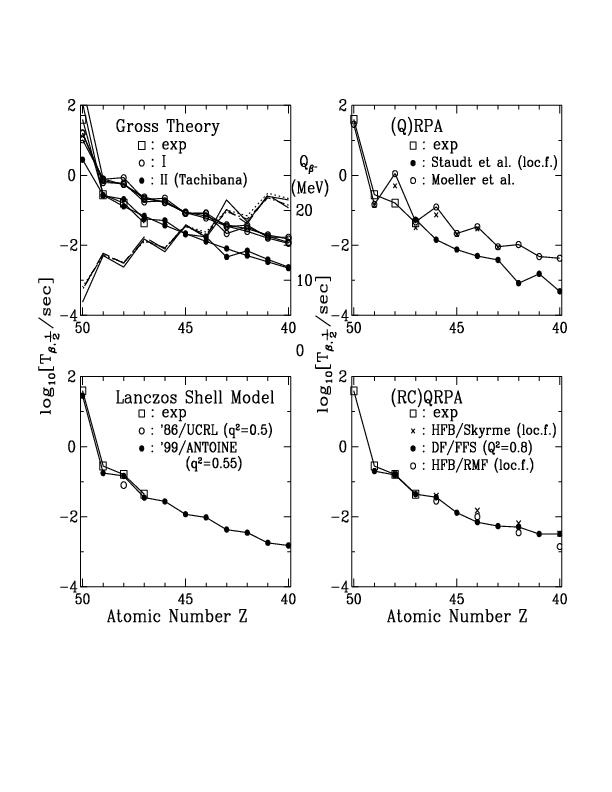}}
\vskip-3.8cm
\caption{Comparison of
 theoretical $\beta^-$ decay half-lives of $N = 82$ isotones in the Zr ($Z=40$) to Sn 
($Z=50$) range (dots) and the measured values \cite{audi03a} (open squares).
The upper two panels concern models that have been used to predict the $\beta$-decay
 properties of thousands of neutron-rich nuclei, whereas the lower panels are for more 
self-consistent models applied to a quite limited number of cases only. The label (loc.f.)
 indicates that a parameter fit has been performed locally in the chart of nuclides, and at 
$^{130}$Cd in particular.
{\it Top Left:} Predictions from the original (I) and the GT2 version (II) \cite{tachibana05}
 of the
Gross Theory with various $Q_{\beta^-}$\ values indicated by the black lines using the scale 
on the right-hand side (see text as for the used mass formula);
{\it Top Right:} Values from \cite{staudt90} (denoted Staudt et al.), and from
 \cite{moeller97} (denoted Moeller et al.) with the modifications from \cite{moeller03} 
(crosses);
{\it Bottom Right:} Predictions from the self-consistent HFB/Skyrme (QRPA) of \cite{engel99},
 the DF/FFS (CQRPA) of \cite{borzov05} with the adoption of the overall quenching factor
 $Q^2 = 0.8$ and the HFB/RMF (RQRPA) of \cite{niksic05};
{\it Bottom Left:} A test calculation for $^{130}$Cd \cite{takahashi86} (denoted 86/UCRL),
 and the more systematic results for the $N = 82$ isotones  \cite{martinez99} 
(noted 99/ANTOINE). The indicated values of $q^2$ refer to the shell-model GT quenching 
factor (see text)}
\label{fig:beta_n82}
\end{figure}

Some limited comparisons between experimental and calculated half-lives and between the 
predictions of different models are presented in Figs.~\ref{fig:beta_cuge} and 
\ref{fig:beta_n82}.
Figure~\ref{fig:beta_cuge} puts the experimental $\beta^-$-decay half-lives of the isotopes 
of the elements Cu to Ge near the neutron magic number $N=50$ in perspective with the 
predictions of three models ranging from macroscopic to self-consistent microscopic types 
(note that none of the calculated values results from a fit to those measurements). It 
appears in particular that the FRDM+RPA approximation predicts by far a too large odd-even 
staggering. This is corrected by the CQRPA model.   

Figure~\ref{fig:beta_n82} compares the $\beta^-$ decay half-lives of the $N = 82$ 
neutron-magic isotones  from various models, and displays available measured values.  Let us
 briefly discuss the data:\\  
(1) {\it Top left panel:} The results of the original Gross Theory (I) are obtained with the
 $Q_{\beta^-}$\ values from HFB9 (Sect.~\ref{nuc_static}), the semi-empirical DM-based mass
 formula of \cite{hilf76},  the FRDM table of \cite{moeller97} (supplementing known 
experimental data), and the mass formula of \cite{duflo95}. A superior agreement to 
measurements (accuracy) is reached by the GT2 version (with the use of the first two mass 
formulae) \cite{tachibana05}.  These calculations include the contributions of  the allowed 
 (essentially GT)  and first-forbidden  transitions;\\
(2) {\it Top Right:} The QRPA results \cite{staudt90} with $Q_{\beta^-}$ from the
 droplet-model masses of \cite{hilf76} and a local parameter  fit of the GT force to the 
measurements are compared with the results of a global fit based on the FRDM approximation 
\cite{moeller97}. An attempt to include the first-forbidden transitions into the latter model
 is made, but with the use of the Gross Theory \cite{takahashi71}. Even though
 the merits of such  a `hybrid' model \cite{moeller03}  are hard to decipher, the
  predictions are added (crosses) for the sake of comparison;\\
(3) {\it Bottom Right:} The good results of the CQRPA model (labelled DF/FFS) \cite{borzov05} 
are  obtained, as in  Fig.~\ref{fig:beta_cuge}, with global parameter values. The GT quenching 
factor $Q$ is universal in the sense of the FFS theory (i.e. it does not depend on the model
 space, such that the the GT coupling constant $G_{\rm A}$ is re-normalised to 
$G_{\rm A}Q$), and is set to  $Q^2 = 0.8$. The first-forbidden transitions are included
 in a consistent way. The GT decay rates predicted for a few even-even nuclei by a 
Skyrme-based HFB plus QRPA model \cite{engel99}, and by a HFB plus RQRPA/RMF model 
\cite{niksic05} are also displayed. The strength of the $T=0$ pairing (particle-particle
 force) has been locally, and sometimes wildly, varied so as to reproduce the observed 
half-life of $^{130}$Cd used as a renormalisation point;\\
(4) {\it Bottom Left:} The Lanczos shell-model calculations \cite{martinez99} with the
 ANTOINE code \cite{caurier05} succeed in reproducing to a satisfactory level the general
 trend and absolute magnitudes of the known half-lives if the shell-model quenching factor 
(see below) $q^2 = 0.55$ is adopted. A former calculation performed  by \cite{takahashi86}
 for $^{130}$Cd with the UCRL code \cite{hausman76} in a very limited model space  and with 
a small number of iterations is inserted to illustrate the progress made in shell model 
calculations over the last two decades thanks to enhanced computational capabilities.
 
 Let us now try to evaluate the merits of the models discussed above, particularly in 
predicting the $\beta$-decay properties of unknown nuclei.  In a global sense, the Gross 
Theory, especially in its GT2 version, appears to be reasonably accurate. This is in 
particular so near closed-shells  where, normally, microscopic approaches are thought to be
 superior  (e.g. \cite{tachibana95}). The success of the Gross Theory-type of models is 
especially remarkable in view of their simplicity. The ease with which $\beta$-decay 
probabilities can be re-computed,  in particular with different $Q_{\beta^-}$ values and 
concomitant new parameter fits, is clearly an additional very pleasing feature.  
 
As already remarked in relation with Figs.~\ref{fig:beta_cuge} and \ref{fig:beta_n82}, 
the oft-used QRPA-type predictions based on the FRDM approximation \cite{moeller97} seem 
to suffer from a too strong even-odd staggering. It is speculated \cite{borzov05} that this
 problem may result from an asymmetric treatment of pairing in the ground and excited states.
 In fact, the $\beta$-decay rates from that model seem to be too closely correlated to the 
even-odd effects in the adopted $Q_{\beta^-}$ values. Another QRPA $\beta$-decay evaluation
 \cite{staudt90} makes use of local adjustments of the force parameter values. This practice
may deserve some criticism, as it makes long-range extrapolations especially unreliable.  
It now remains to be seen whether the CQRPA model with global parameter values 
\cite{borzov05} can successfully reproduce the $\beta$-decay data for a large body of nuclei,
 including well-deformed ones.

As stated in \cite{martinez99}, systematic large-scale shell model calculations may suffer not
 only from computational limitations, but more fundamentally from the lack of the
spectroscopic information that is needed for the proper construction of effective
 nucleon-nucleon interactions. The catch is that the experimental determination of  
$\beta$-decay half-lives has a chance to precede the acquisition of such a spectroscopic 
information! Many models assume some level of GT quenching. In the truncated shell-model, 
the parameter $q^2$ reflects the GT strength that is expected in a certain energy range for
 a given interaction and limited configuration space. It is far from being obvious, and it is
 even doubtful, that a single value of  the quenching factor really applies to all of the
  individual transitions to low-lying states. A universal quenching factor $Q^2$ in relation
 to the Ikeda sum rule can be justified only when the very remote continuum  energy region 
is considered, as in the full-basis CQRPA, or in the so-called no-core shell model (so far
 applicable to very light nuclei only).
 
Some remarks are in order at this point. First,  the adoption of $Q_{\beta^-}$ values from
 mass predictions raises some problems in the RPA framework. In fact, if a fully
 self-consistent treatment is not applied to the parent and daughter nuclei, 
to choose an appropriate $Q_{\beta^-}$ that is not in conflict  with the energetics dictated by 
the RPA is not a trivial matter, as explicitly discussed by \cite{engel99} in a rare 
occasion.  In principle, this difficulty concerning the proper choice of $Q_{\beta^-}$ value 
 is avoided in shell model calculations where both initial and final nuclear states are 
diagonalised. Unfortunately, the current accuracy of the shell-model  predictions of
 $Q_{\beta^-}$ is insufficient for the half-life computations, as noted by \cite{martinez99}.
 For this reason, the predictions from a mass model \cite{duflo95} are used in \cite{martinez99}.
Second, the practice of introducing local adjustments of the models parameters is not the 
monopoly of some QRPA calculations (see above). The local empirical renormalisation technique
 is also adopted in self-consistent models \cite{engel99,niksic05}.  In fact, this procedure
 is especially regrettable in this framework. With the additional quenching corrections that
 are applied as well, one has to acknowledge that  the seemingly  high-minded philosophy that
 is claimed to motivate those models is clearly betrayed.

In conclusion of this section, one may acknowledge that the macroscopic Gross Theory models
 with global parameter values perform remarkably well in their accuracy to reproduce 
experimental data. They also have the important advantage of providing with very limited
 computing efforts all the $\beta$-decay data that are needed in the modelling of the 
r-process, including the contributions of both the allowed and first-forbidden transitions.
 The importance of this has not to be underestimated. Of course, as it is the case for all 
the approximations of the macroscopic type, one may wonder about the reliability of the 
predictions very far from the valley of nuclear stability. As claimed in relation with the
 calculation of nuclear masses (Sect.~\ref{nuc_static}), the reliability of more microscopic
 models may be expected to be higher. It remains to be seen, however, if this statement 
applies in the field of $\beta$-decay studies. In contrast to the situation encountered in
 mass predictions, the level of the accuracy of the currently available microscopic models
 is still far from being satisfactory. This problem is sometimes blurred by local adjustments
 of parameters, which makes the evaluation of the merits of the models difficult, and a global
 comparison between models highly risky. In addition, microscopic models are still a very long
 way from producing the $\beta$-decay half-life data needed in the r-process studies, 
including the due consideration of allowed and first-forbidden transitions and the proper 
inclusion of nuclear deformation. Awaiting better times, the macroscopic models have currently
 to be considered as a good choice. We stress, however, that this is in no way meant to
 discourage further (and wishfully more concerted) effort to develop microscopic descriptions
 of the $\beta$-decay of exotic nuclei.
 
\subsubsection{The $\beta$-decays of neutron-rich nuclei at high temperatures}
\label{beta_annex}

So far, we have been concerned with the $\beta$-decays of neutron-rich nuclei in their 
ground-states. If a high-temperature environment is envisioned for the r-process, significant
 deviations  from the terrestrial $\beta$-decay rates might result from the decays of the 
thermally-populated excited states \cite{AK99}. If the ground and excited states are in 
thermal equilibrium at temperature $T$, the `astrophysical' $\beta$-decay rate including 
these contributions is given by

\begin{equation}
 \lambda_{\beta}^*\ =\ \frac{\sum_\mu\ G_\mu\ \lambda_{\beta,\mu}}{\sum_\mu\ G_\mu}\ \ \
{\rm with}\ \ \ G_\mu\ =\ (2J_\mu+1)\ {\rm exp}(\ -\ \frac{\varepsilon_\mu^*}{kT}),
\label{eq:beta_excite}
\end{equation}

where $\lambda_{\beta,\mu}$ is the $\beta$-decay rate from the $\mu$-th excited state 
(with $\mu = 0$ referring to the ground state),  $k$ is the Boltzmann constant, and $G_\mu$ 
is the equilibrium population of state $\mu$ with total spin $J_\mu$ and excitation energy
 $\varepsilon_\mu^*$.

\begin{figure}
\center{\includegraphics[width=0.8\textwidth,height=0.65\textwidth]{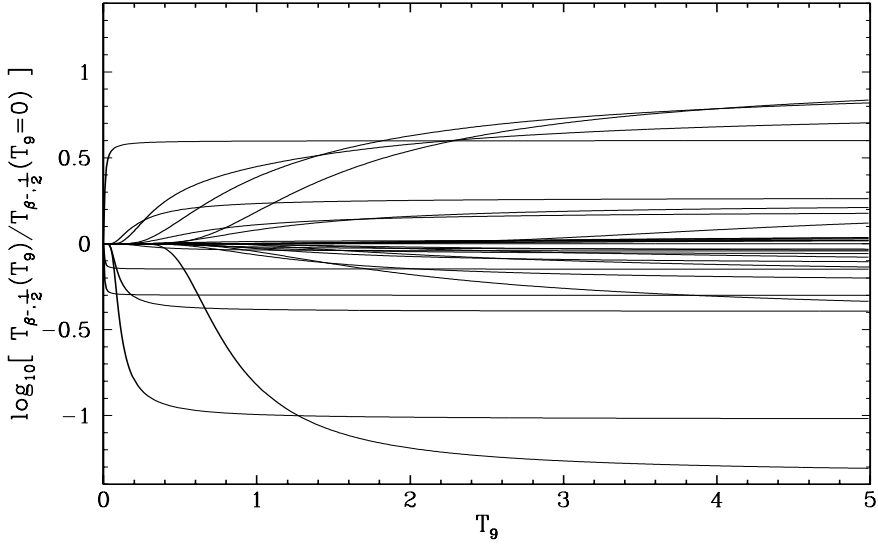}}
\vskip-1.7cm
\caption{Crudely estimated (see text) temperature-dependent $\beta^-$-decay half-lives 
$T_{\beta^-,\frac{1}{2}f}(T_9)$ normalised to their ground state values 
$T_{\beta^-,\frac{1}{2}1/2}(T_9=0) < 1$ sec, where $T_9 = 10^9$ K. The only considered 
 excited states are the known long-lived isomers \cite{audi03a}}
\label{fig:beta_excite}
\end{figure}  

Figure~\ref{fig:beta_excite} illustrates the possible effects of temperature on the
 half-lives of some heavy neutron-rich nuclei with experimentally known (ground-state)
 half-lives shorter than 1 sec. For simplicity, only known long-lived isomeric states 
(one or two per case) are included in the calculation of Eq.~\ref{eq:beta_excite} under
 the assumption that they are in thermal equilibrium with the ground state (which may not 
be true when dealing with isomeric states in certain temperature regimes). As a result,
  it may well be that the temperature effects are much exaggerated. 
(This is just acceptable for the mere purpose of illustration.)  Qualitatively speaking, one
 may identify the following four situations possibly leading to extreme temperature effects:
 (1) a fast-decaying high-spin isomer, (2) a slow-decaying high-spin isomer, (3)
 a fast-decaying low-spin isomer, and (4) a slow-decaying low-spin isomer. These cases lead
respectively to a much-enhanced, a reduced, an enhanced and a nearly-unchanged decay rate.

Equation~\ref{eq:beta_excite} makes clear that $\lambda_{\beta}^*  = \lambda_{\beta}$ 
if $\lambda_{\beta,i} =  \lambda_{\beta} (\equiv \lambda_{\beta,0}$) for all $i$. 
This is indeed nearly so in the Gross Theory, which averages the $\beta$-strength 
distributions over all spins and parities. The influence of the increased energy windows for
 the transitions from excited states is expected to be rather weak since $Q_{\beta^-} \gg kT$ 
in general.    
To evaluate $\lambda_{\beta}^*/\lambda_{\beta}$ is a very hard task for microscopic models 
(e.g. \cite{martinez99}). One may wishfully expect, however,  that the temperature effects
 on the $\beta$-decays of very neutron-rich nuclei is not as wild as those expected  in some
 cases near the line of stability \cite{takahashi87} even at temperatures much lower than 
those considered in some r-process models. When dealing with exotic nuclei, the $Q_{\beta^-}$ 
values are high enough, such that (i) it is likely that potentially strong $\beta$-transitions 
from any given initial states do find appropriate final states, and that (ii) ionisation effects,
which may be dramatic when very low $Q_{\beta^-}$ values are involved \cite{takahashi87},
are negligible.

In passing we note that $\beta$-decay is not an adiabatic process, so that their effect on
the energetics of a phenomenon, and in particular on the thermodynamics of r-process sites,
 has in principle to be taken into account  \cite{sato74,hillebrandt76}.  

\subsection{Fission barriers and spontaneous fission probabilities}
\label{fission}

Any mass model that gives the binding energy of a nucleus as a function of its deformation can
 be applied, in principle,  to the evaluation of fission barriers. In practice, only a few sets of  
calculations of the  barriers of very neutron-rich nuclei that enter the r-process modelling have
 been published so far.  
One of them relies on a droplet-type model \cite{hm80}.  Recently, however, a doubt has been
 cast on fission barrier estimates based on such macroscopic-microscopic approaches
 \cite{msi04}. Another set of calculations relies on the ETFSI high-speed approximation
 to the Skyrme-HF method, with pairing handled in the BCS approximation. ETFSI has been
 developed initially for the calculation of a complete set of masses
 (Sect.~\ref{nuc_static}).  Its extension to the evaluation of fission barriers is 
made by \cite{mprt98,mprt01}. 
Some striking differences are observed between the ETFSI and droplet model predictions for
 the fission barriers of very neutron-rich nuclei, especially in the vicinity of the 
$N=$ 184 magic number. Also note that  \cite{ms99} have proposed a fission barrier formula
 relying on the main trends of a zeroth-order Thomas-Fermi approximation, complemented with
 a prescription for shell corrections. Broadly speaking, the barriers derived in such a way
 lie much closer to those of  \cite{hm80} than to those of \cite{mprt01}.

\begin{figure}
\center{\includegraphics[scale=0.4,angle=90]{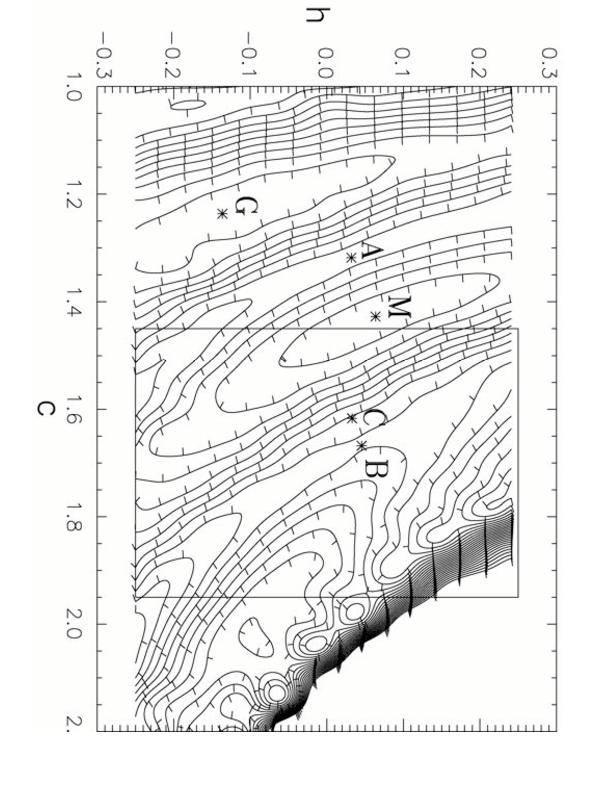}}
\vskip-0.4cm
\caption{ $^{240}$Pu energy surface in the $(c,h)_{\alpha=0}$ plane. The contour lines are
spaced by 1~MeV.   Small tick marks along each contour point in the downhill direction. 
G indicates the ground state location, M the shape isomer, and A, B the reflexion-symmetric 
inner and outer saddle-points, respectively.  The square containing B corresponds to the
 $(c,h)$ plane considered for the computation of 
the local $\alpha\neq 0$ 3D energy-surface with its saddle point C}
\label{fig_fispu240}
\end{figure}

\begin{figure}
\center{\includegraphics[scale=0.45]{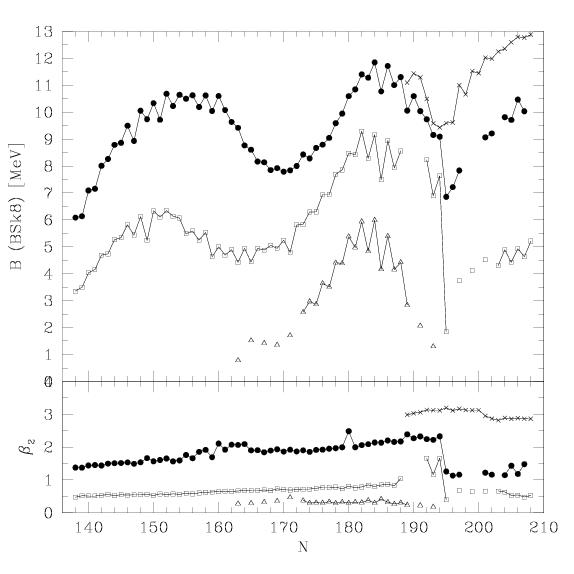}}
\caption{Calculated axial and reflexion symmetric barrier heights for each Pu isotope
 lying between the valley of stability up to the neutron-drip line (upper panel), and
their corresponding dimensionless deformation $\beta_2$ (lower panel). See text for more details}
\label{fig_fispu240e} 
\end{figure}

Pairing correlations and shell effects clearly play a crucial role in the determination of 
fission barriers. This is why  microscopic self-consistent models are required for the estimates of 
the fission properties of exotic neutron-rich nuclei that can be involved in the r-process. 
The  HFB method
corrected for the restoration of broken symmetries  has been used recently to predict not only
  nuclear masses (Sect.~\ref{nuc_static}), but also fission barriers 
\cite{sam05,sam05b}. The calculations are based on the BSk8  Skyrme interaction constructed 
to predict masses to a very satisfactory accuracy (Sect.~\ref{nuc_static}).   

The fission path is conveniently followed in a HFB energy surface in a deformation space 
that is characterised by the parameters  $(c,\alpha,h),$ 
 where $c$  relates to the elongation, $h$ to the necking, and $\alpha$ to the left-right
 asymmetry \cite{bd72}. 
With this parametrisation it is possible to generate from an initial spherical configuration a 
continuous sequence of axially-symmetric deformed shapes of a given nucleus up to and beyond 
the break-up into two separated fragments \cite{sam05}. Such a procedure is illustrated  in
 Fig.~\ref{fig_fispu240}, which displays the reflexion-symmetric energy surface for
 $^{240}$Pu in the left-right symmetric $(c,h)_{\alpha=0}$ plane.   The $\alpha = 0$ fission
 barriers of the Pu isotopes  are shown in Fig.~\ref{fig_fispu240e}.  For each fission 
barrier,  the corresponding dimensionless elongation parameter 
$\beta_2= \sqrt{5\pi}Q_2/(3AR_0^2)$ is also displayed ($Q_2$ is the quadrupole 
moment in fm$^2$ with a choice of the nuclear radius 
 $R_0=1.2 A^{1/3}$ fm). As seen in  Fig.~\ref{fig_fispu240e}, the 
topology of the HFB fission barriers is rather different from the one traditionally predicted
by the macroscopic-microscopic approach (e.g \cite{hm80}). A significant shell effect 
appears around $N=150$ and $N = 184$ for the inner (lower $\beta_2$) as well as outer (higher $\beta_2$) 
barriers. A third, 
low (up to 6~MeV) inner barrier appears before reaching the $N=184$ shell-closure, while for 
$N>184$, a new and rather high (up to 13~MeV) outer barrier at deformations of about 
$\beta_2=3$ starts to dominate. Note, however, that these results are obtained assuming the 
reflexion symmetry. 

\begin{figure}
\center{\includegraphics[scale=0.5]{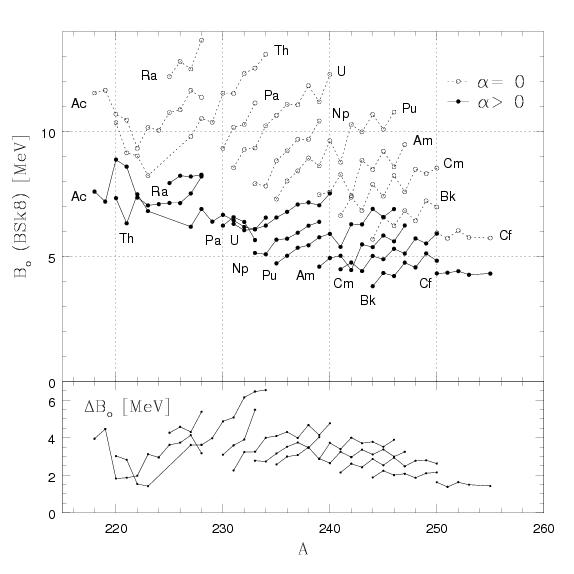}}
\vskip-0.5cm
\caption{Upper panel: Calculated outer barrier heights $B_o$ with or without the inclusion of
 the reflexion asymmetry; lower panel: difference between the reflexion-symmetric and 
asymmetric outer barriers $B_o(\alpha=0)-B_o(\alpha > 0)$. The calculations are based on the
 HFB-8 model}
\label{fig_fisasym}
\end{figure}
 
The predicted fission barriers remain very uncertain in general, and for exotic neutron-rich
 nuclei in particular. One source of uncertainties relates to $\alpha$, the value of which  
 is known to influence the estimated outer barrier heights. This is illustrated in
 Fig.~\ref{fig_fisasym}, where the outer barrier of the $88\leq Z \leq 98$ nuclei is seen to
 be lowered by a few MeV when  a possible asymmetric shape (i.e $\alpha >0$) is considered.
The topology of the energy surface illustrated for $^{240}$Pu in  Fig.~\ref{fig_fispu240} also 
depends on the effective nucleon-nucleus and pairing interactions.  In contrast to the simple
 double-humped picture provided by the macroscopic-microscopic models, a complex path in the 
deformation plane is found, and  often exhibits three or even four barriers.  Another 
illustration of the uncertainties in the barrier predictions is given in Table~\ref{tab_fis}. 
 For some neutron-rich nuclei near the $N=184$ shell-closure, it compares  the primary (i.e.
 the highest) fission barriers calculated with the HFB-8 model \cite{gor04,sam05},  the ETFSI
 approximation \cite{mprt01}, and the microscopic-macroscopic approaches of  \cite{ms99} and 
 \cite{hm80}. Significant differences are seen between model predictions. In particular, the
 low barriers at $N=184$ found by \cite{hm80,ms99} are not supported by the mean-field models.
 One  should acknowledge, however,  that none of the existing large-scale calculations of
 fission barriers includes triaxiality as a shape degree of freedom, although it may play a
 role in reducing the barrier heights \cite{dutta00,bqs04}. 
 
\begin{table*}
\caption{Primary (i.e. highest) fission barriers of some neutron-rich
nuclei near the $N=184$ shell-closure calculated with the HFB-8 (BSK8) model \cite{sam05} 
and with the ETFSI approximation \cite{mprt01}.  
 The reflexion symmetry is assumed in HFB-8, but not for ETFSI.  The asymmetry
 parameter $\tilde{\alpha}=\alpha~c^3$  is given if non zero. The symbols (i) and (o) refer 
to the inner and outer barriers. The earlier predictions of \cite{ms99} (MS) and \cite{hm80} (HM) are added for comparison}
\vskip 0.3cm
\begin{center}
\begin{tabular}{cccccc}
\hline
 Z & N & BSk8 & ETFSI & MS & HM \\ 
\hline
       84 &        170 &          28.27 {\rm (o)} &         26.85 {\rm (o; ${\tilde{\alpha}=.45}$)} &        15.40 &        6.84   \\
        84 &        184 &          37.14 {\rm (o)} &         39.01 {\rm (o)} &        20.08 &        6.21  \\
        92 &        170 &          11.48 {\rm (o)} &          5.25 {\rm (o; ${\tilde{\alpha}=.47}$)} &        4.55 &        3.36 \\
        92 &        184 &          16.77 {\rm (o)} &         17.67 {\rm (o)} &        7.12 &        3.80  \\
        92 &        194 &          15.39 {\rm (o)} &         10.89 {\rm (o)} &        5.06 &        - \\
       100 &        170 &           4.24 {\rm (o)} &          2.18 {\rm (i)} &        1.72 &        2.61  \\
       100 &        184 &           5.34 {\rm (o)} &          5.97 {\rm (i)} &        1.86 &        2.58 \\
       100 &        194 &           3.24 {\rm (o)} &          1.63 {\rm (i)} &        1.36 &        - \\
       100 &        210 &           6.42 {\rm (o)} &          7.30 {\rm (i)} &        3.59 &        - \\
\hline
\end{tabular}
\end{center}
\label{tab_fis}
\end{table*}

\begin{figure}
\begin{center}
\includegraphics[scale=0.6]{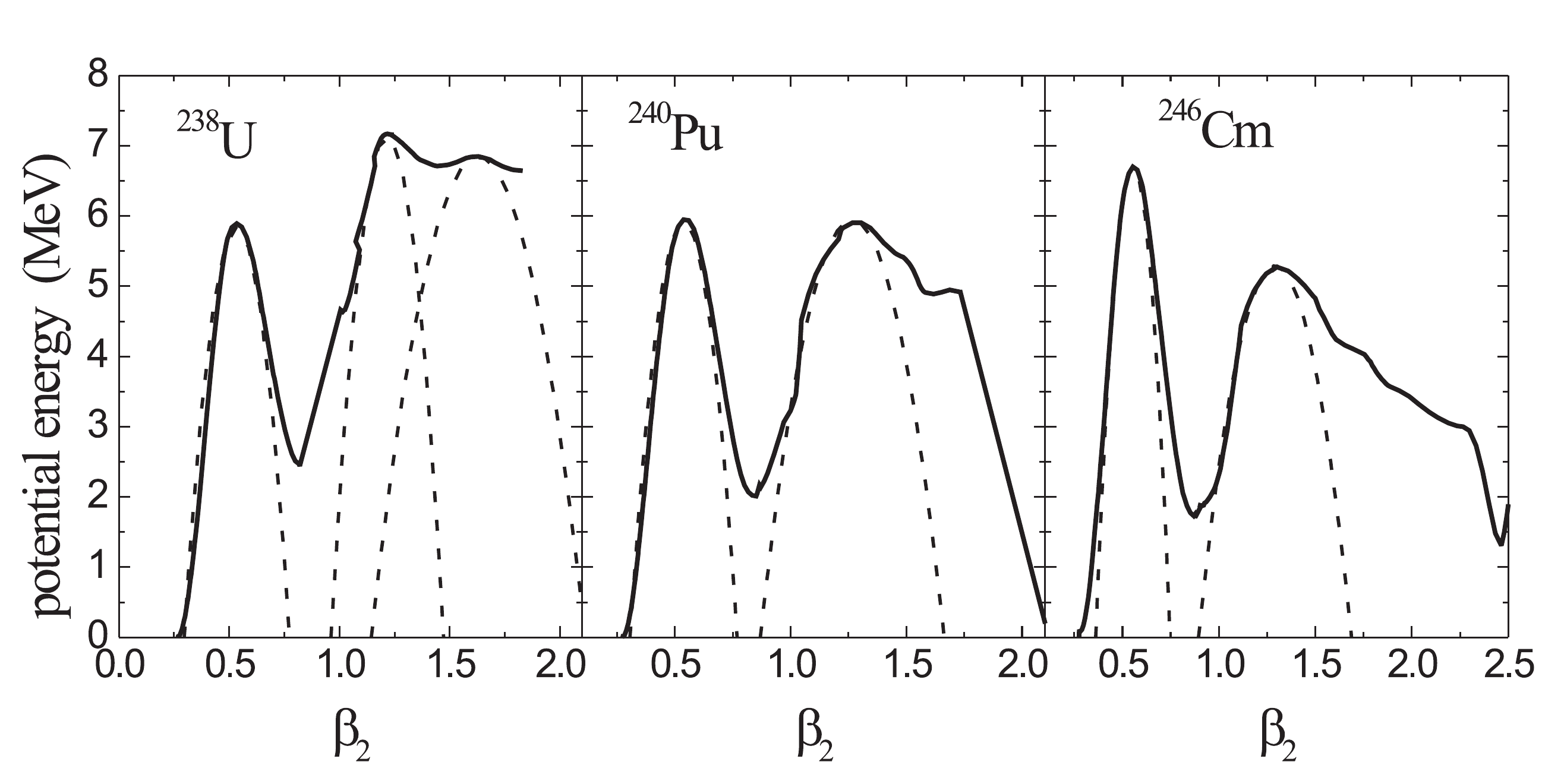}
\caption{HFB-8 static fission paths (black lines) for $^{238}$U, $^{240}$Pu and $^{246}$Cm. 
The dashed lines correspond to the inverted parabola fitted to the inner and outer barriers}
\label{fig_fispath}
\end{center}
\end{figure}

\begin{figure}
\begin{center}
\includegraphics[scale=0.4]{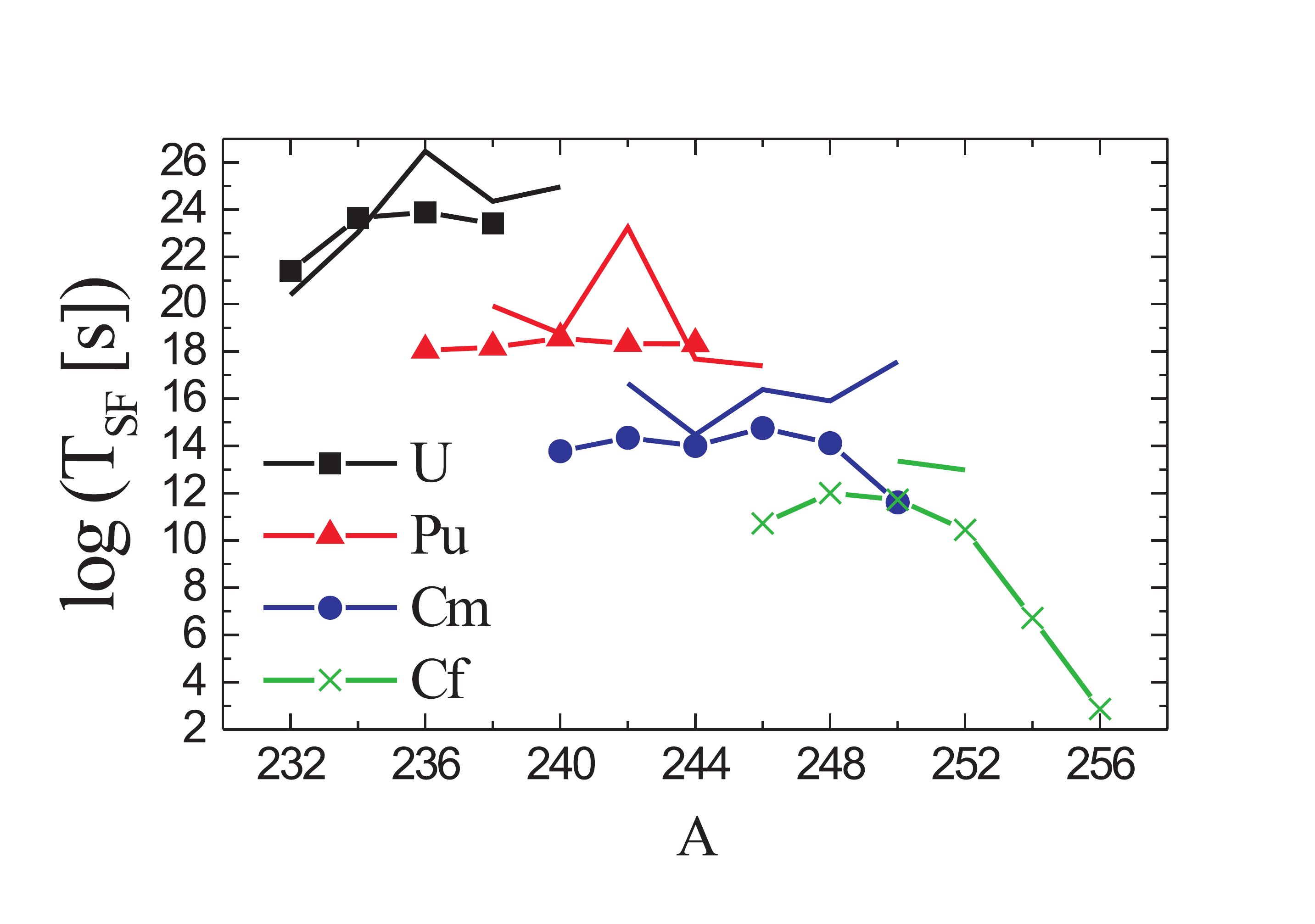}
\vskip-0.8cm
\caption{Comparison between experimental (symbols) \cite{fires96} and theoretical
 (solid lines) T$_{SF}$ of even-even nuclei  obtained with the HFB barriers and widths}
\label{fig_sfis}
\end{center}
\end{figure}

In summary, microscopic models predict energy surfaces that significantly differ from the classical
 double-humped barriers such as suggested  by droplet-type models. The determination within
 a microscopic framework of  the fission barriers along with the full fission paths for all 
the exotic nuclei of relevance to the r-process remains as a considerable challenge that is 
far from being met at the present time.

Another major challenge concerns the calculation of fission half-lives. These are classically
 evaluated from the Hill-Wheeler formulation \cite{bd72}, which is obtained by approximating
 the barrier shapes by  inverted parabolas. In this approach, a barrier is assigned a height 
$B$ and a width $\hbar\omega$ which is related to the curvature of the parabola at the top of 
the barrier.   
The spontaneous fission half-lives $T_{\rm SF}$ evaluated in the Hill-Wheeler framework are
 extremely sensitive to uncertain $B$ and $\hbar\omega$ values for their exponential 
dependence on these quantities. For instance, an uncertainty of approximately 1 MeV on $B$ 
is responsible for an uncertainty  of about four orders of magnitude on $T_{\rm SF}$  if
 $\hbar\omega$ = 0.6 MeV.  Further uncertainties in $T_{\rm SF}$ arise from the fact that the
 Hill-Wheeler approximation fails in general dismally for the second barriers, as illustrated 
in Fig.~\ref{fig_fispath}. Last but not least,  complications arise from the presence of a
 third barrier for some nuclei at large deformations, as it is the case for $^{238}$U. Quite 
clearly, a reliable and accurate evaluation of $T_{\rm SF}$ still remains close to impossible.

The current  (in)capacity to  provide a satisfactory global prediction of $T_{\rm SF}$  is illustrated
 in Fig.~\ref{fig_sfis}, where experimental data for some even-even actinides are compared with
 predictions based on the approximation of HFB-8 barriers  by an inverted parabola
 \cite{dem05}. For $^{236,238}$U, the deviations between theory and experiment do not exceed
one or two  orders of magnitude, and the consideration of the third barrier (Fig.~\ref{fig_fispath}) 
leads to a good agreement between experimental and  predicted $T_{\rm SF}$.  
For $^{242}$Pu and $^{250}$Cm, the discrepancies amount to several orders of magnitude.

\subsection{Nuclear reaction rates}
\label{reac}

A large variety of nuclear reactions come into play in the r-process, their 
precise nature and number depending on little-known astrophysical conditions in which the 
process may develop. There is not the slightest doubt that  neutron captures always play a 
leading role. They are most likely accompanied by photo-reactions of the $(\gamma$,n) type.
 Some models also call for the operation of proton or $\alpha$-particle captures and 
associated photo-disintegrations.

In spite of a concerted effort devoted in the last decades to measurements of 
reaction cross sections for astrophysical purposes (e.g. \cite{AK99}),  experiments on all 
but some of the very exotic neutron-rich nuclei involved in the r-process will remain 
unfeasible for a long time to come. Theory has thus to supply the necessary data, which also
 represents a major challenge. Concomitantly, specific stellar plasma effects come into play,
 like the contribution of target excited levels to the reaction mechanisms, which develops 
at high temperatures, and can become significant in r-process conditions.

\subsubsection{Reaction rate calculations: general framework of a statistical approach}
\label{th_rates-general}

So far, all r-process calculations have made use of nuclear reaction rates evaluated
within the Hauser-Feshbach statistical model \cite{hau52}.  It relies on the fundamental 
assumption (Bohr hypothesis) that the capture process takes place through the
intermediary production of a compound system that can reach a state of thermodynamic 
equilibrium. This compound system is then classically referred to as the compound nucleus
 (CN). The formation of  CN is usually justified if its level density at the excitation 
energy corresponding to the projectile incident energy is large enough. If so, the reaction
$I^{\mu} + j\rightarrow L + l$ of capture of a nucleon or $\alpha$-particle ($j$) on target  $I$
in its state $\mu$ leaving the residual nucleus $L$ and particle or photon $l$ has a cross
section at centre-of-mass energy $E$ given by 

\begin{equation}
\sigma_{jl}^{\mu} (E) =
     \pi \lambdabar^2_j \frac{1}{(2J_I^{\mu}+1)(2J_j+1)} \sum_{J^\pi} (2J+1)
     \frac{T_{j}^{\mu}(J^\pi) T_{l}(J^\pi)}{T_{{\rm tot}}(J^\pi)}\,,
\label{eq_sig}
\end{equation}

where $J_I$ and $J_j$ are the target and projectile spins, and $T_j(J^\pi)$ is the
transmission coefficient measuring the probability for forming the CN in its state $J^{\pi}$ 
obtained from all possible combinations of the orbital and channel
spins. Similarly, $T_l(J^{\pi})=\sum_{\nu} T_l^{\nu}(J^{\pi})$ is the transmission coefficient
for the decay of the compound nuclear state into the pair $L+l$, all states $\nu$ of $L$ which
can be populated in the reaction being taken into account.
$T_{\rm tot}(J^{\pi})=\sum_{i,\lambda}T^{\lambda}_i(J^{\pi})$ is the total
transmission coefficient for the decay of the compound state $J^{\pi}$ into any
combination $i$ of nucleus and particle which can be formed from all its possible decay
modes $\lambda$ (including $I+j$ and $L+l$).  Note the hypothesis of an equilibrium CN 
underlying Eq.~\ref{eq_sig} implies that its formation and decay  are 
independent except for  the basic requirements of conservation of energy, and of the relevant 
quantum numbers.  This may not be fully satisfied, particularly in cases where a few
strongly and many weakly absorbing channels are mixed. As an example, Eq.~\ref{eq_sig} is
known to fail when applied to the elastic channel for which the
transmission coefficients for the entrance and exit channels are identical, and hence
correlated. To account for these deviations, width
fluctuation corrections can be introduced in the Hauser-Feshbach formalism by different
 approximate expressions \cite{tepel74,hofman80}. 
 
Each transmission coefficient is estimated from the sum  over all levels with
experimentally-known energies, spins and parities. At excitation energies for which the required
data are not available, this sum is replaced by an integral 
by folding with a nuclear level density, $\rho$,   so that

\begin{equation}
T_i(J^\pi) = \sum_{\nu=0}^{\omega}\, T_i^{\nu}(J^\pi) +
\int_{\varepsilon^{\omega}}^{\varepsilon^{\rm max}} \int_{J^{\nu},\pi^{\nu}}
T_i^{\nu}(\varepsilon^{\nu},J^{\pi})~\rho({\varepsilon^{\nu},J^{\nu},\pi^{\nu}})
~d\varepsilon^{\nu}d\pi^{\nu}dJ^{\nu}\,,
\label{eq_trans}
\end{equation}

where $\varepsilon^{\omega}$ is the energy of the highest experimentally known 
bound excited state $\omega$ (or more precisely the state up to which the knowledge of
the energy spectrum is considered to be reasonably complete), and
$\rho({\varepsilon^{\nu},J^{\nu},\pi^{\nu}})$ is the density per unit energy interval of $L$
states with spin $J^{\nu}$ and parity $\pi^{\nu}$ at the excitation energy
$\varepsilon^{\nu}$. Similar formulae apply to the other transmission coefficients in
Eq.~\ref{eq_sig}.

A thermodynamic equilibrium holds locally to a very good approximation in stellar interiors
(e.g. \cite{cox68}). Consequently, the energies of both the targets and projectiles, as well
as their relative energies $E$, obey Maxwell-Boltzmann distributions corresponding to the
temperature $T$ at that location. In such conditions, the rate of $I^{\mu} + j\rightarrow L +
l$ per pair of particles in the entrance channel and at $T$ is obtained by integrating the
cross section given by Eq.~\ref{eq_sig} over a Maxwell-Boltzmann distribution of energies
$E$ at the given temperature. In addition, in hot astrophysical plasmas, a target nucleus
exists in its ground as well as excited states. In a thermodynamic equilibrium situation, 
the relative populations of the various levels of nucleus $I^\mu$ with excitation energies
$\varepsilon^\mu_I$ ($\mu = 0$ for the ground state) obey a Maxwell-Boltzmann distribution.
The effective stellar rate of  $I + j\rightarrow L + l$ per pair of particles in the entrance
channel at temperature $T$ taking due account of the contributions of the various target
excited states is thus expressed in a classical notation 
(in cm$^3$~s$^{-1}$~mole$^{-1}$) as

\begin{equation}
 N_{\rm A}\langle\sigma v\rangle^*_{jl}(T) = \Bigl( \frac{8}{\pi m} \Bigr) ^{1/2}
                  \frac{N_{\rm A}}{(kT)^{3/2}~G_I(T)}
                  \int_{0}^{\infty} \sum_{\mu} \frac{(2J_I^{\mu}+1)}{(2J_I^{0}+1)}
                  \sigma_{jl}^{\mu}(E)E\exp \Bigl(-\frac{E+\varepsilon^{\mu}_I}{kT}\Bigr) dE,
\label{eq_rate}
\end{equation}

where $k$ is the Boltzmann constant,
$m$ the reduced mass of the $I^0 + j$ system, $N_{\rm A}$ the Avogadro number, and
$G_I(T) =\sum_{\mu} {(2J_I^{\mu}+1)}/{(2J_I^{0}+1)}\exp(-\varepsilon^{\mu}_I/kT)$ the 
temperature-dependent normalised partition function. Reverse reactions can also be estimated
with the  use of the reciprocity theorem \cite{hol76}. In particular, the stellar
photo-dissociation rates (in s$^{-1}$) 
are classically derived from the reverse radiative capture rates by

\begin{equation}
\lambda_{(\gamma,j)}^*(T) =
\frac{(2J_I^{0}+1)(2J_j+1)}{(2J_L^{0}+1)}~\frac{G_I(T)}{G_L(T)}
 \Bigl( \frac{m kT}{2 \pi \hbar^2} \Bigr)^{3/2} \langle\sigma v\rangle^*_{(j,\gamma)}
~ {\rm e}^{-Q_{j\gamma} /kT},
\label{eq_inverserate}
\end{equation}

where $Q_{j\gamma}$ is the Q-value of the $I^0(j,\gamma)L^0$ capture reaction.
 Note that, in stellar conditions, the reaction rates for targets in thermal equilibrium 
obey reciprocity since the forward and reverse channels are symmetrical, in contrast to the
situation which would be encountered for targets in their ground states only \cite{hol76}. 

For nuclei relatively close to the valley of $\beta$-stability characterised by a relatively 
large nuclear level density at the neutron separation energy, the uncertainties involved in 
any Hauser-Feshbach cross-section calculation are not related to the model of formation and 
de-excitation of the compound nucleus itself (except through the width fluctuation
 correction), but rather to the evaluation of the nuclear quantities necessary for the 
calculation of the transmission coefficients entering  Eqs.~\ref{eq_sig}-\ref{eq_rate}.
 Clearly, the knowledge of the ground state properties (masses,
deformations, matter densities) of the target and residual nuclei is indispensable.  When not
available experimentally (the usual situation for nuclei involved in the r-process), this 
information has to be obtained from nuclear mass models (Sect.~\ref{nuc_static}). The
excited state properties have also to be known. Experimental data may be scarce above some
excitation energy, and especially so for nuclei located far from the valley of nuclear
stability. This is why a frequent resort to a level-density prescription is mandatory. The
transmission coefficients for particle emission are calculated by solving the Schr\"odinger
equation with the appropriate optical potential for the particle-nucleus interaction. Finally,
the photon transmission function is calculated assuming the dominance of dipole $E1$
transitions (the $M1$ transitions are usually included as well, but do not contribute
significantly. They will not be discussed further here). Reaction theory relates the
$\gamma$-transmission coefficient for excited states to the ground-state photo-absorption
strength distribution under the assumption of the Giant Dipole Resonance (GDR) to be built on each excited-state (see
 Sect.~\ref{th_strength}).  Ideally, these various necessary ingredients (properties of nuclei
 in their ground or excited states, nuclear
level densities, optical potentials, $\gamma$-ray strength functions) are 
 to be derived from {\it global}, {\it universal} and {\it microscopic} models, in the same way as
 nuclear masses can nowadays be derived from mean-field models (Sect.~\ref{nuc_static}). 
The situation for each of these ingredients is described below. With this input physics, and
 provided its basic assumptions are satisfied, the Hauser-Feshbach model has proved its 
ability to predict reliably and accurately the reaction cross sections for medium-mass and 
heavy nuclei
nuclei (Sect.~\ref{exp_th_hfrate}). 

\subsubsection{Reaction rate calculations: a direct-capture framework}
\label{th_rates-general_direct}

There are situations in which the compound system cannot reach an equilibrium CN within the 
reaction time. This is in particular the case for reactions with very exotic neutron-rich 
nuclei leading to compound systems with a number of available states that is small enough for
 the validity of the Hauser-Feshbach model to be questioned. In this case, the (neutron) 
radiative captures might be dominated by direct electromagnetic transitions to bound final 
states rather than by  the formation of a CN. These so-called direct radiative capture (DC) reactions
 are known to play an important role for light or closed-shell systems for which no resonant 
 states are available. 

The DC rates have been re-estimated for exotic neutron-rich nuclei by \cite{go97,go98} using a
modified version of the potential model to avoid the uncertainties affecting the
single-particle approach based on the one-neutron particle-hole configuration. The neutron DC
 cross section at energy $E$ is expressed as

\begin{equation}
\sigma^{\rm DC}(E) = \sum_{f=0}^x C_f^2 S_f \sigma_f^{\rm DC}(E) 
+ \int_{E_x}^{S_{\rm n}} \sum_{J_f,\pi_f} \langle C^2 S  
\rangle  \rho(E_f,J_f,\pi_f)\sigma_f^{\rm DC}(E)dE_f ~,
\label{eq_dc}
\end{equation}

 where $x$ corresponds to the last experimentally known level $f$ of the final
 nucleus  with excitation energy $E_x$ (smaller than 
the neutron separation energy $S_{\rm n}$), $S$ is the spectroscopic factor,  $C^2$ is the isospin
 Clebsch-Gordan coefficient, and  $\sigma_f^{DC}$ is the DC cross section to each final state.
  Above $E_x$, the summation is replaced by an integration over the spin ($J$)- and parity
 ($\pi$)-dependent level
density $\rho$, and the product $C_f^2 S$ is replaced by an average quantity $\langle C^2 S
\rangle$. The  cross section $\sigma_f^{DC}$  is calculated within the potential model of
 \cite{go97}, in which the wave functions of the initial and final systems are obtained  by 
solving
the respective Schr\"odinger equations with the use of the same analytic form of the potential.

In addition to the ingredients required to estimate the Hauser-Feshbach rates,  the 
spectroscopic properties (energy, spin, parity, spectroscopic factor, deformation) of the 
low-lying states in (very) exotic neutron-rich nuclei need to be estimated as well. The 
information of this type adopted in the calculation of  Eq.~\ref{eq_dc}, as well as the
 evaluation of the level densities entering this equation, are discussed in
 Sect.~\ref{th_leveldens}. The evaluation of  $\langle C^2 S \rangle$ 
 is very difficult. Experimental systematics  suggest in a first approximation an 
energy-independent value  $\langle C^2 S \rangle \simeq 0.06$ at energies lower than $S_{\rm n}$ 
\cite{go97}.  More details on the sensitivity of the predicted DC rates to the uncertainties
 on the necessary spectroscopic properties can be found in  \cite{go97}.

Both the DC and CN mechanisms may contribute to the radiative capture of neutrons.
 The total capture
rates are often taken as the simple sums of both contributions, neglecting all possible
interferences. However, the Hauser-Feshbach statistical model might  overestimate the 
resonant capture by the most exotic nuclei, since the number of levels available to the 
incident nucleon  in the compound system is too low to ensure a continuum superposition of
 resonances. Special attention should therefore be paid when extrapolating the statistical 
predictions to nuclei close to
the neutron-drip line. One way to account in an approximate way for such an overestimate of the
 Maxwellian-averaged rates at temperature $T$ is to dump artificially the statistical 
contribution following
 
\begin{equation}
\langle\sigma v \rangle^{\rm CN}= \langle\sigma v
\rangle^{\rm HF} ~ {1
\over {\displaystyle 1+(N_{\rm sp}^* /N_{\rm sp}(S_{\rm n}))^\delta}}.
\label{eq_hfdamp}
\end{equation}

where $N_{\rm sp}(S_{\rm n})$ is the number of levels available to {\it s}- and 
{\it p}-neutrons in an energy interval of $2kT$ around $U = S_{\rm n} + kT$ in the CN, and $\delta$ is
 a dumping parameter characterising the disappearance of the CN contribution for nuclei with
 $N_{\rm sp}(S_{\rm n}) < N_{\rm sp}^*$. $N_{\rm sp}(S_{\rm n})$ is calculated with the nuclear level density
formula used in Eq.~\ref{eq_trans} for the spin and parity of the target nucleus.
In Eq.~\ref{eq_hfdamp}, 
$N_{\rm sp}^* \sim 2$ along with $\delta \sim 5$ would be a reasonable choice so as
to  obtain a full Hauser-Feshbach contribution for nuclei with more than two
 $sp$-resonances in the $2kT$ energy interval, and a negligible contribution when less. Values 
$N_{\rm sp}^* \gsimeq 10$ probably exaggerate the dumping of
the CN contribution, while $N_{\rm sp}^* \lsimeq 1$ or $\delta
\lsimeq 5$ would overestimate the resonance contribution for nuclei with few $sp$-resonances. 

Equation~(\ref{eq_hfdamp}) provides at best an extremely schematic approximation. A more
 detailed description of the resonant capture rate when only a few states of the compound 
system are available (for example, in a R-matrix or  Breit-Wigner approach) may be desirable.
 
\subsubsection{Nuclear level densities}
\label{th_leveldens}

Although reliable microscopic models (mainly based on statistical or combinatorial approaches)
have been developed over the last four decades to estimate nuclear level densities (NLDs),
only approaches based on the Fermi gas approximation have been used until recently for
practical applications (e.g. \cite{ripl}). This results mainly from the fact that they
provide a mean to estimate NLDs through the use of simple analytic
formulae. They are, however, obtained at the expense of drastic approximations
concerning in particular the description of the shell, deformation, pairing and collective 
effects. As an illustration, the formulations of the back-shifted Fermi-gas (BSFG) type,
 which are of most common use today,  introduce some phenomenological improvements to the original
prescription of \cite{bethe37}. In particular, use is made of a highly simplified energy
dependence of the key `NLD parameter' $a$ which cannot account properly for the nuclear
excitation spectra. In addition, the complex pairing effects are just
described in terms of an energy shift. In such conditions, it is not surprising that
the existing analytic NLD prescriptions are unable to match the experimental data with a
reasonable level of accuracy. This shortcoming is cured to some extent by adjustments of a
more or less large number of free parameters. Such a procedure introduces, however, a
substantial unreliability if predictions have to be made when experimental data are
scarce or non-existent, as it is very often the case in certain, sometimes extended,
ranges of excitation energies, or
 for nuclei far from stability of importance in the modelling of the r-process and 
in a large variety of other applications. The lack of
measured level densities still constitutes the main problem faced in the NLD modellings and 
the parameter-fitting procedures they require, even though the number of analyses of slow-neutron
resonances and of cumulative numbers of low-energy levels grows steadily. This concerns in
particular the s-wave neutron resonance spacings $D$ at the neutron separation energy $S_{\rm n}$.
For a nucleus $(Z,A + 1)$ resulting from the capture of a low-energy s-wave neutron by a
 target $(Z,A)$ with spin $J_0$, $D$ is given by

\begin{eqnarray}
&D&=\frac{2}{\rho(S_{\rm n},J_0+1/2)+\rho(S_{\rm n},J_0-1/2)} \qquad\qquad {\rm for}~~J_0 >0 \cr
&&=\frac{2}{\rho(S_{\rm n},1/2)}\qquad\qquad \qquad\qquad \qquad\qquad \qquad {\rm
for}~~J_0 =0, 
\label{eq_spacing}
\end{eqnarray}

the factor of 2 in the numerator relating to the classical assumption of equal
probabilities of both parities $\pi$ at all energies.
%

\begin{figure}
\center{\includegraphics[scale=0.5]{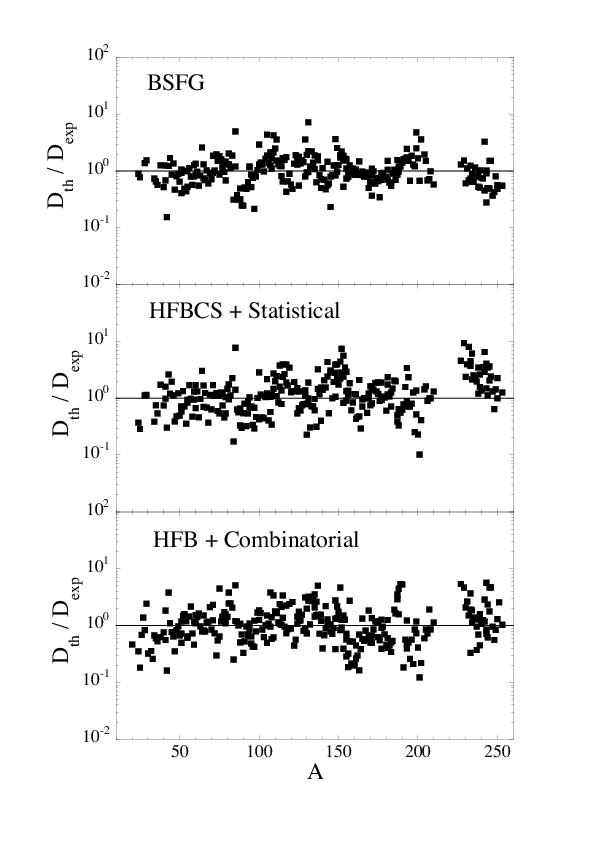}}
\vskip-0.8cm
\caption[]{Comparison between experimental s-wave neutron resonance spacings $D_{\rm exp}$ at
 the neutron separation energy $S_{\rm n}$ and predicted values $D_{\rm th}$ derived from
the use of a HFBCS \cite{dem00} (middle panel) or HFB model \cite{hi05} (bottom). The results from the
 global BSFG approximation of \cite{ndst02} are also shown (top)}
\label{fig_nld}
\end{figure}

\begin{figure}
\centerline{\epsfig{figure=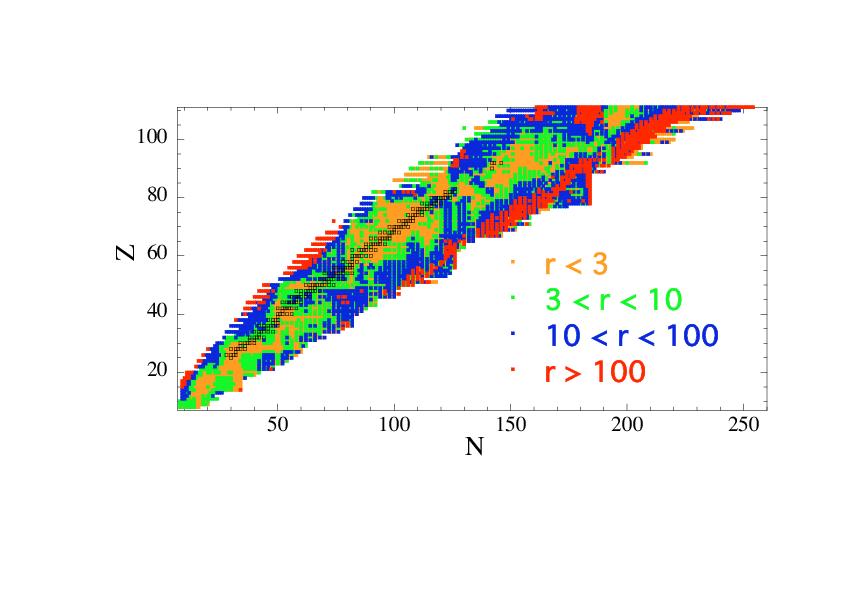,height=14cm,width=17cm}} 
\vskip -3.0cm
\caption[]{Comparison between the state densities 
$\omega (U) = 2\sum_J (2J +1) \rho(U,J.\pi)$ 
at the neutron separation energy $U = S_{\rm n}$ calculated by the HFBCS and BSFG models already
selected in Fig.~\ref{fig_nld}. The factor of 2 in the definition of $\omega(U)$
relates to the assumed equal probabilities of both parities $\pi$ at all energies. The values
of $r$ displayed for all nuclides in the ranges $N$ and $Z \ge 8$ and $Z \le 110$ located
between the proton- and neutron-drip lines are defined as $r = 10^{|\log(\omega_{\rm
HFBCS}/\omega_{\rm BSFG})|}$. Its values are coded as indicated in the figure}
\label{fig_nldtot}
\end{figure}
%
The separation $D$ is thus a direct measure of the spin-dependent level density $\rho(U,J)$ in
the nucleus $(Z,A+1)$ at the excitation energy $U = S_{\rm n}$.  Other sources of information have
also been suggested, such as the analysis of spectra of evaporated particles and  coherence
widths of cross section fluctuations. However, most of these data are affected by
systematic errors resulting from experimental uncertainties, as well as from the use of
approximate models.  

The NLD predictive power can be largely enhanced if the requirement of analytic
formulations is relaxed, which allows to duly take quantitatively into account various
nuclear properties, and in particular the discrete structure of the single-particle spectra
derived from realistic effective nuclear interactions. Such a `microscopic' approach has
the major advantage of being able to treat shell, pairing and deformation effects in a consistent way
 when evaluating the thermodynamic quantities that enter the NLD calculations. It is, however, not
free from some problems related to the very choice of the nuclear and pairing interactions.
In this spirit, \cite{dem00} calculates NLDs by using the single-particle level scheme and
pairing strength of the HFBCS model (Sect.~\ref{nuc_static}). As illustrated in 
Fig.~\ref{fig_nld}, the HFBCS-based NLD formula reproduces experimental data (neutron 
resonance spacings at the neutron separation energy) with an accuracy of typically a factor 
of about 2, which is comparable to the one obtained with a free parameter fit by the 
phenomenological BSFG formula. The work demonstrates the interest of microscopic 
evaluations of the NLDs.  It also provides a reliable extrapolation of the NLDs at low
energies, as seen from a confrontation with available experimental data.\footnote{The
microscopic HFBCS-based model has been re-normalised on experimental neutron-resonance
spacings and low-lying levels of a variety of nuclei in order to enhance the quality of the
predictions concerning the energy dependence of the NLD for these nuclei. This information is
needed in various applications} In spite of the good aforementioned agreement between the
BSFG and HFBCS-based level density predictions when experimental data are available,   
Fig.~\ref{fig_nldtot} shows that large differences may exist between them for nuclei
located far from the line of nuclear stability.  

A new development has been achieved within the combinatorial approach based on
the mean-field single-particle characteristics \cite{hi01}. In such an approach, parity, angular
 momentum, pairing correlations as well as collective enhancements are explicitly treated. 
 A recent study of the NLD based on the Skyrme-HFB single-particle and pairing data
of \cite{hi05} has shown that this approach can reproduce the experimental s- and p-wave 
spacings with an accuracy similar to the one obtained within the HFBCS model 
(Fig.~\ref{fig_nld}). This approach presents the advantage of accounting for the deviation from 
the statistical Fermi-gas limit at the low energies of the residual nuclei of relevance in 
DC cross section calculations, but also the parity-dependence of the NLD that the statistical
 approach fails to predict. For these reasons, the combinatorial model 
may be considered as the approach of choice for the calculations
of the DC process for exotic neutron-rich nuclei.

\subsubsection{Neutron-nucleus optical potentials}
\label{th_pot}

Phenomenological optical potentials (OPs) of the Woods-Saxon type(e.g \cite{koning02}) may 
not be well suited for applications involving exotic nuclei in general, and for the modelling of
the r-process in particular. It is considered profitable to use more microscopically-based
potentials, whenever possible. A semi-microscopic OP, usually referred to as the JLM
potential \cite{jlm77}, is available for the description of the neutron-nucleus case of
 relevance here, but also for the proton-nucleus interaction of importance in e.g. the 
modelling of the p-process \cite{arnould03}. It is
derived from the Br\"uckner--Hartree--Fock approximation (BHF) based on a Reid's hard core
nucleon--nucleon interaction. This OP has been revised recently for nucleons
interacting with spherical or quasi-spherical nuclei with masses $40\le A\le 209$ at energies
ranging from the 1 keV to 200 MeV \cite{bdg01}. The resulting version, referred to as the
JLM-Bruy\`eres or JLMB potential, features in particular a renormalisation to an extensive
set of nucleon scattering and reaction data. It is
 characterised in particular by an isovector [$(N-Z)/A$-dependent]
 component of its imaginary part, which is as much as 50\% larger than the original JLM value.
It remains to be seen whether such an enhancement would apply to larger neutron excesses,
where no experimental data are available. If it indeed remains valid far from the line of 
stability, it would strongly reduce the imaginary component by largely cancelling out the
 isoscalar component. The neutron absorption channel, and consequently the neutron capture cross
sections, would be reduced accordingly \cite{goriely02}. 

 The weakness of the present BHF approach lies in the fact that the asymmetry component of
the JLMB semi-microscopic model  is obtained by differentiating a symmetric nuclear BHF
matter calculation with respect to the asymmetry parameter. Further BHF calculations of the 
asymmetric nuclear matter, such as \cite{zuo99}, would be most useful to test this
crucial effect at large neutron excesses.  Even though  differences remain
in some cases among the rates calculated with different OPs, the predictions based on the
semi-microscopic potentials referred to above give a globally satisfactory agreement with
experimental data. Some improvements would be most welcome,
however, especially in the low-energy domain and in the treatment of deformed or exotic
nuclei.

\subsubsection{$\alpha$-particle-nucleus optical potentials}
\label{th-alphapot}
The situation for the $\alpha$-particle-nucleus OPs is much less satisfactory,
and one still has to rely on phenomenological potentials.  Most of the 
proposed OPs are derived from fits to elastic $\alpha$-nucleus
scattering data at energies $E \gsimeq 80~{\rm MeV}$ or, in some cases, to (n,$\alpha$) cross
sections at lower energies (e.g. \cite{fadden,nolt,avr94} for details). However,
the OP, and in particular its imaginary component, is known to be strongly
energy-dependent at energies  below the Coulomb barrier. As a consequence, its
extrapolation to sub-Coulomb energies is even more insecure than in the case
of nucleons. The development of a global $\alpha$-nucleus
OP to describe scattering and reaction cross sections at energies $E \lsimeq 20$ MeV of more
 relevance to astrophysics has been attempted by adopting a Woods-Saxon
\cite{gr98} or a double-folding (DF) \cite{mohr00,Dem02} component of the real
part. In both cases, a phenomenological form of the imaginary OP and of its
energy dependence is adopted.  More specifically, \cite{Dem02} base their DF model for the
real part of the OP on a realistic nucleon-nucleon interaction. From this, three different
types of imaginary potentials are constructed from the assumption of volume or surface
absorption, or from the adoption of the  dispersion relation linking the real
and imaginary parts of the OP. The three corresponding OPs are constrained in order to
reproduce at best scattering and reaction data. In the case of the OP with a purely volume
imaginary term (OP I of \cite{Dem02}), this is done through the fitting of nineteen free
parameters. This number is decreased to ten in the case of a volume plus surface imaginary
potential (OP II of \cite{Dem02}), with a concomitant reduction of the ambiguities in
deriving the OP from the data. This is even more so when the dispersive relation is used (OP III of  \cite{Dem02}).

The three  global $\alpha$-nucleus OPs derived in such a way are able to reproduce the bulk of
the existing experimental data at sub-Coulomb energies \cite{Dem02}. Experimental data are 
scarce, however, particularly in the $A > 100$ mass range. This limits
dramatically the predictive power of any of the OPs referred to above, especially in view of
their high impact on cross section estimates. At the energies of astrophysical relevance,
these are found to vary in some cases by more than one order of magnitude just
as a result of a different choice among the potentials constructed by \cite{gr98,Dem02}.  
Clearly, additional experimental data extending over a wide mass range, and especially to low-energy
radiative captures by $A \approx 100$ to 200 nuclei, are of importance to further
constrain the construction of, and enhance the reliability of,  global low-energy
$\alpha$-particle-nucleus OPs. Much remains to be done in this field, although such 
uncertainties when applied to the $\alpha$-capture process taking place before the 
neutron-capture r-process within the neutrino-driven wind (Sect.~\ref{r_dccsn}) are somehow
 reduced because of  the high temperatures and short timescales of relevance in such a scenario.

\subsubsection{$\gamma$-ray strength function}
\label{th_strength}

As noted in Sect.~\ref{th_rates-general}, the total photon transmission
coefficient in need for radiative captures is 
dominated by the $E1$ component. The calculation of the $E1$-strength function necessitates
 the knowledge of the low-energy tail of the GDR in the compound system formed in the
reaction process. The photon transmission coefficient is  most frequently described in the
framework of the phenomenological generalised Lorentzian model \cite{go98,mc81,ko90}. In this
 approximation,

\begin{equation}
T_{\rm E1}(\varepsilon_{\gamma})= \frac{8}{3} \frac{NZ}{A} \frac{e^2}{\hbar c}
\frac{1+\chi}{m_{\rm n}c^2} ~\frac{\varepsilon_{\gamma}^4~\Gamma_{\rm GDR}(\varepsilon_{\gamma})}
{(\varepsilon_{\gamma}^2-E_{\rm GDR}^2)^2+\Gamma_{\rm GDR}^2(\varepsilon_{\gamma}) 
\varepsilon_{\gamma}^2},
\label{eq_tg}
\end{equation}

where $E_{\rm GDR}$ and $\Gamma_{\rm GDR}$ are the energy and width of the GDR, $m_{\rm n}$ is
the nucleon mass and $\chi \simeq 0.2$ is an exchange-force contribution to the dipole sum
rule.  This description has been most widely used for practical applications, and more specifically
when global predictions are requested for a large set of nuclei.

The Lorentzian GDR approach suffers, however, from shortcomings of various sorts. On the one
hand, the location of its maximum and its width remain to be predicted from some
underlying model for each nucleus. For astrophysical applications, these properties have
often been obtained from a droplet-type of model \cite{my77}.
In addition, the Lorentzian model is unable to predict the enhancement of the $E1$ strength at
 energies below the neutron separation energy demonstrated by nuclear resonance fluorescence 
experiments. This
departure from a Lorentzian profile may manifest itself in various ways, and especially in the
form of a so-called pygmy $E1$ resonance \cite{gov98,zil02}, which is observed in $fp$-shell
nuclei, as well as in heavier spherical nuclei near closed shells (Zr, Mo, Ba, Ce, Sn, Sm and
Pb).   Calculations \cite{va92,ca97} predict  that the existence of a neutron mantle could
 introduce a new type of collective mode corresponding to an out-of-phase vibration of the 
neutron-proton core
against the neutron mantle. The restoring force for this soft dipole vibration is predicted
 to be smaller than that of the GDR, and could consequently influence  the low-energy $E1$ 
strength of relevance in capture cross-section calculations.

\begin{figure}
\center{\includegraphics[scale=0.5]{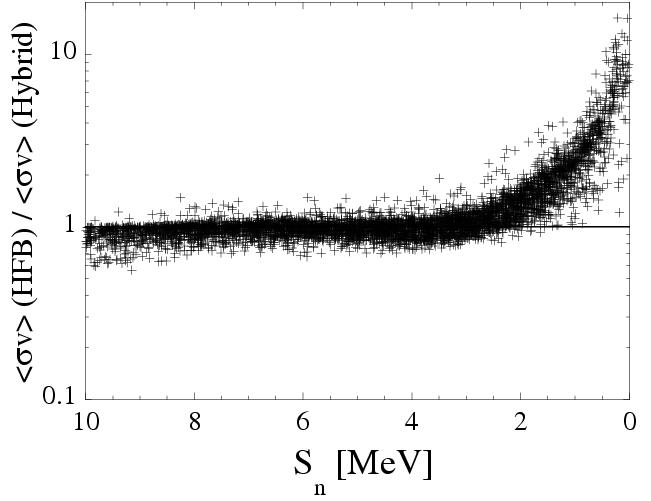}}
\caption{Ratio of the Maxwellian-averaged (n,$\gamma$) 
rate at the temperature  $T = 1.5 \times 10^9$~K obtained with the HFB+QRPA $E1$ strength
\cite{kh04} to the one using the Lorentz-type Hybrid formula \cite{go98} as a function of
the neutron separation energy $S_{\rm n}$ for all nuclei with $8\le Z \le 110$. The rate is
estimated within the Hauser-Feshbach model}
\label{fig_gamma}
\end{figure}

In view of this situation, combined with the fact that the GDR properties and low-energy
resonances may influence significantly the predictions of radiative capture cross sections, it is 
 of substantial interest to develop models of the microscopic type
which are hoped to provide a reasonable reliability and predictive power for the
$E1$-strength function. Attempts in this direction have been conducted within 
the QRPA model based on a realistic Skyrme
interaction.  The QRPA $E1$-strength functions obtained within the HFBCS \cite{kh01} as
well as HFB framework \cite{kh04} have been shown to reproduce satisfactorily the
location and width of the GDR and the average resonance capture data at low energies.
These QRPA calculations have been extended to all the $8\le Z\le 110$
nuclei lying between the neutron- and proton-drip lines. In the neutron-deficient region as
 well as
along the valley of $\beta$-stability, the QRPA distributions are very close to a
Lorentzian profile. However, significant departures are found for
neutron-rich nuclei. In particular, the QRPA calculations of \cite{kh01,kh04} show that the
neutron excess affects the spreading of the isovector dipole strength, as well as the
centroid of the strength function. The energy shift is found to be larger than predicted
by the usual $A^{-1/6}$ or $A^{-1/3}$ dependence given by the liquid-drop
description (e.g. \cite{my77}).
 In addition, some extra strength is predicted to be
located at sub-GDR energies, and to increase with the neutron excess.  Even if it
represents only about a few percents of the total $E1$ strength, as shown in
Fig.~\ref{fig_gamma}, it can be responsible for an increase by up to an order of
magnitude of the radiative neutron capture rate by exotic neutron-rich nuclei with
respect to the rate obtained with Lorentz-type formulae (for more detail, see
\cite{kh01,kh04}). It should, however, be kept in mind that the statistical model might not
 be valid for such exotic nuclei, as discussed in Sect.~\ref{th_rates-general_direct}.

\subsubsection{Fission channel}
\label{fission_ch}
In a similar way as for the particle or photon channels, the probability for the compound
 nucleus to fission can be obtained by estimating the corresponding transmission coefficient.
 The probability  for tunnelling through a barrier of height $B_{\rm f}$ and width $\hbar \omega_{\rm f}$ 
for a compound nucleus with excitation energy $E$ is given by the Hill-Wheeler expression 
(e.g. \cite{wag91})

\begin{equation}
P_{\rm f}(E)=\frac{1}{1+{\rm exp}[-2\pi (E-B_{\rm f})/\hbar\omega_{\rm f}]}\, .
\label{eq_fis1}
\end{equation}

At the saddle-point deformation, the nucleus may be in its corresponding ground state, which 
forms the top of the barrier, or in an excited state. These are called transition states. The 
fission transmission coefficient depends on the penetrability not only through the ground-state
 barrier for a given excitation energy, but also through the barriers associated with these 
transition states and a number of available such states. The total transmission coefficient
 can therefore be expressed as

\begin{equation}
T_{\rm f}(E)=\int_0^{\infty} \frac{1}{1+\exp[-2\pi (E-B_{\rm f}-\varepsilon) / \hbar
 \omega_{\rm f}]}~ \rho_{\rm f}(\varepsilon) ~{\rm d}\varepsilon,
\label{eq_fis2}
\end{equation}

where $\rho_{\rm f}(\varepsilon)$ is the NLD at the saddle-point deformation and at excitation 
energy $\varepsilon$. 
In the case of double-humped barriers and for  excitation energies below the largest of the
 barriers, the non-negligible {\it sub-barrier} effects are taken  into account by means of 
the picket-fence model \cite{Lyn80,sin06}. For higher energies, Eq.~\ref{eq_fis2} is 
generalised to an effective transmission coefficient that is classically calculated from

\begin{equation}
T_{\rm eff}=\frac{T_A~T_B}{T_A+T_B},
\label{eq_fis3}
\end{equation}

where $T_A$ or $T_B$ are the transmission coefficients through barrier A or B given by 
Eq.~\ref{eq_fis2}.  Such an approximation does not take into account either the microscopic
 features of the energy surface, or the complex dynamical processes at play during the fission
 process. More elaborate treatments within a full dynamical and microscopic framework (e.g.
 of the mean-field type) are eagerly awaited.

The fission barrier heights and widths, and the saddle-point NLD necessary for the evaluatation of  
Eq.~\ref{eq_fis2}  remain very uncertain, as already discussed in Sects.~\ref{fission} and 
\ref{th_leveldens}. In particular, no direct experimental data on the NLD at the fission 
saddle-point are available. In such conditions, the predictions of neutron-induced fission 
rates of relevance to the r-process rely on highly-parametrised input data that are adjusted 
to reproduce experimental cross-sections at best. As a consequence, the predictive power of 
such evaluations remains extremely low. 

\subsubsection{Comparison between theoretical and experimental neutron capture rates}
\label{exp_th_hfrate}

\begin{figure}
\centerline{\epsfig{figure=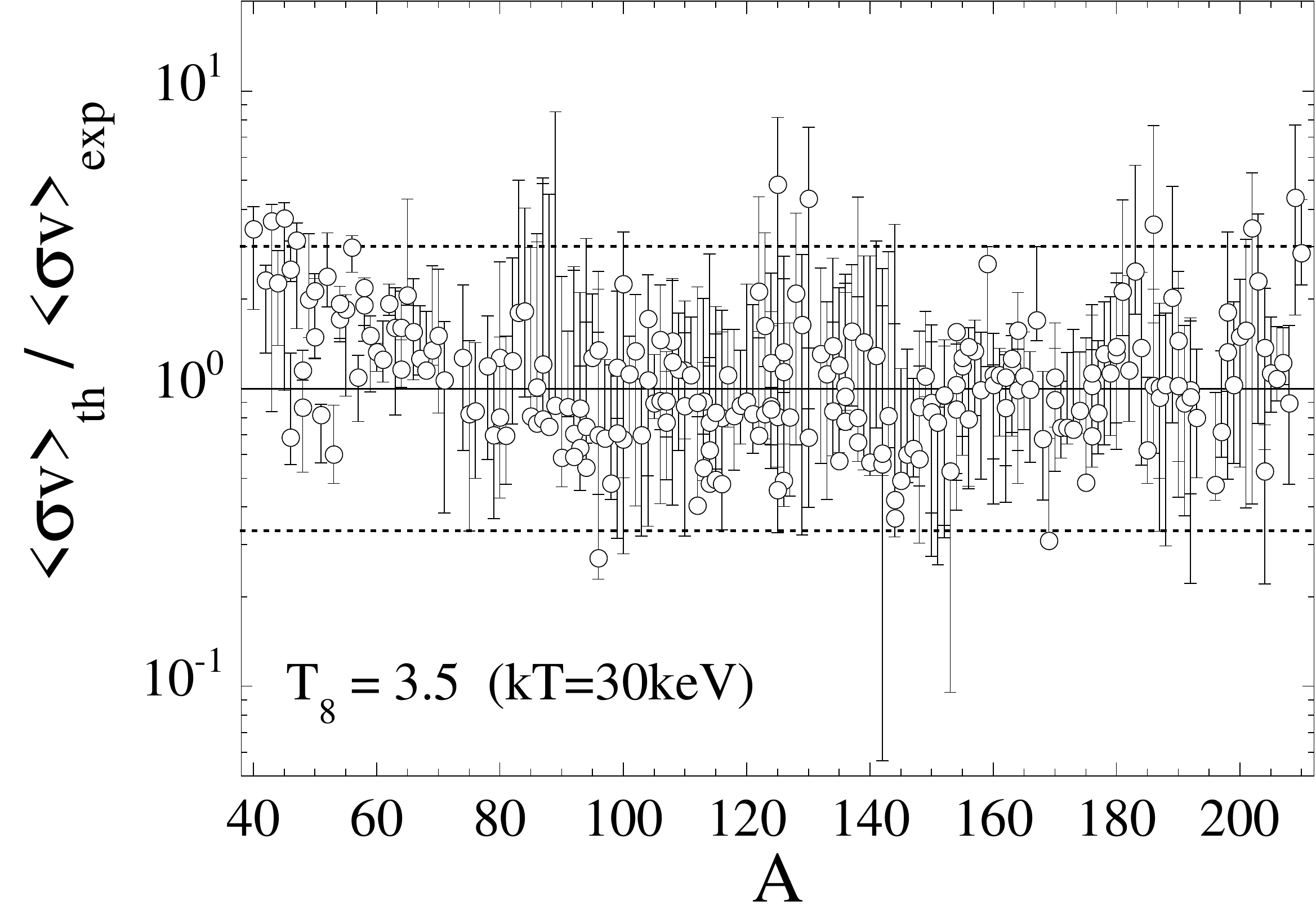,height=7.5cm,width=12cm}}
\caption[]{Comparison of MOST Maxwellian-averaged (n,$\gamma$) rates $\langle\sigma v
\rangle_{\rm th}$ with experimental values \cite{bao00} at $T=3.5 \times 10^8{\rm K}$. The
open dots correspond to the `standard' MOST rates. The vertical bars define the variations in
 the values of the displayed
ratios when the 13 other combinations of nuclear model inputs defined in \cite{arnould03}
are used for the MOST calculations}
\label{fig_ng_exp}
\end{figure}

{\it (1) Radiative neutron captures}. In order to evaluate the overall quality of the reaction
 rate predictions, this section
presents a comparison between  experimental radiative neutron-capture rates and
 Hauser-Feshbach results (the DC contribution being negligible for stable heavy nuclei)
based on Eq.~\ref{eq_rate} in which just the ground-state
contribution ($\mu = 0$) is taken into account. 
(The  consideration of target excited states is  irrelevant in the laboratory  conditions.)
 For astrophysical applications, such Hauser-Feshbach
calculations have been performed with the help of a computer code referred to as MOST
\cite{bruslib1} (see also \cite{bruslib2}). Although many other statistical model codes have been developed 
(e.g. those known as Talys \cite{talys}, Empire \cite{empire} or Non-Smoker
\cite{roro01}), MOST is unique in its 
effort to derive all nuclear inputs from global microscopic models, a feature of particular
 importance as  stressed earlier (Sect.~\ref{nucphys_general}). 
In addition, MOST calculates Eq.~\ref{eq_rate} with the
provision that various sets of models predicting the necessary nuclear input can be selected.
As described in \cite{arnould03}, a clear picture of the model uncertainties affecting the
 prediction of the reaction rates emerges from MOST calculations conducted with fourteen 
different combinations of the global models describing the ground-state properties 
\cite{frdm,sg02}, nuclear level density \cite{ripl,dem00,roro01}, nucleon- and
 $\alpha$-optical potentials \cite{koning02,jlm77,bdg01,Dem02} and $\gamma$-ray strength
 \cite{mc81,go98,kh01}. It should be emphasised that the uncertainties related to the choice
 of different physical models for a given set of nuclear input quantities
 (Sects.~\ref{nuc_static}-\ref{th_strength}) by far exceed the uncertainties associated with
 small variations of the parameters within a given model. 
Figure~\ref{fig_ng_exp} compares the MOST predictions of the Maxwellian-averaged (n,$\gamma$)
 rates $\langle\sigma v\rangle$ at $T = 3.5 \times 10^8$ K with experimental data for some 
230 nuclei heavier than $^{40}{\rm Ca}$ included in the compilation of \cite{bao00}. 

{\it (2) Neutron-induced fission}.  For r-process applications, a compilation of 
 fission barriers for some 2300 nuclei with $78\le Z \le 120$ derived within the ETFSI method 
\cite{mprt01} (see Sect.~\ref{fission}) has been prepared, along with the corresponding NLDs
 at  saddle point deformations predicted by the microscopic statistical model described in 
Sect.~\ref{th_leveldens} \cite{ripl}. 
The predicted neutron-induced fission cross sections based on this nuclear input have been 
compared by \cite{Dem03} with 18 experimentally available values in the 50 to 200 keV range. 
The measured and calculated cross sections globally differ by a factor of about ten, which is 
representative of our poor ability to {\it predict} neutron-induced fission cross sections
 nowadays. Use of highly-parametrised models like the BSFG 
approximation for the nuclear level density (Sect.~\ref{th_leveldens}) does not improve the accuracy, and may not be 
reliable for extrapolations to neutron-rich nuclei of interest in the r-process modelling.
 This unsatisfactory situation clearly calls for more work dedicated to the improvement of 
fission-rate predictions in a coherent and reliable microscopic framework.

\begin{figure}
\center{\includegraphics[scale=0.5]{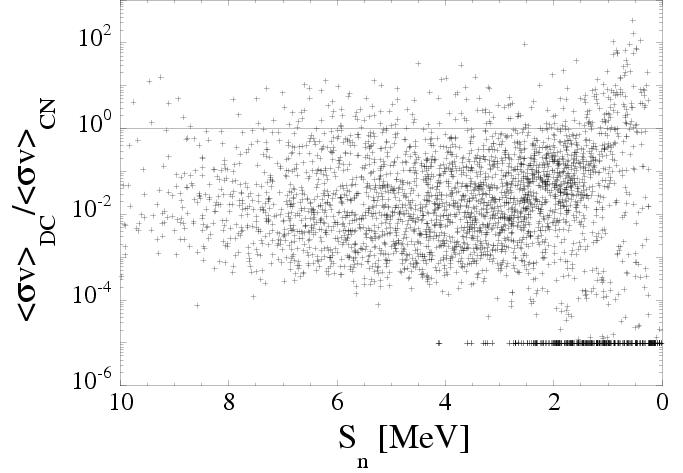}}
\caption{Comparison of the DC and CN $(n,\gamma)$ rates for all nuclei with $20 \le Z \le
92$ located between the valley of stability and the neutron-drip line. A lower limit of
$10^{-5}$ is imposed on the displayed rate ratio. The CN rates  from the BRUSLIB compilation 
\cite{bruslib2} are adopted}
\label{fig_dc}
\end{figure}

\subsubsection{Comparison between CN and DC contributions to radiative neutron capture rates}
\label{sect_dcrate}

 As discussed in Sect.~\ref{th_rates-general_direct}, it is necessary to include the DC 
contribution to radiative neutron captures by exotic neutron-rich nuclei, where the validity
 of the Hauser-Feshbach model is questionable.  Figure~\ref{fig_dc} compares the DC rates
 calculated as described there with the corresponding CN contributions (Sect.~\ref{th_rates-general}).
 Note that no dumping effect of the CN contribution, as described by Eq.~\ref{eq_hfdamp}, is included here.
It appears that the DC contribution exceeds that from the CN for many neutron-drip nuclei. The lower
limit imposed in Fig.~\ref{fig_dc} on the DC/CN rate ratio also indicates that  the DC rates 
can become negligible for some low $S_{\rm n}$ targets. This occurs when the selection rule forbids 
 $E1$ transitions to any of the available levels. 
The absence of DC contribution to supplement diminished CN rates can have a substantial 
influence on the predictions of some r-process models (e.g \cite{go97}).

\subsubsection{Beta-delayed processes}
\label{beta_delayed}
 
If $Q_{\beta^-}(Z,A)$ (Sect.~\ref{beta}) exceeds the neutron separation energy
  $S_{\rm n}(Z+1,A)$ of the daughter nucleus, neutron emissions follow the $\beta^-$ 
transitions to final excited states located above the neutron emission threshold, so that the parent
 nucleus $(Z,A)$ transforms to the residual nucleus $(Z+1,A-1)$.  Experimental data on delayed-neutron  (as well 
as delayed-proton) emissions have accumulated over the years, including a few cases of 
multiple emissions.
Delayed-neutron emission has been recognised already in the early days of the development of
 the r-process models \cite{BBFH57}  (see also \cite{cameron70}) as a mechanism able to help 
 smoothing the even-oddness often found in the computed r-abundance patterns to a point where
 it can compare favourably to e.g. the SoS one, as shown by \cite{kodama73,blake73}.

On the other hand,  $\beta$-delayed fissions of high-$Z$ nuclei may occur if $\beta$-decays
 can feed levels of the daughter nuclei that lie higher than their fission barriers  
\cite{berlovich69,skobelev72}.
This process has attracted attention because of its possible impact on the synthesis
of those nuclei which could possibly be the precursors of the nucleo-cosmochronometers
 $^{232}$Th, $^{235}$U, $^{238}$U, or $^{244}$Pu  (e.g. \cite{wene75,thielemann83}; see also 
Sect.~\ref{chronometry_general}).

The rates of delayed neutron emissions (d.n) or fissions (d.f) can be written as
 \cite{kodama75}

\begin{equation}
 \lambda_{\rm{d.n/d.f}}\approx \sum_{\Omega} \int_{-Q_{\beta^-}}^0
\frac{G_{\Omega}^2}{2 \pi^3} S_{\Omega}
\frac{\Gamma_{\rm n/f}}{\Gamma_{\rm n}+\Gamma_{\rm f}+\Gamma_\gamma} f_{\Omega}(-E){\rm d} E,
\label{eq:beta_delay}
\end{equation}

where $\Gamma_{\rm n}, \Gamma_{\rm f}$, and $\Gamma_\gamma$ are the neutron emission, fission,
 and $\gamma$ de-excitation widths, respectively. The other symbols have the same meaning as 
in Eq.~\ref{eq:beta_rate}.

\begin{figure}
\center{\includegraphics[width=1.0\textwidth,height=0.85\textwidth]{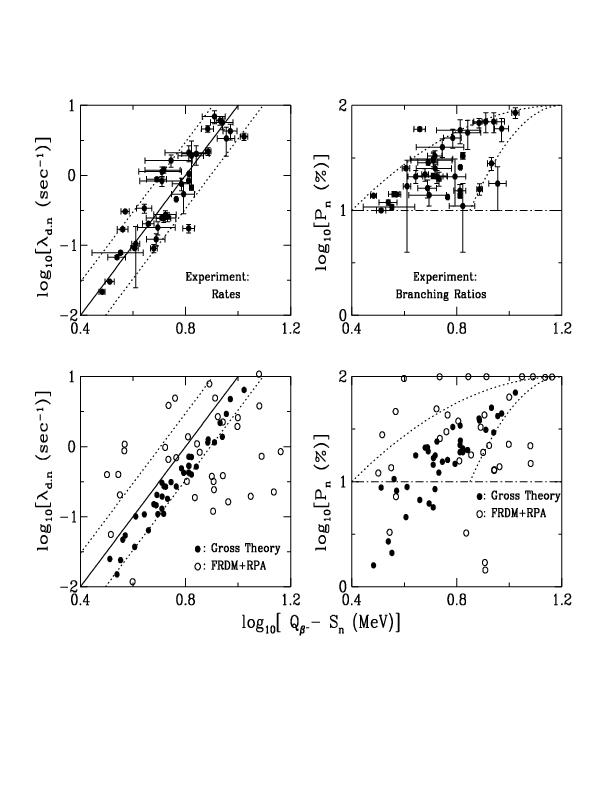}}
\vskip-2.4cm
\caption{Beta-delayed neutron emission rates $\lambda_{\rm d.n}$  and their branching
 ratios to the total $\beta$-decay rates $P_{\rm n} = \lambda_{\rm d.n}/\lambda_\beta$. The
 upper panels display experimental values for the $P_{\rm n} > 10$ \% cases. The lower panels
 show the corresponding predictions from the original Gross Theory as in 
Fig.~\ref{fig:beta_tq} (solid circles), and from the FRDM+RPA model \cite{moeller97} (open 
circles). In the left panels, the solid lines correspond to the straight line in
 Fig.~\ref{fig:beta_tq} with $(Q_{\beta^-}-S_{\rm n})$ replacing $Q_{\beta^-}$, and represent 
$\lambda_{\rm d.n} = 10^{-4} [(Q_{\beta^-} - S_{\rm n})$ in MeV]$^5$ s$^{-1}$. The dashed 
lines indicate deviations by a factor of three from this formula. The delineations by dotted 
lines in the right panels are just for guiding the eye}
\label{fig:beta_dn}
\end{figure}  

Except at energies close to the neutron emission threshold $(Q_\beta^- - S_{\rm n}$), 
$\Gamma_{\rm n} \gg \Gamma_\gamma$ generally holds. In the absence of fission channels, this 
reduces the calculation of $\lambda_{\rm d.n}$  (Eq.~\ref{eq:beta_delay}) solely to the 
evaluation of the relevant $\beta$-strength functions and nuclear masses.  A complication 
arises, however, in the  case of multiple-neutron emissions, as discussed by \cite{kodama75}.
 Figure~\ref{fig:beta_dn} displays experimental delayed-neutron emission rates and their 
ratios to the total $\beta^-$ decay rates, compared with model predictions. The
 $\lambda_{\rm d.n}$ rates are seen to correlate well with $[Q_{\beta^-} - S_{\rm n}]$, but 
the $P_{\rm d.n}$-values do not \cite{takahashi72}. This situation results from the generally
 quite irregular behaviours of 
the $\beta^-$  strengths to the final states at low excitation energies. The qualitative
 reproduction of the observed $\lambda_{\rm d.n}$ by the Gross Theory and  the improvements 
of its  $P_{\rm d.n}$ value predictions with increasing window energies are apparent. The
 performances of the global FRDM+RPA model of \cite{moeller97} are in general poorer. In 
this approximation, the `microscopic' character of the $\beta^-$ strength distributions is 
in fact exaggerated.

The fission channel can be neglected except for very heavy nuclei for which fission barriers
are low. In such cases,  fissions and neutron emissions could competitively occur from
 the excited states above $S_{\rm n}$ in the $\beta^-$-decay daughter. In general, therefore,
 $\Gamma_{\rm n}/\Gamma_{\rm f}$ ratios have to be estimated \cite{kodama75,meyer89}.
Seemingly, the early claims (e.g. \cite{wene75,thielemann83}) of an extreme importance of the
 $\beta$-delayed fissions in the r-process does not hold \cite{meyer89,cowan91}.
However, the final word in this matter has to await experimental information on the properties
 of very heavy neutron-rich nuclei that could make possible a calibration of the theoretical 
predictions of  masses, $\beta$-decay strength distributions, and especially fission barriers
 (e.g. \cite{panov05}). So far, only a handful of $\beta$-delayed fission branches have been 
measured, mostly in  neutron-deficient nuclei \cite{audi03a}.

\subsubsection{Neutrino captures}
\label{beta_annex2}

The interaction of neutrinos with matter may lead to a large variety of processes that 
possibly play a pivotal role in supernova explosions (Sects.~\ref{explo_1D} and 
\ref{explo_multiD}), as well as in various nucleosynthesis mechanisms, including the
 r-process (Sect.~\ref{r_wind} and Fig.~\ref{fig_apro7}). In addition to scatterings on
 electrons or nucleons, and coherent scatterings on complex nuclei that affect the neutrino 
opacity, charged-current $\bar\nu_{\rm e}$- or $\nu_{\rm e}$-captures or 
neutral-current scatterings by a nucleus $(Z,A)$ may lead to the release of a nucleon or 
an $\alpha$-particle. Some cases of special interest concern $\nu$ scatterings off 
\chem{4}{He} \cite{meyer95}.

Neutrino captures via the charged-current can be formulated analogously to ordinary 
$\beta^-$ decay. The major difference lies in the widening of the energy range of the 
$\beta^-$ strength distributions due to the possibly high energies of the neutrinos. 
Therefore, the knowledge of the global behaviour of the strength functions 
(Sect.~\ref{beta_sub2}) becomes more important than that of their detailed structure.
The capture cross section can be written as  

\begin{equation}
 \sigma(E_\nu) \approx \sum_{\Omega} \int_{-Q_{\beta^-}}^{E_\nu}
\frac{G_{\Omega}^2}{\pi}  S_{\Omega} p_{\rm e} W_{\rm e} F(Z,W_{\rm e}) {\rm d} E,
\label{eq:beta_nucross}
\end{equation}

where $E_\nu$  is the neutrino incident energy, which determines the electron energy
$W_{\rm e}$ as $ E_\nu =  E + (W_{\rm e}-1)m_{\rm e}c^2$. The local neutrino flux at distance
 $R$ from a source with neutrino luminosity  $L_\nu$  is given by  

\begin{equation}
 \phi = \frac{L_\nu}{4\pi\ R^2}\frac{1}{\langle E_\nu\rangle} \frac{E_\nu^2\ f_\nu(E_\nu)}
 {\int_0^\infty\ E_\nu^2 f_\nu(E_\nu) {\rm d}E_\nu},
\label{eq:beta_nuflux}
\end{equation}

where $E_\nu^2 f_\nu(E_\nu) {\rm d}E_\nu$ represents the neutrino spectral distribution 
function, and  $\langle E_\nu \rangle$ is the neutrino energy averaged over this distribution. The capture rate is obtained by integrating $\phi \sigma(E_\nu)$, which leads to

\begin{equation}
 \lambda_\nu  = \frac{L_\nu}{4\pi\ R^2}\ \frac{\langle \sigma_\nu \rangle}{\langle E_\nu \rangle}\ \
\approx 5 \times \left[ \frac{L_\nu}{10^{51} {\rm erg\ s}^{-1}}\right]\ \left[\frac{{\rm MeV}}{\langle E_\nu \rangle}\right]\
\left[\frac{100 {\rm km}}{R}\right]^2\ \left[\frac{\langle \sigma_\nu \rangle}{10^{-41} {\rm cm}^2}\right]\ \ {\rm s}^{-1}, 
\label{eq:beta_nurate}
\end{equation}

the last expression allowing easy order-of-magnitude estimates. The CQRPA estimates of 
$\langle \sigma_\nu \rangle$ \cite{borzov00} supplemented with the neutrino emission rates of
 \cite{hoffman92}  for nuclei lighter than Kr are used in Sect.~\ref{r_wind} to evaluate the 
possible consequences of neutrino captures on the r-nuclide abundance calculations. 

Finally, we add that there have been many speculations on the possible effects of neutrino
 captures on the r-process yields, and in particular on the role of  fissions possibly 
 induced by high-energy neutrino captures (e.g. \cite{langanke03,qian03,kolbe04}). The cross
 sections of these fissions cannot be calculated at present with any degree of reliability
 because of the quite poor knowledge of the neutrino fluxes combined with large uncertainties 
in fission barriers and fission probabilities of highly neutron-rich actinides 
(Sect.~\ref{fission}). We will not discuss this mechanism further here, waiting for better
 times.

\section{Site-free parametric high-temperature r-process models}
\label{high_t_param}
 
Already in their seminal work on the theory of nucleosynthesis, \cite{BBFH57} proposed the 
r-process (for rapid neutron-capture process) to result from the availability of neutron
 concentrations that are so high that neutron captures (especially of the radiative type) 
are faster than $\beta$-decays, at least for a substantial number of neutron-rich nuclides 
that are not located too far from the valley of nuclear stability. This is in marked contrast
 to the situation envisioned for the s-process (for slow neutron-capture process). 
Such conditions clearly provide a natural way
 to transform any pre-existing material into very neutron-rich species, viewed as the 
progenitors of the r-nuclides, as already schematised in Fig.~\ref{fig_srp}. The hypothesised
 high neutron affluence has been the framework adopted by the vast majority of studies of the 
r-process. In many cases, the consequences of such an assumption have been scrutinised only 
from a purely nuclear physics point of view by considering that one astrophysical site or
 the other, and in particular the inner regions of massive-star supernova explosions, could be
 the required neutron provider. A classical additional hypothesis has been that
the otherwise-unspecified stellar location is hot enough to allow ($\gamma$,n) photo-disintegrations to
 counteract to a more or less large extent the action of the inverse radiative neutron 
captures. Finally, it is supposed that a decrease of temperature that allows the `freezing' 
 of the photo-disintegrations occurs concomitantly with a decrease of the neutron density to 
values that are low enough to freeze the 
neutron captures.\footnote{A transformation is said to be `frozen' if its typical mean 
lifetime gets longer than a typical evolutionary timescale of the considered astrophysical
 site}
%

\begin{figure}
\centerline{\epsfig{figure= 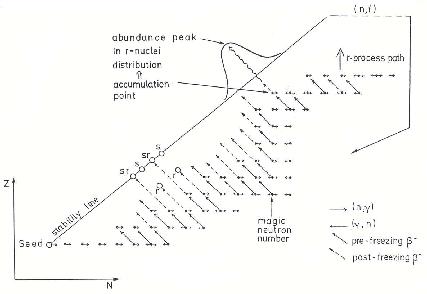,height=8cm,width=14.0cm}} 
\caption{Schematic representation of the r-process in the $(N,Z)$-plane}
\label{fig_nzplane}
\end{figure}

The aforementioned requirements on neutron concentration and temperature suffice to fix 
qualitatively several of the main features of the nuclear flow associated with the r-process
 and to identify the involved nuclear physics.  Figure~\ref{fig_nzplane} depicts the situation
 very schematically. In the course of the transformation of a given seed into more neutron-rich
 isotopes by a series of (n,$\gamma$) reactions, ($\gamma$,n) photo-disintegrations have the rates 
increasing with the neutron excess or, equivalently, with the associated decrease of the neutron
 separation-energy $S_{\rm n}$ (see Eq.~\ref{eq_inverserate}). At some point, the nuclear flow may 
proceed to  higher $Z$ elements through the intervening fast $\beta$-decays. 
In this picture, the flow takes a special
 character at neutron closed-shells.  The especially low $S_{\rm n}$ values just past a neutron magic 
number indeed hinders the flow to proceed to more neutron-rich species, so that
$\beta$-decays drive the material closer to the valley of stability following a path
with increasing $Z$ at practically constant $N$. The $\beta$-decays of the corresponding 
relatively less exotic nuclei become less probable and tend
to slow down the nuclear flow. As a consequence, some material accumulates at nuclei with a 
magic neutron number. However, when the path gets close enough to the stability line, $S_{\rm n}$ 
eventually becomes large enough to allow (n,$\gamma$) reactions to become more rapid than the 
$\beta$-decays, and to proceed without being counteracted by the ($\gamma$,n)
 photo-disintegrations. The flow then resumes normally until a new neutron magic-number is 
reached. In this picture, the accumulation of matter at neutron closed-shell nuclei due to 
the relatively slow $\beta$-decay bottlenecks provides a natural explanation of the SoS 
r-process peaks (Fig.~\ref{fig_srp}), as suggested in Fig.~\ref{fig_nzplane}.

If the nuclear flow towards increasing $Z$ values reaches the actinides or transactinide 
region, it is stopped by neutron-induced or $\beta$-delayed fissions, which lead to a recycling
 of a portion of the material to lower $Z$ values.  At freezing of the neutron captures or inverse 
photo-disintegrations, mainly $\beta$-decays  but also spontaneous or
$\beta$-delayed fissions and single or multiple $\beta$-delayed neutron emissions drive
the neutron-rich matter towards the valley of stability. These post-freezing transformations
 are included in Fig.~\ref{fig_nzplane} in a schematic way.

The evolution of the abundances dictated by the pre- and post-freezing transformations is 
obtained by solving a set of coupled nuclear kinetic equations of the form\footnote{Neutrino 
interactions with nuclei have to be considered in some r-process scenarios (see
 Sect.~\ref{r_wind}). They are neglected here}
%
 
\begin{eqnarray}
\frac{{\rm d}N(Z,A)}{{\rm d}t} 
& = & 
N(Z,A-1)~\lambda_{{\rm n}\gamma}^{Z,A-1}  + N(Z,A+1)~\lambda_{\gamma{\rm n}}^{Z,A+1} \nonumber\\ 
& + & 
N(Z-1,A)~\lambda_{\beta 0}^{Z-1,A} + \sum_{k} N(Z-1,A+k)~\lambda_{\beta k{\rm n}}^{Z-1,A+k} \nonumber \\
& + & N(Z+2,A+4) ~\lambda_{\alpha}^{Z+2,A+4} \nonumber \\  
& - & N(Z,A) \left\lbrack \lambda_{{\rm n}\gamma}^{Z,A} + \lambda_{\gamma{\rm n}}^{Z,A}  +\lambda_{\beta}^{Z,A}  \right\rbrack  \nonumber \\  
& - & N(Z,A) \left\lbrack \lambda_{\rm f}^{Z,A} +  \lambda_{\rm nf}^{Z,A}+\lambda_{\alpha}^{Z,A} \right\rbrack \nonumber \\ 
& + & \sum_f q_{Z_f,A_f}(Z,A)~\lambda_{\rm f}^{Z_f,A_f}~N(Z_f,A_f)  \nonumber \\
& + & \sum_f q_{Z_f,A_f}^\beta(Z,A)~\lambda_{\beta {\rm f}}^{Z_f-1,A_f}~N(Z_f-1,A_f)  \nonumber \\
& + & \sum_f q_{Z_f,A_f}^n(Z,A)~\lambda_{\rm nf}^{Z_f,A_f-1}~ N(Z_f,A_f-1)~,
\label{eq_net}
\end{eqnarray}

where $N(Z,A)$ is number density of nucleus $(Z,A)$ and the $\lambda$s refer to the rates of the
following reactions: $\lambda_{{\rm n}\gamma}=N_{\rm n}\langle\sigma v\rangle$ for radiative neutron 
captures, namely the product of the neutron number density $N_{\rm n}$ and the quantity
$\langle\sigma v\rangle$ given by Eq.~\ref{eq_rate} applied to the considered captures; 
$\lambda_{\gamma{\rm n}}$ for the inverse photo-disintegration given by Eq.~\ref{eq_inverserate}; 
$\lambda_{\beta 0}$ for $\beta$-decays followed by no delayed neutron or fission;
 $\lambda_{\beta k{\rm n}}$ for  $\beta$-decays followed by the delayed emission of $k$
neutrons; $\lambda_{\beta {\rm f}}$ for $\beta$-delayed fissions; 
$\lambda_{\beta}=\lambda_{\beta 0}+\sum_{k}\lambda_{\beta k{\rm n}}+\lambda_{\beta {\rm f}}$ for
the total $\beta$-decay rate; $\lambda_{\alpha}$ for $\alpha$-decay, and 
$\lambda_{\rm f}^{Z,A}$ and $\lambda_{\rm nf}^{Z,A}$ for spontaneous and neutron-induced fissions.
The last three terms reflect the feedback due to fissions of the synthesised heavy elements. 
 The factor $q_{Z_f,A_f}(Z,A)$ is the probability for the spontaneously-fissioning nucleus
($Z_f,A_f$)  to produce a $(Z,A)$-fragment. Similar fragmentations can also result from 
$\beta$-delayed or from neutron-induced fissions.
 Fission of  $Z<80$ nuclei does not play any role, so that neutron captures,
photo-disintegrations and $\beta$-decays dominate for these nuclei in Eq.~\ref{eq_net}.

In order to derive abundances from the system of equations of Eq.~\ref{eq_net}, the
 evolution of the thermodynamic quantities (temperature, density) has to be specified, along
 with the initial abundances. Such information may be obtained through detailed stellar 
models (Sects.~\ref{explo_1D} - \ref{compact_general}), or from more or less highly simplified
 approximate prescriptions described below.

\subsection{The canonical r-process model (CAR) and the waiting point (WP) approximation}
\label{canonical}

The early works of \cite{BBFH57} and \cite{seeger65} have proposed  the simplest and most
 widely used form of the r-process scenario, referred to as the canonical r-process (CAR)
 model. It assumes that pre-existing material located in the valley of nuclear stability is 
driven by neutron captures into a location of the neutron-rich region determined by the
 neutron supplies and by the highly temperature-sensitive reverse photo-dintegrations.  
Although this canonical model does not make reference to any specific astrophysics scenario, 
but builds on nuclear properties only, it has 
 greatly helped paving the way to more sophisticated approaches of the r-process.

The canonical model relies on the following assumptions:\\
(1) the neutron density $N_{\rm n}$ remains constant over the whole timescale $\tau$, and is high
 enough for the (n,$\gamma$) captures by any neutron-rich nucleus to be faster than its 
$\beta$-decays;\\
(2) the temperature $T$ is high enough for the ($\gamma$,n) photo-disintegrations to be faster
 than the $\beta$-decays;\\
 (3)  the neutron irradiated material is made initially of pure \chem{56}{Fe}. This assumption
 is validated by the fact that the r-process has been suspected already by 
\cite{BBFH57,seeger65} to develop in the Fe-rich inner core of exploding massive stars 
(Sect.~\ref{explo_1D} and \ref{explo_multiD}); \\
(4) the transmutations appearing in Eq.~\ref{eq_net} that are not listed above are neglected.
 Some provision for fission is, however, considered in a very approximate way by 
\cite{seeger65}. The corresponding `long-time solution' is not discussed here.

Under assumptions (1) - (3), Eq.~\ref{eq_net} takes the simple form ($Z \geq 26$)

\begin{equation}
{{\rm d}N(Z,A) \over {\rm d}t} =  \lambda_{\gamma,{\rm n}}^{Z,A+1} N(Z,A+1) - \langle\sigma v\rangle_{Z,A} N(Z,A)
 N_{\rm n}~;
\label{eq_abond_canonical}
\end{equation}

(5) in addition to conditions (1) - (4), an equilibrium between the (n,$\gamma$) and
 ($\gamma$,n) reactions holds during the whole timescale $\tau$ for all isotopes of each of 
the $Z \geq 26$ elements. In such conditions, it follows from Eq.~\ref{eq_inverserate} 
that (e.g. \cite{seeger65})

\begin{eqnarray}
{N(Z,A+1) \over N(Z,A)} &=& {\langle\sigma v\rangle_{Z,A} \over
\lambda_{\gamma,{\rm n}}^{Z,A+1}} \,  N_{\rm n} \cr
&=&{G^*(Z,A+1) \over 2 G^*(Z,A)} \left( {2 \pi \hbar^2N_{\rm A} \over m k T}
\right)^{3/2} N_{\rm n}\exp{\left[ \frac{S_{\rm n}(Z,A+1)}{kT} \right]}~,
\label{wpa2}
\end{eqnarray}

where the reduced mass $m$ is well approximated by the nucleon mass $m_{\rm n}$ for
the heavy nuclei of our current interest. 
The partition functions $G^*$ relate to the normalised values $G$ of
 Eq.~\ref{eq_inverserate} through $G^* = (2J^0 + 1) G$, $J^0$ being the ground state spin of 
the considered nucleus. All the other symbols have the same meaning as in
 Eq.~\ref{eq_inverserate}. Assumption (5) is classically known as the `waiting point 
approximation'. The origin of this name is as follows:\\
(i) for a given $T$ and $N_{\rm n}$, Eq.~\ref{wpa2} indicates that the abundance of a given
 element $Z$ is almost entirely concentrated on its isotope (and closest neighbours) with a neutron 
separation energy $S_{\rm n}(Z,A)$ approaching the value

\begin{equation}
 S_{\rm a}^0 [{\rm MeV}] = \left( 34.075
- \log N_{\rm n} [{\rm cm}^{-3}]+ {3 \over 2}~ \log T_9 \right) {T_9 \over 5.04}~, 
\label{eqSa}
\end{equation}

where $T_9$ is the temperature in $10^9$~K. The locus of the isotopes of the $Z \geq 26$
 elements for which $S_{\rm n} \approx S_{\rm a}^0$ defines the r-process path for the considered $T$ and 
$N_{\rm n}$;\\
(ii) under the aforementioned assumptions, each isotopic  chain $Z$ in a state of
 $(n,\gamma)-(\gamma,n)$ equilibrium has to {\em wait} the $\beta$-decays of the constituting 
isotopes to transform into the next $Z + 1$ chain. The rate of this $Z  \to Z + 1$ 
transformation is given by 

\begin{equation}
\lambda_{\beta}^Z=\sum_A~ \lambda_{\beta}^{Z,A}~ \frac{N(Z,A)}{N(Z)} 
\label{eq_lamZ}~,
\end{equation}

 where $N(Z)=\sum_A N(Z,A)$ is the abundance of element $Z$, $N(Z,A)$ being derived from
Eq.~\ref{wpa2}. 
 
In such conditions,  the evolution of the elemental abundance $N(Z)$ is given by 

\begin{equation}
{{\rm d}N(Z) \over {\rm d}t} = N(Z-1) \lambda_{\beta}^{Z-1} - N(Z)  \lambda_{\beta}^Z~;
\label{eq_can}
\end{equation}

 (6) at time $\tau$, $T$ and $N_{\rm n}$ are supposed to go to zero abruptly, so that the
 (n,$\gamma$) and ($\gamma$,n) reactions are frozen suddenly. Each of the unstable nuclides 
produced during the irradiation (mainly those located on the r-process path defined by 
Eq.~\ref{eqSa}) then transforms into a stable r- or sr-nuclide through a $\beta$-decay
 cascade.

\begin{figure}
\centerline{\epsfig{figure= 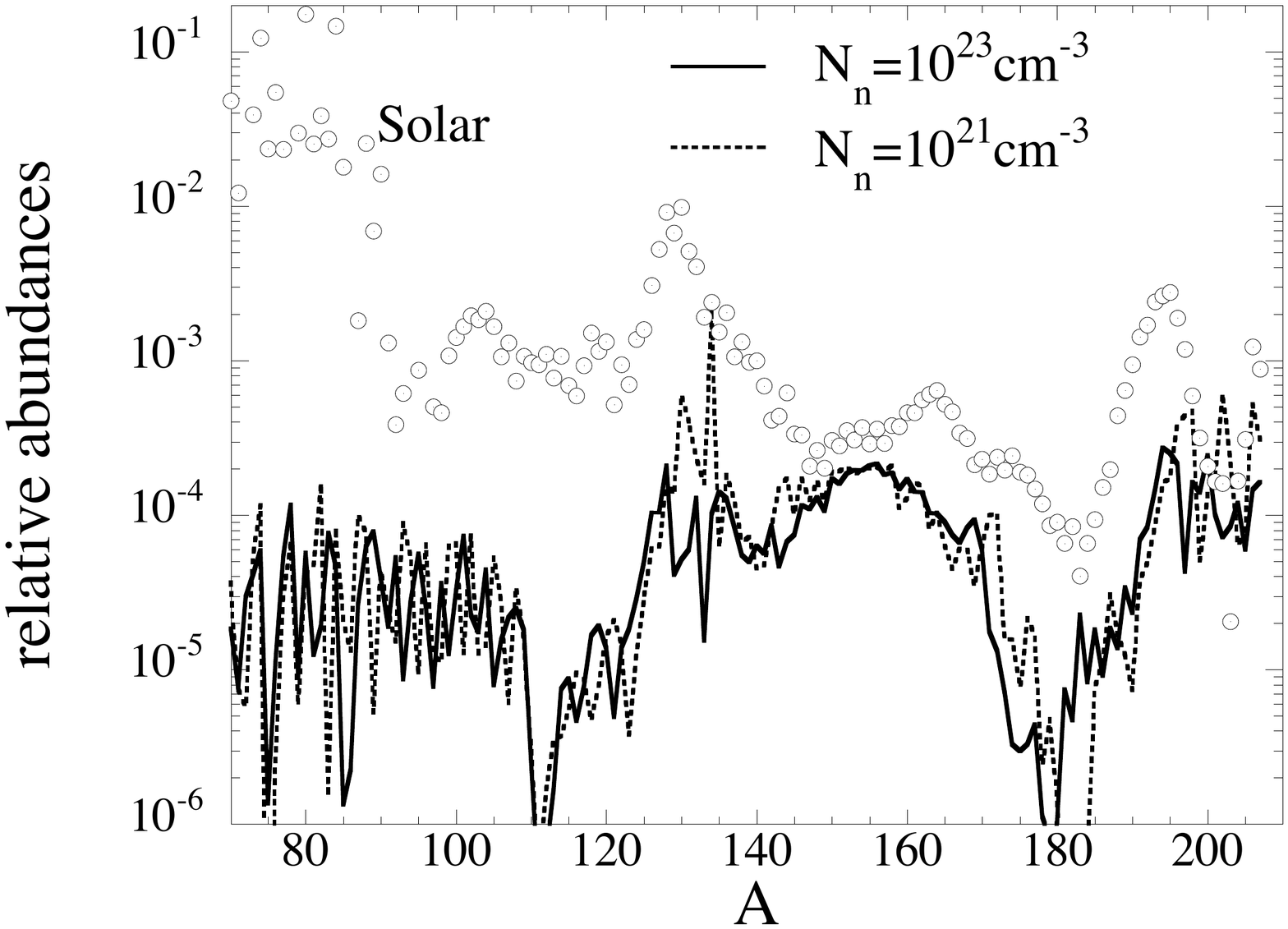,height=8cm,width=14.0cm}} 
\vskip-0.5cm
\caption{Abundances predicted by the steady-flow canonical model for  $T_9=1.2$ and
 $N_{\rm n}=10^{21}$ or $10^{23}~{\rm cm}^{-3}$. The $S_{\rm n}$ values are derived from the HFB-9 
nuclear masses (Sect.~\ref{nuc_static}),  and the GT2 $\beta$-decay rates are used 
(Sect.~\ref{beta}). The solar abundances are shown merely for illustrative purposes}
\label{fig_steady1}
\end{figure}

The waiting point approximation can be simplified further by assuming that a steady state is
 reached between production of each isotopic chain and its destruction. In such conditions,
 Eq.~\ref{eq_can} leads to 

\begin{equation}
 N(Z) = N_0~\tau_{\beta}^{Z}
\label{eq_steady1}
\end{equation}

where $N_0$ is a normalisation constant and 
$\tau_{\beta}^Z=1/\lambda_{\beta}^Z$ (Eq.~\ref{eq_lamZ}). An example of elemental abundances
 derived under the steady flow approximation is given in Fig.~\ref{fig_steady1} for a 
temperature $T_9=1.2$ and two $N_{\rm n}$ values. The abundance peaks corresponding to the N=82 and
 N=126 neutron shell-closures already emerge from such a simple model.

Even when  the steady-flow approximation is discarded, the canonical model in its waiting-point
 approximation version still offers a quite easily tractable mathematical (set of 
Eqs.~\ref{eq_can}) and physical framework for the r-process. In particular, abundance 
evaluations only request the knowledge of nuclear masses, partition functions and $\beta$-decay rates,
 and are fully determined by the choice of the 
three parameters: temperature $T$, neutron number density $N_{\rm n}$, and duration $\tau$.
  In the following, a 
\{$T, N_{\rm n}, \tau$\} set of constant values will be referred to as a `canonical event (CEV)',
 the additional adoption of the waiting point approximation leading to a so-called CEV+WP. 
Equivalently, a CEV+WP may be characterised by $S_{\rm a}^0$ and  the number $n_{\rm cap}$ of neutrons
 captured by \chem{56}{Fe} seed. It is given by adapting Eq.~\ref{eq_ncap} to the CEV+WP 
situation, leading to  $n_{\rm cap}\equiv n_{\rm cap}(\tau)=\sum_{Z,A} A N(Z,A)(\tau)56 N(26,56)(t=0)$,
 where $N(26,56)(t=0)$ is the initial amount of \chem{56}{Fe} seeds.
 
The substantial simplifications introduced by the CEV+WP approach have secured its popularity.
  Since the early work of \cite{seeger65}, it has been used in countless attempts to fit the 
SoS r-nuclide abundance distribution. This line of research has occcasionally gone so far (too far indeed) 
that  the quality of the fit obtained in this highly schematic r-process framework has been 
used as a measure of the quality of the input nuclear physics (restricted in the CEV+WP 
approximation to masses and $\beta$-decays). The attempt to fit the SoS r-abundances has led
 \cite{seeger65} to the conclusion that a single CEV+WP could not reproduce correctly the 
peaks observed in the r-nuclide distribution (Fig.~\ref{fig_srp}). It was concluded in this 
early work that two CEV+WPs were required instead for that purpose. Many subsequent works 
have followed the line set by \cite{seeger65}, and have concluded that three, four,  or some more
 CEV+WPs are necessary to obtain a satisfactory fit. Considering the highly schematic nature
 of the CEV+WPs, this kind of fitting exercise is not likely to set as meaningful 
constraints on the underlying nuclear physics as is sometimes asserted.

\begin{figure}
\centerline{\epsfig{figure= 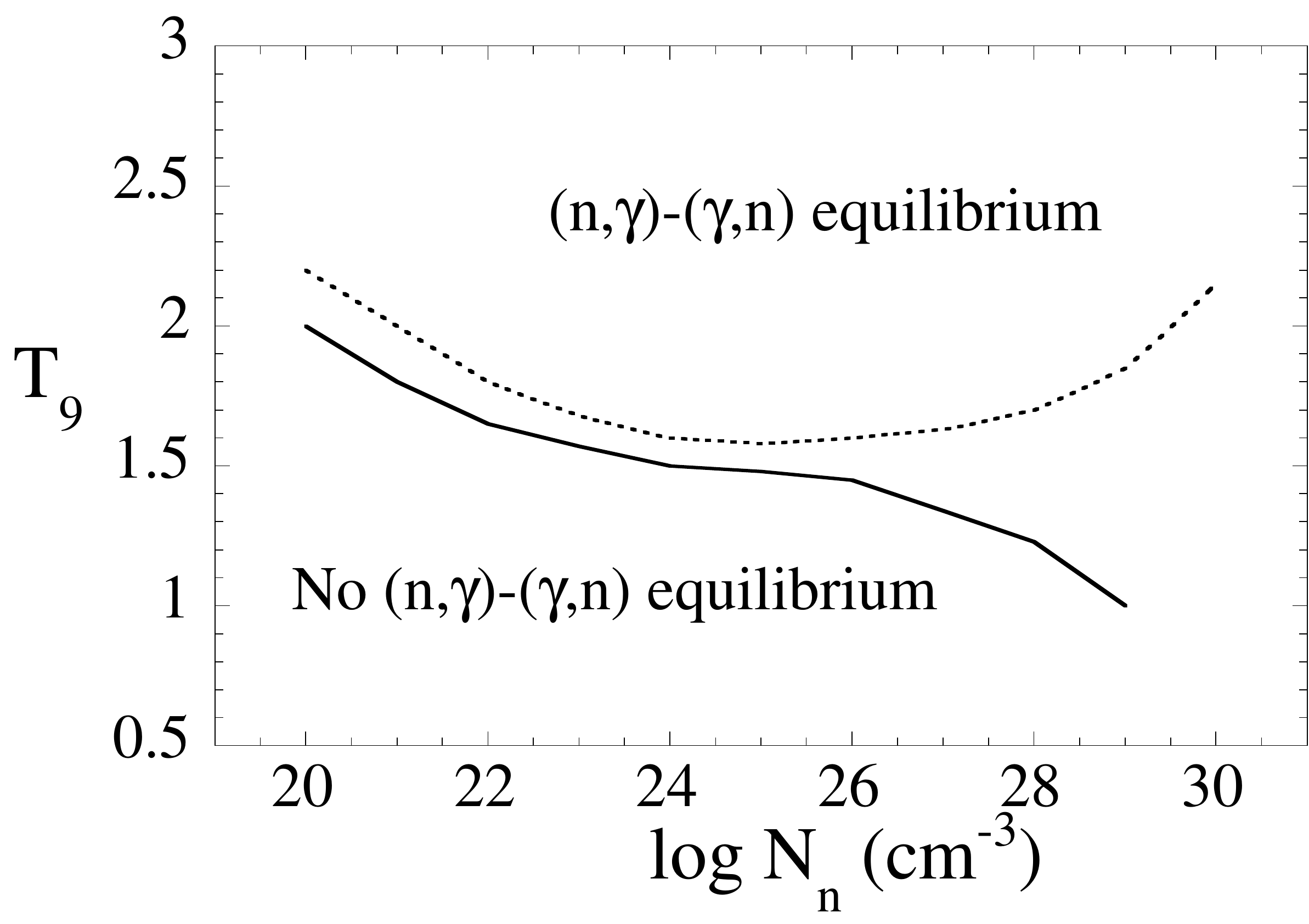,height=7cm,width=12.0cm}} 
\caption{Validity of the CEV + WP approximation in the ($T_9, N_{\rm n}$) plane.  The solid line is 
obtained with the $(n,\gamma)$ and $(\gamma$,n) rates estimated from the
Hauser-Feshbach calculations of \cite{goriely96} (see also Sect.~\ref{th_rates-general}). The
 dotted line is obtained when direct captures (see Sect.~\ref{th_rates-general_direct}) and 
some other properties specific to exotic neutron-rich nuclei are taken into account \cite{go97}. The rate 
estimates are based on the HFB-2 nuclear masses. The GT2 $\beta$-decay rates are used. For the
 adopted nuclear physics input, the CEV+WP approximation is invalid below the displayed lines}
\label{fig_wpa}
\end{figure}

\begin{figure}
\centerline{\epsfig{figure= 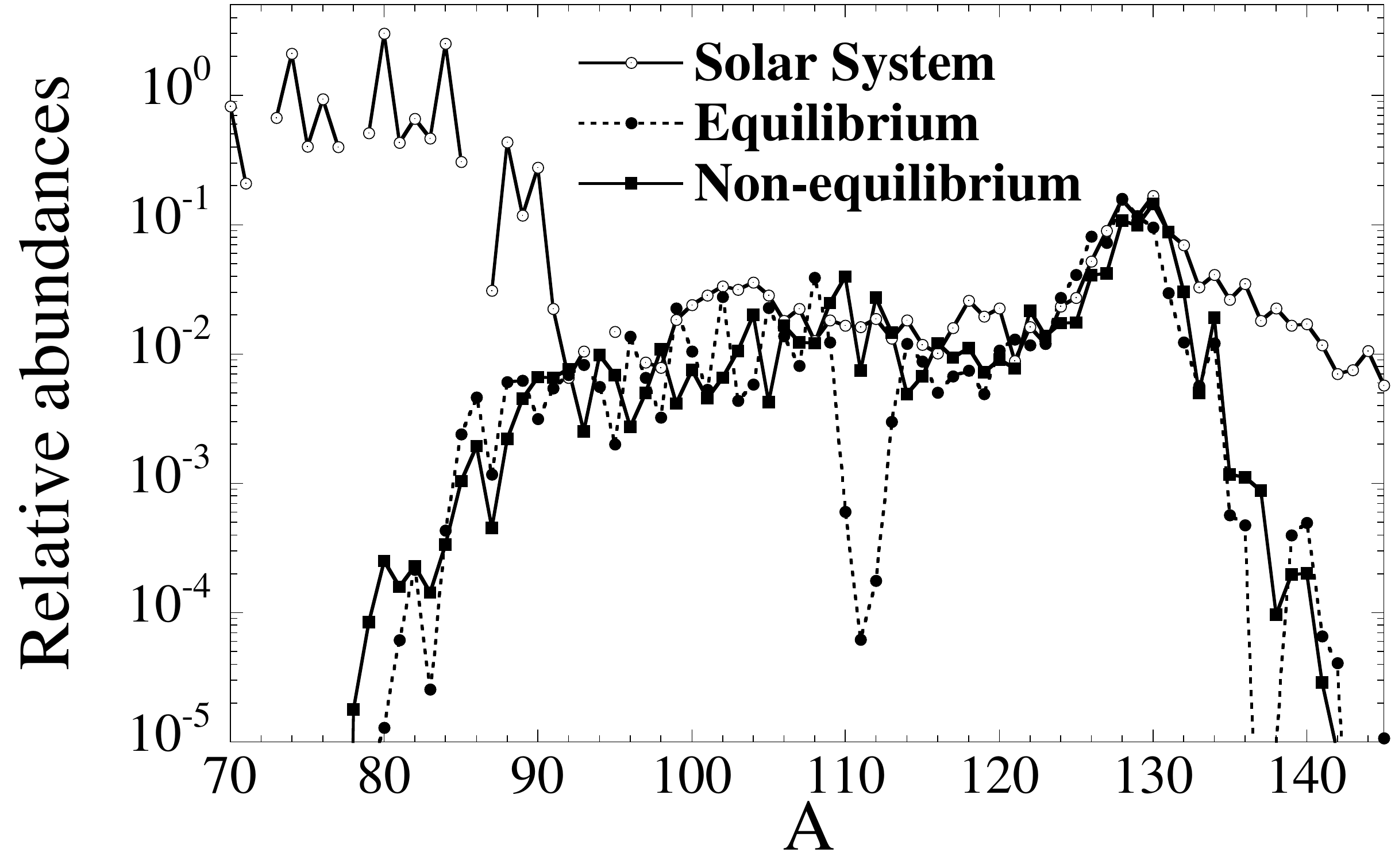,height=8cm,width=14.0cm}} 
\caption{Abundances from a canonical event \{$T_9 = 1.2, N_{\rm n} = 10^{21}$ cm$^{-3}$, $\tau = 1$
 s\} ($S_{\rm a}^0 = 3.1$ MeV) when the waiting-point approximation is 
inappropriately applied (dashed line) or not (solid line). 
The nuclear input is the same as in
 Fig.~\ref{fig_wpa} . The solar abundances are shown just for illustrative purposes}
\label{fig_canonical}
\end{figure}

An illustration of the danger of adopting the CEV+WP framework without restriction is 
provided by Fig.~\ref{fig_wpa}. For this approximation to be valid,  $T$ and $N_{\rm n}$ have to be 
such that the basic assumption of a (n,$\gamma$) - ($\gamma$,n) equilibrium  can be reached
 within time $\tau$ for each isotopic chain. This is not the case below the lines shown in
 Fig.~\ref{fig_wpa}.  In the corresponding conditions, abundances have to be obtained from 
the solution of a reaction network coupling for each nuclide its $\beta$-decays, radiative 
neutron captures and photo-disintegrations.  The limits of the the CEV+WP validity are 
obviously sensitive to the uncertainties in the rates of the transmutations involved 
\cite{goriely96,go97}, as illustrated in part in Fig.~\ref{fig_wpa}.
 Figure~\ref{fig_canonical} gives a measure of the errors introduced by the adoption of a 
CEV+WP in conditions for which it is invalid, as predicted by Fig.~\ref{fig_wpa}.  These
 errors are substantial in some cases.
 
\subsection{The multi-event r-process model (MER)}
\label{MER}

A parametric approach of the r-process differing from the one pioneered by \cite{seeger65}
 has been developed by \cite{goriely96,bouquelle96}. In its formulation, it is identical to
 the model MES used for the s-process in the decomposition  between the SoS s- and
 r-abundances (Sect.~\ref{sr-splitting}).

As MES, the multi-event r-process model, referred to as MER, relies on a superposition of a 
given number of CEVs, but the waiting point approximation is not imposed for any of the 
considered CEVs.  
In order to avoid unnecessary confusions, it may be worth noting that the term
"multi-event" may not only refer to {\it numerous stars}  responsible for the
production of r-nuclides. If massive star explosions are indeed possible r-process sites 
(see Sect.~\ref{r_dccsn}), one may conclude from a rough estimate that about as many as
 $10^7$ supernovae may have contributed to the r-nuclide contamination of the solar system. 
As there is no serious observational or theoretical reason to assume the global universality 
of the r-process (see Sects.~\ref{galaxy_universality} and \ref{MER_universality}), the SoS
 r-nuclide composition may indeed be the result of a quite large variety of different events.
 In addition, the name `multi-event' may also relate to a {\it suite of thermodynamic conditions} 
ready for the r-process that can likely be encountered {\it in a single object} (see
 Sect.~\ref{r_dccsn}).  In addressing those aspects statistically,  
a `multi-event' approach is likely more realistic than the consideration of a few  events. 

\begin{figure}
\centerline{\epsfig{figure= 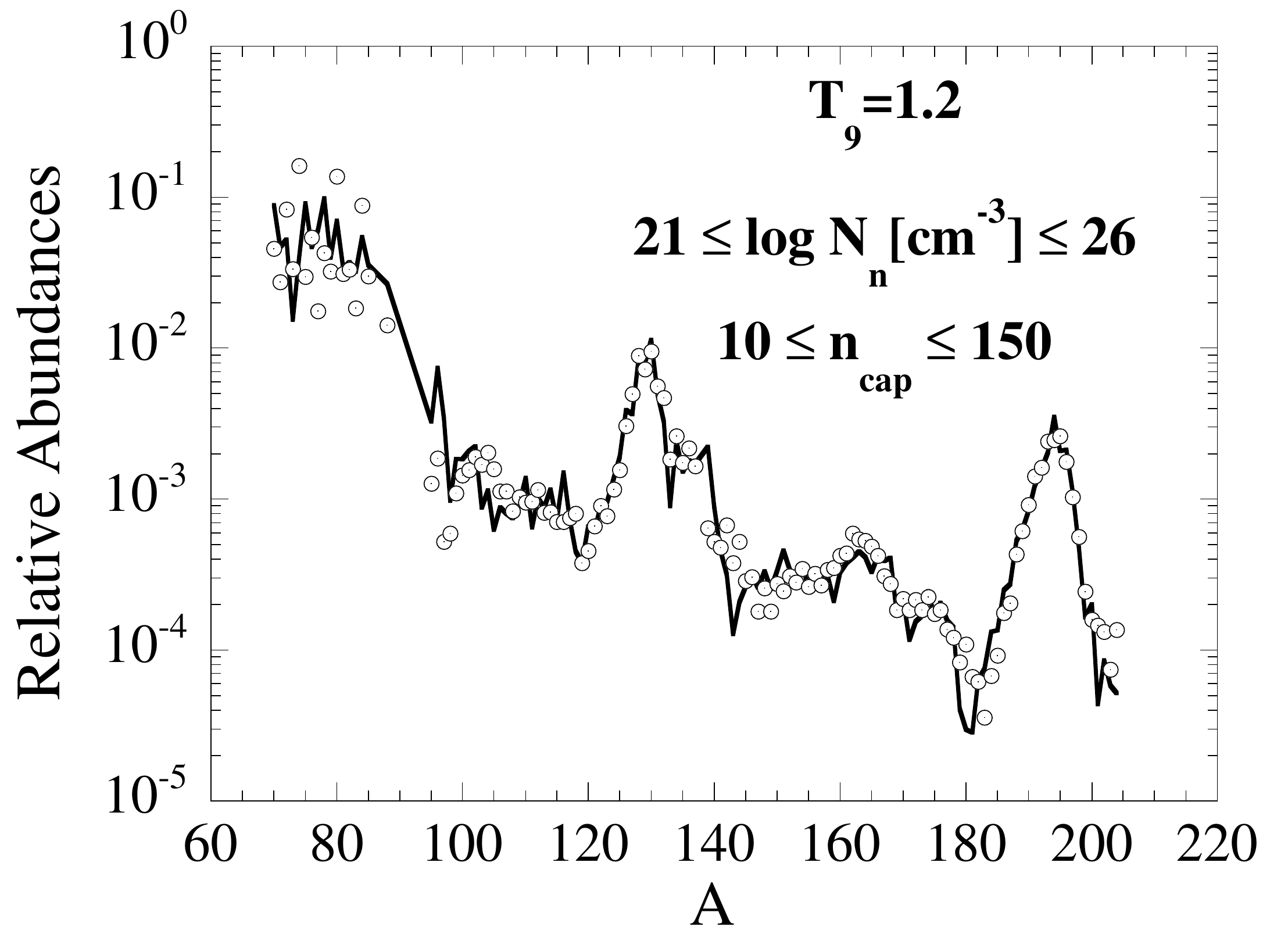,height=8cm,width=13.0cm}} 
\caption{Comparison between the SoS r-abundances (Fig.~\ref{fig_rsol_isot}; the uncertainties 
are not shown) and a MER fit obtained with the $T_9 = 1.2$ CEVs shown in 
Fig.~\ref{fig_multie2} and located in the indicated ranges of $N_{\rm n}$ and $n_{\rm cap}$. The
 adopted nuclear input is the same as in Fig.~\ref{fig_wpa} (from \cite{goriely96})}
\label{fig_multie1}
\end{figure}

\begin{figure}
\centerline{\epsfig{figure= 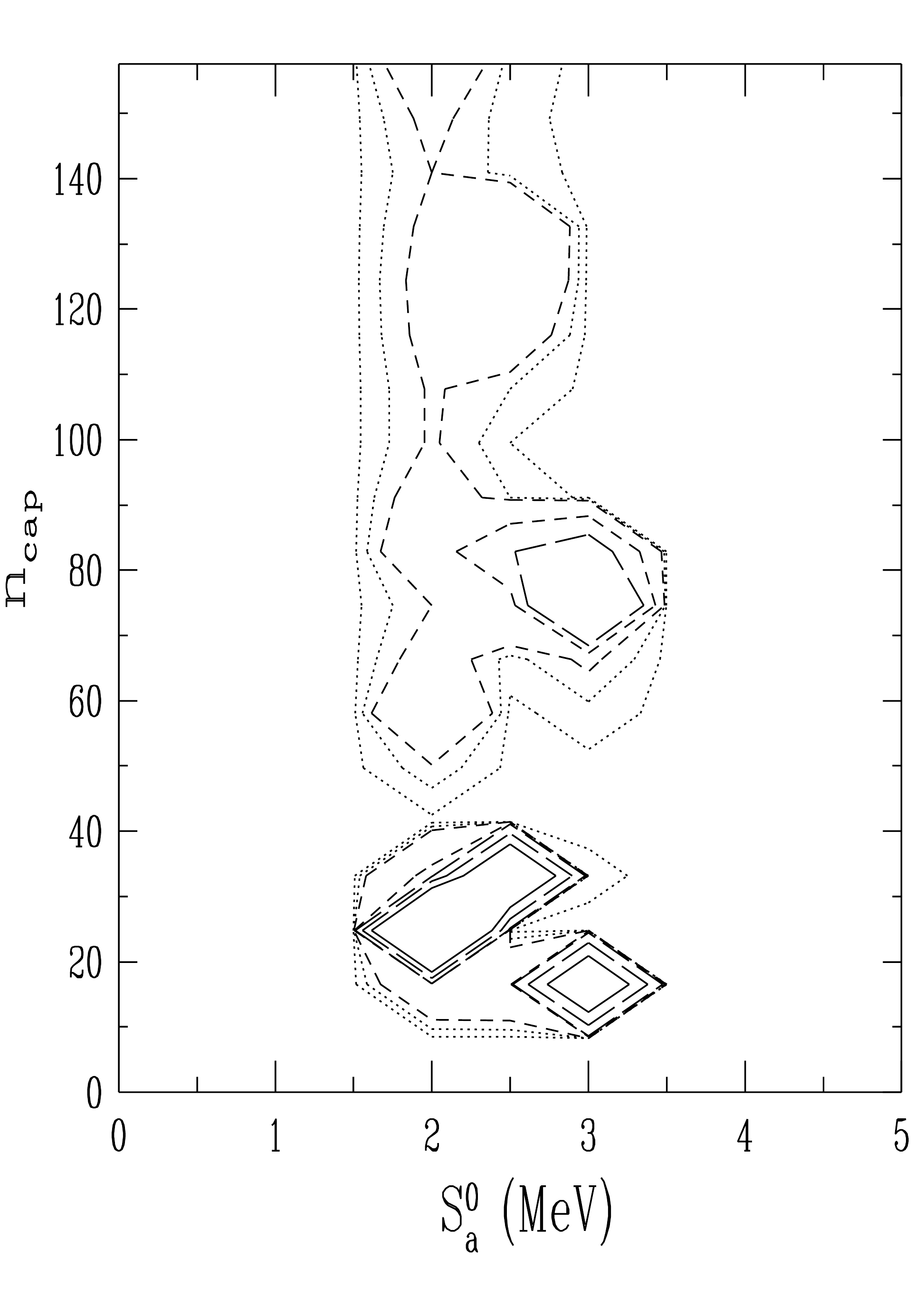,height=9.01cm,width=10.01cm}} 
\vskip-0.2cm
\caption{Statistical distribution in the ($S_{\rm a}^0$,$n_{\rm cap}$) plane of the CEVs
leading to the abundance distribution of Fig.~\ref{fig_multie1}. Dotted contours
correspond to identical statistical weights of  $5~10^{-4}$ and $10^{-3}$. Contours
of weight 10, 100 and 1000 times larger are represented by short-dashed, long-dashed
and solid lines, respectively (from \cite{goriely96})}
\label{fig_multie2}
\end{figure}

MER, similarly to MES (Sect.~\ref{sr-splitting}), relies on an iterative inversion
procedure in order to find the ensemble of CEVs and their corresponding statistical weights
 which provide for a given nuclear input the best fit to a given abundance distribution, and
 in particular the one in the SoS. In this procedure, each nuclide is given a weight that is 
inversely proportional to the uncertainty found to affect its r-abundance (see
 Fig.~\ref{fig_rsol_isot}). Figure~\ref{fig_multie1} illustrates the type of fit to the SoS
 r-abundances that can be obtained from MER. The CEVs required for that fitting are
shown  in Fig.~\ref{fig_multie2}. 

 As MES, MER is a unique and efficient tool to carry out a
systematic study of the impact on the CEV characteristics or to yield predictions of 
 uncertainties of nuclear physics nature \cite{goriely99}, any change in the nuclear input
 translating indeed into a different set of CEVs fitting best a given observed r-nuclide
 distribution.  In this respect, it may be argued that MER has a real
drawback because it can mask nuclear structure effects by introducing spurious CEVs. 
It might be so, but a problem of concurring with this criticism right away is that there 
is no way at this point to distinguish spurious CEVs from real ones. 
This ambiguity is made even more serious as MER as
 well as the classical canonical approach make use of the CEV oversimplification. So far, we 
thus have to live with our inability to make a clear distinction between 
astrophysical- and  nuclear-physics-related deficiencies of the r-process model \cite{goriely97a}. 
 
A final remark is in order here. MER retains all the basic assumptions of the CAR 
model but the WP approximation. In particular, the CEV approach is maintained [assumptions (1),
 (2) and (6) of Sect.~\ref{canonical}]. This CEV approach is clearly questionable, as the
 high-temperature r-process is most likely associated with highly dynamical situations
 (Sect.~\ref{r_dccsn}). In addition,  iron is still the presumed 
 seed for the r-process [assumption (3)
 of Sect.~\ref{canonical}]. This may not be justified, as illustrated by the neutrino wind
 (Sect.~\ref{r_dccsn}) or neutron-star merger (Sect.~\ref{compact_general}) models. In these 
cases, the composition of the material from which the r-process may eventually develop is 
governed by thermodynamic equilibrium either at very high temperature, or at very high
 density. In the neutrino wind model, this corresponds to an initially nucleon-dominated 
composition from which $\alpha$-particles, and subsequently complex nuclei gradually build 
up through charged-particle induced reactions, possibly up to masses as high as about $A = 90$.
 In such conditions, the fit of the SoS r-nuclide abundance curve in the 
$A \lsimeq 90$ region with the help of CEVs (Fig.~\ref{fig_multie1}) is of course just meaningless.

\subsection{MER and the r-nuclide abundance convergence between Ce and Os}
\label{MER_universality}

\begin{figure}
\center{\includegraphics[scale=0.5]{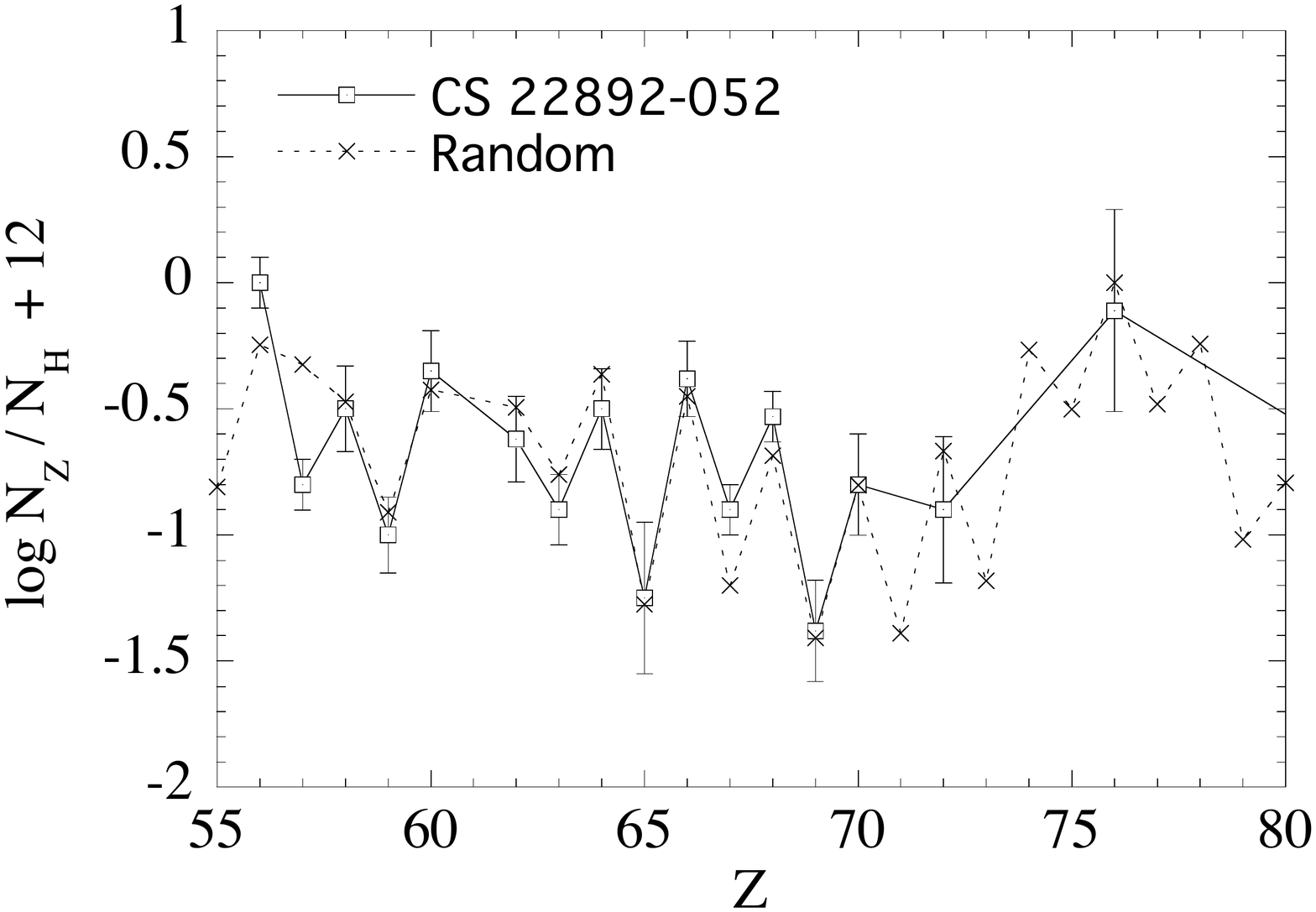}}
\vskip-0.7cm
\caption{Elemental abundances  $N_Z$ (relative to the H abundance $N_H$) in the 
$55 \le Z \le 80$ range obtained from a random superposition of  40 CEVs with characteristics
 provided in Fig.~\ref{fig_cs22892b} (for more details, see \cite{goriely97a}). The observed 
abundances are from \cite{sneden96}} 
\label{fig_cs22892a}
\end{figure}

\begin{figure}
\center{\includegraphics[scale=0.5]{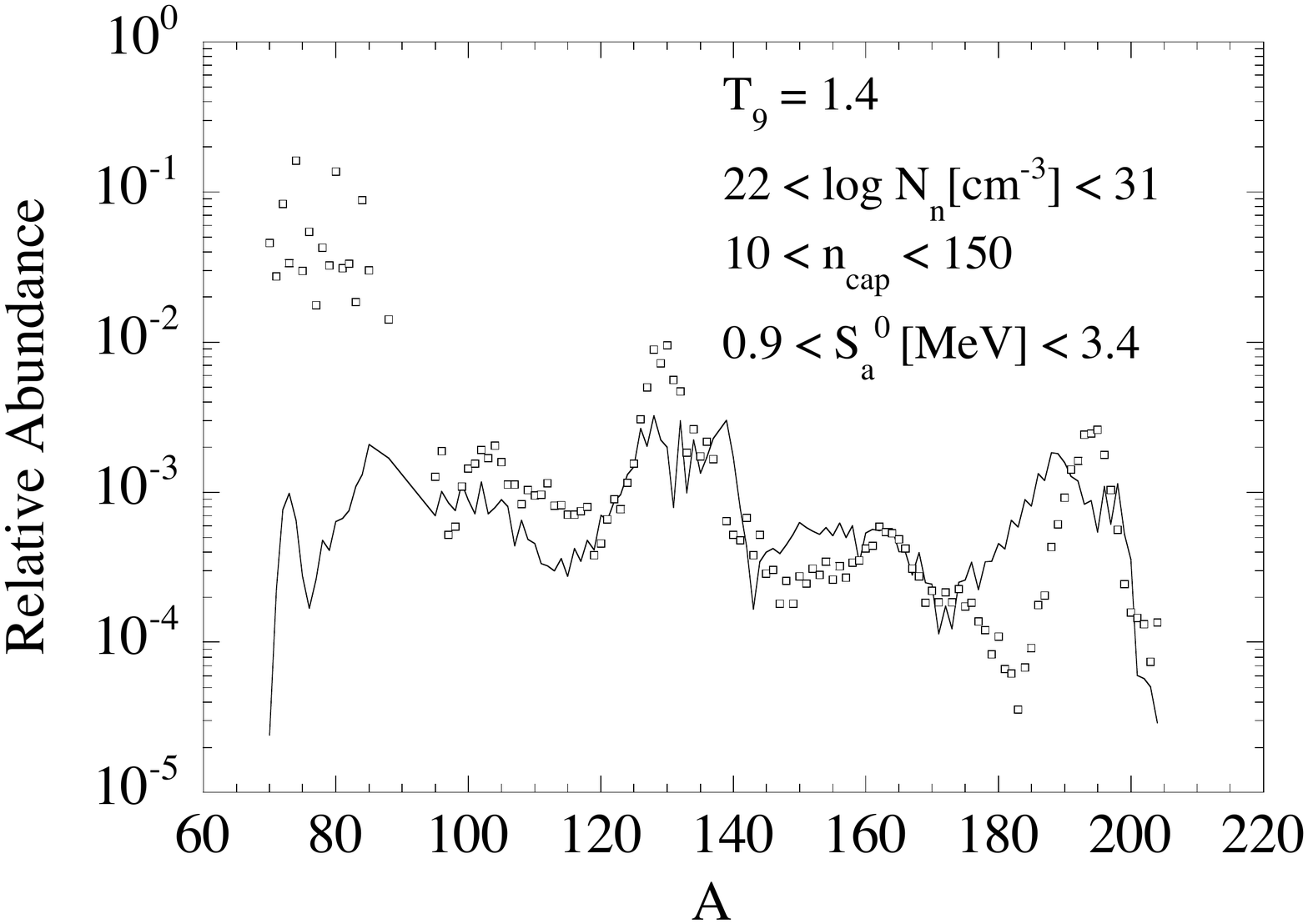}}
\vskip-0.7cm
\caption{Comparison between the SoS r-process abundance curve proposed by 
\cite{kappeler89} and the distribution predicted by the random superposition of CEVs leading
 to Fig.~\ref{fig_cs22892a}. Both distributions are arbitrarily normalised (from 
\cite{goriely97a})}
 \label{fig_cs22892b}
\end{figure}

\begin{figure}
\center{\includegraphics[scale=0.5]{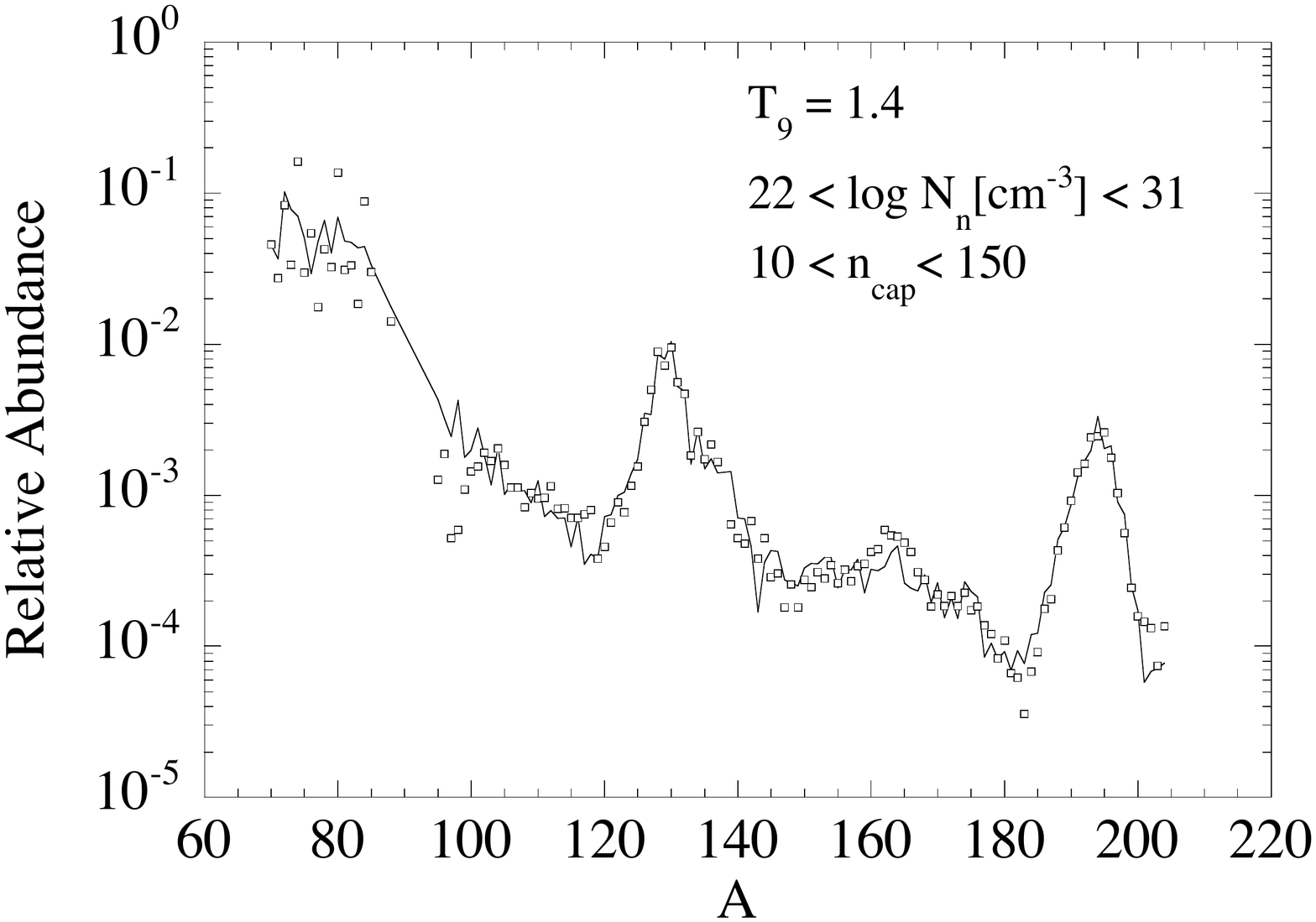}}
\vskip-0.7cm
\caption{Same as Fig.~\ref{fig_cs22892b}, but for the selection of CEVs that reproduce at 
best the whole SoS r-nuclide abundance curve (from \cite{goriely97a})} 
\label{fig_cs22892d}
\end{figure}

\begin{figure}
\center{\includegraphics[scale=0.5]{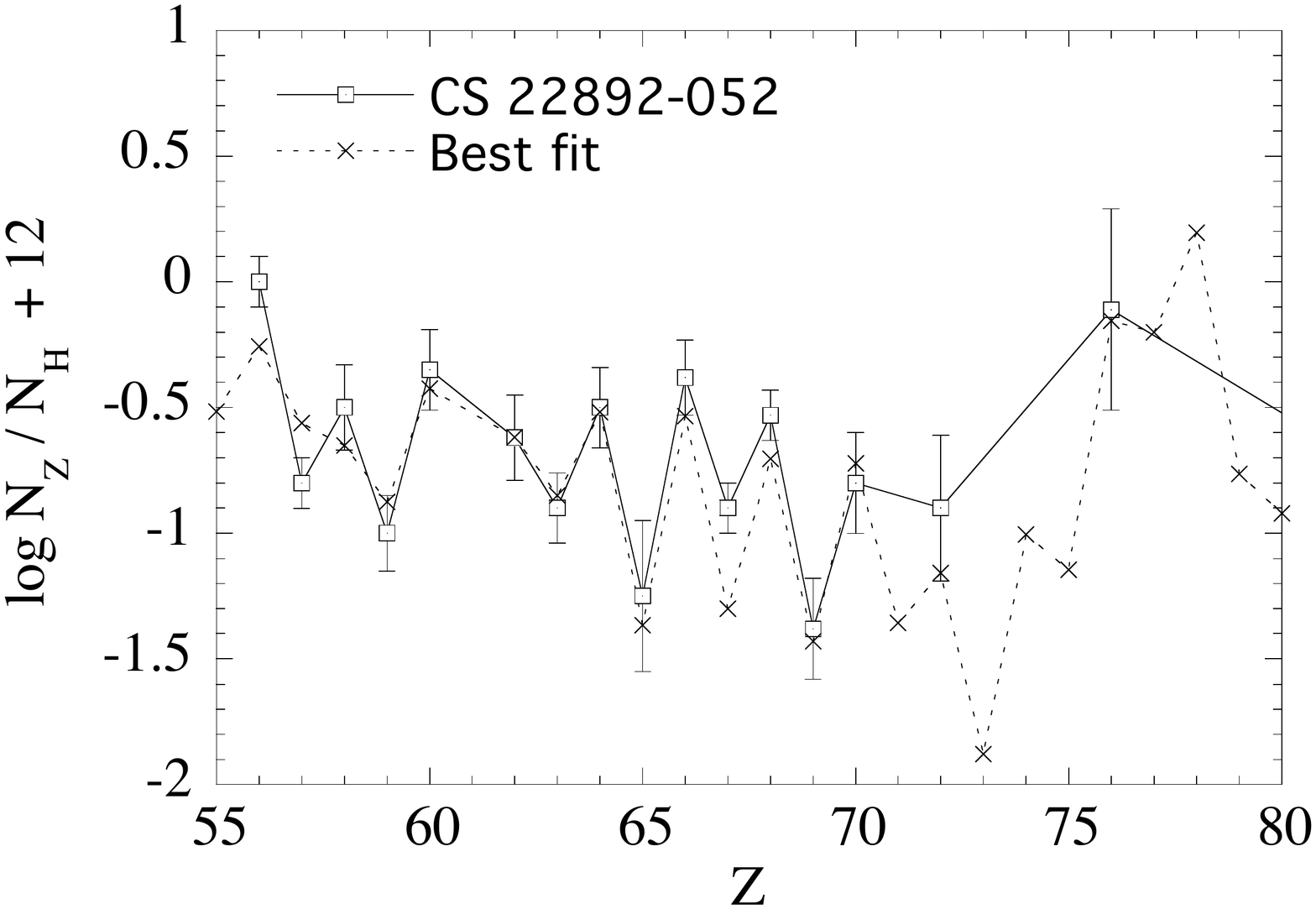}}
\vskip-0.7cm
\caption{Same as Fig.~\ref{fig_cs22892a}, but for the CEVs of Fig.~\ref{fig_cs22892d} (from 
\cite{goriely97a})} 
\label{fig_cs22892c}
\end{figure}

An interesting application of MER concerns the quite striking observation that the patterns of
 abundances of heavy neutron-capture elements in the Ce to Os range observed in r-process-rich
 metal-poor stars show a remarkable similarity to the one found in the SoS, as stressed in 
Sect.~\ref{galaxy_universality}. This has led to a quite frequent claim in the literature 
that the r-process is `universal'.
 
An early word of caution concerning this sort of claim has been given on theoretical grounds 
by \cite{goriely97a}  and  reinforced (see \cite{yushchenko05}  for
 references). A critical evaluation of the universal nature of the r-process relies on the 
interpretation by MER  of the observed r-nuclide content of the star CS 22892-052, and of 
its similarity with the SoS composition in the $58 \leq Z \leq 76$ range (see 
Figs.~\ref{fig_galaxy_specific1} and \ref{fig_galaxy_specific2}). This study demonstrates 
that a {\em random} superposition of CEVs
that is selected to fit nicely the CS 22892-052 abundance pattern in the Ba to Os range 
(see Fig.~\ref{fig_cs22892a})  is  unable to account satisfactorily for the SoS r-nuclide 
content. Neither the position, nor the width or height of this abundance distribution are
 reproduced (see Fig,~\ref{fig_cs22892b}). This mismatch is  not surprising for the 
SoS can be satisfactorily fitted by a specific selection of CEVs (e.g. \cite{goriely96}). As 
a complement, it is shown that a superposition of CEVs that fits best the whole SoS r-nuclide
 abundance curve (Fig.~\ref{fig_cs22892d}) is also able to fit the CS 22892-052 Ba to Os 
abundances (Fig.~\ref{fig_cs22892c}).

The main conclusions drawn by \cite{goriely97a} from these results are that (1) the pattern 
of abundances in the Ba to Os range is mainly governed by nuclear physics properties (and in
 particular by the fact that even $Z$ elements have more stable isotopes that can be fed by 
the r-process), so that a possible universality in this $Z$ range does not tell much about
 specific astrophysical conditions,  and (2) the convergence of abundances in the above 
mentioned range does not provide any demonstration of any sort of a more global universality 
involving lighter and heavier elements. As already stressed in 
Sect.~\ref{galaxy_universality}, these reservations have received mounting support from 
observation. Section~\ref{chronometry_lowz} discusses more specifically the consequences of
 the likely non-universality of the actinides production on the reliability of attempts to 
develop galactic chronologies from the observed actinides content of very metal-poor stars.  

\subsection{Dynamical r-process approaches (DYR)}
\label{DYR}

In associating the r-process with supernova explosions, several attempts to go beyond the CAR
 model (Sect.~\ref{canonical}) have been made by taking into account some evolution of the 
characteristics of the sites of the r-process during its development. The earliest of these
 models,  coined as `dynamical'  (DYR) in the following as a reminder  of the time variations 
of the thermodynamic state of the r-process environment, have been proposed by 
\cite{sato74,hillebrandt76,cameron70,delano71,schramm73}. These DYR models 
by and large do not rely on any specific explosion model. They consider 
 a material that is initially hot enough
 for allowing a nuclear statistical equilibrium (NSE) to be achieved expands and cools in a 
prescribed way on some selected timescale. This evolution is 
in general highly parametrised, with a notable exception of the hydrodynamical
 treatment by \cite{hillebrandt76}.

With the requirement of charge and mass conservation, and if the relevant nuclear binding 
energies are known, the initial NSE composition is determined from the application of the 
nuclear Saha equation (e.g. Sect. 7-2 of \cite{clayton68} for a general presentation) for an
 initial temperature and density  (or, equivalently, entropy, following 
Eqs.~\ref{eq:wind_srad}), and electron fraction (or net electron number per baryon) 
$Y_{\rm e}$, that are free parameters in a site-free r-process approach. The evolution 
of the abundances during expansion and cooling of the material from the NSE state is derived
 by solving an appropriate nuclear reaction network. The freeze-out of the charged-particle
 induced reactions might be followed by an r-process during which the abundances are 
calculated from the set of equations of the form of Eq.~\ref{eq_net}. 

\begin{figure}
\center{\includegraphics[width=0.85\textwidth,height=0.75\textwidth]{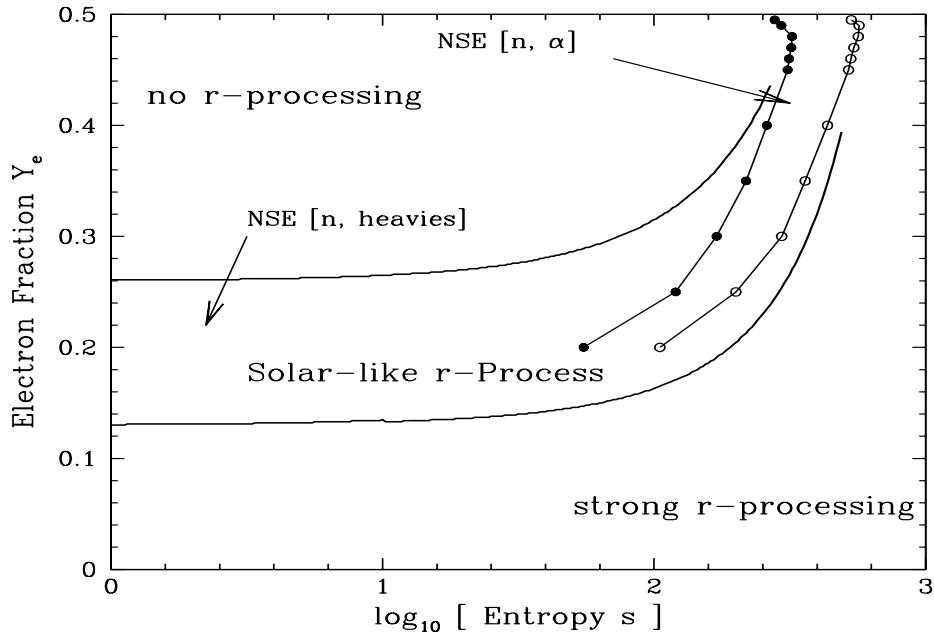}}
\vskip-2.1cm
\caption{The likelihood of a DYR r-process for given combinations of the electron fraction
$Y_{\rm e}$ and the entropy per baryon $s$ (hereafter in units of $k$). A SoS-like r-process is expected for a
 suitable superposition of  conditions between the black lines. The results inferred from an
 initial NSE phase at low $s$ (see Fig.~18 of \cite{ruffert97}) are smoothly connected to 
those of various nuclear network calculations \cite{witti94a,witti94,hoffman97} for  high 
$s$ values. In the latter cases, the assumed expansion timescales imply that the freeze-out 
of the charged-particle induced reactions is reached for dynamical timescales 
$\tau_{\rm dyn}$ (see Eq.~\ref{eq:wind_time}) in excess of about 50 - 100 ms. The two
 lines with dots represent the contours of successful r-processing for
 $\tau_{\rm dyn} = 50$ ms (black dots) and 100 ms (open dots) (see \cite{hoffman97} for 
details)}
\label{fig:wind_yes} 
\end{figure}
 
As temperature, density and $Y_{\rm e}$ are free parameters in a site-free approach, many
 choices of initial NSE compositions may clearly be made, involving a dominance of light or 
heavy nuclides, as illustrated in Fig.~\ref{fig:wind_yes}. However, in view of its relevance
 to the supernova models reviewed in Sect.~\ref{explo_1D} and \ref{explo_multiD}, we limit 
ourselves here to the consideration of an initial NSE at temperatures of the order of 
$10^{10}$ K which favours the recombination of essentially all the available protons into 
$\alpha$-particles (the region noted NSE [n,$\alpha$] in Fig.~\ref{fig:wind_yes}). The 
evolution of this initial composition to the stage of charged-particle induced reaction 
freeze-out has been analysed in detail by \cite{meyer98a}, and we just summarise here some of 
its most important features that  are of relevance to a possible subsequent r-process:\\
(1) at some point in the course of the expansion and cooling of the initially $\alpha$-rich 
material, full NSE breaks down as the result of the slowness of a fraction of the 
charged-particle reactions relative to the expansion timescale. The formation of
 quasi-equilibrium (QSE) clusters results. In this state, the intra-QSE composition still
 follows the NSE Saha equation, but the relative inter-cluster abundances do not, and depend
 on the kinetics of the nuclear flows into and out of the QSE clusters. To be more specific,
 the QSE phase is dominated in its early stages by a light cluster made of neutrons and
 $\alpha$-particles and traces of protons, and by a heavy cluster made of \chem{12}{C} and 
heavier species. The population of the latter is determined mainly by the $\alpha + \alpha + 
n$ reaction, followed by
  $^9{\rm Be}(\alpha,n)^{12}{\rm C}(n,\gamma )^{13}{\rm C}(\alpha,n)^{16}{\rm O}$, as first 
noticed by \cite{delano71};\\
(2) as the temperature decreases further, the QSE clusters fragment more and more into 
smaller clusters until total breakdown of the QSE approximation, at which point the
 abundances of all nuclides have to be calculated from a full nuclear reaction network. In
 the relevant $\alpha$-particle-rich environment, the reaction flows are dominated by
 $(\alpha,\gamma)$ and $(\alpha,n)$ reactions with the addition of radiative neutron captures.
 Nuclei as heavy as Fe or even beyond may result. For a low enough temperature, all
 charged-particle induced reactions freeze-out, only neutron captures being still possible.
 This freeze-out is made even more efficient if the temperature decrease is accompanied by 
a drop of the density $\rho$, which is especially efficient in bringing the operation of the 
$\rho^3$-dependent $\alpha + \alpha + n$ reaction to an end.

In the following, the process summarized above, which develops in a medium that is both neutron-rich and $\alpha$-rich at freeze-out of the charged-particle induced reactions, will be referred to
 as the $\alpha$-process for simplicity, and for keeping the terminology introduced by 
\cite{woosley92} in order to avoid further confusion (the $\alpha$-process of \cite{BBFH57}
 refers to a different nuclear process).

The composition of the material at the time of freeze-out depends on the initial
 $Y_{\rm e}$, on the entropy  $s$ (see \cite{meyer98a} for a detailed discussion), as well
 as on the dynamical timescale $\tau_{\rm dyn}$. The heavy nuclei synthesised at that
 moment may have on average neutron numbers close to the $N = 50$ closed shell, and average 
mass numbers around $A = 90$.
 These nuclei can be envisioned to be the seeds for a subsequent 
r-process, in replacement of the iron peak assumed in the CAR model (Sect.~\ref{canonical}). 
For a robust r-process to develop, favourable conditions have to be fulfilled at the time of
 the $\alpha$-process freeze-out. In particular, the ratio at that time of the neutron 
concentration to the abundance of heavy neutron-rich seeds has to be high enough for allowing
  the heaviest r-nuclides to be produced. As an example, $A = 200$ nuclei can be 
produced if an average of 110 neutrons are available per $A = 90$ nuclei that could emerge
 from the $\alpha$-process. The availability of a large enough number of neutrons per seed
 can be obtained under different circumstances: (i) at high enough entropies (high enough 
temperatures and low enough densities), even in absence of a large neutron excess, as it is 
the case if  $Y_{\rm e}$ is close to 0.5  \cite{woosley92}, (ii) at lower entropies if
 $Y_{\rm e}$ is low enough, and/or (iii) if the temperature decrease is fast enough for 
avoiding a too prolific production of heavy seeds. Figure~\ref{fig:wind_yes} sketches in a
 semi-quantitative way the conclusions of the discussion above concerning the likelihood of
 development of a successful r-process in terms of entropy and $Y_{\rm e}$. 

Some predictions from a DYR-type of r-process based on dynamical conditions inspired by 
supernova simulations will be presented in Sect.~\ref{r_dccsn}.
 
\section{A site-free high-density  r-process scenario (HIDER)}       
\label{HIDER}

\begin{figure}
\centerline{\epsfig{figure= 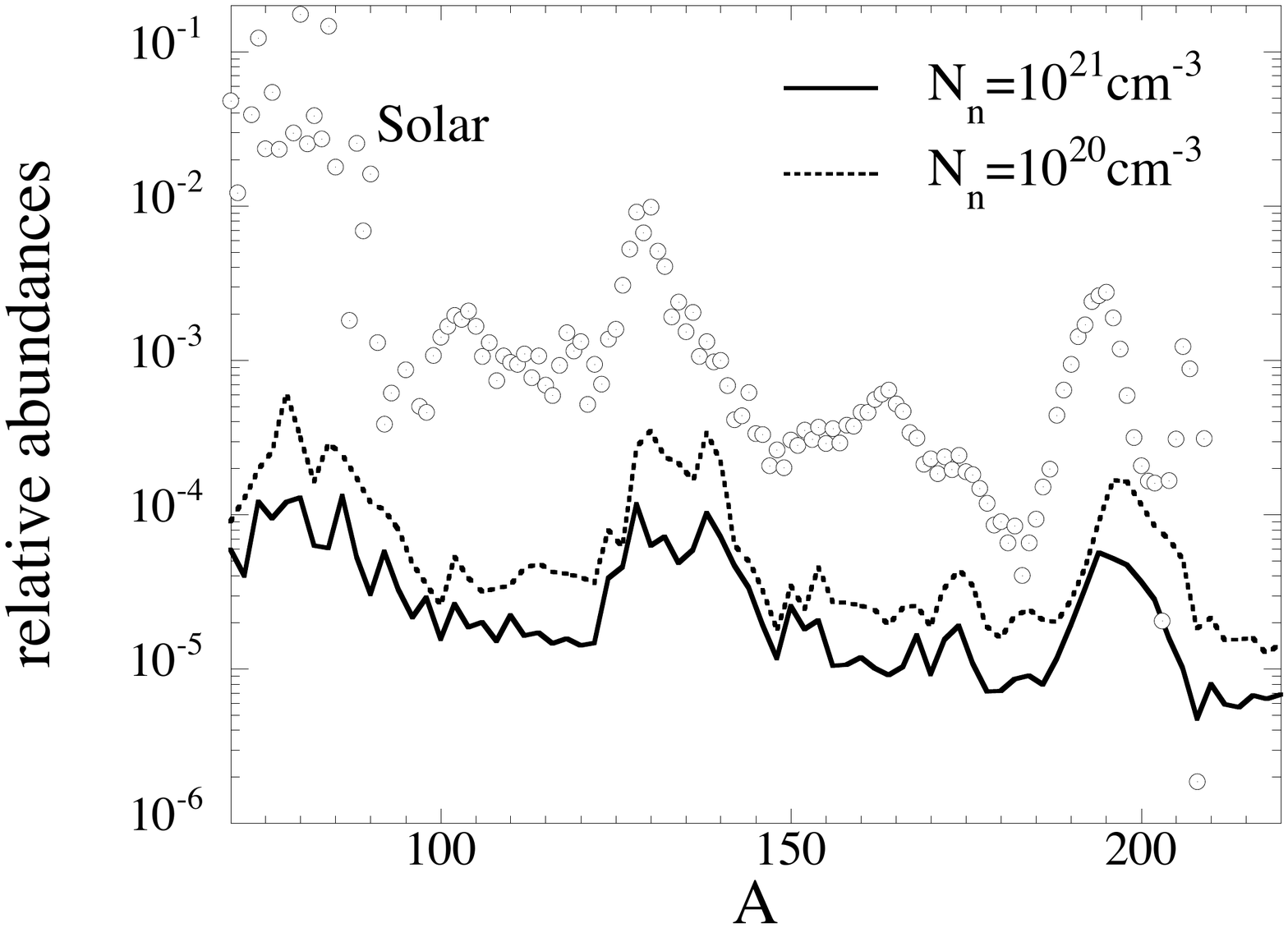,height=8cm,width=14.0cm}} 
\vskip-0.5cm
\caption{Abundance distributions predicted by the steady flow HIDER for $N_{\rm n}=10^{20}$ and 
$10^{21}~{\rm cm}^{-3}$. The neutron capture and $\beta$-decay rates are derived from a
 Hauser-Feshbach calculation, and from the GT2 approximation, the HFB-9 nuclear masses being 
adopted throughout. The solar abundances are shown for illustrative purposes}
\label{fig_steady2}
\end{figure}

Early in the development of the theory of nucleosynthesis, an alternative to the high-T
 r-process canonical model (Sects.~\ref{canonical} and \ref{MER}) has been proposed 
\cite{tsuruta65}. It relies on the fact that very high densities (say
 $\rho > 10^{10}$ gcm$^{-3}$) can lead material deep inside the neutron-rich side of the
 valley of nuclear stability as a result of the operation of endothermic free-electron
 captures (e.g. \cite{AK99} for a short review), this so-called `neutronisation' of the 
material being possible even at the $T = 0$ limit. The astrophysical plausibility of this
 scenario in accounting for the production of the r-nuclides has long been questioned, and
 has remained largely unexplored until the study of the composition of the outer and inner 
crusts of neutron stars \cite{baym71} and of the decompression of cold neutronised matter 
resulting from tidal effects of a black hole on a neutron-star companion \cite{lattimer77}. 
The decompression of cold neutron star matter has recently been studied further 
(Sect.~\ref{compact_general}).
 
In view of the renewed interest for a high-density r-process, a simple steady-flow model,
 referred to in the following as HIDER, may be developed.  Irrespective of the specific 
details of a given astrophysical scenario, it allows to follow in a very simple and
 approximate way the evolution of the composition of an initial electron-degenerate 
neutronised matter under the combined effect of $\beta$-decays and of the captures of free
 neutrons that are an important initial component of the considered material. These are the
 only two types of transformations that have to be considered if fissions are disregarded, 
and if any heating of the material resulting from the $\beta$-decay energy deposition is
 neglected, so that photo-disintegrations can be ignored (this assumption has been made by 
\cite{lattimer77} as well). Under such assumptions, the evolution of  the abundance of 
nucleus $(Z,A)$ follows

\begin{eqnarray}
\frac{{\rm d}N(Z,A)}{{\rm d}t} 
& = & \lambda_{{\rm n}\gamma}^{Z,A-1} N(Z,A-1)  +  \lambda_{\beta}^{Z-1,A}~ N(Z-1,A) \nonumber \\  
& - &  \lambda_{{\rm n}\gamma}^{Z,A} ~N(Z,A) -  \lambda_{\beta}^{Z,A}~ N(Z,A),
\label{eq_steady2a}
\end{eqnarray}

where the symbols have the same meaning as in Eq.~\ref{eq_net}, at least if it is assumed that
the neutrons obey a Maxwell-Boltzmann distribution. (Otherwise, $ \langle\sigma v\rangle$.
in Eq.~\ref{eq_rate} would have to be obtained by integrating over a Fermi-Dirac distribution.)
For non-relativistic neutrons (the only situation of practical interest here) this  limits
the applicability of Eq.~\ref{eq_steady2a} to temperatures in excess of about 
$(N_{\rm n}[\rm {cm^{-3}}]/10^{20})^{2/3}$ K.
This constraint is met in all cases of practical interest (Sect.~{\ref{compact_general}). 
 By summing over all the isobars of
 a given isobaric chain $A$, Eq.~\ref{eq_steady2a} transforms into 

\begin{equation}
\frac{{\rm d}N(A)}{{\rm d}t}   =  \lambda_{\rm n}^{A-1} N(A-1)  -  \lambda_{\rm n}^{A} ~N(A), 
\label{eq_steady2b}
\end{equation}

where $N(A)=\sum_{Z \geq Z_{\rm min}(A)} N(Z,A)$ and $\lambda_{\rm n}^{A}=\sum_{Z \geq Z_{\rm min}(A)} 
\lambda_{\rm n}^{Z,A} N(Z,A)/N(A)$, with $Z_{\rm min}(A)$ being the minimum $Z$ contributing to the isobaric
 chain $A$.

The ratio $N(Z,A)/N(A)$ represents the contribution of nucleus $(Z,A)$ to the isobaric chain
 $A$. It is essentially a function of the branching ratio between its $\beta$-decay and 
free-neutron capture probabilities, so that it may be approximated by 
$N(Z,A)/N(A)=\lambda_{\beta}^{Z,A}/[\lambda_{\rm n}^{Z,A}+\lambda_{\beta}^{Z,A}]$. In such an 
approximation, the steady-state solution of Eq.~\ref{eq_steady2b} takes the form

\begin{equation}
N(A)   = N_0  \frac{1}{\lambda_{\rm n}^A} =N_0 \Big( \sum_Z \frac{ \lambda_{\beta}^{Z,A}
  \lambda_{\rm n}^{Z,A}}{ \lambda_{\beta}^{Z,A} + \lambda_{\rm n}^{Z,A}} \Big)^{-1},
\label{eq_steady2c}
\end{equation}

where $N_0$ is a normalisation constant.

The predictions of a steady-flow HIDER are illustrated in  Fig.~\ref{fig_steady2}. From a 
comparison with Fig.~\ref{fig_steady1}, this very simple model is seen to do roughly as well in 
reproducing the three SoS abundance peaks as a steady state CE for comparable neutron
densities. It also shows that a high-$T$ environment is not a must in order to account
  either for the location, or for the width of the observed SoS r-abundance peaks.

\section{Some generalities about the evolution of massive stars}
\label{evol_massive}
 
Very early in the development of the theory of nucleosynthesis, and in particular following 
the predictions of the canonical high-temperature r-process model (Sect.~\ref{canonical}), 
the inner regions of massive stars undergoing a supernova explosion have been considered as 
a viable r-process site. With time, a better characterisation of the supernova mechanisms 
that could provide suitable conditions for the r-process has been attempted. The result is 
that the most-likely progenitor stars are those that could develop pre-explosively an iron 
core or a massive enough O-Ne core that could eventually lead to a core collapse supernova 
(CCSN). In the case of single stars, this restricts the relevant objects to those in the
 approximate $9 \lsimeq M \lsimeq$ 100 $M_\odot$ mass range (these limiting masses are still
 quite uncertain, and probably depend in particular on metallicity and pre-explosion mass-loss
 rates, not to mention rotation).  Observationally, these massive star CCSNe may be of
 Type I (SNI) or II (SNII). A SNII exhibits H-lines in its spectrum at maximum light, as it
 is expected when the star has retained its H-envelope up to the time of the explosion. 
In contrast, a SNI spectrum is devoid of H lines. Depending upon additional features, SNI 
 a, b or c subtypes are defined.
  SNIb/c explosions are classically considered to result from the explosion of massive stars having
 lost their H-envelopes through steady winds. While SNIa are generally not viewed as CCSN 
events, some rare SNIa explosions could well be of this type (See Sect.~\ref{r_aic}). 

One has to stress at this point that the pre-supernova structure of the $9 \lsimeq M \lsimeq$ 
100 $M_\odot$ stars remain very uncertain. The treatment of convective instabilities, rotation,
 magnetic fields and mass loss, even when
 treated in a one-dimensional approximation, bring their share of intricacies, not 
to mention  binarity. The complications clearly culminate when
 multi-dimensional simulations are attempted. As stressed by  \cite{meakin06} (and references
 therein), due consideration of internal-wave physics and of different symmetry-breaking 
mechanisms (density perturbations induced by turbulence, wave interactions between burning 
shells, rotationally-induced distortions) should be included in the modelling of stars, and 
might deeply affect the predicted pre-supernova models. The still quite uncertain knowledge of
 the stellar models that define the initial conditions for supernova simulations
reduces further the reliability of the current modellings of the explosions themselves.

\subsection{A one-dimensional perspective of the explosive evolution of massive stars}
\label{explo_1D}

At the end of their nuclear-powered evolution, stars between about 10 and 100 $M_\odot$ develop a
 core made of nuclides of the iron group (`iron core') at temperatures in excess of about $4 
\times 10^9$ K. As these nuclides have the highest binding energy per nucleon, further nuclear
 energy cannot be released at this stage, so that the iron core contracts and heats up. This 
triggers endothermic photo-disintegrations of the iron-group nuclides down to 
 $\alpha$-particles, and even nucleons. The corresponding energy deficit is accompanied by
 a pressure decrease which can be  responsible of the acceleration of the contraction into a
 collapse of the core. Endothermic electron captures can make things even worse. To a first 
approximation, this gravitational instability sets in near the classical Chandrasekhar mass 
limit for cold white dwarfs, $M_{\rm Ch} = 5.83\,Y_{\rm e}^2$, $Y_{\rm e}$
 being the electron mole fraction. In the real situation of a hot stellar core,  collapse may
 start at masses that differ somewhat from this value, depending on the details of the core 
equation of state.

The gravitational collapse of the iron core does not stop before the central densities exceed
 the nuclear matter density  $\rho_0\approx 2.5\times 10^{14}\,$g$\,$cm$^{-3}$ by about a 
factor of two. At this point, the innermost ($M \lsimeq$ 0.5 $M_\odot$) material forms an 
incompressible, hot and still lepton-rich `proto-neutron' star (PNS) whose collapse is stopped
 abruptly. A shock wave powered by the gravitational binding energy released in the collapse
 propagates supersonically into the infalling outer layers. For many years, 
it has been hoped that this shock could be sufficiently strong for ejecting explosively most of the 
material outside the core, producing a so-called `prompt core collapse supernova' (PCCSN) 
with a typical kinetic energy of 1--$2\times 10^{51}\,$ergs, as observed. The problem is that
 the shock is formed roughly half-way inside the iron core, and looses a substantial fraction
 of its energy in the endothermic photo-disintegrations of the iron-group nuclei located in 
the outermost portion of the core. The shock energy loss is aggravated further by the escape
 of the neutrinos produced by electron captures on the abundant free protons in the
 shock-heated material. Detailed one-dimensional hydrodynamic simulations conclude that the
 initially outgoing shock wave transforms within a few milliseconds after bounce into an 
accretion shock. The matter behind the shock continues to accrete on the PNS. The bottom line 
is that no recent simulation is able to predict a successful PCCSN  for a Fe-core progenitor 
star ($M \gsimeq$ 10 $M_\odot$). This failure is illustrated in Fig.~\ref{ccsn_fail} for a 15 
$M_\odot$ star.

\begin{figure}
\center{\includegraphics[scale=0.6,angle=270]{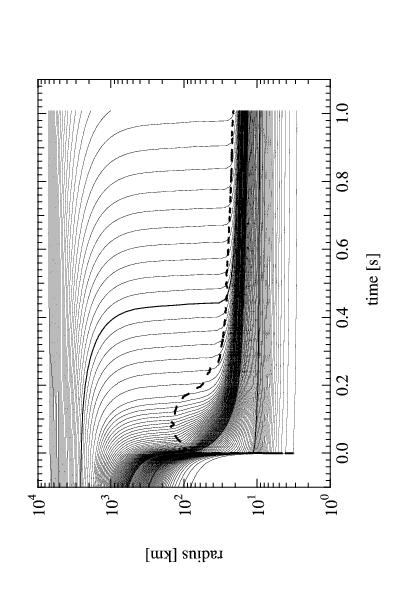}}
\vskip-0.6cm
\caption{Radial trajectories of several mass elements of the core of a 15 $M_\odot$ star
 versus time after bounce. The trajectories are plotted for each 0.02 $M_\odot$ up to
 1 $M_\odot$, and for each 0.01 $M_\odot$ outside this mass. The thick solid lines correspond
 to 0.5, 1.0, and 1.5 $M_\odot$. The thick dashed line indicates the location of the shock 
wave. The prompt shock stalls within 100 ms after reaching 150 km, and recedes down to below
 100 km. No sign of a revival of the shock that possibly leads to a successful D(elayed-)CCSN is seen 
either, even after 300 ms. Instead, a stationary accretion shock forms at several tens of km. 
A PNS is seen to form, reaching 1.6 $M_\odot$ around 1 s after bounce (from \cite{sumiyoshi05})}
\label{ccsn_fail}
\end{figure}

Even so, some hope to get a CCSN of a non-prompt type has been expressed if there is a way to
 `rejuvenate' the shock efficiently enough to obtain an explosive ejection of the material 
outside the PNS. This rejuvenation remains a matter of intensive research. Neutrinos might
 well play a pivotal role in this matter. They are produced in profusion from the internal 
energy reservoir of the PNS that cools and deleptonises hundreds of milliseconds after bounce,
 and their total energy might amount to several $10^{53}\,$ergs, that is about 100 times the
 typical explosion energy of a SNII. The deposition of a few percent of this energy would thus
 be sufficient to unbind the stellar mantle and envelope, and provoke a `delayed' CCSN (DCCSN)
 (these qualitative statements assume that a black hole is not formed instead of a PNS; see 
below). 
Many attempts to evaluate the precise level of neutrino energy deposition have 
been conducted over the last decades, based on more or less controversial simplifications of 
the treatment of the neutrino transport (e.g. \cite{liebendorfer05} for a recent re-analysis 
of the problem, which is made even more complex by the due consideration of neutrino flavor 
mixing). In fact, theoretical investigations and numerical simulations performed with
 increasing sophistication over the past two decades have not been able to come up with a clearly
 successful CCSN for a Fe-core progenitor ($M \gsimeq$ 10 $M_\odot$). This conclusion is 
apparently robust to changes in the highly complex physical ingredients (like the neutrino 
interactions, or the equation of state), and in the numerical techniques (e.g. 
\cite{liebendorfer05}). In fact, the neutrino-energy deposition should have to be 
significantly enhanced over the current model values in order to trigger an explosion. 
An illustration of a failed DCCSN is shown  in Fig.~\ref{ccsn_fail}. 

\begin{figure}
\center{\includegraphics[width=0.75\textwidth]{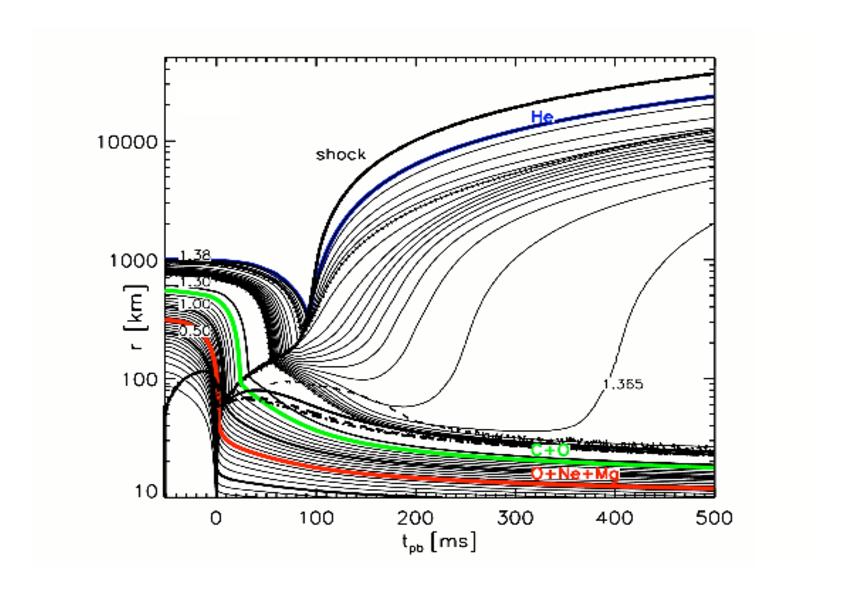}}
\vskip-0.3truecm
\caption{Simulation of an electron-capture supernova following the collapse of an O-Ne core. 
The time evolution of the radius of various mass shells is displayed with the inner boundaries
 of the O+Ne, C+O and He shells marked by thick
 lines. The inner core of about
 0.8 $M_\odot$ is mainly made of Ne at the onset of collapse (\cite{nomoto87}, and references
 therein). The explosion is driven by the baryonic wind caused by neutrino heating around the
 PNS.
 The thick solid, dashed, and dash-dotted lines mark the neutrino spheres of
 $\nu_{\rm e}$, $\bar\nu_{\rm e}$, and heavy-lepton neutrinos, respectively. The
 thin dashed line indicates the gain radius which separates the layers cooled from those heated 
 by the neutrino flow. The thick 
line starting at $t = 0$ is the outward moving supernova shock (from \cite{kitaura06})} 
\label{fig_onecore}
\end{figure}

This adverse circumstance may not mark the end of any hope to get a DCCSN, however. In the case
 of single stars considered here, one might just have to limit the considerations to the
 $\sim$9 to 10 $M_\odot$ stars that possibly develop O-Ne cores instead of iron cores at the 
termination of their hydrostatic evolution. Efficient endothermic electron captures could 
trigger the collapse of that core, which could eventually transform into a so-called 
electron-capture supernova that may be of the SNIa or SNII type, depending upon the extent
 of the pre-explosion wind mass losses.\footnote{The range of initial masses of single stars
 which could experience an electron-capture instability is still quite uncertain, and depends 
in particular on a subtle competition between the growth of the stellar cores resulting from 
thermal pulses developing during the Asymptotic Giant Branch evolution and their erosion 
resulting from steady mass-losses. Other stars in the approximate 8 to 12 $M_\odot$ mass 
range might end up as O-Ne white dwarfs or experience of Fe core collapse instead of 
experiencing an electron-capture supernova resulting from the collapse of the O-Ne core 
(\cite{siess06} for a review). Binary systems might offer additional opportunities of
 obtaining electron-capture supernovae (see Sect.~\ref{explo_multiD})}
 It was once claimed that these explosions could even be of the PCCSN type (e.g.
  \cite{hillebrandt84}). As illustrated in Fig.~\ref{fig_onecore}, this is not confirmed by 
recent 1D simulations \cite{kitaura06}. However, and in contrast with the conclusions drawn
 for more-massive stars, a successful DCCSN is obtained. The neutrino heating is efficient
 enough for rejuvenating the shock wave about 150 ms after bounce, and mass shells start 
being ablated from the PNS surface about 50 ms later, leading to a so-called `neutrino-driven
 wind'.\footnote{Unless otherwise stated, neutrino-driven winds refer to transonic as well as
 subsonic winds, the latter being referred to as breeze in Sect.~\ref{r_dccsn}} 
 No information is provided by the current simulations on the conditions at times much later 
than a second after bounce. Note that the predicted successful delayed electron-capture 
supernova is characterised by a low final explosion energy (of the order of $0.1 
\times 10^{51}$ ergs, which is  roughly ten times lower than typical SN values), and by just a
 small amount of ejected material (only about 0.015 $M_\odot$). These features might suggest
 a possible connexion with some sub-luminous SNII events and with the Crab nebula
 \cite{kitaura06}. This successful electron-capture DCCSN simulation is in qualitative 
agreement with earlier calculations \cite{mayle88}, even if some important differences exist,
owing in particular to differences in the micro-physics (e.g. the neutrino-transport algorithm).
 Some of these differences concern the properties of the neutrino-driven wind, which has been
 envisioned as a possible site of the r-process in DCCSNe (Sect.~\ref{r_dccsn}). They may 
thus have an important impact on the nucleosynthesis questions of direct relevance here. 
It has also to be recalled that the structure of the progenitors of the electron-capture 
supernovae remains especially uncertain (e.g. \cite{siess06}), which endangers any conclusion 
one may draw on these SN types.

Note that the outcome of a failed CCSN is the transformation of the PNS into a black hole
 through the fall back onto the neutron star of the material that cannot be shock-ejected.
 A black hole is even expected to form `directly' instead by fall back in
 $M \gsimeq$40 $M_\odot$ non-rotating stars, at least under the assumption of no strong mass
 losses.  In fact, this assumption is likely to be invalid for a large fraction at least of
 the not-too-low metallicity $M \gsimeq$ 40 $M_\odot$ stars which transform through strong 
steady mass losses  into Wolf-Rayet stars (e.g. \cite{maeder96}) that might eventually 
experience a CCSN.

\subsection{A multi-dimensional perspective of the explosive evolution of massive stars}
\label{explo_multiD}

A major effort has been put recently in the development of simulations of explosions that go
 beyond the one-dimensional approximation. This is motivated not only by  the difficulty of
 obtaining successful CCSNe in one-dimensional simulations, but also by the mounting 
observational evidence that SN explosions deviate from spherical symmetry, not to mention
 the possible connexion between the so-called soft long-duration gamma-ray bursts, and grossly
 asymmetric explosions accompanied with narrow jets of relativistic particles, referred to as 
JetSNe.  The multi-dimensional extension of the simulations opens the potentiality to treat 
in a proper way different effects that may turn out to be essential in the CCSN or JetSNe
 process. As briefly reviewed by e.g. \cite{mezzacappa05},  they include fluid instabilities,
 or rotation and magnetic fields on top of the neutrino transport already built into the 
one-dimensional models. Acoustic power may be another potential trigger of CCSNe 
\cite{burrows06}.

Fluid instabilities deeply alter the one-dimensional views about the neutrino reheating
 scenario. Multi-dimensional simulations indicate that the outer PNS layers and the region 
behind the shock wave may exhibit convective instabilities supported by the neutrino heating.
 They affect in one way or another the PNS or neutrino-driven wind properties. In particular,
 the convection that is predicted to develop just behind the shock (referred to as `hot bubble 
convection') may have a strong impact on the explosion itself by enhancing an already
 efficient neutrino reheating of the shock.  An instability of the diffusive type leading to 
the development of so-called `lepto-entropy fingers' has also been identified. Its role in
 the CCSN phenomenon remains to be scrutinised further. Finally, a new hydrodynamic 
non-radial instability of the stalled accretion shock wave, referred to as `stationary 
accretion shock instability',  has been discovered recently in numerical simulations 
\cite{blondin03} after its analytic prediction. This instability is now considered as a
 serious active participant to the generation of CCSNe, and may also be held responsible for
 the observed neutron star kicks (e.g. \cite{scheck06} for references). It might even lead to
 bipolar explosions of the JetSN type even in absence of rotation, to the polarisation of the 
light, or anisotropy and patchy composition structure observed in many SN remnants, and
 especially in the most extensively studied SN1987A (see \cite{wang02}, and references 
therein).

An alternative CCSN mechanism has recently been proposed by \cite{burrows06} based on 2D 
simulations of the post-collapse phases of a non-rotating 11 $M_\odot$ star. It relies on the
 generation in the core and the propagation into the outer layers of strong sound waves that 
are a very efficient means to transport energy and momentum, sound being almost 100\% absorbed
 in the matter. Interestingly enough, the mechanism is found to be efficient well after (more 
than  550 milliseconds after bounce in the simulations of \cite{burrows06}) the neutrino 
reheating mechanism and stationary accretion-shock instability have failed to provoke a 
successful explosion. In fact, the acoustic powering of the shock remains efficient as long 
as the accretion onto the PNS continues, and consequently plays a regulating role as the
 accretion stops at the time of an eventual explosion. On the other hand, an extreme breaking
 of the spherical symmetry is predicted, the initial phase of the explosion being in fact
 found to be unipolar. Last but not least, high entropies are found in a fraction of the 
ejecta. This may have important consequences for the r-process (see Sect.~\ref{r_dccsn}). 
    
Several calculations show that rotation can significantly influence in various ways the CCSN 
mechanism in both its pre-bounce and post-bounce phases (e.g. \cite{warren04} for 3D 
simulations, and references therein). In particular, (1) the collapse along the equator is 
slowed down, and the density of the stellar core at bounce may be lowered, (2) the accretion 
flow through the stalled accretion-shock onto the PNS becomes non-spherical, (3) the 
post-bounce neutrino flux is enhanced in the direction of the rotation axis, with a possible
 significant impact on the shock revival, (4) rotation provides an additional source of
 internal energy that may augment the energy supplied by the neutrinos, (5) the development 
of the fluid instabilities is altered, and (6) the growth of magnetic fields and their 
topology may be significantly affected (see below). 

Multi-dimensional simulations also suggest that, much like rotation, magnetic fields may
 have a significant impact on the development of magneto-hydrodynamically-driven CCSNe 
(referred to as MHD CCSNe in the following). In particular, a large magnetic pressure may 
influence the core-collapse and post-bounce phase, the development of fluid instabilities,
 as well as the internal energy budget through viscous dissipation. In fact, sufficiently 
strong magnetic fields possibly generated by magneto-rotational instabilities or other 
mechanisms (e,g, \cite{sawai05}) might provide a substantial fraction of the energy powering
 an explosion, even rivalling the neutrino luminosity (e.g. 
\cite{sawai05,akiyama03,moiseenko06}, and references therein). The stochastic nature of
 these instabilities might also lead to supernova shapes of different forms. An extreme
 configuration is of the JetSN type for dipole magnetic configurations, but a  preferential
 ejection in the equatorial disc that forms perpendicular to the rotation axis is also
 possible for a quadrupole-like field. 

JetSN-like MHD CCSNe might lead to the formation of pulsars and account for their observed
 kicks. Depending upon the progenitor properties, magnetars could also result (see
 Sect.~\ref{magnetar_model}). These explosions open the further possibility of formation of
 a magnetised `supramassive' PNS collapsing into a black hole through the loss of angular 
momentum \cite{vietri98}, leading to the generation of so-called `supranovae' (another
 supranova channel associated with the formation of a supra-massive neutron star through
 accretion of mass and angular momentum from a companion in a low-mass X-ray binary is 
proposed by \cite{vietri99}. Note that the term `faint supernova' has also been introduced in
 the literature, and is associated by e.g. \cite{nomoto06} to a DCCSN producing a non-rotating
 black hole). Supernovae with final kinetic energies typically of an order of magnitude larger
 than the typical values for CCSNe, referred to as `hypernovae',  have also attracted much 
attention, and are interpreted (e.g. \cite{nomoto06}) as being associated with the formation 
of rotating black holes. A fraction at least of the supranovae or hypernovae are likely to be
 associated with accretion discs. Some portion of these discs can be wind-ejected through 
magnetic centrifugal forces or viscosity. Part of the innermost disc material may also be 
collimated into ultra-relativistic jet-like structures with properties close to those of 
certain gamma-ray bursts, especially of the long duration type (\cite{piran05} for a review).

\begin{figure}
\center{\includegraphics[width=0.70\textwidth]{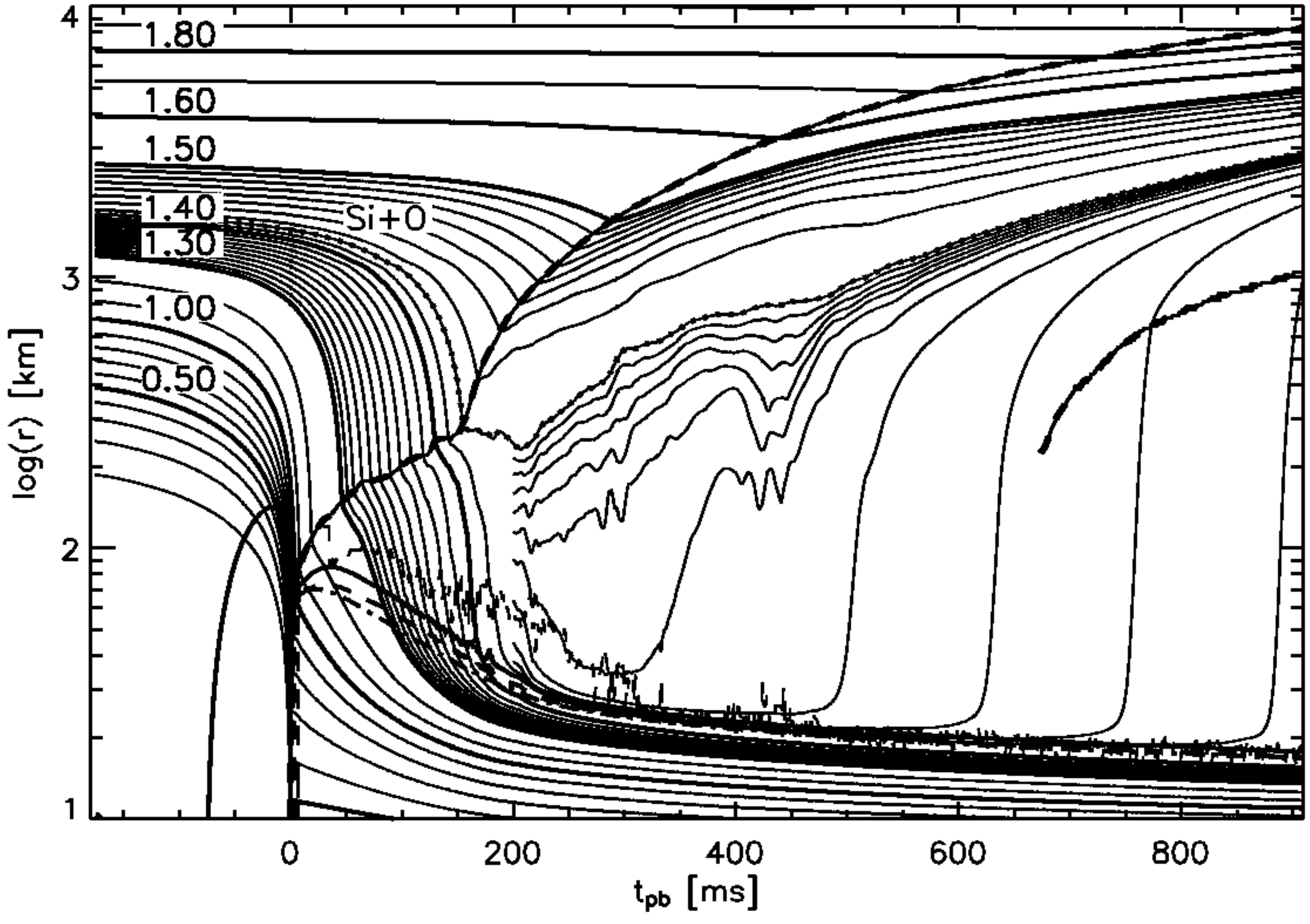}}
\caption{Mass trajectories obtained in a simulation of a 15 $M_\odot$ star CCSN \cite{buras05}.
 The explosion is caused by the convectively supported neutrino-heating 
mechanism, but the neutrino energy deposition behind the shock is artificially enhanced in
 order to get a successful explosion. The calculations are performed in 2D until about 470 ms 
after shock formation (at $t = 0$), and are reduced to 1D at later times. The outward moving
 shock is indicated by a thick black line located at increasing radii as time passes.  The 
extra thick line starting at $\sim\,$650$\,$ms indicates the separation between the fast 
neutrino-driven wind and the slower expanding supernova ejecta (from \cite{pruet05})}
\label{fig_15msunexplosion}
\end{figure}

As far as neutrinos are concerned, there is little doubt that they continue to play an 
important role in the CCSN physics by transporting energy and momentum, and by driving 
convective instabilities. 
In so doing, they possibly combine with other mechanisms cited above 
which could act as rejuvenating agents of the stalled shock wave. In spite of clear progress
 made in the description of neutrino transport, all schemes used up to now in
 multi-dimensional simulations suffer from some shortcomings (e.g. \cite{cardall03}). Some of 
them are of pure physics nature (like the due treatment of neutrino flavor mixing 
\cite{strack05}), but many are enforced by the algorithmic complexity and by the computational 
demands. The first generation of 2D simulations, as well as more recent 2D and 3D models, are,
 in fact,  either parametric studies with some imposed neutrino luminosities (e.g. 
\cite{scheck06}), or are computed with a simplified treatment of the neutrino transport 
 (e.g. \cite{warren04}). Doubts about the robustness of the obtained explosions against 
improvements in the neutrino transport have been brought forward, and have been confirmed by 
recent 2D simulations with improved neutrino transport schemes \cite{buras05,buras05a}. 
Previously predicted one-dimensional explosions of 15 $M_\odot$ or more massive progenitors 
\cite{buras03} have not been confirmed, unless the neutrino-matter coupling in the region 
just behind the shock is artificially enhanced by roughly a factor of two in the case of a 
15 $M_\odot$ progenitor. This artificially obtained successful CCSN is illustrated in 
Fig.~\ref{fig_15msunexplosion}.  

A nice illustration of the mutual support of neutrinos and rotation in generating a successful
 explosion is provided by the simulation of the collapse and eventual explosion of a rotating
 O-Ne white-dwarf resulting from the evolution of massive Asymptotic Giant Branch stars in a
 binary system. Through accretion of mass and angular momentum from a non-degenerate 
companion, it might experience a so-called accretion-induced collapse (AIC) of the 
electron-capture supernova type (Sect.~\ref{explo_1D}). This event might lead to a rare type
 of SNIa event that remains to be identified in nature (e.g. \cite{dessart06} for references).
 As described by \cite{dessart06}, the AIC progenitor is a rotating O-Ne white dwarf. Its 
collapse triggered by electron captures is followed in 2D. The shock generated at bounce is
 unable to produce a PCCSN. At later times, a neutrino-driven wind develops, and leads to a 
successful explosion. Its energy is very modest ($5-10 \times 10^{49}$ ergs), and only a few
 10$^{-3}$ $M_\odot$ of material is ejected, leaving a neutron star of about 1.4 to  1.9
 $M_\odot$ (note that about one hundred times more mass is predicted to be ejected in the
 AIC simulation of \cite{fryer99}). 

In summary, there are obviously many crucial questions that remain to be answered before one
 can hope putting together a clear and coherent picture of the CCSN fate of massive stars. 
 The structure of the pre-supernova stars remains uncertain in many important aspects which
 may have a significant impact on the properties, and even the very existence, of the 
explosive fate of the massive stars.  This concerns in particular the mass loss rates, 
angular-momentum distributions, couplings to magnetic fields, chemical mixing, not to mention
 multi-dimensional effects. The simulations of CCSNe and of JetSNe face crucial 
problems of micro- and macro-physics nature. 
Aborted model explosions are currently commonplace. PCCSNe appear to be excluded, and so are 1D
 DCCSNe. Multi-dimensional simulations leave some hope through the interplay between fluid 
instabilities, acoustic waves, rotation, magnetic fields and neutrinos. Mild or weak 
explosions of stars developing O-Ne cores or of accreting and rotating O-Ne white dwarfs 
have been obtained thus far, sometimes at the expense of an artificial enhancement of the 
neutrino luminosity. Detailed three-dimensional simulations are most  needed in order
 to clarify the role of various mechanisms listed above, and their precise couplings.  

\section{The neutrino-driven DCCSNe: a site for the r-process?}
\label{r_dccsn}

As briefly reviewed in Sects.~\ref{explo_1D} and \ref{explo_multiD}, neutrino-driven
 winds might naturally accompany, if not trigger, many DCCSNe possibly resulting 
from the collapse of an O-Ne core, an iron core, or an accreting white-dwarf in a 
binary system.  These winds originate from the ablation of the PNS surface layers 
by the deposition of the energy of the neutrinos streaming out of the cooling PNS. This
 ejected material is hot (temperatures of typically in excess of $10^{10}$ K) and of
 relatively low-density (i.e. relatively high entropies, or, in other words, rather high 
photon-to-baryon ratios). 
Figure~\ref{fig_15msunexplosion} illustrates that the wind phase starts about half
 a second after the bounce of the iron core of a 15 $M_\odot$ progenitor, 
which persists  for more than about 10 seconds. The mass lost through the wind during 
this period  is determined mainly by the properties (radius, mass, neutrino 
emission) of the PNS, and to a lesser extent by the properties of the progenitor
 star and of the explosion. 
 
The neutrino-driven winds are certainly interesting from a purely hydrodynamical
 point of view. In addition, their nucleosynthesis has been scrutinised in detail,
 especially following the excitement raised by the hope that they could provide a 
natural site for an $\alpha$-process and for a subsequent dynamical r-process
(Sect.~\ref{DYR}, and e.g. \cite{woosley94} for early calculations). This hope has gained 
support from a one-dimensional DCCSN simulation of an iron-core progenitor that predicted
 that entropies as high as about 400 (/baryon/k) could be attained in the wind more than 10 seconds after
 bounce (e.g. \cite{woosley94}). Such a high entropy allows the development of a robust 
r-process for a large variety of values of  the neutron excess or $Y_{\rm e}$ and the 
dynamical timescale $\tau_{\rm dyn}$ (see Sect.~\ref{DYR} and Fig.~\ref{fig:wind_yes}).
 However, another  one-dimensional iron-core DCCSN model has
predicted about five times lower entropies, so that the development of an extended
 r-process was severely endangered  \cite{takahashi94}. The subsequent studies
have confirmed that this r-process scenario could only be recovered at the expense 
of some twists. Beside an artificial increase of the entropy \cite{takahashi94}, they include
 some ad-hoc decrease of $Y_{\rm e}$ and/or of $\tau_{\rm dyn}$ from the values 
obtained by the one-dimensional numerical simulation used by \cite{takahashi94} (see also 
\cite{witti94}), which were in excess of 0.45, and of 100 ms, respectively. These alterations
 are inspired by the situation depicted in
  Fig.~\ref{fig:wind_yes} in order to obtain a successful r-process.
 In general, they are difficult to justify. For instance, the deviations of $Y_{\rm e}$
 from 0.5 results from the asymmetry of the reactions involving neutrinos and anti-neutrinos 
(see Eq.~\ref{eq_ye}), making any manoeuvring difficult.

On top of the neutrino-driven wind, the ejecta from the DCCSN deep layers also contains the
 more external hot-bubble material that is driven convectively unstable by neutrino heating
 just behind the SN shock. The bubbles have entropies that are typically lower 
($s \lsimeq 50$) than those of the neutrino-driven wind, and are most likely not of direct
 relevance in r-process studies. In the following, we will thus be concerned only with the 
neutrino-driven wind material. 
 
The failure of exploding one-dimensional neutrino-driven DCCSNe of
 iron-core progenitors is generally agreed upon nowadays, at least for $M >$ 10 $M_\odot$ 
stars (Sect.~\ref{explo_1D}). In 2D simulations, this type of explosion mechanism has been
 found to have some viability, but so far at the expense of an artificial
 enhancement of the neutrino-energy deposition behind the shock, as discussed in 
Sect.~\ref{explo_multiD} and illustrated by Fig.~\ref{fig_15msunexplosion}.
Nonetheless, its not being the key trigger of DCCSNe does not necessarily mean that the very
 existence of neutrino-driven winds is categorically denied. Nor is their possible roles as
 r-process agents. Accordingly, some effort has been put in the understanding  of the physics  of
 neutrino-driven winds through the development of (semi-)analytic  models, some 
aspects of which may be inspired by (failed) explosion simulations. 
These models confirm that the wind nucleosynthesis critically depends on $Y_{\rm e}, s$, and 
on $\tau_{\rm dyn}$, as in the $\alpha$-process discussed in Sect.~\ref{DYR}. 
The wind mass-loss rate d$M$/d$t$ is influential as well. Ultimately, the 
quantities acting upon the synthesis in the neutrino-driven DCCSN model depend 
crucially on the details of the interaction of neutrinos with the innermost 
supernova layers, as  well as on the mechanisms that might aid the DCCSN scenario. Their 
relative importance remains to be quantified in detail (Sect.~\ref{explo_multiD}).  

\subsection{Analytic models of a spherically-symmetric steady-state wind}
\label{anal}

Several wind models of analytical nature exist. They differ in their level of physical
 sophistication and in their way to parametrise the wind characteristics. In all 
cases, the wind is assumed to be spherically symmetric, which appears to be a 
reasonable first approximation even in two-dimensional simulations, at least at
 late enough times after core bounce (see \cite{pruet05} and 
Fig.~\ref{fig_15msunexplosion}). In addition, the wind is generally treated as a 
stationary flow, meaning no explicit time dependence of any physical  quantity at a
 given radial position $r$, so that $\partial x/\partial t = 0$, let $x$ be the velocity,
 temperature, density, internal energy, pressure, entropy, or composition.
The validity of this approximation is discussed in \cite{thompson01}, where
 it is concluded that stationarity may be reasonably assured, even though
 some caution is warranted. Newtonian and post-Newtonian descriptions of a spherically-symmetric 
stationary  neutrino-driven winds emerging from the surface of a PNS have  been developed  
with a preceding long history in the background of the studies of the solar wind and of
 accretion onto black holes.

The radial wind flow with the inclusion of special- and general-relativistic effects is 
basically described by the following equations (e.g. \cite{thompson01}:
 
\begin{equation}
        \frac{1}{vy} \frac{{\rm d}(vy)}{{\rm d}r} + \frac{1}{\rho} 
        \frac{{\rm d}\rho}{{\rm d}r} +\frac{2}{r} = 0\ \ \ 
        {\rm (continuity\ equation)},
\label{eq:wind_continuity}
\end{equation}

\begin{equation}
        \frac{(\varepsilon + p)}{\rho} \frac{1}{y} \frac{{\rm d}y}{{\rm d}r}
 + \frac{1}{\rho} \frac{{\rm d}p}{{\rm d}r}
          = 0\ \ \ {\rm (Euler\ equation)},
\label{eq:wind_euler}
\end{equation}
and

\begin{equation}
         \frac{{\rm d}\varepsilon}{{\rm d}r} -
        \frac{(\varepsilon + p)}{\rho} \frac{{\rm d}\rho}{{\rm d}r}
         + \rho \frac{\dot{q}}{vy}  = 0\ \ \ {\rm (energy \ equation)}.
\label{eq:wind_energy}
\end{equation}

In the above equations, $\rho, \varepsilon, p$ and $\dot{q}$ are the rest-mass density,  the
 mass-energy density, the pressure, and the rate of energy deposition per unit mass. The
 quantity $y$ is defined as $[(1 - 2GM_*/rc^2)/({1-v^2/c^2})]^{1/2}$, where  $M_*$ is
 the PNS mass, which is taken to be constant (the effect of the wind mass loss is neglected 
here), $G$ is the gravitational constant, and $v$ is the radial outflow velocity as measured
 by a local, static observer. For the sake of later discussions, a dynamical timescale  
 
\begin{equation}
    \tau_{\rm dyn} \equiv 
 \frac{1}{v~y}\left|\frac{1}{\rho}\frac{\partial\rho}{\partial{\rm d}r}\right|^{-1}
\label{eq:wind_time}
\end{equation}

is defined in terms of  the e-folding time of the density (e.g. {\cite{thompson01}).

Before summarising the discussion of the wind properties based on 
Eqs.~\ref{eq:wind_continuity}-\ref{eq:wind_energy}, let us introduce
 some simplifications to these equations in order to make the following discussion of the
 viability of the r-process in the neutrino-driven wind more transparent. 

\subsubsection{A Newtonian, Adiabatic, Steady-State (NASS) wind model}
\label{NASS}

An analytic Newtonian, adiabatic and steady-state wind model,  referred in the following to
 as NASS has been sketched by \cite{takahashi97}. It is aimed at providing  a simple, 
fully-analytic description of the dynamics of the wind at relatively late times or sufficiently 
far away from the PNS surface. 
 
 NASS relies on the general assumptions listed above, complemented with those of a Newtonian 
PNS gravitational potential,  and of an adiabatic expansion. In addition,  all the elementary
 $\nu/\bar{\nu}$ and e$^-$/e$^+$ weak interaction processes are frozen out,
nuclear $\beta$-decays do not affect $Y_{\rm e}$ or $s$, and possible deviations from nuclear
 equilibrium with regard to strong and electromagnetic interactions have no influence on the
 thermodynamical properties of the wind. Under such simplifying assumptions, NASS cannot 
predict  any time variation of $s, Y_{\rm e}$, or of the wind mass loss rate 
${\rm d}M/{\rm d}t,$
 which are thus treated as input, constant parameters.
 
The basic NASS equations can be derived easily, in particular from
 Eqs.~{\ref{eq:wind_continuity}-\ref{eq:wind_energy}. By setting $v \ll c$, and $y = 1$ at 
sufficiently large $r$, the  integration of Eq.~\ref{eq:wind_continuity} defines the 
(constant) mass loss rate ${\rm d}M/{\rm d}t =  4 \pi r^2 \rho v $. With the aid of
 Eq.~\ref{eq:wind_continuity}, Eq.~\ref{eq:wind_euler} transforms into the familiar form 
$ v{\rm d}v/{\rm d}r + G M_*/r^2 + {\rm d}p/\rho{\rm d}r = 0$,
provided that $v$ and $[(\varepsilon+p)/2\rho]^{1/2}$ are much smaller than $c$. Finally,
 $\dot{q} = 0$ for an adiabatic expansion and no composition changes, which simplifies 
Eq.~\ref{eq:wind_energy}. The wind motion in the regime under consideration, and in
 particular for high enough entropies, is thus approximated by 

\begin{equation}
\frac{1}{2}v^2 - \frac{G M_*}{r} + N_{\rm A} k T s_{\rm rad} = E,
\label{eq:wind_motion}
\end{equation}

where the total energy per unit mass $E$ may be obtained by setting a boundary condition, and
 the `radiation entropy' $s_{\rm rad}$ is given by

\begin{equation}
s_{\rm rad} = s_{\rm rad}^{(0)} \Bigl[ \frac{4}{11} + \frac{7}{11} 
f_{\rm e} \Bigr]\ \ \ {\rm with}\ \ \ s_{\rm rad}^{(0)} =
 \frac{11\pi^2}{45 \rho N_{\rm A}} \Bigl( \frac{kT}{\hbar c}  \Bigr) ^3, 
\label{eq:wind_srad}
\end{equation}

with $f_{\rm e}$ being unity in the  high-$T$ limit, and decreasing with $T$ for high
 $s_{\rm rad}^{(0)}$-values. Within the $f_{\rm e} = 1$ approximation, 
Eq.~\ref{eq:wind_motion} can
 be scaled with the aid of Eq.~\ref{eq:wind_continuity} as \cite{takahashi97}
 
\begin{equation}
    \frac{1}{2}\ \hat{v}^2 
    - \frac{2}{\hat{r}} + \frac{3}{\hat{r}^{2/3}\ \hat{v}^{1/3}} = 
                   \frac{3}{2}\ f_{\rm w},
\label{eq:wind_scaled}
\end{equation}

where $\hat{v}=v/v_{\rm s}$, $\hat{r}=r/r_{\rm s}$, $f_{\rm w}=E/E_{\rm wind}$ with
 $E_{\rm wind}=3v_{\rm s}^2/2$. Indeed, $v_{\rm s}$ now equals the local adiabatic sound
 speed, and is given by $v_{\rm s}=1.01\times 10^3~\alpha^{1/3}~\beta^{-2/3}$ km~s$^{-1}$
 in terms of $\alpha \equiv s^4~{\rm d}M/{\rm d}t\ M_\odot$/s and
 $\beta \equiv M_*/$(1.5 $M_\odot$), and $r_{\rm s}=9.74\times 10^4~\alpha^{-2/3}~\beta^{7/3}$ km.

\begin{figure}
\center{\includegraphics[width=0.80\textwidth,height=0.80\textwidth]{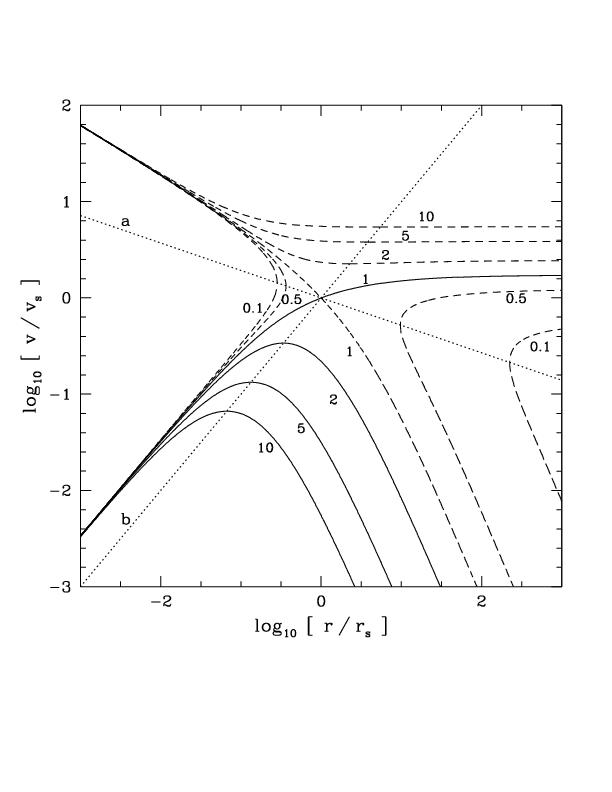}}
\vskip-2.2cm
\caption{Scaled velocity profiles derived by solving  Eq.~\ref{eq:wind_scaled}.
 The dotted  lines $a$ and $b$ are obtained by setting  $A = 0$ and $B = 0$ in 
its derivative form $A$d$v/$d$r = B$, respectively.
The curves are labelled with the selected $f_{\rm w}$ values. Solid lines
 correspond to the physically-meaningful `wind' ($f_{\rm w} = 1$) and 'breeze'
 ($f_{\rm w} > 1$)  solutions}
\label{fig:wind_vr} 
\end{figure}

The possible types of solutions of Eq.~\ref{eq:wind_scaled} are drawn in 
Fig.~\ref{fig:wind_vr} for a few selected $f_{\rm w}$ values. If $f_{\rm w}=1$, the 
transonic wind solution is obtained, which crosses the singularity at $\hat{v}=\hat{r}=1$. 
For  $\hat{r}\rightarrow\infty$, 
$\hat{v}(\hat{r})\rightarrow\sqrt{3}$, so that $\rho(\hat{r})$ and $T(\hat{r})$ decrease as
 $\hat{r}^{-2}$ and $\hat{r}^{-2/3}$. For $f_{\rm w}>1$, the solutions are of subsonic wind
 type, with $T(\hat{r})$ and $\rho(\hat{r})$ reaching constant values at infinity. In the 
following, the transonic and subsonic solutions will be referred to as 
 the `wind solution' and `breeze solution', respectively.\footnote{The solutions above can
 be cast into the approximate but useful form
 
\begin{equation}
    \bigl[\hat{v}(\hat{r})\bigr]^{-1}\approx a_{-1}\hat{r}^{-1}+a_{-1/2}\hat{r}^{-1/2}+a_0+a_{1}\hat{r}+a_{2}\hat{r}^2 
\label{eq:wind_vrapp}
\end{equation}
%
with  $a_{-1}=8/27, a_{-1/2}=(19-9\sqrt 3)/27, a_{0}=1/\sqrt 3,
 a_{1}=a_{2}=0$ if $f_{\rm w}=1$ {\it and} $\hat{r}\gsimeq 0.28$.
Otherwise $a_{-1}=8/27, a_{-1/2}=0, a_{0}=2f_{\rm w}/3,
 a_{1}=f_{\rm w}^2/2, a_{2}=(f_{\rm w}/2)^3$.
The errors are large for $f_{\rm w}$-values just above unity.
The scaled time $\hat{t}(\hat{r})$ can be obtained approximately by
 integrating Eq.~\ref{eq:wind_vrapp} over $\hat{r}$} 

\begin{figure}
\center{\includegraphics[width=0.80\textwidth,height=0.75\textwidth]{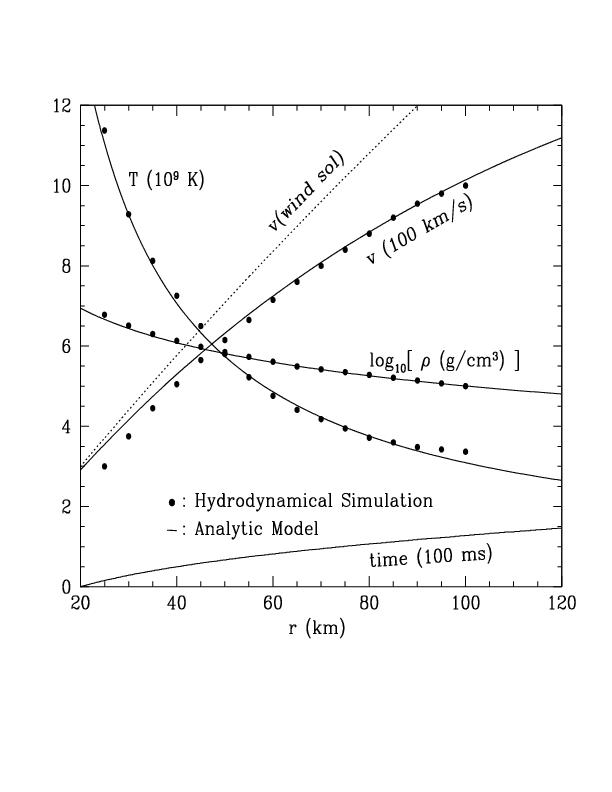}}
\vskip-2.0cm
\caption{Temperature $T$, density $\rho$ (in logarithm), and velocity $v$ profiles
 as functions of the distance $r$ from the centre of the PNS, in respective units as
 labelled. The dots are from a hydrodynamical simulation of neutrino wind material 
from a PNS with $M_*=1.63 M_\odot$ \cite{janka93}, which shows nearly constant
$s_{\rm rad} \approx 97.5$ and d$M/$d$t \approx 6.5 \times 10^{-6}$  $M_\odot$/s
in the approximate  $r\gsimeq 30 $ km range. These values are used as input for the
 fully-analytic NASS model along with the choice of $f_{\rm w}=3.7$, this meaning
 that the simulation most likely produced a breeze solution. The dotted line labelled
 $v$(wind sol) shows the corresponding NASS wind ($f_{\rm w}=1$) velocity  profile. The
 curve labelled `time' displays the time elapsed from the starting point at $r_0=20$ km, 
as derived from NASS}
\label{fig:wind_fit}
\end{figure}

Once $v(r)$ is known for a given set of $M_*, s, {\rm d}M/{\rm d}t$,
 and $f_{\rm w}$, then $\rho(r)$ and  $T(r)$ can be derived for adiabatic motions. 
 Figure~\ref{fig:wind_fit} demonstrates that the NASS model reproduces closely the late 
variations predicted by a 
hydrodynamical simulation \cite{janka93}. A slight discrepancy is observed in the velocity
 very close to the PNS surface, where adiabaticity is not obtained. 

\begin{figure}
\center{\includegraphics[width=0.80\textwidth,height=0.77\textwidth]{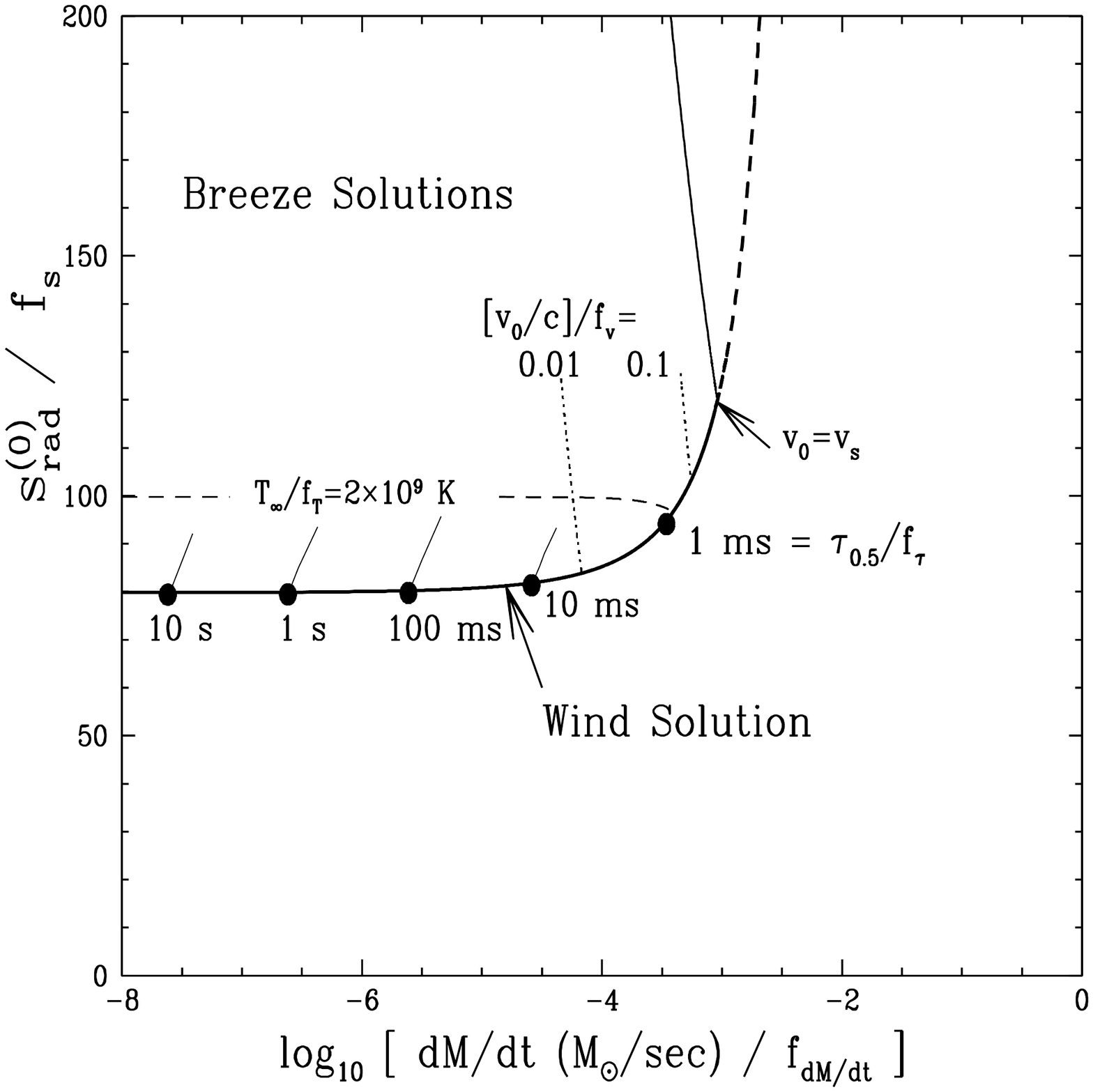}}
\vskip-1.8cm
\caption{The wind and breeze solutions of the NASS model as mapped onto the d$M$/d$t \equiv
{\rm d}M/{\rm d}t$  (in 
logarithm) -- $s_{\rm rad}^{(0)}$ plane. The material is assumed to leave the PNS of mass
 $M_*$ at the distance $r_0$  with temperature $T_0$, the standard values of which are
 $M_* =$ 1.5 $M_\odot$, $r_0 = 30$ km, and $T_0 = 10^{10}$ K. For different sets of values, 
the following scaling relations on the coordinates apply, with $T_{0,10} \equiv 
[T_0/10^{10}{\rm K}]$, $r_{0,30} \equiv [r_0/30{\rm km}]$, and $M_{*,1.5} \equiv 
[M_*/$1.5 $M_\odot]$: $f_{{\rm d}M/{\rm d}t} = [r_{0,30}]^{+5/2}~[M_{*,1.5}]^{-1/2}~[T_{0,10}]^{+4}$, 
and $f_s = [r_{0,30}]^{-1}~[M_{*,1.5}]^{+1}~[T_{0,10}]^{-1}$. The physically meaningful 
solutions are located in the upper-left region enclosed by solid lines.  The quantity
 $\tau_{{\rm dyn},0.5}$ defines the dynamical timescale (Eq.~\ref{eq:wind_time}) at the time
 when $kT$ gets down to 0.5 MeV. The contours of $\tau_{{\rm dyn},0.5}/f_\tau$ with $f_{\tau}
 = [r_{0,30}]^{3/2}~[M_{*,1.5}]^{-1/2}$ are drawn (thin solid lines) for some selected values.
 The point at which the velocity $v_0$ at $r_0$ equals the sonic velocity $v_{\rm s}$ is 
indicated. The wind solution beyond this point (thick dashed line) is unphysical, as it
 corresponds to $r_0$ being located beyond the sonic point. The breeze solutions with 
$s_{\rm rad}^{(0)}/f_s \gsimeq 120$ are also unphysical since 
${\rm d}v/{\rm d}r (r_0) < 0$. The contours of velocities $v_0/f_v$ (at $r_0$) 
equalling $c/10$ and $c/100$ are drawn by dotted lines, where 
$f_v=[r_{0,30}]^{-1/2}~[M_{*,1.5}]^{+1/2}$. Above (below) the near-horizontal dashed
 line, the asymptotic temperature $T_\infty$ of the breeze solutions at large distances
 become higher (lower) than $2 \times 10^{9} f_T$ K, where $f_T = 
[T_{0,10}]$}
\label{fig:wind_sm}
\end{figure}
 
In order to ease the comparison with hydrodynamical simulations, it may be of interest to 
express the NASS model in terms of  inner and outer boundary conditions. Denote $r_0$ the
innermost radial  position at which adiabaticity and steady state are likely to hold. Numerical
 simulations suggest that this occurs at $r_0 \approx 30$ km from the PNS centre, at which 
point the temperature $T_0 \approx 10^{10}$ K. In the following, these values along with
 $M_* = 1.5$ M$_\odot$ are adopted as standard.  Figure~\ref{fig:wind_sm} maps the NASS wind 
and breeze solutions onto the log$_{10} {\rm d}M/{\rm d}t$ -- $s_{\rm rad}^{0}$ plane.  

Figure~\ref{fig:wind_sm} may be used to evaluate the chance of having a successful r-process
 developing under the conditions dictated by the NASS breeze or wind solutions. Let us first 
assume that the standard $T_{0,10}=r_{0,30}=M_{*,1.5}=1$ conditions hold.  Most breeze 
solutions  at high entropies have asymptotic temperatures in excess of $2 \times 10^9$ K, and
 thus lead to an $\alpha$-process that has time to produce 
too much heavy seeds per neutron to allow for
 a well-developed r-process. As a result, only relatively low entropies
 in the $80 \lsimeq s \lsimeq 100$ range with tightly correlated $\tau_{\rm dyn}$ and 
mass-loss rates are left to be considered. For a reasonable choice of $Y_{\rm e} \gsimeq
 0.4$, $\tau_{\rm dyn}$ must be much shorter than 50 ms,  and the mass-loss rates
${\rm d}M/{\rm d}t$  are 
limited to a narrow range around 10$^{-4}$ $M_\odot$/s. This conclusion remains essentially the
 same even when the wind solutions are considered since the high-entropy wind solutions would 
require unrealistically high velocities already near the PNS surface. The situation would be
 a bit less constrained with lower-than-standard $r_0$ and larger-than-standard $M_*$ 
values, so that $s$ would be increased by  $f_s$ and the dynamical timescale would be
 decreased by $f_\tau$. This explains the apparent `success'  of some r-process scenarios that
 call for a very massive and  compact PNS (see Sect.~\ref{RESSW}).
 
 \subsubsection{The General-Relativistic Steady-State Wind (RESSW) solution}
 \label{RESSW}
 
 We briefly review here an analytic model developed by \cite{thompson01} for the
 stationary and spherically symmetric {\sl wind} solution of 
Eqs.~\ref{eq:wind_continuity}-\ref{eq:wind_energy} . It is the most sophisticated version to
 date of this type of models. In particular, it 
generalises the NASS model by including neutrino interactions, the effects of general (and
 special) relativity, and brings a solution to some uncertainties and ambiguities resulting 
from other analytic wind models (e.g. \cite{qian96,meyerbs97}). It is referred to in the 
following as the Relativistic Steady-State Wind (RESSW) model.

The wind solutions are provided for different PNS masses and luminosities from 
shortly (about 1 s) after bounce, at which time the wind starts emerging from a 
still hot, highly luminous, and quite extended (radius of the order of 30 to 50 km)
 PNS, to about 10 s, at which point the PNS has had time to cool and contract 
quasi-hydrostatically (note that the cooling timescale depends in particular on 
the still-uncertain PNS equation of state). Some of the main RESSW results which may have an 
impact on the wind nucleosynthesis may be 
summarised as follows:\\
(1) the electron fraction $Y_{\rm e}$ reaches a value close to asymptotic
 $Y_{\rm e}^{\rm a}$ very near (within about 10 km) the PNS surface. This value is largely 
determined by the weak interactions with $\bar\nu_{\rm e}$ and $\nu_{\rm e}$, as well as by
 electron- and positron-captures on free nucleons. In other words, it depends on the PNS 
$\bar\nu_{\rm e}$ and $\nu_{\rm e}$ luminosities and average neutrino energies, and is 
predicted to increase as the total PNS neutrino luminosity decreases in time. As shown 
by  \cite{qian96}, $Y_{\rm e}^{\rm a}$ is well approximated by
 
 \begin{equation}
\frac{1}{ Y_{\rm e}^{\rm a}}\simeq  \frac{\Gamma_{\nu_en}+\Gamma_{\bar{\nu}_ep}}{\Gamma_{\nu_en}}
 \simeq 1+ \frac{L_{\bar{\nu}_e}}{L_{\nu_e}}\frac{\langle\varepsilon_{\bar{\nu}_e}\rangle-2\Delta+1.2\Delta^2/\langle\varepsilon_{\bar{\nu}_e}\rangle}{\langle\varepsilon_{\nu_e}\rangle+2\Delta+1.2\Delta^2/\langle\varepsilon_{\nu_e}\rangle} 
 \label{eq_ye}
  \end{equation}
%
 where $\Gamma_{\nu_en}$ ($\Gamma_{\bar{\nu}_ep}$) is the $\nu_e+n\rightarrow p+e^-$ 
($\bar{\nu_e} +p \rightarrow n + e^+$) reaction rate and $\Delta = m_{\rm n} - m_{\rm p}\simeq 
1.293~{\rm MeV}$ is the neutron-proton mass difference. $L_{\nu_e}$ ($L_{\bar{\nu}_e}$) and
 $\langle\varepsilon_{\nu_e}\rangle$ ($\langle\varepsilon_{\bar{\nu}_e}\rangle$) are the 
neutrino (anti-neutrino) luminosity and mean energy. The ratio 
$\langle\varepsilon_{\bar{\nu}_e}\rangle / \langle\varepsilon_{\nu_e}\rangle$ typically 
ranges from 1.1 to 1.4, and $L_{\bar{\nu}_e} / L_{\nu_e}$ from 1.0 to 1.4 in supernova
 simulations \cite{thompson01}, leading to typical values of $Y_{\rm e}^{\rm a}$ in
 the approximate 0.46-0.49 range \cite{thompson01}. Other effects, like the formation of 
$\alpha$-particles from free nucleons, are also expected to affect 
$Y_{\rm e}^{\rm a}$, as reviewed by \cite{thompson01}, but are neglected in 
Eq.~\ref{eq_ye}.  Values substantially lower than about 0.40 are estimated to apply 
at best to a very small fraction of the ejected wind. The difficulty of reaching low  
$Y_{\rm e}^{\rm a}$ values is confirmed by the artificially triggered DCCSN of a 
15 $M_\odot$ progenitor shown in Fig.~\ref{fig_15msunexplosion} \cite{pruet05};\\
(2) the entropy is found to be enhanced in a general-relativistic rather than
 Newtonian framework. Even so,  asymptotic values $s_{\rm a}$ not exceeding about
 80 to 140 are found, and are obtained already very close to the PNS surface. 
Higher entropies that have been often claimed to be required in order to obtain a 
robust r-process could be reached through an increase of the ratio of the PNS 
mass to its radius over the value that is viewed as typical for 
a 1.4 $M_\odot$ PNS with a radius of about 10 km. General-relativistic effects help
 increasing this ratio. It is constrained, however, by observations of neutron-star
 binaries and by the high-density equation of state. An increased $s_{\rm a}$
 could also be obtained through an enhanced neutrino-energy deposition rather close 
to the PNS, an artifact that, in fact,  also helps getting a successful DCCSN 
(see Fig.~\ref{fig_15msunexplosion}). The situations considered by 
\cite{thompson01} do not allow, however, to obtain entropies in excess of about
 200;\\
(3) any simple parametrisation of the dynamical timescale $\tau_{\rm dyn}$ is found
 to be an oversimplification, as non-uniform variations are obtained at least for 
some of the considered neutrino luminosities. 
At large distances (in excess of about  100 to 200 km) from the PNS, 
$\tau_{\rm dyn}$ increases monotonically up to values between about 0.01 and 0.05 s
 at about 700 km from the PNS. General-relativistic corrections tend to increase 
$\tau_{\rm dyn}$ somewhat. In contrast, shorter $\tau_{\rm dyn}$ could be
 obtained along with somewhat higher entropies [see (2) above] for an artificial
 neutrino-energy deposition.  

 From the RESSW wind predictions summarised above, it is concluded by 
\cite{thompson01} that the r-process advertised by \cite{woosley94} is extremely 
unlikely in the context of a wind at late times (say in excess of about a few 
seconds) after bounce. By the time the wind evolves to high entropy,
 $\tau_{\rm dyn}$ gets too long and $Y_{\rm e}^{\rm a}$ too high
 for allowing a robust r-process. Anyway, the predicted wind mass-loss rates at 
those  times are found to be so small that any significant r-nuclide ejection would
 be precluded. This conclusion is in line with those reached by
 \cite{takahashi94,qian96}. Instead, \cite{thompson01} examines the possibility of
 an early r-process episode just a second or two after bounce, when the PNS has 
completed its contraction phase.  At such early times, only relatively modest 
entropies ($s_{\rm a} \lsimeq 150$), short dynamical timescales 
($\tau_{\rm dyn} \lsimeq 1.5$ ms), and rather high electron fractions 
($0.46 \lsimeq Y_{\rm e}^{\rm e} \lsimeq 0.50$) could allow a level of 
r-processing leading to the formation of the Pt r-process peak.  This conclusion 
is in line with some previous results \cite{hoffman97,otsuki00}. In addition, at 
the considered early times, the wind mass-loss rates would still be high enough 
for ejecting an amount of r-nuclides (roughly $10^{-6}$ to 10$^{-5}$ $M_\odot$) that
 can be significant at the galactic scale (see Sect.~\ref{galaxy_obs_th}). 
A major difficulty of this early-time r-process  is that highly luminous and very
 compact PNSs (e.g. $M \gsimeq$  2 $M_\odot$ with a radius less than about 9 km) are
 called for. This, perhaps physically unhealthy, trick is consistent with the conclusions
 drawn from the NASS model and based on Fig.~\ref{fig:wind_sm} (see Sect.~\ref{NASS}).

\begin{figure}
\center{\includegraphics[width=0.77\textwidth,height=0.68\textwidth]{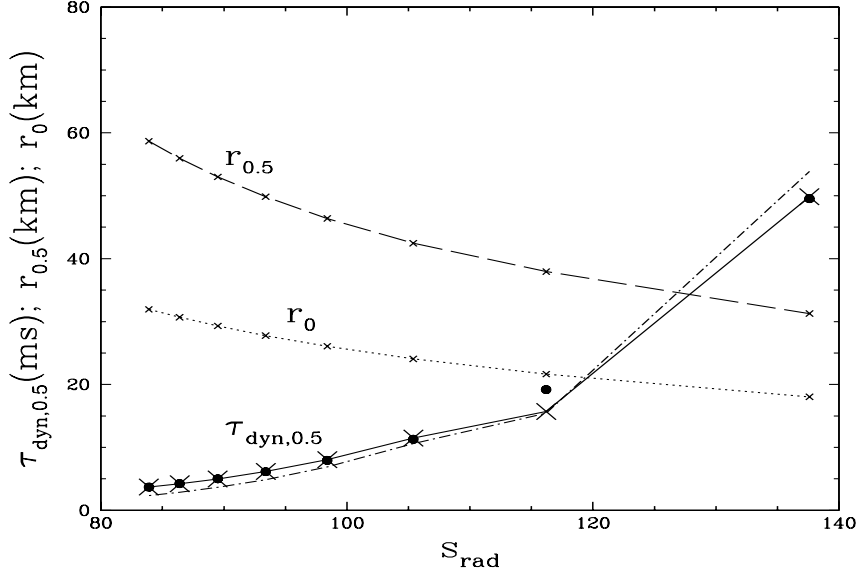}}
\vskip-1.8cm
\caption{Comparison between eight RESSW $\tau_{{\rm dyn},0.5}$ values (dots) with 
$M_* =$ 1.4 $M_\odot$, but different values for the total neutrino luminosity (Table I  of 
\cite{thompson01}) and the NASS wind solutions obtained with the RESSW $s_{\rm a}$ and
${\rm d}M/{\rm d}t$ 
 values.  The RESSW results are well reproduced by NASS (large crosses connected by
 a solid line) if an effective PNS mass $M_{*,{\rm eff}} =$ 1.55 $M_\odot$ is adopted. The 
corresponding NASS inner boundaries $r_0$ at which $T = 10^{10}$ K, and the radial positions
 $r_{0.5}$, at which the $\tau_{{\rm dyn},0.5}$ values are obtained are shown by dotted and 
dashed lines, respectively. The dash-dot line is obtained for a uniform expansion
$\hat{v} \sim (3/2)^3~\hat{r}$ which is expected to hold during the very early phase of the
NASS evolution (see Fig.~\ref{fig:wind_vr})}
\label{fig:wind_tau}
\end{figure}
 
It may be instructive to compare the NASS and RESSW wind solutions. The predictions for
 $\tau_{{\rm dyn},0.5}$ are seen in  Fig.~\ref{fig:wind_tau} to agree nicely if NASS 
makes use of the RESSW $s_{\rm a}$ and d$M/$d$t$ values, and if an effective PNS mass 
$M_{*,{\rm eff}} =$ 1.55 $M_\odot$ larger than the RESSW value $M_* =$ 1.4 $M_\odot$ is adopted.
 This increase is consistent with the neglect in NASS of any general-relativistic effect 
(see also \cite{qian96}).\footnote{Slightly different values of $M_{*,{\rm eff}}$
 are needed to mimic the RESSW evolutions at much later times than those of relevance for 
the r-process}

All in all, it is suggested \cite{thompson01} that the neutrino-driven wind from a PNS may not
 be the primary site for the r-process.  A possible remedy to the PNS compactness is discussed 
in Sect.~\ref{simulation_breeze}.

\subsection{A numerical approach to the wind or breeze problem}
\label{simulation_breeze}
  
The (transonic) wind and (subsonic) breeze are both allowed in the NASS model, while only the
 wind solution is discussed in detail in the RESSW framework, even though some qualitative 
considerations are presented by \cite{thompson01} on the breeze regime. It now remains to be 
seen if one of these types of solutions may be favoured by the DCCSN physics. This question is
 far from being just academic, as it is likely that its answer may have some impact on the
 predicted development of the r-process. It is quite intricate as well.
 
One difficulty arises as the neutrino-driven material is likely not to flow 
unperturbed to infinity in a variety of DCCSN situations. The wind indeed 
catches up with the slower hot bubble material, and may also interact with matter 
and radiation in that portion of the star through which the SN shock has already 
passed. This interaction is likely to depend, among other things, on the pre-SN 
structure, and is more limited as the mass of the outer layers decreases as one
 goes from massive SNII progenitors to SNIb/c events whose progenitors 
(Wolf-Rayet stars) have lost their extended H-rich envelope prior to the explosion, or to 
envelope-free accretion-induced DCCSNe scenarios (Sec.~\ref{explo_multiD}). 

The interaction of the material ablated from the PNS and the outer SN layers has
 several important consequences. It may give rise to a reverse shock responsible
 for the fallback of a more or less large amount of material onto the PNS, and 
whose properties (location and strength)  alter more or less deeply the 
characteristics of the neutrino-ejected material.\footnote{See \cite{wang06} for the
 search of an observational evidence of this fallback phenomenon} If the energy 
of the reverse shock is large enough for it to propagate to the sonic point, the 
whole region between the PNS surface and the shock would be brought in sonic 
contact. The wind would thereby transforms into a breeze. It may also be that the 
reverse shock is unable to disrupt the wind interior to the sonic point. In a 
steady state situation, it results that, across the shock, the velocity decreases
 the density increases so as to maintain the mass loss rate, while the temperature 
increases \cite{thompson01}. 

Some numerical simulations attempt to answer some of the 
intricate questions concerning the wind or breeze nature of the PNS-ablated 
material, as well as the validity of the steady-state approximation adopted in the NASS
 model, or in other (semi-)analytic models. A calculation performed  by  \cite{janka95}
 in a Newtonian approximation shows that a strong reverse-shock propagating toward the 
PNS results when the SN shock crosses the Si-O-rich layers that surround the iron core of
a 15 $M_\odot$ progenitor, causing a slowing down of the velocity of the wind from 
about $2 \times 10^9$ to only a few times $10^8$ cm~s$^{-1}$, and even possibly 
its disruption (see also the simulation by \cite{janka93} displayed in
 Fig.~\ref{fig:wind_fit}).  A general-relativistic simulation by \cite{sumiyoshi00}
 which adopts 
approximate PNS structure and neutrino luminosities and spectra shows that the
 general-relativistic corrections increase the entropy (see also Sect.~\ref{RESSW}), while the 
dynamical timescale  decreases  with respect to the Newtonian treatment.\footnote{Note 
that \cite{sumiyoshi00} define a dynamical timescale as the e-folding time at $kT = 0.5$ 
MeV of the temperature instead of the density. This leads to a value that is about one-third
 of the timescale provided by Eq.~\ref{eq:wind_time}}
An important result  concerns the sensitivity of the dynamics of the PNS-ablated
material on adopted boundary conditions which are meant to mimic the pressure just 
behind the shock. Adopting the model by \cite{sumiyoshi00}, this dependence has been
 studied further by \cite{terasawa02}, and some results are shown in Fig.~\ref{wind_boundary}. 
This model shows that the asymptotic temperature in the PNS ejecta decreases with 
the boundary pressure, while the neutron-to-seed ratio correspondingly increases. 
Following \cite{terasawa02}, an r-process could develop under such conditions.  So, the
 requirement of a high-mass compact PNS which makes
 the r-process in the RESSW model unlikely appears to be circumvented by a certain
 choice of the wind boundary conditions. It remains to be demonstrated, however, 
if the imposed boundary conditions could be obtained in less schematic supernova 
environments than the ones envisioned by the simulations of 
\cite{sumiyoshi00,terasawa02}. 

\begin{figure}
\center{\includegraphics[scale=0.50]{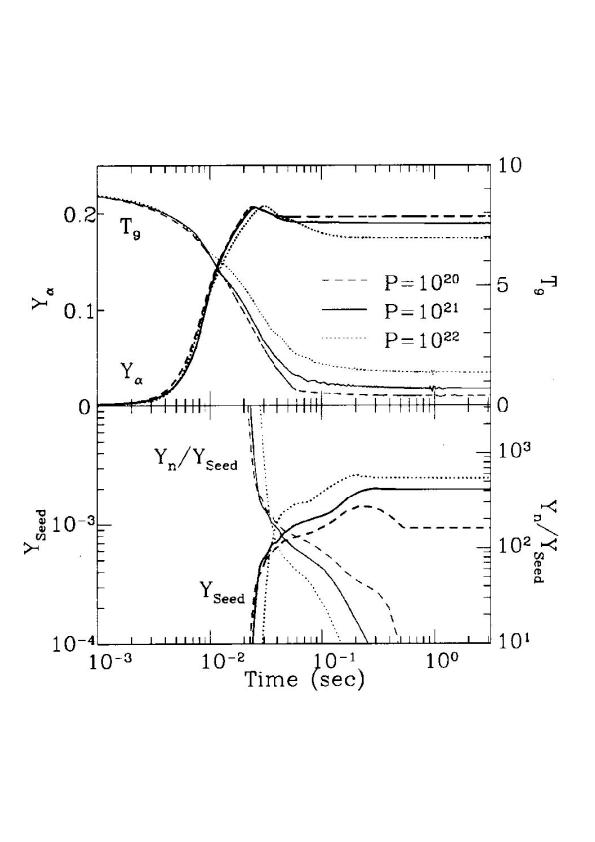}}
\vskip-2.3truecm
\caption{Some properties (temperature $T_9$, $\alpha$-particle abundance 
$Y_\alpha$, neutron-to-seed ratio $Y_{\rm n}/Y_{\rm seed}$, and seed 
abundance $Y_{\rm seed}$) of the neutrino-driven wind versus time (time zero 
refers to the moment when $T_9 = 9$ is reached) derived from the model of 
\cite{sumiyoshi00} for a 1.4 $M_\odot$ and 10 km radius PNS, and three selected
values for the boundary pressure $P$ (in dyn cm$^{-2})$ (from \cite{terasawa02})}
\label{wind_boundary}
\end{figure}

We note that the approach to the breeze adopted in these simulations with a prescribed 
boundary pressure is in fact equivalent to the way NASS  parametrises the breeze solutions in
  terms of $f_{\rm w}$. In Eq.~\ref{eq:wind_motion}, $E$ approaches $N_{\rm A}kTs_{\rm rad}$ 
at large distances in the case of breeze solutions, and thus the 'outer' pressure, as adopted in 
the numerical calculations, is equal to $N_{\rm A}kTs_{\rm rad}\rho/4$ in the 
high-entropy limit. This pressure is approximately proportional to the fourth power of the
 asymptotic temperature $T_\infty^4$. As it is the case in order to reconcile the RESSW and
 NASS results (Fig.~\ref{fig:wind_tau}),
 the NASS parameters have to be adjusted differently at different times to 
mimic the relativistic effects taken into account in the simulations leading to
 Fig.~\ref{wind_boundary}.
 
\subsection{The r-process in the neutrino-driven wind and breeze regimes}
\label{r_wind}

As already stressed in Sects.~\ref{NASS} and \ref{RESSW}, it appears difficult, if not 
impossible, for the wind or breeze from a PNS to provide conditions allowing the development
 of a successful r-process, at least if one restricts the discussion to conditions inspired 
by numerical simulations. It is true that this conclusion is largely based on various 
simplifications allowing the construction of analytic, spherically-symmetric models. 
However, numerical simulations do not lead to much more optimism, at least if 
multi-dimensional effects are not considered, in which case it is very difficult to get
 a successful explosion (see Sect.~\ref{explo_1D}). It is also true that the 1D DCCSN 
simulations suffer from many shortcomings. In particular, none of the calculations of the
 winds at late times incorporate a reliable description of the energy-dependent 
neutrino physics. This leaves quite uncertain key quantities like the luminosities 
and spectral shapes of the neutrinos of different flavors, and, as a consequence, 
quantities which largely determine the possibility for a successful r-process, like
 the time variation of $Y_{\rm e}$. The impact of the reverse shock and of 
material fallback on the wind properties for different DCCSN progenitors remain 
unexplored. This whole situation clearly weakens any conclusion one may try to draw on the
 development of the r-process in DCCSN environments, not to mention multi-dimensional 
 effects associated with rotation or magnetic fields (see Sects.~\ref{explo_multiD} and 
\ref{r_others}).

\begin{figure}
\center{\includegraphics[scale=0.45]{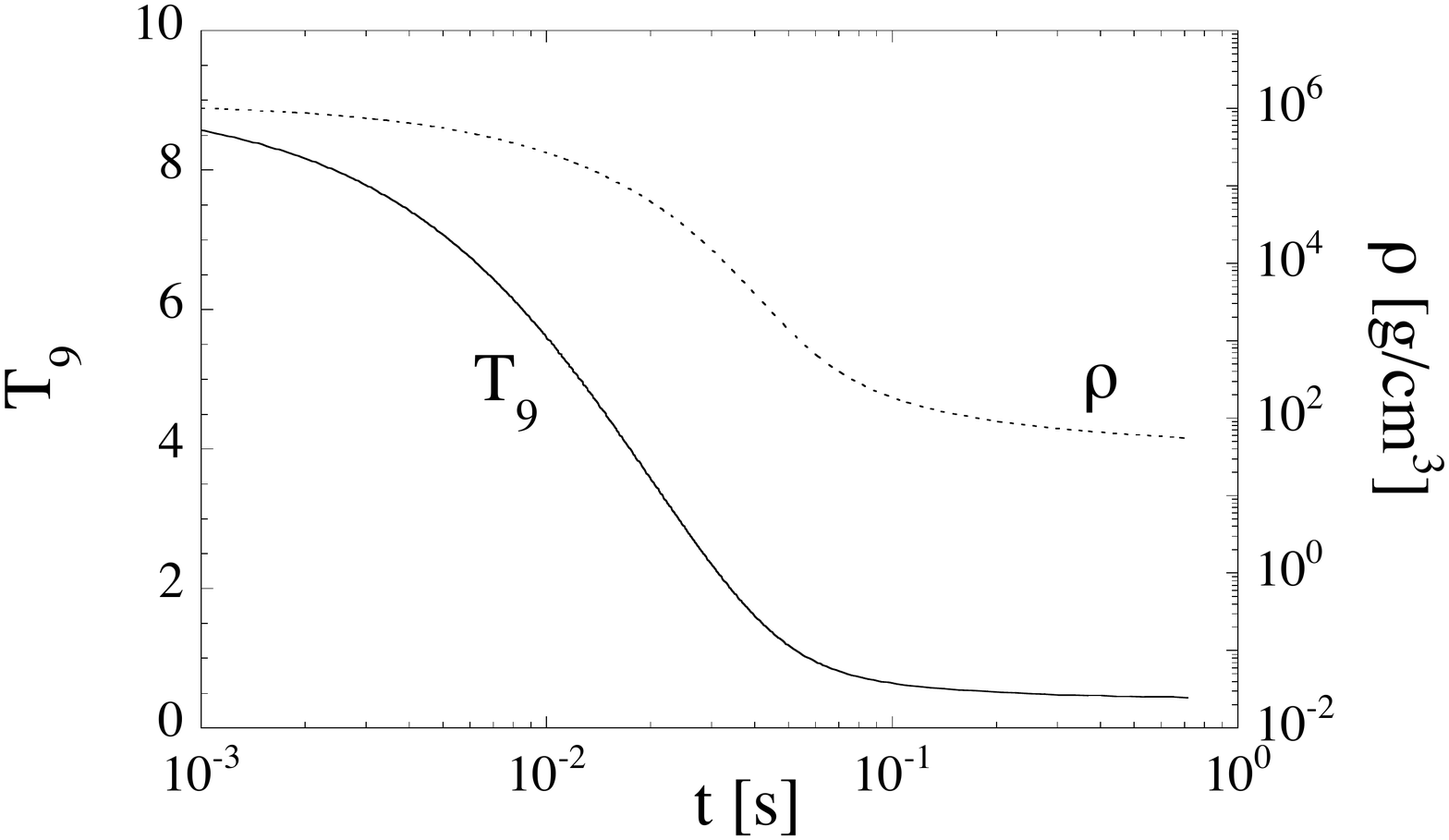}}
\vskip-1.2cm
\caption{Evolution of density $\rho$ and of temperature (expressed in $10^9$K) from an 
initial ($t = 0$) value $T_9 = 9$ for the NASS breeze solution obtained with 
$M_*=$ 1.5 $M_\odot$, $s_{\rm rad} = 200$,  ${\rm d}M/{\rm d}t=$
0.6 $\times$ 10$^{-5}$ $M_\odot$/s), and $f_{\rm w} = 3$. The corresponding initial radial position is
 $R_0 = 13.8$~km.} 
\label{fig_apro_tak1}
\end{figure}

\begin{figure}
\center{\includegraphics[scale=0.45]{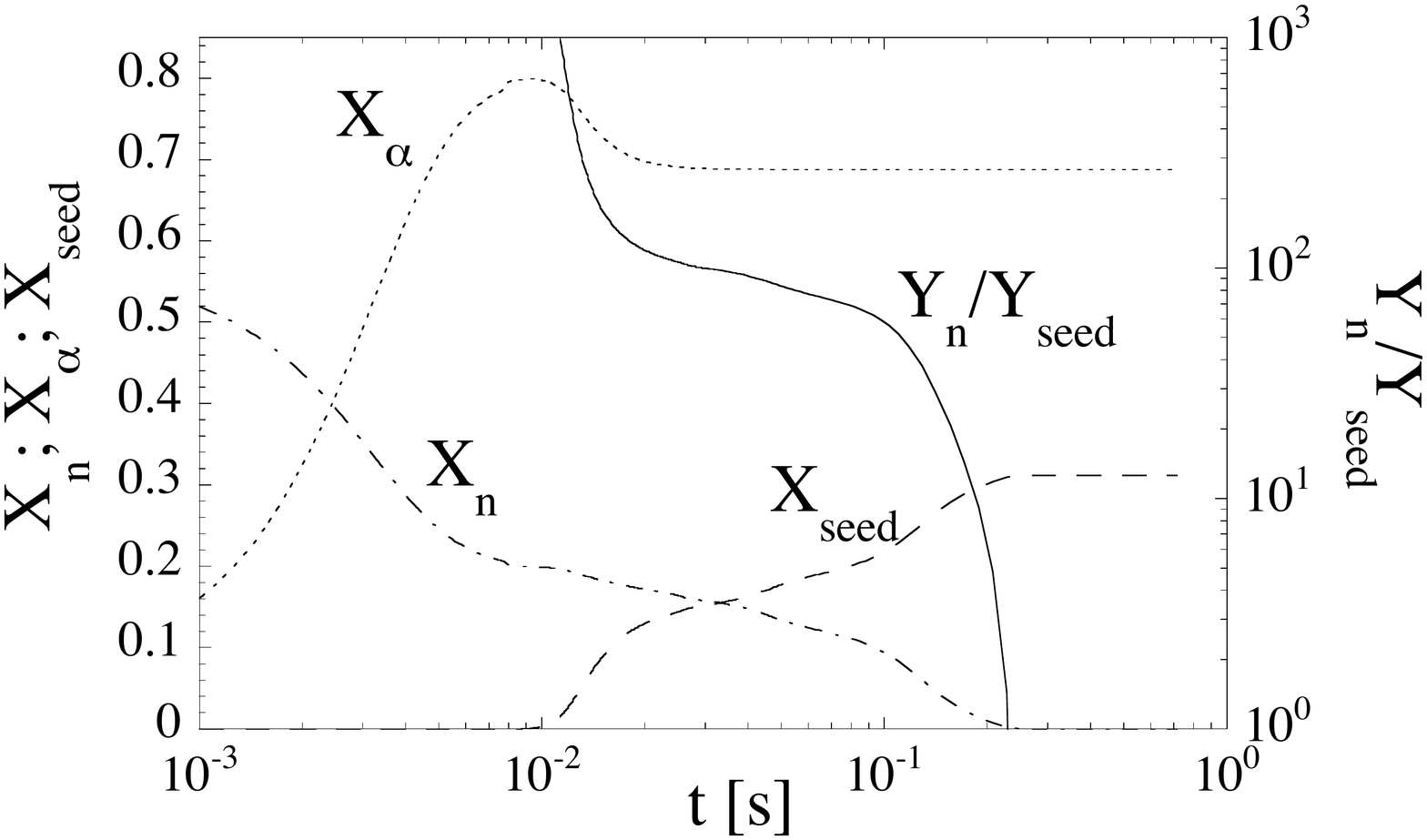}}
\vskip-0.9cm
\caption{Evolution of the mass fraction of the neutrons ($X_{\rm n}$),  $\alpha$-particles
 ($X_{\alpha}$) and  seed ($A>4$) nuclei ($X_{\rm seed}$), as well as of  the
 neutron-to-seed number ratio ($Y_{\rm n}/Y_{\rm seed}$)
for the breeze model shown in Fig.~\ref{fig_apro_tak1},  and
 with the choice $Y_{\rm e} = 0.40$} 
\label{fig_apro_tak2}
\end{figure}

In such a situation (and at least if one wants to proceed in the quest of suitable 
astrophysical sites of the r-process!), the best one can do is to introduce artificial  
modifications to the wind or breeze solutions that seem to be most 
realistic on grounds of the still uncertain
 numerical simulations. It is  sometimes attempted to relate these changes
 to special astrophysical sites.   Optimistically enough, one might hope to orientate in such
 a way realistic simulations in the quest of the proper r-process conditions in DCCSNe. This
 strategy is widely used, and it is the one adopted in the illustrative calculations of the
 r-process yields described below relying on some NASS wind or breeze solutions.

In these illustrations, the abundances are obtained from the solution of an extended nuclear 
reaction network including all nuclei with $0\le Z \le 92$ lying between the valley of 
stability and the neutron-drip line. All n-, p- and $\alpha$-capture reactions as well as 
$\beta^{\pm}$-decays and $\beta$-delayed neutron emissions
 are taken into account. The nuclear-physics input used is described in
 Sect.~\ref{nucphys_general} and includes the HFB-2 nuclear
 masses, the GT2 $\beta$-decay rates (Sect.~\ref{beta}) and the Hauser-Feshbach reaction 
rates (Sect.~\ref{reac}) without any direct-capture contribution.
 
Figure~\ref{fig_apro_tak1} displays the run of temperature and density in one of the NASS
 breeze solutions that is expected to provide suitable conditions for the development of the
 r-process: 
 $s_{\rm {rad}}^{(0)}/f_s = 82.8$ and log~d$M$/d$t = -4.19$ in Fig.~\ref{fig:wind_sm}. 
The evolution of the neutron, $\alpha$-particles and heavy ($A > 4$) seeds, as 
well as of the neutron-to-seed ratio in the material obeying this breeze solution is displayed
 in Fig.~\ref{fig_apro_tak2}. The expansion of the matter leads to a decrease of the 
temperature from its initial value $T = 9 \times10^{9}$~K, and to the development of an 
$\alpha$-process (see Sect.~\ref{DYR}). Below $T \simeq 9 \times 10^{9}~{\rm K}$, neutrons
 and protons start to recombine into $\alpha$-particles. Between  about $7 \times 10^9$ and
 $5 \times 10^9~{\rm K}$, part of these $\alpha$-particles and neutrons combine to form 
heavier nuclei through the nuclear bottleneck
$\alpha + \alpha + n \rightarrow \, ^9{\rm Be}(\alpha,n)^{12}{\rm C}(n,\gamma )^{13}{\rm C}(\alpha,n)^{16}{\rm O}$.
 Subsequent $(\alpha,\gamma)$, $(\alpha,n)$ and $(n,\gamma)$ reactions produces nuclides in the 
$50 \lsimeq A \lsimeq 100$ range.  Snapshots of the nuclear flow leading to the build-up of these
 heavy nuclei are shown in Fig.~\ref{fig_alpha_flow}. The $\alpha$-process freezes out when the 
temperature drops below about $T = 2 \times 10^9$~K (or $t \gsimeq 0.03$~s; see
 Fig.~\ref{fig_apro_tak1}).   At this time, the material is dominated by $\alpha$-particles and
 by roughly equal mass fractions of neutrons and heavy seeds, leading to a number density ratio 
$Y_{\rm n}/Y_{\rm seed}  \approx 100$, as the most abundant seeds have mass numbers $A$ 
around 100. This allows the development of a successful r-process, as illustrated by the
 $t = 0.40$~s flow of Fig.~\ref{fig_alpha_flow}.

\begin{figure} 
\center{\includegraphics[width=1.00\textwidth,height=0.80\textwidth]{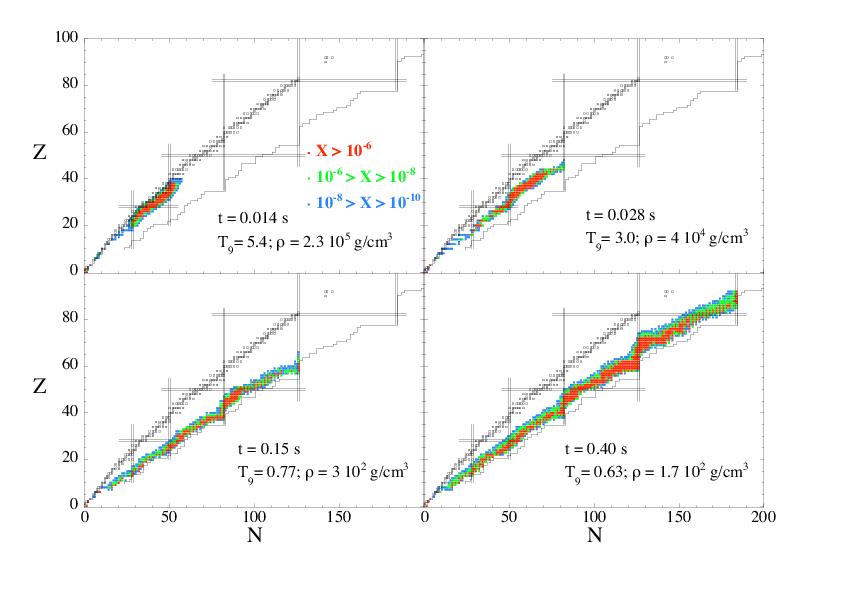}}
\vskip-0.8cm
\caption{Snapshots of the nuclear flows calculated for the NASS breeze solution adopted in 
Fig.~\ref{fig_apro_tak1}. The first three panels ($t \le 0.15$~s) describe the gradual 
build-up of heavy nuclei by the $\alpha$-process. These act as the seeds for the r-process
 that  develops after the $\alpha$-process freeze-out, as shown in the last panel. The mass
 fractions are colour-coded as defined in the upper left panel} 
\label{fig_alpha_flow}
\end{figure}

\begin{figure}
\center{\includegraphics[scale=0.45]{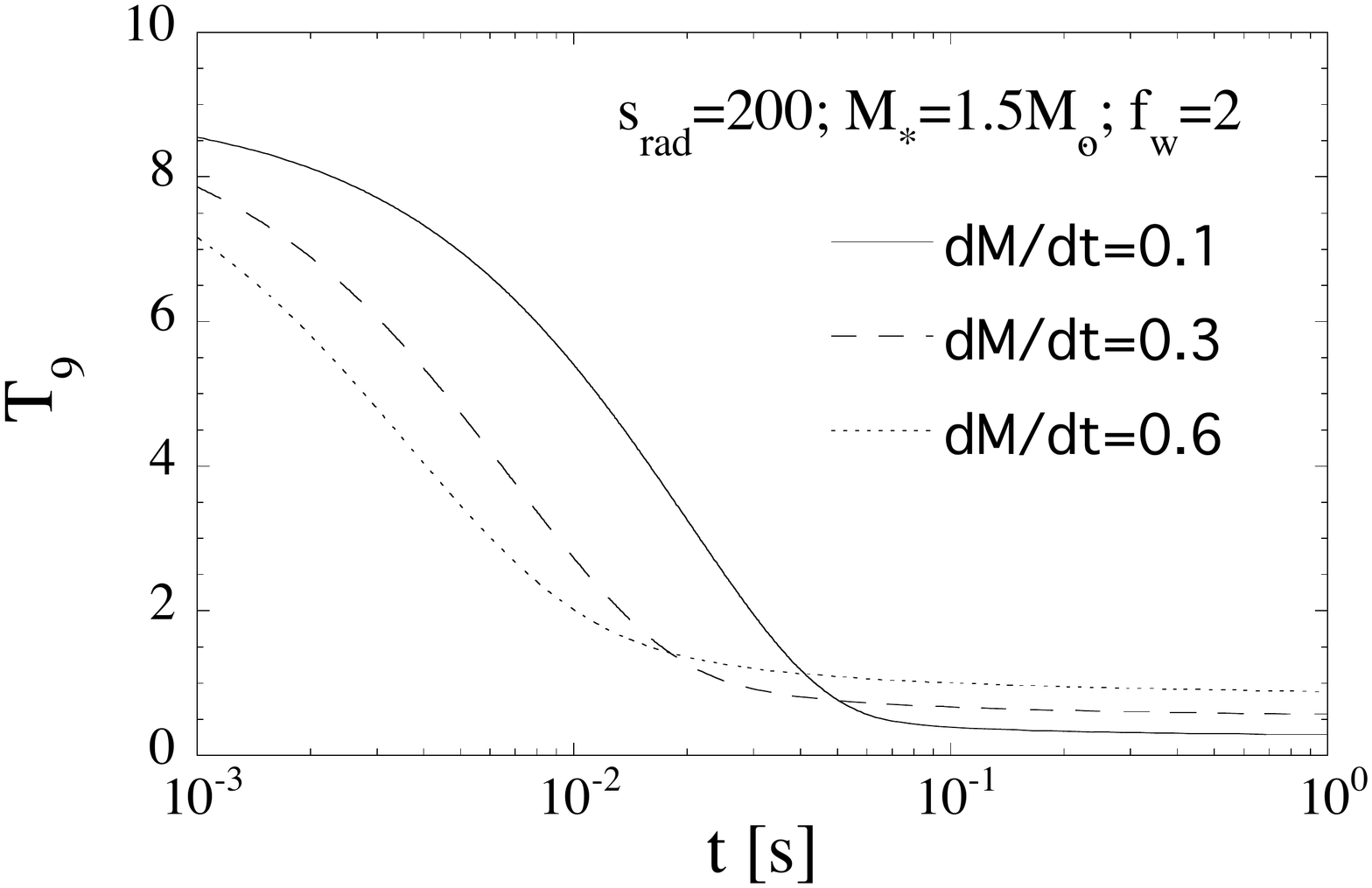}}
\vskip-0.8cm
\caption{Evolution of temperature (in $10^9$ K) from an initial ($t = 0$) value $T_9 = 9$ in 
the NASS breeze solution corresponding to $f_{\rm w}=2$ and to different values of 
${\rm d}M/{\rm d}t$ (in units of $10^{-5}~M_\odot$/s).  The adopted values of 
$M_*=1.5~M_\odot$
 and of $s_{\rm rad} = 200$ are the same as in Fig.~\ref{fig_apro_tak1}} 
\label{fig_apro1a}
\end{figure}

\begin{figure}
\center{\includegraphics[scale=0.45]{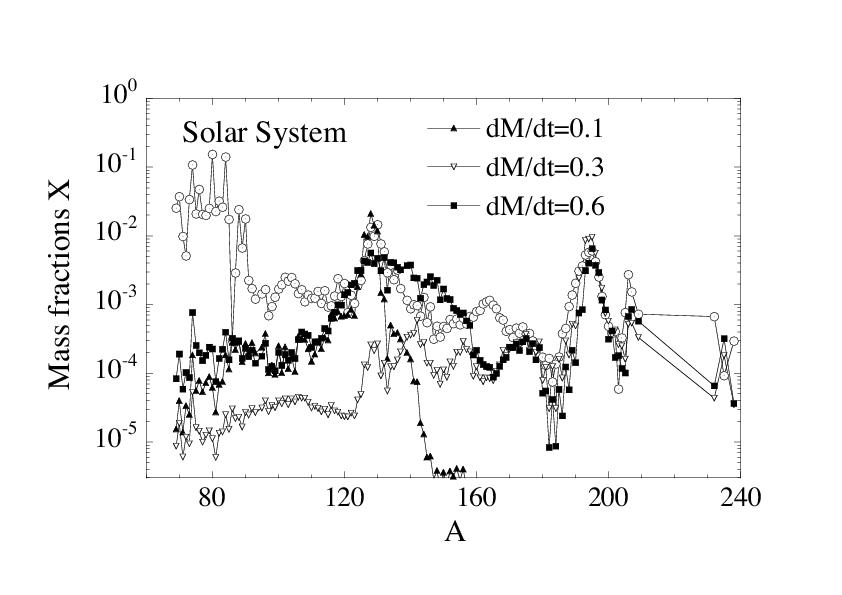}}
\vskip-0.7cm
\caption{Distribution of the r-nuclide abundances obtained with the wind characteristics 
from Fig.~\ref{fig_apro1a} (the values of ${\rm d}M/{\rm d}t$ are in units of 
10$^{-5}$ $M_\odot$/s) and an initial electron fraction $Y_{\rm e}=0.48$. The nuclear physics 
input is based on the microscopic models described in Sect.~\ref{th_rates-general} 
\cite{bruslib1,bruslib2}, except that the direct contributions to the radiative neutron 
captures are neglected, and the fission processes are included only at the very final stage of
 the r-process. The $\beta$-decay and $\beta$-delayed processes are estimated within the GT2 
model (Sects.~\ref{beta} and \ref{beta_delayed}). The upper curve corresponds to the SoS 
r-nuclide abundances normalised to $\sum_{i} X_i=1$} 
\label{fig_apro2a}
\end{figure}

\begin{figure}
\center{\includegraphics[scale=0.45]{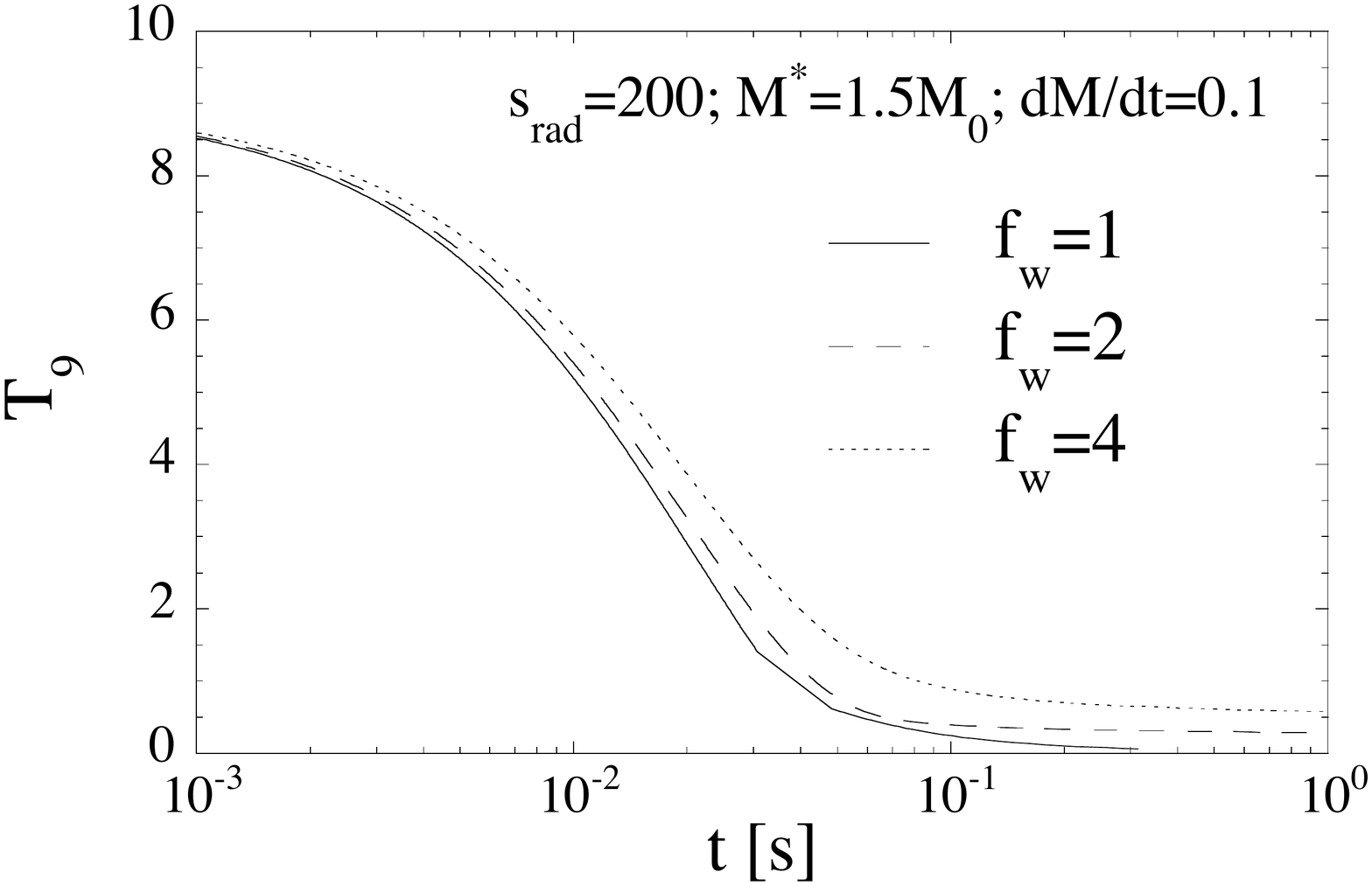}}
\vskip-0.7cm
\caption{Evolution of temperature  (in $10^9$K) from an initial ($t = 0$) value $T_9 = 9$ in
 the NASS model for ${\rm d}M/{\rm d}t = 0.1 \times$ 10$^{-5}$ $M_\odot$/s and different values 
of  $f_{\rm w}$. The $f_{\rm w} = 1$ case corresponds to the wind solution, the other
 ones being of the breeze type. The values $M_* =$ 1.5 $M_\odot$ and $s_{\rm rad} = 200$ are the 
same as in Fig.~\ref{fig_apro_tak1}} 
\label{fig_apro1b}
\end{figure}

\begin{figure}
\center{\includegraphics[scale=0.45]{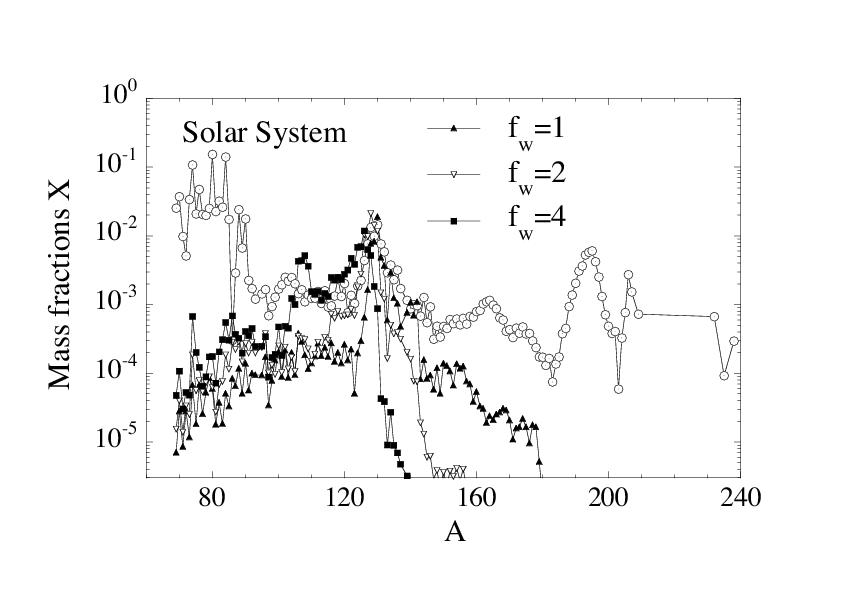}}
\vskip-0.7cm
\caption{Same as Fig.~\ref{fig_apro2a}, but for the wind $(f_{\rm w}=1)$ and
 breeze $(f_{\rm w} > 1)$  solutions of  Fig.~\ref{fig_apro1b}}
\label{fig_apro2b}
\end{figure}

\begin{figure}
\center{\includegraphics[scale=0.5]{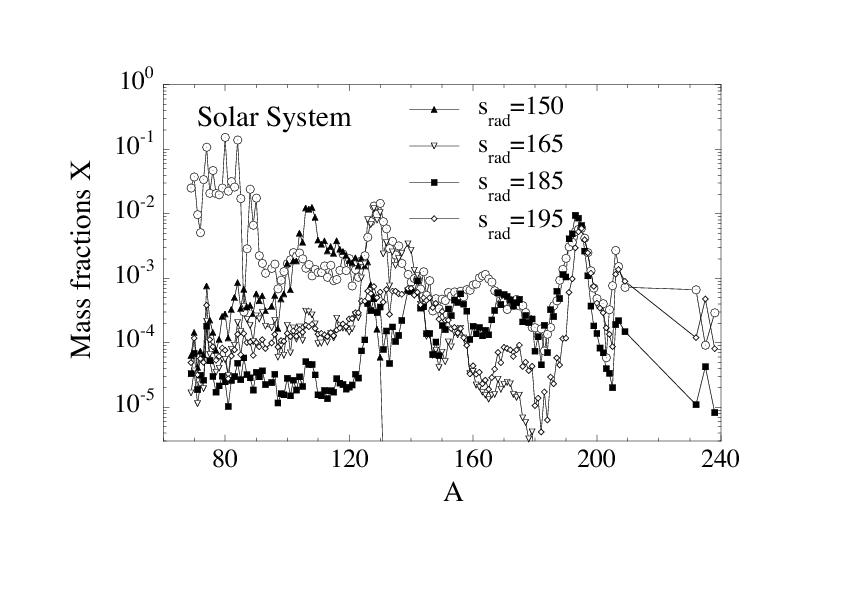}}
\vskip-1.5cm
\caption{Same as Fig.~\ref{fig_apro2a}, but for the breeze solution of Fig.~\ref{fig_apro1a}
 with  ${\rm d}M/{\rm d}t = 0.6 \times 10^{-5}$ $M_\odot$/s and
 four different  values of the entropy $s_{\rm rad}$} 
\label{fig_apro3}
\end{figure}

\begin{figure}
\center{\includegraphics[scale=0.5]{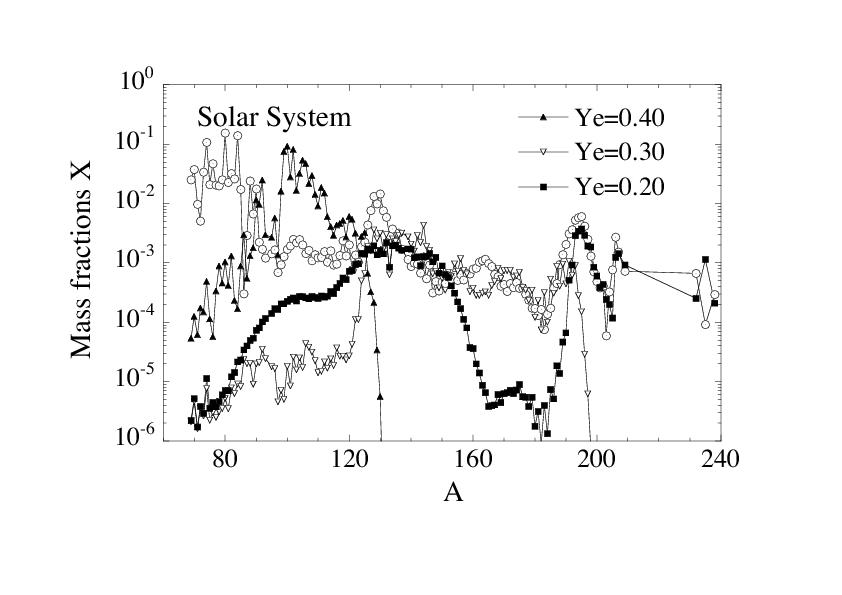}}
\vskip-1.5cm
\caption{Same as Fig.~\ref{fig_apro2a}, but for the breeze solution of Fig.~\ref{fig_apro1a} 
 with ${\rm d}M/{\rm d}t = 0.6 \times$ 10$^{-5}$ $M_\odot$/s, and $s_{\rm rad} = 100$ 
and for  different $Y_{\rm e}$ values} 
\label{fig_apro4}
\end{figure}
 
We now consider several variations to the breeze model adopted above in order to illustrate 
their influence of the r-process: 

 (1) {\it Influence of the expansion timescales}. The wind expansion timescales may
 be influenced by the wind energy through $f_{\rm w}$ and by the mass-loss rate
 ${\rm d}M/{\rm d}t$, these two characteristics having in their turn an impact on the 
time variation of the wind temperature, even at large distances, as illustrated in 
Fig.~\ref{fig_apro1a} and Fig.~\ref{fig_apro1b}. More specifically, the expansion 
timescales decrease with increased mass-loss rates, but increase for higher wind-energies.
  For fast 
expansions, $\alpha$-particles have less time to recombine, so that the number of neutrons per
 seed nuclei at the time of activation of the r-process is increased, as is its efficiency, for
 a given entropy and electron fraction. This is illustrated in Fig.~\ref{fig_apro2a} and 
Fig.~\ref{fig_apro2b}.  In comparison with breeze expansions ($f_{\rm w} > 1$), the wind 
solution ($f_{\rm w} = 1$), through its shorter expansion timescale, favours the development
 of the r-process. Considering a wind rather than a breeze model has, however, a smaller impact
 on the predicted r-abundances than a change in mass-loss rate can have.

 (2) {\it Influence of the entropy}.   For a given expansion timescale and electron 
fraction, an increase in the entropy can have a drastic effect on the predicted abundances. 
This is illustrated in Fig.~\ref{fig_apro3}, which shows that, for  $Y_{\rm e} = 0.48$ and 
a relatively fast expansion (see Fig.~\ref{fig_apro1a}), an entropy $s_{\rm rad} = 165$ 
would allow a substantial  production of the $A \simeq 130$ r-nuclides, this limit being 
pushed up to the $A \simeq 195$ r-process peak for $s_{\rm rad} = 185$, and even to the
 actinides for  $s_{\rm rad} = 195$. 
 
 (3) {\it Influence of the electron fraction}.  The r-process efficiency increases with
 a decrease of $Y_{\rm e}$ in the wind, which favours the recombination of  
$\alpha$-particles into more neutron-rich nuclei. Figure~\ref{fig_apro4} illustrates the 
effect of $Y_{\rm e}$ on the r-nuclide abundance distribution resulting from a breeze with
 a relatively low entropy ($s_{\rm rad} = 100$).  It is seen that values of $Y_{\rm e}$ 
down to 0.20 are needed to produce the actinides, even with relatively short expansion 
timescales. Note that the initial electron fraction of the wind strongly depends on the
 properties of the neutrino flux escaping the PNS. This effect is neglected in constructing
 Figs.~\ref{fig_apro2a} - \ref{fig_apro4}, but is briefly examined below. Also note that for 
large neutron excesses, the nuclear flow comes close to the neutron-drip line and reaches very
 rapidly the fissioning region of the very heavy nuclei. In this case, the final r-abundance
 distribution is affected by the different fission properties of the nuclei produced during
 the r-process. This is observed for the $Y_{\rm e}=0.20$ abundance curve in Fig.~\ref{fig_apro4}.

\begin{figure}
\center{\includegraphics[scale=0.5]{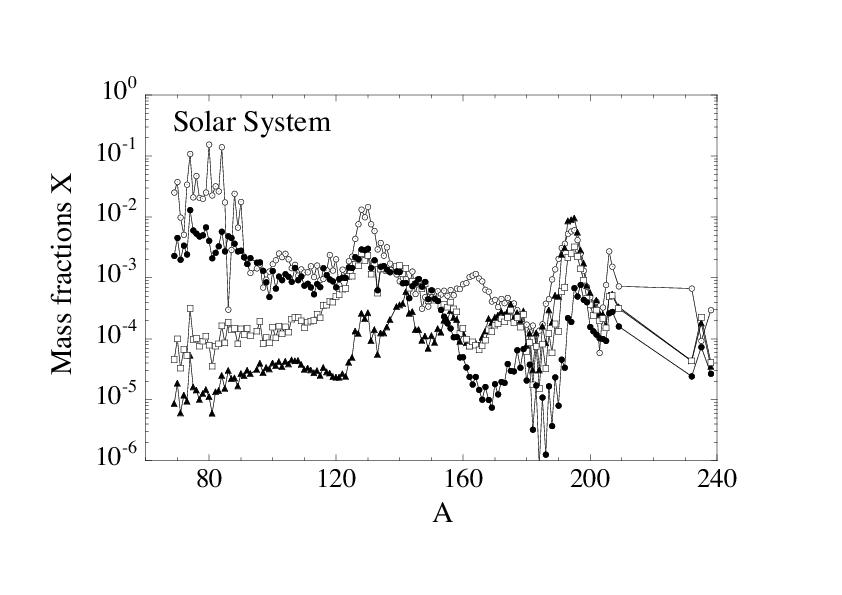}}
\vskip-0.9cm
\caption{Same as Fig.~\ref{fig_apro2a} for the breeze solution corresponding to 
${\rm d}M/{\rm d}t = 0.3 \times10^{-5} M_\odot$/s, but for different neutrino
 luminosities. The solid triangles correspond to the r-abundance distribution without neutrino 
interactions, the open squares to the same case when adopting $\nu_e$ and $\bar{\nu_e}$ 
luminosities  $L_{\nu_e}=10^{51}$ and $L_{\bar{\nu}_e}=1.3 \times 10^{51}$~ergs/s, and $\mu$ 
and $\tau$ (anti)neutrinos luminosities $L_{\nu_i}=10^{51}$~ergs/s. The solid circles 
correspond to a ten-fold increase of the luminosities of all the neutrino flavors.  In all
 cases, the initial electron fraction is set equal to $Y_{\rm e}=0.48$}
\label{fig_apro7}
\end{figure}

 Neutrino interactions on free nucleons, $\alpha$-particles and heavy nuclei play an important 
role in the r-process in the wind. In particular, as already mentioned, the $\nu_e +n 
\rightarrow p + e^- $ and $\bar{\nu_e} +p \rightarrow n + e^+$ captures determine the initial
 $Y_{\rm e}$ value (see Eq.~\ref{eq_ye}). If the luminosity of the anti-neutrinos
exceeds that of neutrinos, the initial material is neutron-rich ($Y_{\rm e} < 0.5$).
 As pointed out by 
\cite{fuller95,mclaughlin96}, (anti)neutrino captures on nucleons also affect $Y_{\rm e}$
 later during the expansion at the time when neutrons and protons have mainly recombined into 
$\alpha$ particles. A this moment, the neutrino captures on the remaining free neutrons increase
 $Y_{\rm e}$.  Finally, neutrino captures on free nucleons also increase the total number of 
seed nuclei available for the r-process, and consequently reduce the neutron-to-seed ratio 
(for more details, see \cite{meyer98}). 

In addition to the captures of $\nu_e$ and $\bar{\nu_e}$ on free nucleons, those on heavy
 nuclei, as well as the captures of neutral current $\mu$- and $\tau$- (anti)neutrinos may affect 
the nucleosynthesis taking place during the expansion (e.g \cite{fuller95,mclaughlin96,meyer98}).
 While the charged-current channel plays a role similar to the weak $\beta^{\pm}$ interaction, 
the neutral-current neutrino spallation reactions affect the nuclear flow in the same way as  
photo-disintegrations. To illustrate the impact of the neutrino effects, the r-process has been 
re-calculated for a NASS breeze solution corresponding to
 $s_{\rm rad}=200$, $Y_{\rm e}=0.48$, $M_*=1.5 M_\odot$, $f_{\rm w}=2$ and
 ${\rm d}M/{\rm d}t=0.3\times$ 10$^{-5}$~$M_\odot$/s, and with the neutrino captures on
 free-nucleons and nuclei duly
 taken into account.  The rates are estimated assuming that the neutrinos streaming out of the
 PNS have a Fermi-Dirac distribution with zero chemical-potential and with 
`(anti)neutrino temperatures'
$kT_{\nu_e}=3.5$, $kT_{\bar{\nu}_e}=4.0$ and $kT_{\nu_i}=6.0$~MeV, where $i$ stands for $\mu$ 
and $\tau$ neutrinos and anti-neutrinos.  The $e$-, $\mu$- and $\tau$- (anti)neutrino interaction
 cross sections are taken from \cite{borzov00,hoffman92,mclaughlin96,woosley90}. 

The resulting abundances obtained with a selection of (anti)neutrino luminosities are compared
 in  Fig.~\ref{fig_apro7} with those which would be obtained in the same wind conditions,
but ignoring the neutrino captures.  As seen in Fig.~\ref{fig_apro7}, the neutrino interaction
 is detrimental to the r-process as a result of the reduction of the number of neutrons 
available per seed nucleus. In addition, the abundance distribution is reshaped by the neutrino
 interactions. Such effects strongly depend on the adopted neutrino luminosities and 
temperatures, which remain rather uncertain. The impact of the neutrino interaction is complex 
and subtle, and also depends on the breeze/wind properties. However, it always tends to reduce 
the r-process efficiency, so that more extreme conditions for the breeze/wind characteristics,
 i.e an even higher entropy, lower $Y_{\rm e}$ or faster expansion, need to be invoked for a 
successful r-process. More details can be found in  \cite{fuller95,mclaughlin96,meyer98}.

\begin{figure}
\center{\includegraphics[scale=0.40,angle=270]{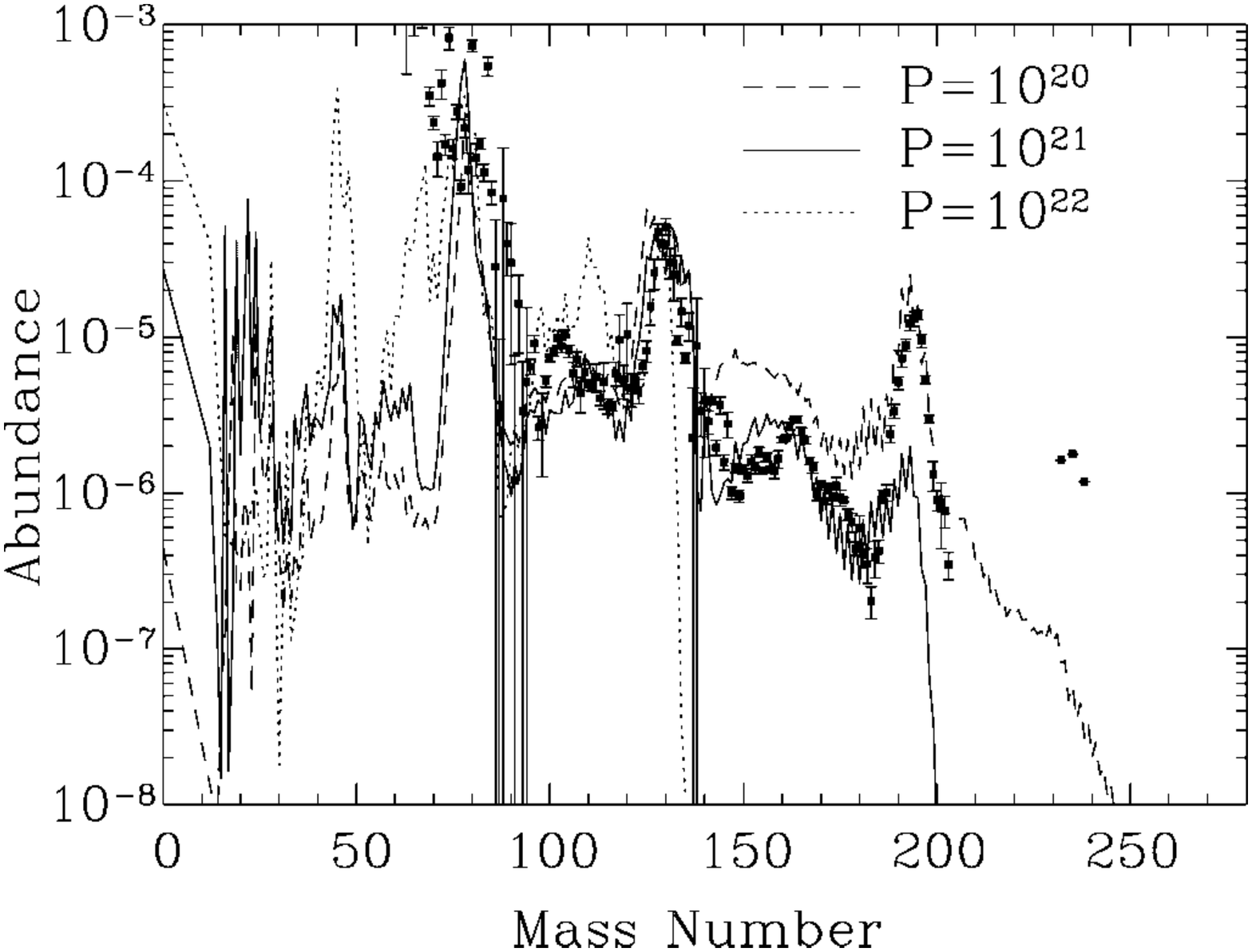}}
\caption{Abundances of r-nuclides calculated for the breeze solutions of 
\cite{terasawa02}, some characteristics of which are displayed in Fig.~\ref{wind_boundary}. The 
dots represent the SoS r-abundances from \cite{kappeler89} (from \cite{terasawa02})}
\label{r_terasawa}
\end{figure}

As a final illustration of the r-process yields in a breeze regime from a PNS, 
Fig.~\ref{r_terasawa} presents an r-process calculation performed by \cite{terasawa02}. It is 
based on numerical breeze simulations (see Fig.~\ref{wind_boundary}). 
These solutions that make use of a parametrised boundary pressure, 
which is in fact equivalent to the NASS parametrisation in terms of $f_{\rm w}$ 
(Sect.~\ref{simulation_breeze}). It is seen that an r-process can develop under the selected
 conditions, the Pt peak being even produced for the lowest of the selected boundary pressures
 ($P = 10^{20}$ dyn cm$^{-2})$. From the (quite limited) information available in 
\cite{terasawa02}, we guess that the model may be reproduced by a NASS breeze solution located
 in Fig.~\ref{fig:wind_sm} at around $s_{\rm rad}^{(0)}/f_s \approx 84$ and
 $\log_{10} ({\rm d}M/{\rm d}t/f_{{\rm d}M/{\rm d}t}) \approx -5$.

Some words are also in order concerning the case of the electron-capture DCCSNe, which could 
result from the electron-capture triggered collapse of the O-Ne cores of stars 
of about 9 to 10 $M_\odot$
(Sect.~\ref{explo_1D}). Some successful explosions with low final explosion
 energies and small amount of ejected material have been modelled (see Fig.~\ref{fig_onecore}). 
 As emphasised in Sect.~\ref{simulation_breeze}, the pre-SN structure may have an important 
impact on the neutrino-driven wind properties. It may thus well be that the situation in this 
respect is different for the electron-capture DCCSN of a compact O-Ne white-dwarf and for the
 explosions of more massive iron-core progenitors discussed above. 

The 1D electron-capture DCCSN simulation of \cite{kitaura06} (Sect.~\ref{explo_1D} and 
Fig.~\ref{fig_onecore}) predicts that the ejected material has entropies $10 \lsimeq s 
\lsimeq 40$ and $Y_{\rm e}$ values ranging between 0.46 and 0.53 during the first second of
 the explosion, which is higher than in former models \cite{mayle88}. This difference is 
ascribed by \cite{kitaura06} to an improved treatment of the neutrino transport. These 
conditions make it unlikely that electron-capture DCCSNe are suitable sites for the r-process, 
at least during the early (up to about one second) phase of the explosion simulated by 
\cite{kitaura06}. This is in line with an earlier conclusion \cite{wanajo03}, but contradicts
 claims that even the Pt r-peak could be produced in the explosion of O-Ne white dwarfs (e.g. 
\cite{wheeler98}). It is considered by \cite{kitaura06} that multi-dimensional effects (see 
Sect.~\ref{explo_multiD}) are unlikely to modify the predicted inability of electron-capture 
DCCSNe to produce r-nuclides. Some uncertainties might relate to the still quite poor knowledge 
of the pre-SN evolution and structure.

In conclusion, the possibility for a strong r-process to develop in the spherically-symmetric
 wind or breeze from a PNS formed following the one-dimensional collapse of an iron core remains
 an open question, in spite of the many studies that have been devoted to this question. The
 optimism in this matter is quite limited, especially in view of the fact that the 
one-dimensional DCCSN simulations not only do not seem to provide the suitable wind or breeze
 conditions, but do not even lead to successful supernova explosions! It has to be acknowledged,
 however, that uncertainties remain in the modelling of the properties of the material leaving 
the PNS, this giving some, if faint, hope, especially when one considers that only a 
relatively modest addition of energy input to  the material ablated from the PNS could not only 
greatly aid an explosion (see Fig.~\ref{fig_15msunexplosion}), but possibly also help developing  an 
r-process. Also note that the neutrino-driven winds could in fact produce proton-rich, instead 
of neutron-rich, species during the PNS contraction phase \cite{frohlich06}, 
which even precedes  the early-time r-process discussed by \cite{thompson01}.  

\subsection{The r-process in DCCSN models with additional physics}
\label{r_rescue}

As discussed in Sect.~\ref{explo_multiD}, the failure of the neutrino-driven explosions briefly
 reviewed in Sect.~\ref{explo_1D} has triggered a flurry of simulations involving in particular
 accretion in binary systems, rotation, magnetic fields, or acoustic waves. These additional 
physical processes have been described in the framework of multi-dimensional simulations. One
 has naturally to wonder about the possibility of development of the r-process under the 
modified conditions predicted by these simulations.

\subsubsection{An r-process in accretion-induced DCCSNe?}
\label{r_aic}

\begin{figure}
\center{\includegraphics[scale=0.5]{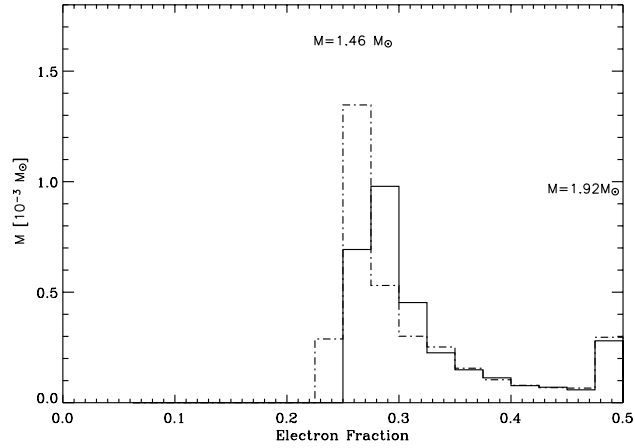}}
\vskip-6.2truecm
\caption{Distribution of the ejected mass versus $Y_{\rm e}$ for 1.46 (dot-dashed line) and 
1.92 (solid line) $M_\odot$  O-Ne white-dwarf AIC progenitors at the end of the simulation 
(slightly over 600 ms after bounce) (from \cite{dessart06})}
\label{fig_aicye}
\end{figure}

\begin{figure}
\center{\includegraphics[scale=0.55 ]{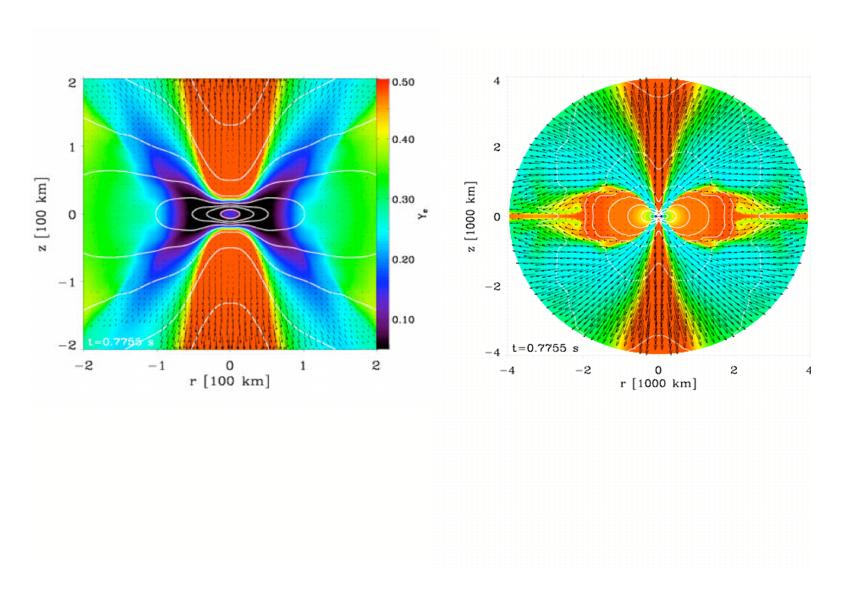}}
\vskip-3.7truecm
\caption{Two-dimensional distribution of $Y_{\rm e}$, 775 ms after bounce of the 1.92 
$M_\odot$ electron-capture AIC. Note the difference in distance scales between the left and
 right panels, the former being a blow-up of the innermost regions of the latter one.
 Superimposed white lines and black arrows represent iso-density contours and velocities,
 respectively (from \cite{dessart06})}
\label{fig_aicye2D}
\end{figure}

As recalled in Sect.~\ref{explo_multiD}, rotating O-Ne white dwarfs might  experience
 a low-energy explosion with a modest mass ejection through the electron capture mechanism
 following an accretion-induced collapse (AIC) in a binary system.  These events are the 
counterparts of the electron-capture DCCSNe briefly reviewed in Sect.~\ref{explo_1D}. 

Figure~\ref{fig_aicye} displays the  overall $Y_{\rm e}$ distribution in the ejecta for two
 progenitor O-Ne white dwarf masses calculated by \cite{dessart06}. This distribution is seen 
to be bimodal. The early ejection of material with $Y_{\rm e} \ge 0.5$ (the calculations 
artificially limit the value of $Y_{\rm e}$ to 0.5) is followed about 200 ms after bounce 
by a neutron-rich $Y_{\rm e} = 0.25 - 0.35$ neutrino-driven collimated wind. Some more details
 on the $Y_{\rm e}$ pattern within the ejecta are displayed in Fig.~\ref{fig_aicye2D}. During
 the expansion of the ejecta, $Y_{\rm e}$ is altered, as in all DCCSN models, by
 $\nu_{\rm e}$ and $\bar\nu_{\rm e}$ captures on nucleons. The resulting asymptotic
 $Y_{\rm e}^{\rm a}$ value is found to be latitude dependent, as illustrated in 
Fig.~\ref{fig_aicye2D} (right panel). The highest $Y_{\rm e}$ values are obtained towards 
the poles, while $Y_{\rm e}$ is close to 0.3 in the equatorial disc at short enough distances,
 but rises again further out. The entropies are decreasing as well away from the poles.  The
 impact of the still very poorly known progenitor evolution and pre-SN structure, the role of 
magnetic fields and possibly associated jets, as well as of 3D effects on the AIC-type of
 supernovae remain to be explored. The possibility of r-processing clearly calls for further
 studies based on new generations of AIC simulations. Could it be that these events are more 
efficient r-producers than their isolated (non-rotating) exploding electron-capture DCCSN white 
dwarf counterparts discussed below ?
 
\subsubsection{Fluid instabilities, rotation, magnetic fields, acoustic waves, and others}
\label{r_others}

As briefly reviewed in Sect.~\ref{explo_multiD}, various fluid instabilities, rotation and
 magnetic fields, as well as the generation of acoustic waves might trigger successful CCSN 
explosions, possibly of the bipolar JetSN type. These different mechanisms might be aided by
 neutrino transport in the production of successful supernovae. They might concomitantly aid 
in increasing the neutrino wind entropy, and consequently in the development or strengthening 
of the r-process. 

The possibility of development of the r-process in the stationary accretion shock instability
or in the acoustically-powered explosion mechanism (Sect.~\ref{explo_multiD}) remains to be 
explored.  Entropies growing in time to large values are predicted by \cite{burrows06} in a
 simulation of the explosion of an $11 M_\odot$ progenitor, as seen in Fig.~\ref{fig_s_acoustic}.
  Entropies even larger than about 300 are obtained at late times. This is in favour of the 
r-process. As discussed by \cite{burrows06}, these high entropies may be related to the adopted
 relatively low-mass progenitor. If this is indeed so, the level of r-processing might decrease
 with increasing mass of the progenitors of acoustically-driven explosions.  These inferences 
clearly deserve further scrutiny.

\begin{figure}
\center{\includegraphics[scale=0.4,angle=270]{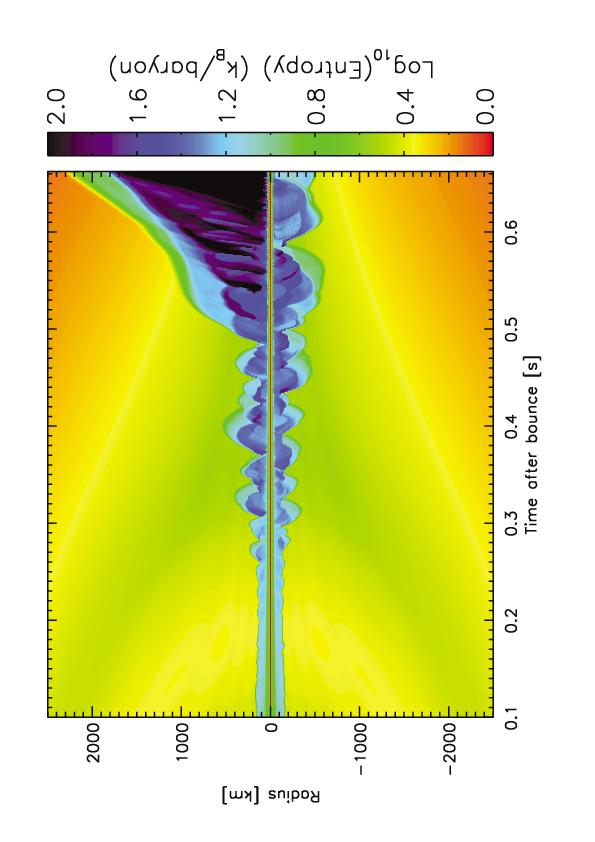}}
\vskip-0.5cm
\caption{Evolution of the entropy after core bounce of an $11 M_\odot$ progenitor along the
 symmetry axis of an acoustically powered supernova, and within the innermost 2500 km (from 
\cite{burrows06})}
\label{fig_s_acoustic}
\end{figure}

\begin{figure}
\resizebox{0.49\hsize}{!}{\includegraphics{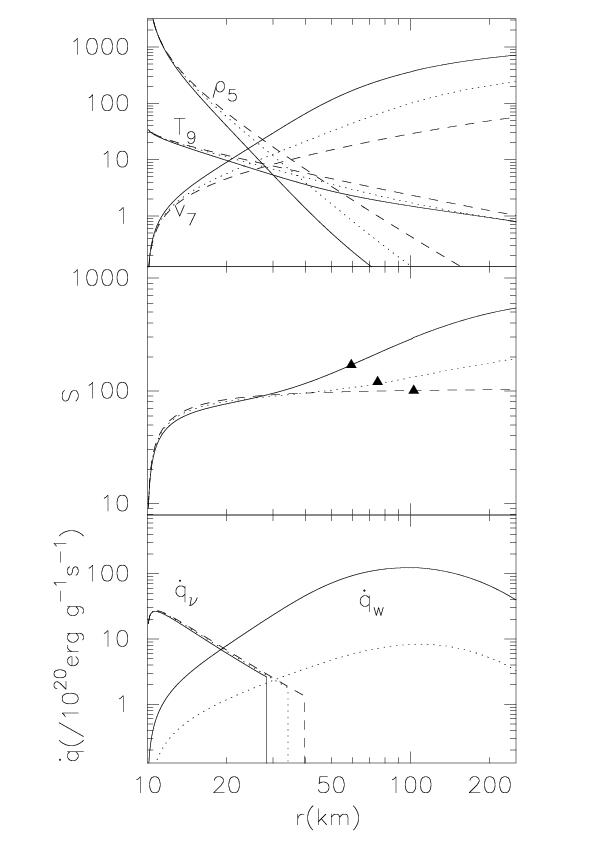}}
\resizebox{0.49\hsize}{!}{\includegraphics{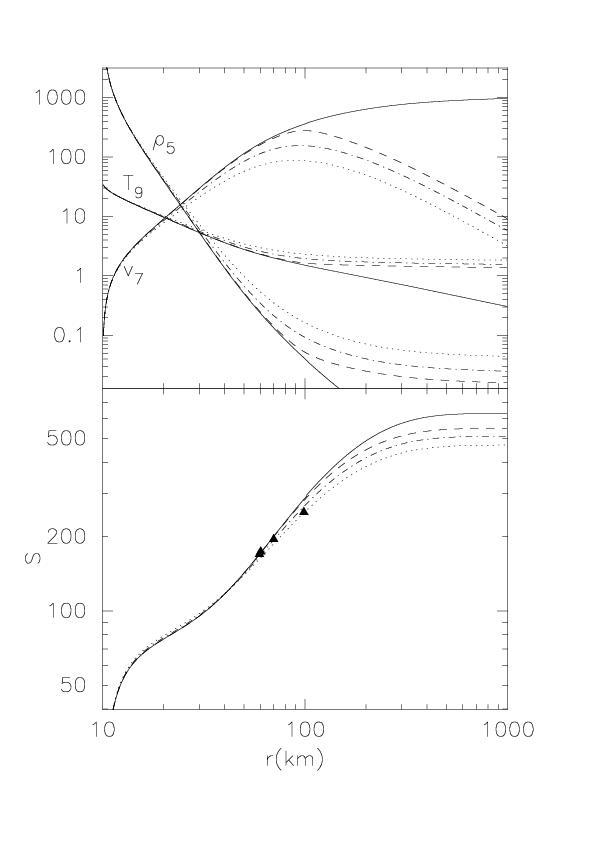}}
\caption{{\it Left panel}: Radial dependence of density $\rho_5$ (in units of $10^5$ gcm$^{-3}$),
 temperature $T_9$ (in units of $10^9$ K), velocity $v_7$ (in units of $10^7$ cms$^{-1}$),
 entropy $s$, heating of the wind by neutrinos $\dot {q_\nu}$ and by Alfv\'en wave dissipation
 $\dot{q}_{\rm w}$ (both in units of $10^{20}$ ergg$^{-1}$s$^{-1}$) of Alfv\'en wave-driven 
transonic winds for magnetic fields at the PNS surface $B_0 = 5 \times 10^{14}$ G (solid lines)
 and $3 \times 10^{14}$ G (dotted lines) and for a simple parametrisation of the wave dumping 
through a dissipation length $l = 10~r_0$, where $r_0$ is the PNS radius, taken to be 10 km. 
For comparison, dashed lines correspond to the absence of Alfv\'en waves. Triangles indicate 
the $kT = 0.2$ MeV location; {\it Right panel}: Comparison between the transonic wind properties 
shown in the two upper left panels for $B_0 = 5 \times 10^{14}$ G and the characteristics 
obtained with the same field and dissipation length in a subsonic case mimicked by simply 
reducing the wind velocity at the PNS surface calculated in the transonic case by 0.1, 1 and 2.5
 \% (dashed, dot-dash, and dotted lines) (from \cite{suzuki05})}  
\label{fig_alfven_wind}
\end{figure}

\begin{figure}
\center{\includegraphics[scale=0.4,angle=270]{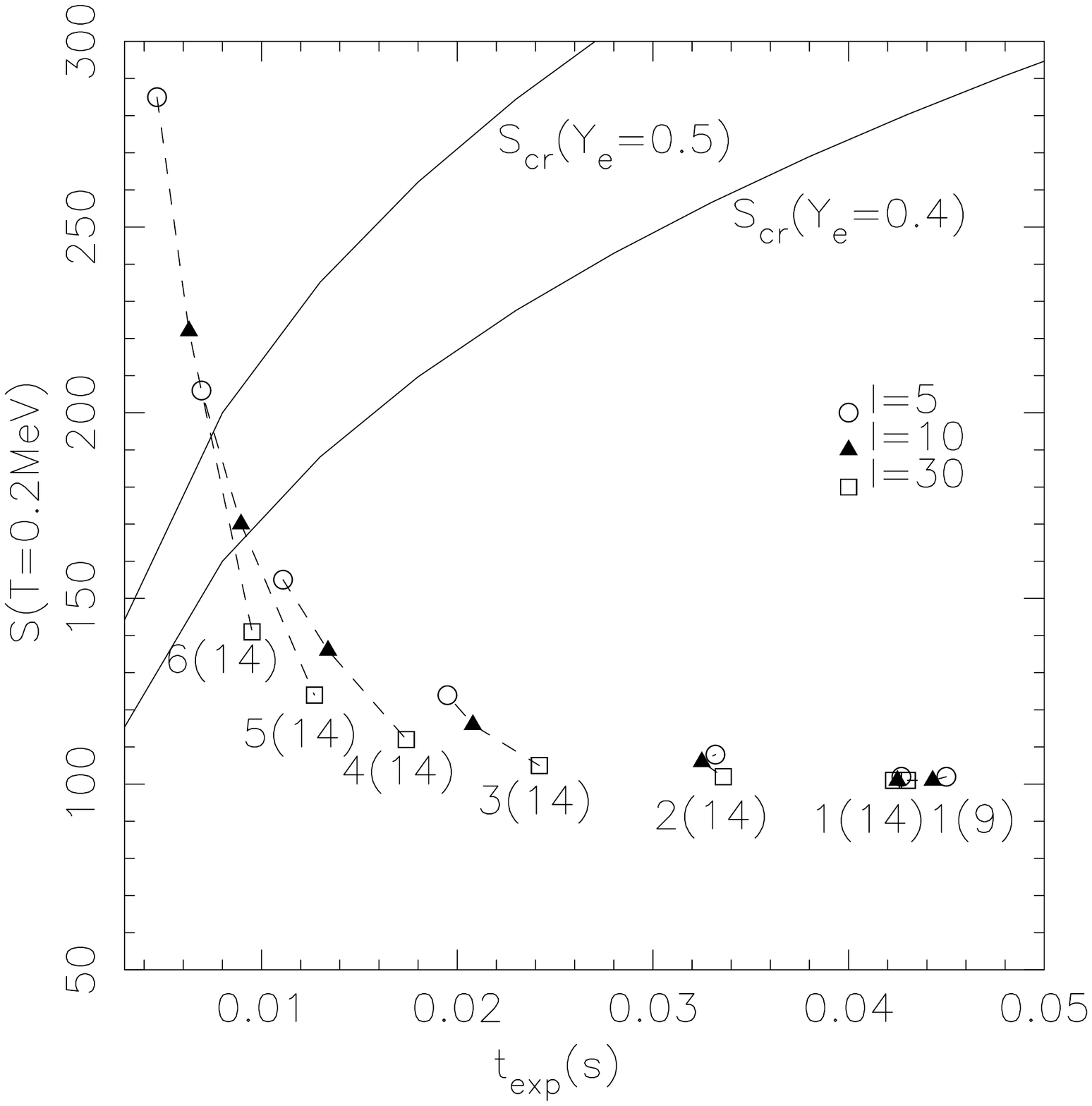}}
\caption{Values of $s$ at $k T = 0.2$ MeV versus expansion timescale $t_{\rm exp}$, defined as 
the time needed for the temperature to decrease from $9 \times 10^9$ to $2.5 \times 10^9$ K for
 various values of $l$ connected by dashed lines if they correspond to the same $B_0$. The 
notation a(b) refers to a field of $a \times 10^b$ G. The curves labelled $S_{\rm cr}$ are 
calculated from $S_{\rm cr} = 2 \times 10^3 Y_{\rm e} t_{\rm exp} {\rm [sec]}^{1/3}$,
 which is the minimum value of the entropy for the r-process to develop, following 
\cite{hoffman97} (from \cite{suzuki05})}
\label{fig_alfven_entropy}
\end{figure}

\begin{figure}
\center{\includegraphics[scale=0.4,angle=270]{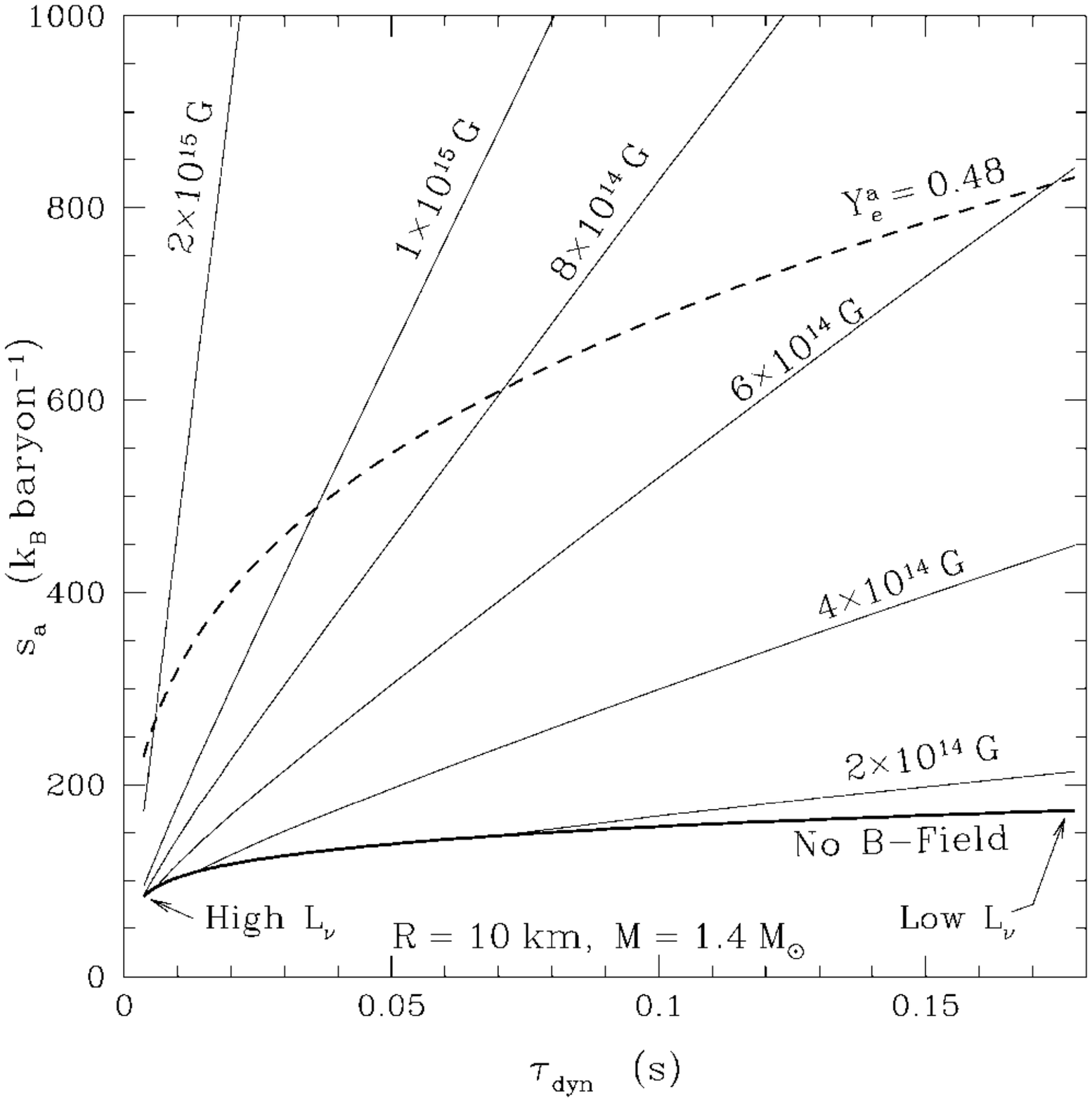}} 
\caption{Wind asymptotic entropy $s_{\rm a}$ versus the dynamical timescale
 $\tau_{\rm dyn}$ defined by Eq.~\ref{eq:wind_time} for a cooling (the neutrino luminosity 
decreases from left to right) $1.4 M_\odot$ PNS with 10 km radius, and for different values of
 the PNS magnetic dipole strength $B_0$. It is assumed that $\tau_{\rm dyn}$ is independent
 of $B_0$. The dashed line shows the value of $S_{\rm cr}$ defined in 
Fig.~\ref{fig_alfven_entropy} for the asymptotic value $Y_{\rm e}^{\rm a} = 0.48$
 (from \cite{thompson03})}
\label{fig_dipole_entropy}
\end{figure}

The extent to which the r-process could be made more probable as a result of the additional
 energy input brought into the neutrino-driven wind by magnetic fields has been studied by 
\cite{suzuki05} using a simple configuration of radially-open magnetic flux tubes in a steady-state 
situation without rotation or general relativistic effects.  It is demonstrated that the
 outward propagation of Alfv\'en waves excited by the surface motions of a PNS heats and
 accelerates its ablated material, at least for field strengths in excess of about $10^{14}$ G,
 which is typical of magnetars (Sect.~\ref{magnetar_model}). The impact of Alfv\'en waves on the
 wind properties is illustrated in Fig.~\ref{fig_alfven_wind} for the transonic and several
 subsonic (breeze) winds. It appears that the wave heating is much more widely distributed 
around the adopted dissipation length $l$ (100 km in the figure) than the neutrino 
heating is. This leads to differences in the wind structure.
The differences get larger and larger beyond a certain distance from the PNS that increases 
with $l$.
  This concerns in particular the entropy, which still
 increases quite significantly at large distances instead of remaining close to constant when
 just neutrino heating is taken into account.  Figure \ref{fig_alfven_entropy} further 
demonstrates that the value of $s$ at $k T = 0.2$  MeV (adopted as it roughly represents 
the condition of freeze-out of the $\alpha$-process, and the start of neutron captures possibly 
leading to an r-process) rises quite substantially with increasing $B_0$ and  decreasing 
dissipation lengths $l$ in the considered $5 \leq l \leq 30$ range, this entropy increase being 
accompanied by a shortening of the expansion timescale.  This translates into better 
possibilities of development of the r-process, as illustrated in Fig.~\ref{fig_alfven_entropy},
 neutron captures being efficient above the curves labelled `$S_{\rm cr}$'  if one relies on 
an approximate criterion proposed by \cite{hoffman97}.

Concerning Alfv\'en-driven subsonic winds,  the right panel of Fig.~\ref{fig_alfven_wind} shows 
that their structure does not differ markedly from the transonic wind solution at distances 
$r \lsimeq l = 10~r_0 = 100$ km, while differences are seen at larger distances.  It is
 speculated by \cite{suzuki05} that the subsonic winds might offer a somewhat better opportunity
 of development of the r-process because of  the slowing down of these winds at large distances 
(Fig.~\ref{fig_alfven_wind}), giving more time for neutron captures to take place. This 
question has clearly to be scrutinised further.

Closed magnetic fields might have a variety of effects \cite{suzuki05}, some of them being 
possibly in favour of an r-process. In particular, \cite{thompson03} shows that matter may become
 trapped by an ordered magnetar-type magnetic field close to the PNS surface before escaping 
as a result of neutrino heating with a substantial entropy increment when compared with
 the  non-magnetic case. This is vividly illustrated in Fig.~\ref{fig_dipole_entropy}, where 
entropies well in excess of $S_{\rm cr}$ (defined in Fig.~\ref{fig_alfven_entropy}) can be
 obtained for large-enough magnetic-dipole strengths. As a result, a robust r-process might develop.

The conclusions that magnetic fields could bring a more or less significant aid to the
 development of the r-process in PNS neutrino-driven winds are challenged by \cite{ito05}
 through numerical simulations making use of  a homogeneous PNS magnetic field of $10^{12}$ to
 $5 \times 10^{15}$ G perpendicular to the radial direction. The wind dynamics and the key
 quantities for the r-process ($s, Y_{\rm e}, \tau_{\rm dyn}$) are indeed found not to be 
significantly different from the non-magnetised case. The question of the impact of magnetic 
fields on the development of the r-process in the PNS wind is clearly worth 
further investigations by relaxing some of the approximations and assumptions made by 
\cite{suzuki05,thompson03}. A detailed solution of the full time-dependent MHD equations has 
clearly to be worked out in order to examine in particular more complicated magnetic field 
topologies, trapping timescales in closed magnetic loops, or the development of magnetic 
instabilities and magnetic re-connexions. The coupling of magnetic fields with rotation or
 convection may additionally lead to JetSNe configurations of r-process relevance.
 
In fact, nucleosynthesis in JetSNe may occur through nuclear burning associated with the 
fallback material, in the material outflowing from the accretion disc (disc outflow), or in the
 jets themselves, which are likely ultra-relativistic, as it is the case in gamma-ray bursts.

{\it Nucleosynthesis in the JetSN jets}. The nucleosynthesis in the jets, and in 
particular the possible r-nuclide enrichment of their material, has been discussed only on 
very qualitative grounds in the framework of the collapsar model of ultra-relativistic jets 
emerging from the vicinity of the central black hole \cite{pruet03,cameron03}. Even if the 
neutron excess in the jets is large, the expansion timescale may be too short, and the density 
at the temperatures allowing the recombination of the nucleons may be too low for complex 
nuclei to form. Some hope remains of producing these nuclei when the jet material starts 
interacting with the overlying stellar material, and in particular with the He-rich to Si-rich
 layers of  possible Wolf-Rayet progenitors of some gamma-ray bursts. It is speculated by 
\cite{pruet03} that most of the neutrons in the jets would have time to be captured by these
 nuclei, and possibly lead to the production of some r-nuclides. It is quite clear that no 
definitive conclusion can be drawn at present on this matter. Multi-dimensional relativistic 
simulations are required, as well as a careful study of the possible role of spallation 
reactions taking place during the interaction between the jets and the overlying material 
\cite{cameron03}.

\begin{figure}
\center{\includegraphics[scale=0.4,angle=270]{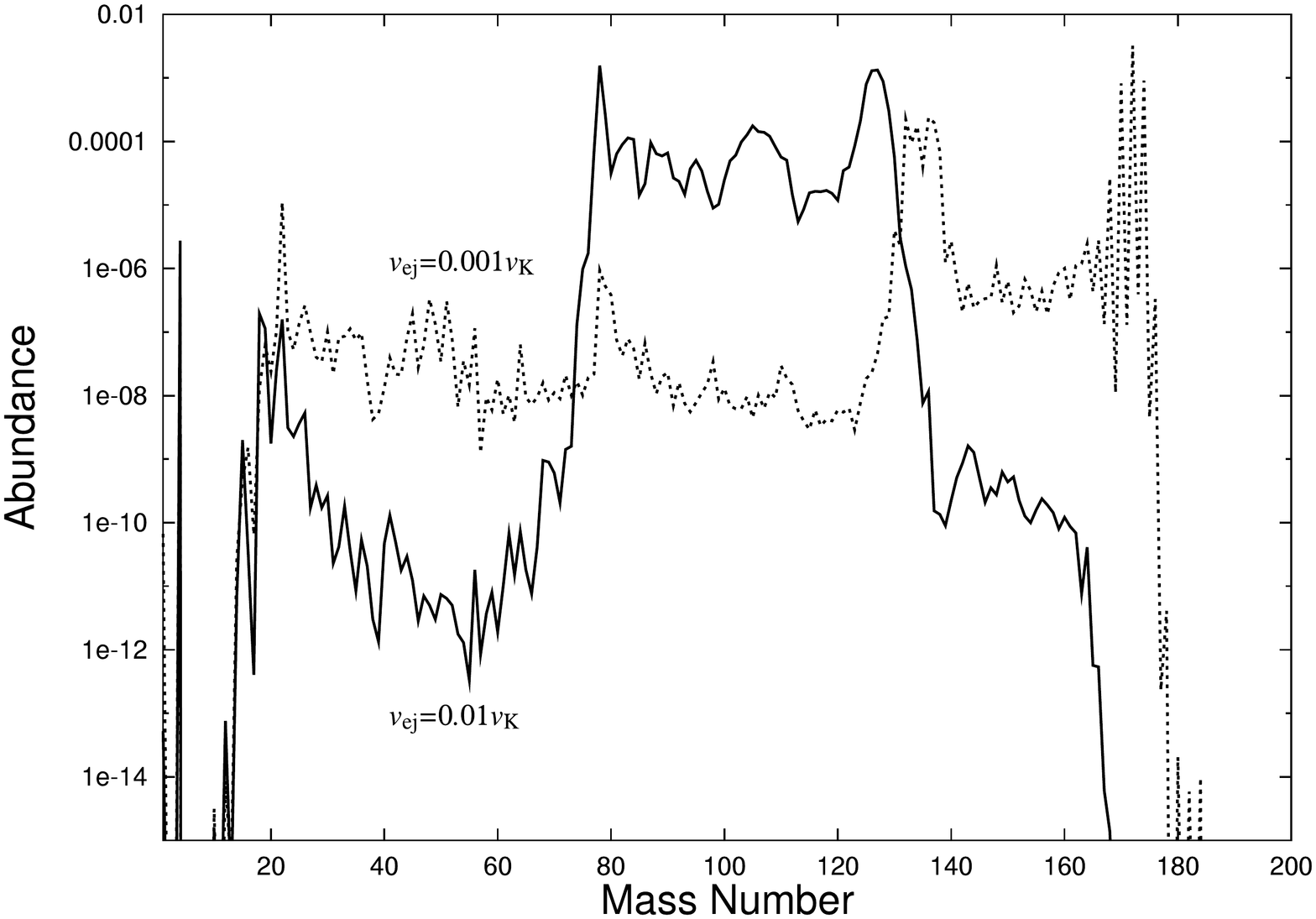}}
\vskip-0.3truecm
\caption{Distribution of r-nuclides in the wind from the inner regions of a collapsar-type
 accretion disc. The wind is adopted to emerge at a distance of ten times the black hole 
Schwarzschild radius ($r \approx 1000$ km) and with constant ejection velocities $v_{\rm ej} =
 0.01$ and 0.001 $v_{\rm K}$, where $v_{\rm K}$ is the Keplerian velocity at the considered 
position in the disc. Initially, $Y_{\rm e} = 0.30$ and $s = 13.1$ are assumed (from 
\cite{fujimoto04})}
\label{fig_r_collapsar}
\end{figure}

 {\it Nucleosynthesis in the JetSN disc outflows}. The possibility of development of an 
r-process in 2D simulations of the MHD CCSN disc outflow of a $13 M_\odot$ progenitor has been
 explored by \cite{nishimura06} under various assumptions concerning pre-SN rotation and magnetic
 fields. It is found that the structure, as well as the nucleosynthesis outcome, depend sensitively
 on the selected initial rotation and magnetic fields. More specifically, the level of r-processing 
is concluded to increase with the extent of collimation of the ejected material,  the Pt 
r-abundance peak being obtained for the best-developed JetSN structure. These results have to be
 taken with some care, however. Worries have been expressed on the ZEUS-2D code used for the MHD
 simulation \cite{falle02}. On the other hand, and perhaps more importantly, neutrinos have been
 neglected at the level of both their transport and their interaction with matter, which may in 
particular affect quite drastically the predicted entropies and $Y_{\rm e}$ values of the ejecta,
 and consequently the level of r-processing.  

The nucleosynthesis in the disc outflow has also been studied in the framework of JetSNe of the 
collapsar type \cite{macfadyen99}, which involves a rapidly accreting black hole at the centre of
 a massive progenitor and associated jets that may account for at least certain gamma-ray bursts.
  A typical collapsar-powered gamma-ray burst involves accretion rates up to typically 
$0.1 M_\odot$/y, leading to an accumulated mass of the order of one to several $M_\odot$, of which
 a  sizable fraction may be ejected. As demonstrated by \cite{pruet03,fujimoto04,surman06}, the 
composition of the disc outflow depends to various extents on the disc initial composition, the 
black-hole mass and rotation velocity, accretion rate, viscosity and neutrino luminosity of, and 
trapping in, the disc. These quantities indeed affect $s$ and $Y_{\rm e}$, which are 
 varying with the distance from the accreting black hole. Magnetic fields may also influence the 
properties of the outflow, as discussed by \cite{daigne02} in the framework of a simplified model 
which in particular freeze $Y_{\rm e}$ to 0.5, so that nucleosynthesis considerations based on 
this model are premature. Note that the disc is optically thin to neutrinos in typical collapsar
 conditions, which is likely not the case in typical merger conditions (Sect.~\ref{coales_model}).

Figure~\ref{fig_r_collapsar} displays the r-abundance distribution calculated by \cite{fujimoto04}
 with a simple one-dimensional model of an adiabatic, steady wind emerging from the inner
 neutron-rich portion of an accretion disc where $Y_{\rm e} \ll 0.5$ may be obtained. It is seen 
that the amount of produced heavy r-nuclides increases substantially with decreasing ejection 
velocities. This results from the longer time available for neutron captures in slower ejecta. For
 $v_{\rm ej} \gsimeq 0.1 v_{\rm K}$, the winds calculated by \cite{fujimoto04} only contain  
light elements up to Be. 

It has to be made clear that the above conclusions are still highly uncertain, and that the 
possibility of r-processing in collapsar-type disc winds remains to be scrutinised further with 
the use of more realistic MHD-driven or viscosity-driven wind models from detailed disc models 
calculated for a variety of pre-collapsar models.

{\it Nucleosynthesis in the JetSN fallback material}. As already encountered at several
 occasions above, some of the material ablated from a PNS may not reach the escape velocities, and
 falls back onto it when the shock wave associated with the explosion decelerates as it moves 
through the star. This slowing  down is able to generate a reverse shock that has implications on 
the dynamics of the neutrino-driven winds (Sect.~\ref{simulation_breeze}). The amount of fallback
 material and the configuration in which it settles depend on many characteristics of the DCCSN. 
 Current simulations predict that the fallback may lead to a total accreted mass of the order of
 0.1 to over $1 M_\odot$. A black hole may in fact result from high enough fallback rates. The view
 has also been expressed that fallback develops only in explosions with total kinetic energies
 lower than the typical values (of the order of $10^{51}$ erg). The fallback is also sensitive to
 the rotation rate of the stellar core, and may settle into a disc with associated jet outflows, 
possibly related in some cases to certain gamma-ray bursts. Some fraction of the fallback material
 may eventually be ejected, the efficiency of this ejection being sensitive to the inner boundary 
conditions (black hole accretion disc, or PNS surface), angular momentum of the infalling material,
 as well as on the neutrino cooling and heating. The ejecta from the fallback expands and cools 
from temperatures in excess of about $10^{10}$ K to values of the order of 2-4 $\times 10^9$ K on 
millisecond timescales. This expansion is, however, slowed down by the interaction of the ejected 
material with the material still falling back. As a result, the temperature drops less
 precipitously. As noted by \cite{fryer06}, the temperature history is highly sensitive to the
 initial conditions of the fallback, and is consequently still highly uncertain.

The explosive nucleosynthesis in the fallback material in a collapsar-type model has been shortly 
discussed by \cite{fujimoto04}. No reference, however, is made to the possible development of an 
r-process.

\begin{figure}
\center{\includegraphics[scale=0.5]{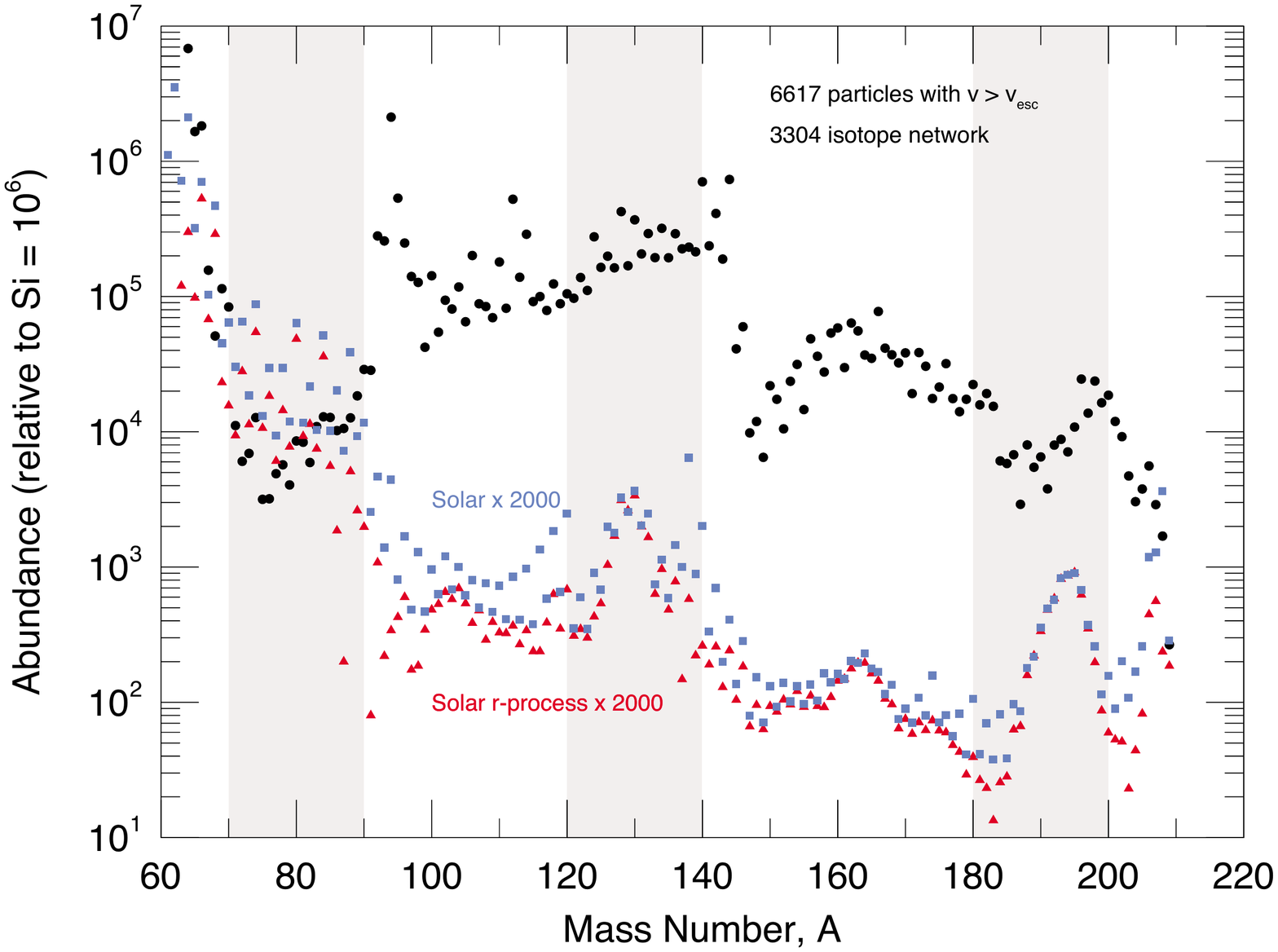}}
\vskip-0.5cm
\caption{Stable nuclide abundances (solid circles) in the ejecta from the fallback material on a 
low-angular momentum $1.4 M_\odot$ PNS with a radius of 10 km. A NSE is assumed initially, and 
$Y_{\rm e}$ is held at the constant value of 0.5. The (normalised) total SoS and r-nuclide 
SoS abundances from \cite{anders89} are shown by
 squares and
 triangles (from  \cite{fryer06}) }
\label{fig_fallback2}
\end{figure}

In contrast, the possibility of having the r-process developing in material falling back onto a
 $1.4 M_\odot$ PNS with a 10 km radius has been explored by \cite{fryer06} in a 2D Smooth Particle
 Hydrodynamics simulation under the condition
that the core angular momentum is low enough for the
 accreting matter not forming a centrifugally supported disc. In this study, some simplifications
 are made concerning in particular the luminosities of the neutrinos. Their interactions are also
 neglected, so that $Y_{\rm e}$ remains constant to assumed initial value of 0.5.
 Figure~\ref{fig_fallback2} displays the composition of the ejected fraction of the fallback 
material for $Y_{\rm e} = 0.5$ and for an initial temperature that is high enough for a NSE to
 hold. It is seen that a significant amount of heavy nuclides are produced. These nuclides are of
 the r-type. The region around the $A = 195$ peak is the result of a three-step operation during
 which the $\alpha$-particle captures first freeze-out before the proton-captures become too slow
 in their turn. The remaining neutrons are then captured on a timescale of milliseconds, driving 
the material on an r-process path. 

The fact that the computed distribution shown in Fig.~\ref{fig_fallback2} does not fit the SoS is
 not a concern at this point, considering the very preliminary nature of the calculation. The
 characteristics of the fallback material and of the ejecta remain to be modelled in detail before 
deriving any firm conclusion on the possibility of development of the r-process in the fallback 
scenario.

\subsection{Neutron captures in exploding He- or C-rich layers}
\label{coco_HeC}

It has long been recognised that neutrons could be released in the He- or C-rich layers of SNIIe 
 heated by the associated outward-moving shock wave.  This has raised the hope that an r-process 
could develop in such locations (e.g. \cite{meyer94} for references). 

\begin{figure}
\center{\includegraphics[scale=0.6,angle=0]{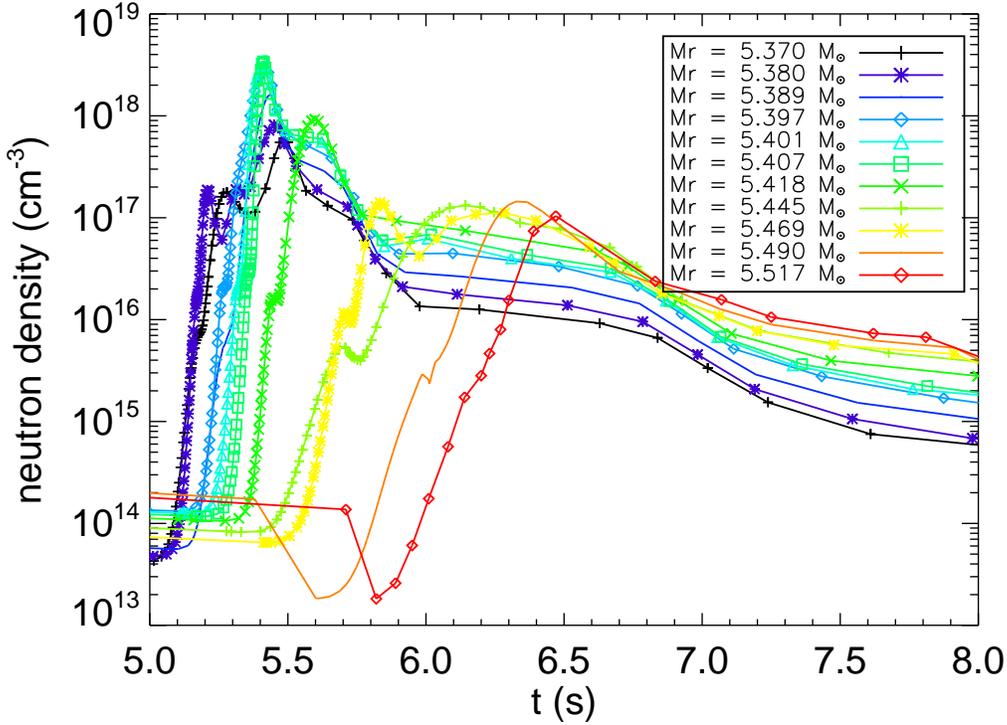}}
\vskip-1.3cm
\caption{Time evolution of the neutron number density in various layers of the
 explosively burning He shell of a 25 $M_\odot$ star. Each curve corresponds to a fixed mass
 coordinate $M_{\mathrm r}$, as labelled.  Time is measured 
since the energy triggering the explosion was put at the stellar centre. 
(Courtesy: L.-S. The, B.S. Meyer; see also \cite{meyer04})}
\label{fig_burst}
\end{figure}

A neutron-capture episode could be encountered in explosive He-burning as a result of the neutrons 
produced by ($\alpha$,n) reactions on pre-existing \chem{22}{Ne} or Mg isotopes. This neutron
 production could be augmented by inelastic scattering on \chem{4}{He} of neutrinos streaming out 
of the PNS at the centre of the exploding star \cite{epstein88}. Parametric studies have 
demonstrated, however, that too little neutron supply could be achieved 
to allow for the production  of a SoS type of r-nuclide abundance pattern. 
 Only a limited redistribution of the pre-explosion heavy nuclide abundances is
 possible. This conclusion is confirmed by the use of recent stellar
 models \cite{rauscher02,meyer04} which indeed predict by far too low neutron densities to generate
 a well-developed r-process (see  Fig.~\ref{fig_burst}). This shortcoming has been shown 
to be circumvented only if totally ad-hoc and astrophysically implausible assumptions were made, 
particularly concerning the initial amount of the relevant Ne or Mg isotopes and/or of the heavy 
seeds for radiative neutron captures (especially Ba). The hope has been expressed that some 
meteoritic r-nuclide anomalies (Sect.~\ref{anomalies}) could emerge from the limited neutron-capture
 process (sometimes referred to as the n-process) accompanying explosive He burning (see
 Sect.~\ref{anomalies_obs_th}).  In addition, it is speculated that some short-lived radio-nuclides
 (\chem{36}{Cl}, \chem{41}{Ca}, \chem{60}{Fe}, \chem{182}{Hf}) might have been injected live from 
the He shell into the early solar system \cite{meyer04}.

Parametric studies of explosive carbon burning along lines similar to those  followed for the
 explosive He burning have also been conducted. In particular, the production by neutron 
captures of the elements up to Zr has been computed by \cite{wefel81}, who conclude that only some
 contribution to the ultra-heavy cosmic-rays or to some isotopic anomalies can be expected
 (Sect.~\ref{anomalies_obs_th}), whereas the bulk SoS r-nuclides just beyond iron cannot be produced 
in such a way.  A very limited production of some light ($A \lsimeq 88$) r-nuclides is predicted 
in the C- and Ne-rich shells 

Finally, it may be worth noting  that the limitations of the neutron-capture efficiency in the explosively
 processed He and C shells derived from the parametric models mentioned above are confirmed by
 calculations performed in the framework of detailed stellar models \cite{blake81,rauscher02}. 

\section{Compact objects: a site for the high-density r-process scenario?}
\label{compact_general} 
 
 As recalled in Sect.~\ref{HIDER}, the decompression of the crust of cold neutron stars (NSs) made 
of a lattice of very neutron-rich nuclei immersed in a gas of neutrons and degenerate electrons has 
long been envisioned as a possible site for the development of a high-density r-process (HIDER). 
This decompression could result from the coalescence of two NSs or of a NS and a black hole (BH)
 in a binary system. It could also result from the ejection of material from magnetars.  
  
\subsection{Neutron star coalescence}
\label{coales_model}

The modelling of the coalescence of two NSs (e.g. \cite{oechslin06}, and references therein) or of 
a NS and a BH (e.g. \cite{rosswog04,rosswog05}, and references therein) has attracted a flurry of 
interest recently. These events are indeed considered to be among the strongest known sources 
of gravitational wave radiation, this emission being in fact responsible for the coalescence after
 typical times of the order of  tens of millions  to billions of years. They are also viewed as 
the likely progenitors of the class of short hard gamma-ray bursts (e.g. \cite{piran05}). The 
available simulations demonstrate that the details of a merging and its ability to account for 
gamma-ray bursts depend more or less drastically on the characteristics of the binary partners, 
like their nature (NS or BH), mass, rotation and magnetic field, as well as on the nuclear NS equation 
of state. 

In the NS + NS case, the main stages of the merging may be briefly sketched as follows:\\
(1) as a result of mass transfer between the two NSs during the last phases of the inspiral 
process and of the centrifugal forces acting at that time, the two NSs develop long tidal arms
 stretching into a disc/torus made of cold material from their crust. Through its expansion, 
this material decompresses while releasing energy via  nucleon recombination in nuclei and
 radioactive decay. An r-process might accompany this decompression, as described in 
Sect.~\ref{r_decompression}. The  energy production may lead a fraction of the tips of the tidal
 arms to escape from
the system prior to the merger, the ejected material remaining unaffected by the
 event. The amount of mass lost in such a way (from about $10^{-4}$ to even more than
 $10^{-2} M_\odot$) depends sensitively on the total angular momentum of the system and is 
largest when the NS spins are aligned along the orbital angular momentum. It also depends on the
 time it takes for the merging to produce a  BH, being possibly prevented if this formation 
occurs on a dynamical timescale [see (2) below];\\
(2) the merging may lead to the formation of a central hyper-massive hot NS that collapses
 immediately or after some delay to a rapidly rotating BH, depending on its mass, rotation rate 
and of the equation of state adopted in the simulations. It is surrounded by a thick neutron-rich
 disc/torus made of material of the two NSs in proportions that vary with their initial mass-ratio.
 This material is heated by neutrinos emitted by the merging NSs, or by viscous dissipation when
 the hot torus starts being swallowed by the newly-born BH. As a result, the torus may experience 
mass outflow in the form of a neutrino-driven wind. This wind shows some similarity with the one 
 described in the case of DCCSNe (Sect.~\ref{r_dccsn}). However, as a result of possibly higher
 disc/torus accretion rates, neutrinos are likely to be trapped much more efficiently in the torus
 than in the disc envisioned in the collapsar-type models. Much larger entropies could also be
 achieved, especially along the rotational axis. It has to be emphasised that these statements are 
qualitative (e.g. \cite{surman06}), no detailed simulation providing a quantitative picture of the
 development of the neutrino-driven wind and of its post-merging evolution. The possibility of 
development of an r-process in the wind-ejected material is described in Sect.~\ref{r_torus};\\ 
(3)  jets of ultra-relativistic material may emerge from the post-merger BH-torus, and establish
 the connexion between this system and the short duration gamma-ray bursts. The jets may be powered
 by the thermal energy released by the annihilation of $\nu\bar\nu$ pairs, or by MHD effects. The
 development of the jets puts constraints on the density and configuration of the material ejected 
through the neutrino-driven wind.  

Note that an alternative, and still quite speculative, scenario for the fate of a strongly
 asymmetric neutron-star binary has been discussed by \cite{sumiyoshi98}. It has to do with the
 transfer by the less massive NS to its companion of enough material to bring its mass just below
 its minimum equilibrium configuration value before disruption into a tidal disc. The stripped star
 might explode after a phase of quasi-static expansion, possibly leading to the production of 
r-nuclides, including the heavy ones \cite{sumiyoshi98}. This possibility is not reviewed further,
 as this r-process scenario has not been discussed in much detail up to now.

The NS + BH merger shares some similarities with the NS + NS case, but also shows important 
specificities (e.g. \cite{rosswog04,rosswog05}), the leading one being that none of the simulated 
NS + BH cases are viable gamma-ray burst progenitors. This is a direct consequence of the fact that
 only low-mass geometrically thin and relatively low-temperature discs develop for the considered 
$M < 18 M_\odot$ BH cases, the difficulty of forming an accretion disc outside the BH Schwarzschild
 radius increasing with the NS spin. The configurations obtained at the end of the simulations are 
made of a BH and a low-mass NS engulfed in a common low-density envelope consisting of material 
decompressed from the NS. From a nucleosynthesis point of view, the decompression of this NS 
material might be quite similar to the one considered in the NS + NS case (see 
Sect.~\ref{r_decompression}). The late-time fate of the predicted `mini NSs' is unknown. The
 situation might well be reminiscent of the one considered by \cite{sumiyoshi98} (see above), 
with the possible NS explosion and r-nuclide production. Even
when this production occurs during the 
decompression or as a result of the NS explosion, it remains to be seen if the r-nuclides can
 escape the system.

An extreme situation is found for BH masses $M \gsimeq 18 M_\odot$ \cite{rosswog05}. In this case,
 most of the NS disappears into the BH without disc formation, the remaining NS debris whose mass
 can be as high as about $0.1 M_\odot$ being spun up by tidal torques and ejected as a half-ring 
of neutron-rich matter. The rapid decompression of this NS matter may lead to a robust r-process 
(Sect.~\ref{r_decompression}).

As emphasised by e.g. \cite{oechslin06}, much remains to be done in the modelling of the NS + NS/BH
 mergers in order to derive truly reliable predictions concerning the possible links with short
 gamma-ray bursts, not to mention the possible r-nuclide yields from these systems. The frequency of
 these events is also an open and important issue for galactic chemical evolution predictions 
 (Sect.~\ref{galaxy_obs_th}).

\subsection{Magnetars}
\label{magnetar_model}

As mentioned in Sect.~\ref{explo_multiD}, magnetars may result from certain JetSNe. Their
 properties are reviewed by  \cite{woods06,harding06}. They are observed as Soft Gamma Repeaters 
emitting sporadically bright bursts of energy over a period that is estimated to be of the order
 of 10000 years. Their most remarkable characteristic is their magnetic fields whose  values are
 typically of the order of $10^{14-15}$ G, that is 100 to 1000 times larger than classical pulsar 
values.   This results in extremely-large magnetic forces that deform the crust and causes the 
magnetic field penetrating the crust to shift and move. Occasionally, the crust and the overlying
 field become catastrophically unstable, leading to observed giant flares and inducing significant
 changes in the crust structure with the possible ejection of baryonic matter, at least if the
 field is of the order of $10^{14}$~G or larger in order to affect the solid NS crust to a 
sufficient extent (this may somehow remind us of the NS `volcanoes' talked about by e.g.
 \cite{lattimer77}).  In the particular case of the giant flare detected from the magnetar SGR 
1806-20 on Dec. 27, 2004,  it has been estimated \cite{gelfand05}  that as much as
 $5 \times 10^{-9}$ to 5 ~ 10$^{-7}$ $M_\odot$ have been ejected with an initial kinetic energy  
$E \gsimeq 10^{44.5}$~ergs.  Note that newly-born NSs could also eject some material before the
 formation of a crust as a result of the so-called dynamo action process induced by the combined
 effect of rotation and convection \cite{thompson03}.  

Even if highly-magnetised NSs are likely capable of ejecting part of their surface matter into 
the interstellar medium, the frequency of such events and the total amount of matter they can expel
 remain unknown. 

\subsection{The r-process in the decompression of cold neutron star matter}
\label{r_decompression}
 
As briefly reviewed in Sect.~\ref{coales_model}, the coalescence of  NS + NS and of certain NS +
 BH binaries may be accompanied by the ejection of decompressed cold NS material. This matter 
can be highly neutron-rich, as illustrated in Figs.~\ref{fig_crust} and \ref{fig_nsbh}. Note that
 the $Y_{\rm e}$ distribution shown in Fig.~\ref{fig_nsbh} closely resembles the NS crust values
 of Fig.~\ref{fig_crust}, this similarity resulting from the fact that the parts of the NS that 
experience $Y_{\rm e}$ changes due to heating during the merging process disappear very quickly
 into the BH.

\begin{figure}
 \begin{center}
\includegraphics[scale=0.5]{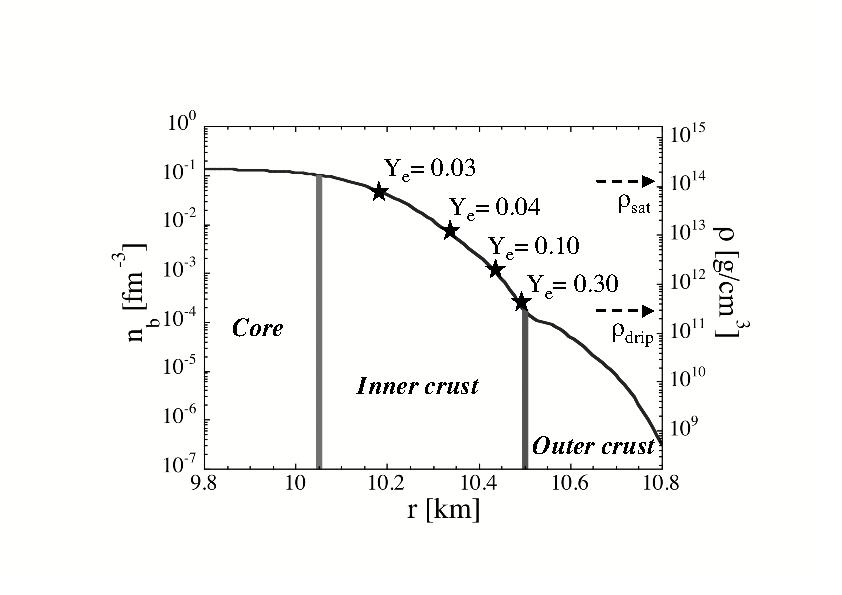}
\vskip -1cm
\caption{Density profile, expressed in baryon number density $n_{\rm b}$ and matter density $\rho$, in 
a typical 1.4 $M_\odot$ NS showing the structure of the crust.  The saturation and drip densities
 are noted $\rho_{\rm sat}$ and  $\rho_{\rm drip}$. The electron fraction $Y_{\rm e}$ at 
$\beta$-equilibrium is calculated as described in the text. Its values are given at
 $\rho=4 \times 10^{11}, 2 \times 10^{12}, 10^{13}$ and $10^{14}~{\rm g/cm^{3}}$}
\label{fig_crust}
\end{center}
\end{figure}

\begin{figure}
\center{\includegraphics[scale=0.57,angle=270]{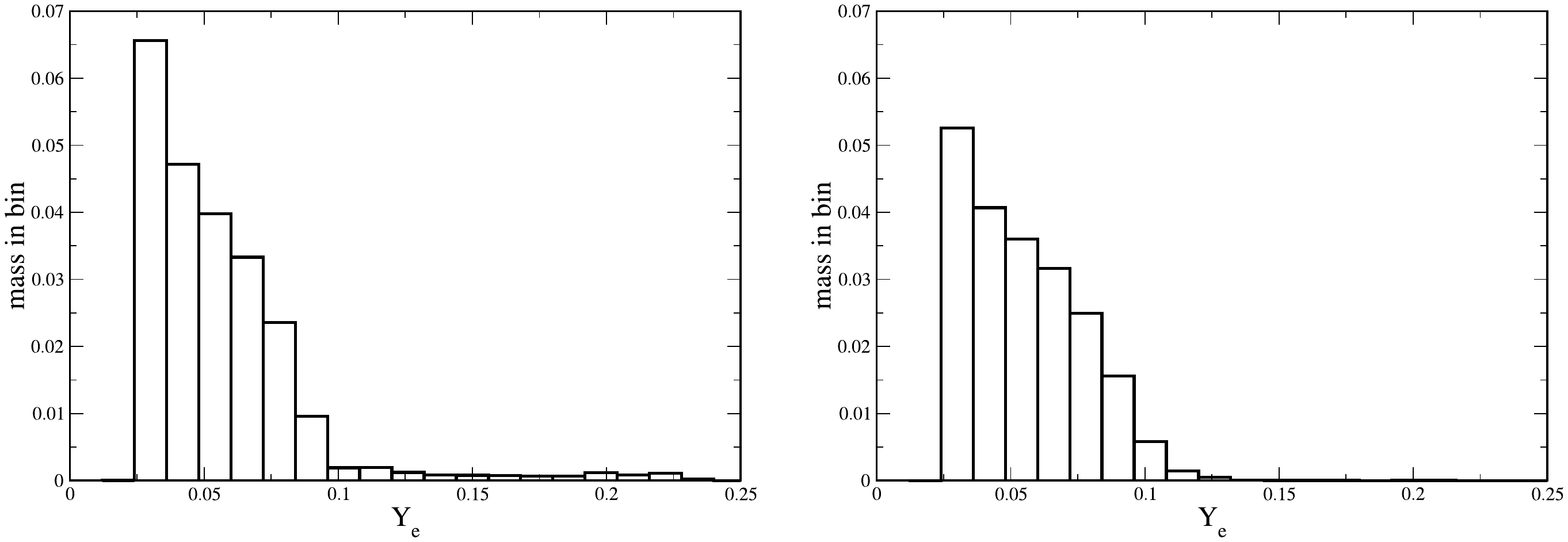}}
\vskip-3.0cm
\caption{Snapshot distribution of the $Y_{\rm e}$ values within the ejected NS debris resulting 
from the merging of a 1.4 $M_\odot$ NS and a 14 $M_\odot$ BH. {\it Left panel}: the NS is assumed 
to co-rotate (Run 2 of \cite{rosswog05}), in which case 0.2 $M_\odot$ of NS material is ejected; 
{\it Right panel}: the NS is not spinning (Run 7 of \cite{rosswog05}), with the ejection of 
0.17 $M_\odot$ of its matter}
\label{fig_nsbh}
\end{figure}

Following the early works of \cite{lattimer77,symbalisty82}, the first detailed calculation of the
 r-process that can accompany the decompression of NS material has been performed by \cite{meyer89a}
 in a systematic parametric study. The expansion has been followed down to densities around
 the neutron-drip density ($\rho_{\rm drip}\simeq 3 \times 10^{11} {\rm g/cm^3}$) only, however. In 
contrast, only densities below this value have been considered by \cite{frei99}, the initial 
composition of the material being assumed to result from a NSE at high temperatures ($T_9\simeq 6$) 
and with $Y_{\rm e}$ viewed as a free parameter. However, these selected  high temperatures cannot
 plausibly be reached in the unshocked NS crust material that eventually gets dynamically stripped 
and expands very quickly from the NS surfaces during a merger. 

An improved model calculation has been performed \cite{go05} of the 
composition of the dynamically ejected material from cold NSs making use of a detailed treatment 
of the micro-physics and thermodynamics during  the decompression. The adopted initial density 
structure of the NS crust is depicted in Fig.~\ref{fig_crust}. The evolution of the matter density 
is modelled by considering the pressure-driven expansion of a self-gravitating clump of NS matter
 under the influence of tidal forces on an escape trajectory.  The expansion is characterised by
the expansion timescale $\tau_{\rm exp}$, defined as  the time needed for the initial density 
to drop by three orders of magnitude.

The final isobaric composition of the outer crust (Fig.~\ref{fig_crust}; initial densities 
$\rho < \rho_{\rm drip}$)  after decompression is almost identical to the one prior to the
 ejection. Only $\beta$-decays (including $\beta$-delayed neutron emission) can change the initial 
composition. This conclusion is drawn from a calculation identical to the
previous one \cite{baym71},
but performed with the updated nuclear physics of Sect.~\ref{nuc_static} and \ref{beta}. The matter 
is assumed to be in thermodynamic equilibrium and a minimisation of the Gibbs free energy
 to estimate the zero-temperature composition. The energy of the body-centred cubic lattice and of
 the relativistic electrons is included. For densities above $ 3 \times 10^9  {\rm g/cm^3}$,  the
 $N=50$ and $N=82$ r-nuclides dominate the composition. More precisely, for
 $10^9 \le \rho[{\rm g/cm^3}] \le 6 \times 10^{10}$, the decompression of the outer crust leads to
 $N=50$ nuclei with $80 \le A \le 86$. At these densities, only those nuclei whose masses 
are experimentally known are involved. This is not the case at higher densities.
The use of the HFB-9 mass table (Sect.~\ref{nuc_static}), complemented with available experimental
 masses, predicts the 
production of $N=82$ nuclei with $120 \le A \le 128$ for $6\times10^{10} \le \rho[{\rm g/cm^3}] 
\le 3\times10^{11}$. Consideration of neutron emissions by $\beta$-delayed processes, as well as the effects of 
finite temperatures on the energy distribution, would certainly
  spread the matter over a wider mass range than the one originally found in the crust.

The situation is quite different in the inner crust (Fig.~\ref{fig_crust}; initial densities 
$\rho>\rho_{\rm drip}$) because of the presence of a neutron sea in which the nuclei are immersed.
 The matter is assumed to be in $\beta$-equilibrium prior to the expansion. This equilibrium is
 estimated on the basis of a Thomas-Fermi equation of state \cite{onsi97} with the BSk9 Skyrme
 force used to build the HFB-9 mass table (Sect.~\ref{nuc_static}). As shown in Fig.~\ref{fig_crust},
 the higher the density, the lower the initial $Y_{\rm e}$ value. 

For the upper part of the inner crust (Fig.~\ref{fig_crust}; 
 $3 \times 10^{11} \le \rho {\rm [g/cm^{-3}]} \le 10^{12}$), the Wigner-Seitz cells at
 $\beta$-equilibrium are made of some 39 protons and $Y_{\rm e}$ lies initially in the
 0.30-0.15 range. In this case, when the density reaches the drip value during the expansion, 
the number of free neutrons available for the r-process is limited, and the neutron-to-seed ratio 
does not exceed 100. For higher densities, $Y_{\rm e}$ is smaller and neutron-to-seed ratios 
up to 1000 can be obtained. For example, at an initial density $\rho\simeq 10^{14}  {\rm g/cm^3}$,
 the Wigner-Seitz cells are characterised initially by $Y_{\rm e}=0.03$, corresponding to a
 $Z=39$ and $N=157$ nucleus. In this case, the neutron-to-seed ratio is as high as 1300 at the 
time the density reaches its drip value.

The expansion is followed all the way to the neutron-drip density as described in \cite{meyer89a}, 
allowing for the co-existence of Wigner-Seitz cells with different proton numbers obtained through
 $\beta$-transitions. These decays are estimated according to \cite{lattimer77},  and are found to
 heat the matter as soon as the electron and nucleon chemical potentials satisfy 
$\mu_{\rm n} - \mu_{\rm p} - \mu_{\rm e}>0$. Depending on the value assumed for the $\beta$ transition probability, 
 the material at the drip density is either of the same charge as that  at $\beta$-equilibrium, or is made of
 nuclides with atomic numbers distributed in a wide $39 \leq Z \leq 70$ range. In both 
cases, the material lying at the neutron-drip line is surrounded by free neutrons with a typical
 number density  $N_{\rm n}\simeq 10^{35}~{\rm cm}^{-3}$.
The expansion can then be followed by a more `classical' r-process reaction network including
a full suite of radiative neutron captures, photo-disintegrations, $\beta$-decays,
$\beta$-delayed neutron emissions, as well as neutron-induced, $\beta$-delayed and
spontaneous fission processes.  The energy deposited by the $\beta$-decays and fission reactions
 can be responsible for a temperature increase, which is evaluated on the basis of an equation of 
state (e.g. the one proposed by  \cite{tim99}).

The final composition of the material ejected from the inner crust is expected to depend on the 
initial density, at least for the upper part of the inner crust at  
$3 \times 10^{11} \le \rho {\rm [g/cm^{3}]} \le 10^{12}$. Figure~\ref{fig_hider3} shows the
 r-nuclide abundances calculated  for four initial densities $\rho < 10^{12}$ g/cm$^3$. Although 
the composition is seen to be density-dependent, it is characterised by peaks with location and 
width similar to those in the SoS.

\begin{figure}
\center{\includegraphics[scale=0.5]{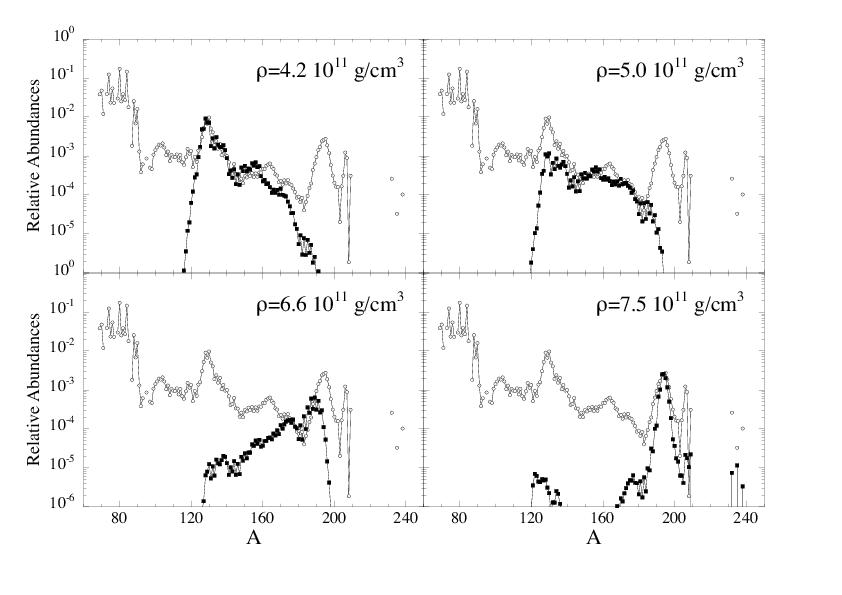}}
\vskip-0.5cm
\caption{Final composition of the ejected NS inner crust material with four different initial 
densities lower than $10^{12}$ g/cm$^3$ (solid squares). The SoS r-abundance distribution is also 
shown} 
\label{fig_hider3}
\end{figure}

For inner-crust layers at  higher densities ($\rho > 10^{12}~{\rm g/cm^{3}}$), large neutron-to-seed
 ratios bring the nuclear flow into the very heavy mass
 region, which leads to fission recycling. In this
 case, the matter can be heated to temperature up to $T_9\approx 0.8$, as seen in 
Fig.~\ref{fig_hider1}.  This figure also displays the associated evolutions of  the neutron 
mass-fraction $X_{\rm n}$ and of the average mass-number of the heavy nuclei $\langle A \rangle$ in 
a clump of material expanding from an initial density $\rho = 10^{14}  {\rm g/cm^3}$ on a timescale
 $\tau_{\rm exp}=6.5$~ms.  In such conditions, all the available neutrons are seen to be 
captured. The fission recycling leads to the displayed oscillations of $\langle A \rangle$. 

\begin{figure}
\center{\includegraphics[scale=0.45]{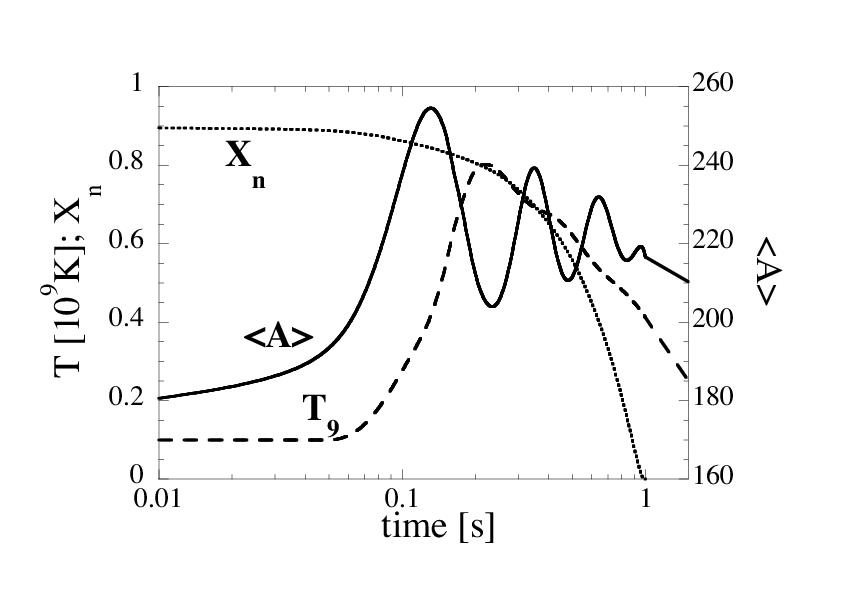}}
\vskip-0.7cm
\caption{Evolution of the temperature, the average mass number $\langle A \rangle$,
 and the neutrino mass fraction $X_{\rm n}$ for a clump of material with
 initial density $\rho = 10^{14}  {\rm g/cm^3}$ expanding on a timescale $\tau_{\rm exp}=6.5$~ms} 
\label{fig_hider1}
\end{figure}

Figure~\ref{fig_hider2} shows the final abundances obtained in the conditions displayed in
 Fig.~\ref{fig_hider1}. For $A>140$, the calculated composition is in relatively good agreement 
with the SoS pattern. In particular, the $A=195$ peak has the right location and width. The 
abundance distribution is now independent of the initial conditions, and in particular of the 
initial density. 

\begin{figure}
\center{\includegraphics[scale=0.4]{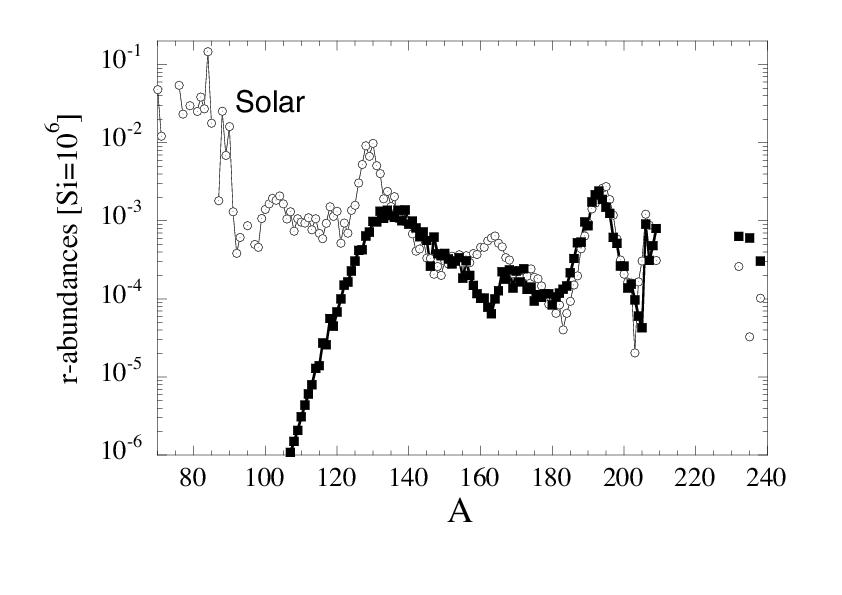}}
\vskip-1.0cm
\caption{Final r-abundance distribution for the same clump of material as in Fig.~\ref{fig_hider1}
 (solid squares). The SoS r-abundance distribution is also shown} 
\label{fig_hider2}
\end{figure}

\begin{figure}
\center{\includegraphics[scale=0.57]{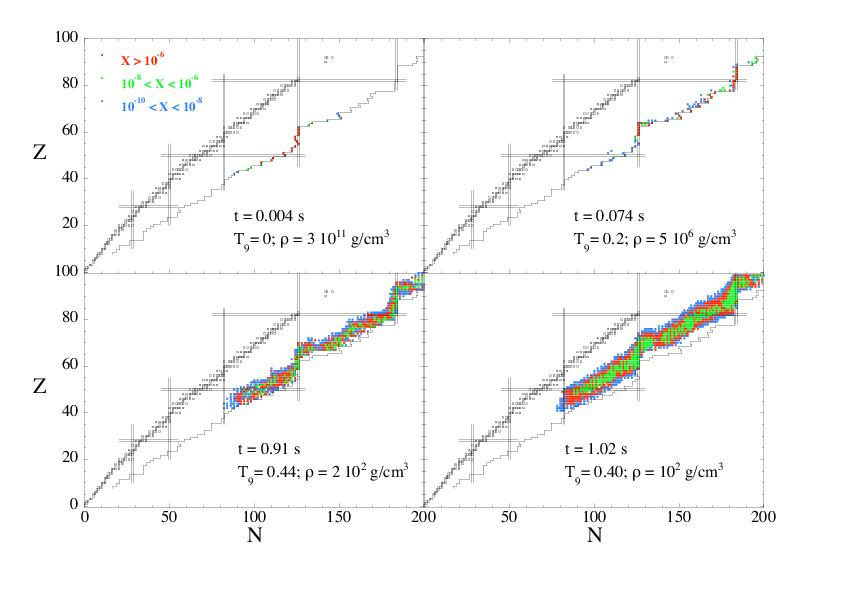}}
\vskip-0.7cm
\caption{Evolution of the r-process flow in the NS coalescence scenario. Strips represent the 
synthesised nuclides between the valley of stability (black squares) and the neutron-drip line 
(stairs). The double lines represent
  the neutron and proton magic-numbers. The initial density is 
 $\rho = 10^{14}$ g/cm$^3$. The final abundance distribution is shown in Fig.~\ref{fig_hider2}. 
The mass fractions are colour-coded, as indicated in the upper left panel}
\label{fig_nsm_flow}
\end{figure}

 As explained in Sect.~\ref{HIDER}, such a r-nuclide distribution results from a sequence of 
nuclear mechanisms that significantly differ from traditional one invoking  the establishment 
of an $(n,\gamma)-(\gamma,n)$ equilibrium followed by the $\beta$-decays of the associated 
waiting-points (Sects.~\ref{canonical} and \ref{MER}). In the present scenario, the neutron
 density is 
initially so high that the nuclear flow follows a path touching the neutron-drip line for the 
first hundreds of ms after reaching the drip density, as illustrated in Fig.~\ref{fig_nsm_flow}.
 Fission keeps on recycling the material. After a few hundreds of ms,
the density has dropped by orders of magnitude, and the neutron abundance falls 
dramatically as a result of efficient captures (Fig.~\ref{fig_hider1}). 
During this period of time, the nuclear
 flow around the $N=126$ region follows the isotonic chain. When the neutron density reaches some 
$N_{\rm n}=10^{20}$~cm$^{-3}$, the timescale of neutron captures by the most abundant $N=126$ nuclei
 becomes larger than a few seconds, and the nuclear flow is dominated by $\beta$-decays back to 
the stability line. 

 In this scenario, photo-disintegrations do not play any major role, so that the calculated  
abundances do not depend sensitively on temperature. For example, a result very similar to the
one displayed in Fig.~\ref{fig_hider2} would be obtained for a constant temperature $T_9 = 0.1$. 
  It is also found to be robust to most of the nuclear uncertainties affecting masses, 
$\beta$-decay rates, or reaction rates. The adopted fission-fragment distribution mainly affects
 the abundance of the $A<140$ nuclei. In fact, the parameter that has the largest impact on the 
calculated abundances appears to be the expansion timescale. As long as the expansion is relatively
 slow, all neutrons have time to be captured. The freeze-out takes place at densities around 
$N_{\rm n}=10^{20}$~cm$^{-3}$, which  leads invariably to an abundance distribution similar to 
the one of Fig.~\ref{fig_hider2}, and close to the predictions of the simple `canonical'
 model described in Sect.~\ref{HIDER} (see Fig.~\ref{fig_steady2}). In contrast, for fast expansions 
 ($\tau_{\rm exp}<3$~ms) of the clumps with initial density of $10^{14} {\rm g/cm^3}$, not all 
the free neutrons are captured, and the neutron density falls proportional to the density. In this 
case, the final distribution becomes sensitive to the expansion timescale, and can differ 
significantly from the SoS pattern.
 
\subsection{The r-process in the outflow from NS + NS discs}
\label{r_torus}

A parametric study of the r-process similar to the one conducted for accretion rates thought to 
be typical for collapsars (Sect.~\ref{r_others}) has been extended by \cite{surman06} to the 
outflow from discs with accretion rates in excess of about $1 M_\odot$/s that are thought to be 
typical of NS mergers. In contrast to the situation encountered for lower accretion rates, neutrinos 
interact with the disc and outflowing material. The result is that $Y_{\rm e}$ values well in 
excess of 0.5 are obtained for moderate accretion rates (about $1~ M_\odot$/s). The
 situation is quite different 
when an accretion rate of $10~M_\odot$/s is considered.
 In this case, the material is driven neutron-rich through 
$\bar\nu_{\rm e}$-captures by protons which  are found to overwhelm the 
$\nu_{\rm e}$-captures by neutrons. This effect increases with decreasing 
outflow accelerations, which gives more time to the neutrinos to interact. The $Y_{\rm e}$ 
values also depend on the entropy. For low enough $s$ values, the material is electron-degenerate
 and has a high $Y_{\rm e}$ regardless of the neutrino or anti-neutrino interactions.

Figure~\ref{fig_outflow1} shows the dependence of the production of the  r-nuclide abundance peaks 
on the outflow acceleration and entropy for the 10 $M_\odot$/s accretion rate. It is seen 
in particular that a strong Pt-peak results for $s \sim 40$ and for low enough accelerations. 
This is confirmed by Fig.~\ref{fig_outflow2}. More secure conclusions on the possibility of 
r-nuclide production in outflows from accretion discs formed in the process of NS + NS coalescence
 has yet to come from detailed hydrodynamical simulations of the merging process.

\begin{figure}
\center{\includegraphics[scale=0.4,angle=270]{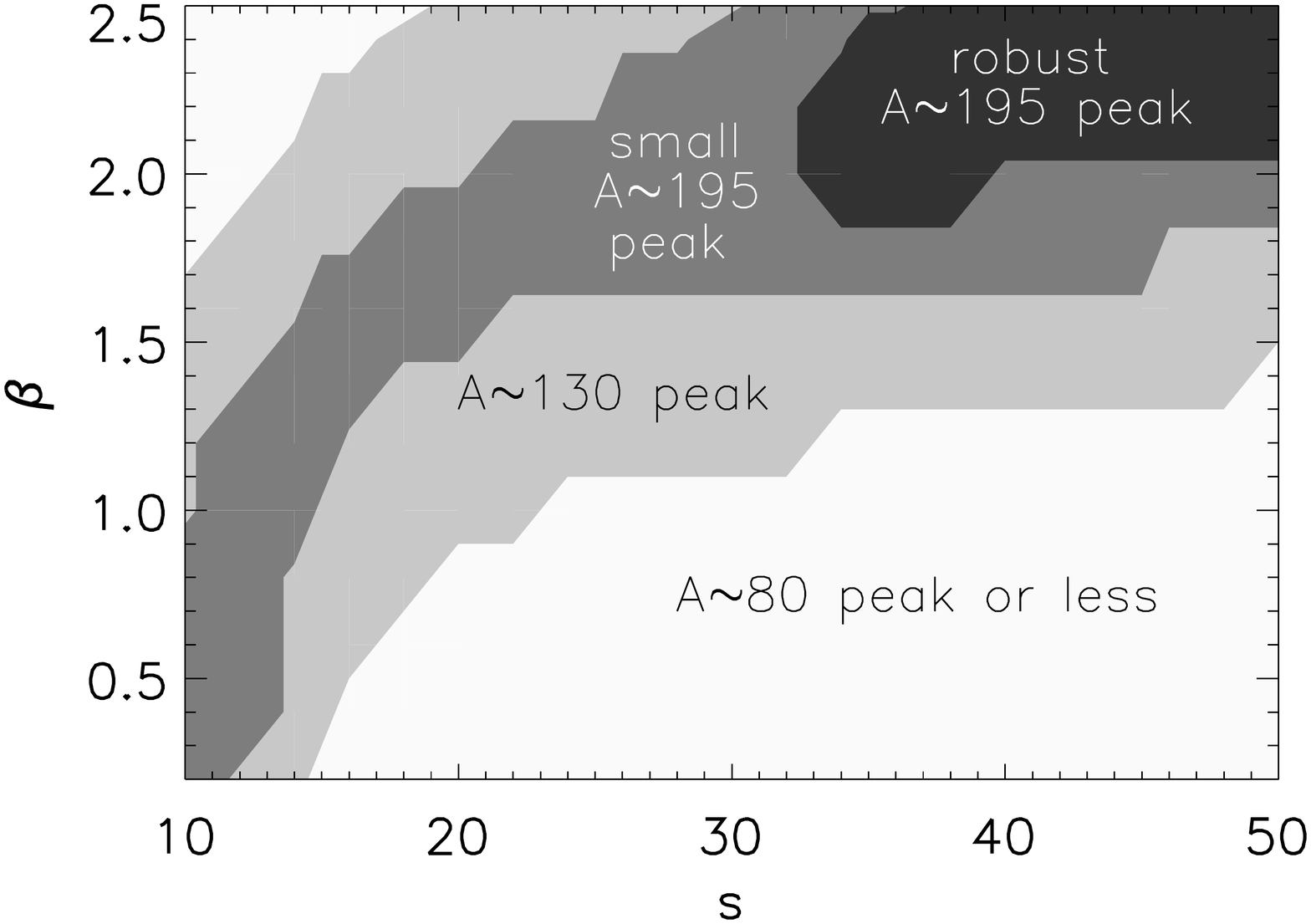}}
\vskip-0.3truecm
\caption{Dependence of the production of the r-nuclide abundance peaks on entropy $s$ and outflow 
acceleration $\beta$ defined by $|u| = v_\infty(1 - R_0/R)^\beta$, where $|u|$ is the absolute 
value of the velocity, whose asymptotic value is $v_\infty$, $R_0$ defines the starting position
 of the disc outflow, and $R = (z^2 + r_{\rm c})^{0.5}$ in regions where the outflow has 
cylindrical symmetry, $r_{\rm c}$ being the radial cylindrical coordinate, or $R = r$ in the 
part of the outflow where spherical symmetry applies. Following its definition, $\beta$ increases 
with decreasing acceleration.  The various labelled regions indicate where the three r-nuclide 
abundance peaks are dominantly produced (from  \cite{surman06})}
\label{fig_outflow1}
\end{figure}

\begin{figure}
\center{\includegraphics[scale=0.4,angle=270]{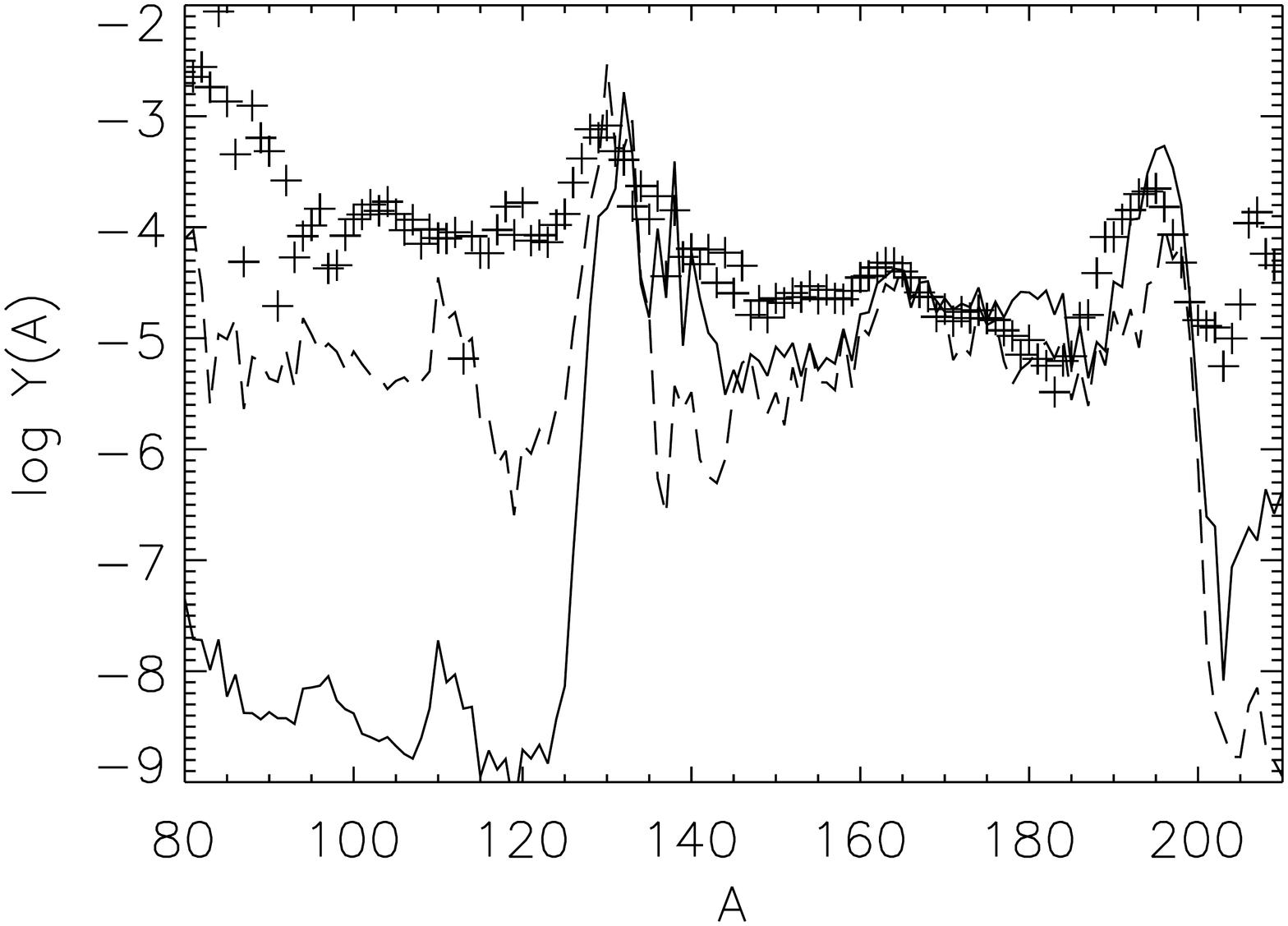}}
\vskip-0.3truecm
\caption{Abundances of r-nuclides produced in an accretion disc outflow for the
 10 $M_\odot$/s accretion rate, $s = 10$ with $\beta = 0.4$ (dashed line), and $s = 50$
 with $\beta = 2.2$ (solid line). A scaled SoS r-abundance distribution (crosses) is displayed 
for comparison (from \cite{surman06})}
\label{fig_outflow2}
\end{figure}

\section{The r-process: confrontation between observations and predictions}
\label{r-process_obs_th}
 
\subsection{Evolution of the r-nuclide content of the Galaxy}
\label{galaxy_obs_th}

A quite natural astrophysicists' dream is to understand the wealth of data on the evolution of
 the r-nuclide content of the Galaxy that are accumulating from very many spectroscopic 
observations,
 a limited sample of which is reviewed in Sect.~\ref{galaxy}. These observations clearly
 demonstrate a huge complexity that will probably keep increasing as new observations become
 available. The best one can attempt at this stage is to explain broad trends which may 
be identified through the analysis of the r-nuclide abundance information. In this exercise, 
one has always to keep in mind that at best more or less reliable {\emph {elemental}} abundances 
are derived from spectral analyses that often rely on approximate classically-used  techniques.
 This necessitates to disentangle the s- and r-process contributions to a given elemental 
abundance. It is generally done by assuming that these two nucleosynthesis contributions are at 
a largely metallicity-independent relative level, and thus do not differ widely from the SoS case.
 This assumption cannot yet be ascertained in any quantitative way, and is in fact not expected to 
hold, at least for the s-process.

With these reservations in mind, one may identify the following main trends: 

(1) the earlier enrichment of the Galaxy with r- than with s-nuclides.
This classical statement has, however, to be taken with some care. As discussed in
 Sect.~\ref{galaxy_s_r_evolution}, the conclusion that can be derived from the evolution of
 La/Eu abundance ratio with metallicity is that there is no clear and unambiguous value of [Fe/H] 
at which the signature of  the production of at least the heavy s-nuclides becomes clearly 
identifiable during the galactic evolution. This is partly related to the quite large abundance 
scatter observed at low metallicity (see (3) below). The situation appears to be much different 
for the light neutron-capture nuclides (Sr in particular), which appear to have had a different 
nucleosynthesis history (see Sect.~\ref{galaxy_s_r_evolution}). This is often interpreted in
 terms of different r-process components producing mainly the light or the heavy r-nuclides, thus
 reviving in some way the original idea of \cite{seeger65} based on the development of the 
canonical high-temperature r-process model (Sect.~\ref{canonical}). The possibility of an 
s-process production of nuclides like Sr early in the galactic history cannot be categorically
 excluded, at least if it can be demonstrated that massive low-metallicity stars can produce the 
light s-nuclides efficiently enough;

 (2) the so-called universality of the relative r-nuclide abundances in the 
$58 \leq Z \leq 76$ range. This is briefly touched upon in Sect.~\ref{galaxy_universality}, 
where it is emphasised in particular that this feature might be preferably termed `nuclear 
physics convergence', as it most likely owes to nuclear physics properties and does not tell
 much about the astrophysics of the r-process;

 (3) a scatter in the r-nuclide abundances (Eu in particular) relative to Fe that significantly 
increases with decreasing [Fe/H]. The classical interpretation of this trend calls for abundance 
inhomogeneities related to the short time spans sampled by the low-metallicity stars early in the
 galactic history. 

There have been several attempts to interpret these general trends in the framework of different
 models for the chemical evolution of the Galaxy. This is an immense task, one perhaps next
 to impossible in the present state of affairs. In this respect, one has to recall 
that no site for the r-process has been identified yet with any reliability (Sects.~\ref{r_dccsn}
and \ref{compact_general}).
This combines with the very schematic nature of the available galactic chemical
 evolution models, not to mention the various intricacies of observational nature. In view of
 this situation, it may be enough to limit ourselves here to a very brief account of the subject,
 waiting for better times to treat the subject in greater depth and provide a more secure
evaluation of the chemical evolution predictions.

Various toy-models of galactic chemical evolution have been constructed, which often focus on the 
evolution of the abundances of two representative elements, Ba and Eu (see e.g. \cite{argast04}
 for references). They adopt different schematic descriptions of the galactic halo and disc, and 
different prescriptions for the physical input quantities to these models. In particular,  fully 
ad-hoc assumptions are made or free parameters are chosen concerning the r-process yields from 
stars of different masses and metallicities.  

\begin{figure} 
\center{\includegraphics[scale=0.75,angle=0]{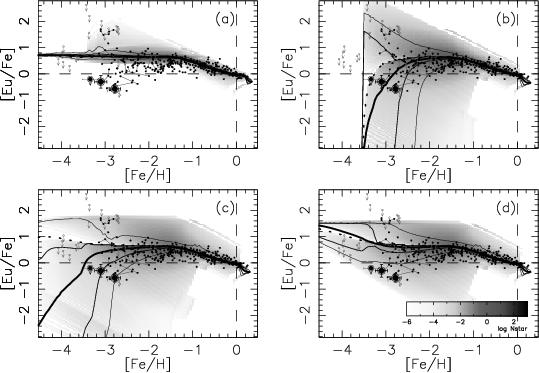}}
\caption{Comparison between observed and predicted [Eu/Fe] ratios. The calculated values are 
obtained with the use of a homogeneous model and under the assumption of a nucleosynthesis 
contribution from supernovae in the following mass ranges (in $M_\odot$) (a)  $ \geq 10$, (b) 
 8 -- 10, (c) 20 -- 25, and (d) $\geq 30$. The thick solid lines indicate the derived average
 abundance ratios, the 50 and 90 \% confidence intervals being delineated by solid and thin-solid 
lines, respectively. Dots represent observations from various sources. The grey scale gives a 
measure of the predicted number density of contributing stars (from \cite{wanajo06})}
\label{fig_chemevol1}
\end{figure}

Figure~\ref{fig_chemevol1} illustrates some predictions by \cite{wanajo06} for [Eu/Fe] derived
 from a homogeneous one-zone model in which it is assumed that stars in prescribed mass ranges
 produce an artificially selected amount of r-nuclides  through the neutrino wind or prompt 
explosion mechanisms (Sect.~\ref{explo_1D} and \ref{explo_multiD}). Quite clearly, the predicted
 [Eu/Fe] ratio is very sensitive to the selected stellar mass range. This result might be 
optimistically considered as providing a way to constrain the site(s) of the r-process from 
observation. Reality is most likely less rosy, as very many uncertainties and severe approximations 
drastically blur the picture.
 
One among the many approximations in use has to do with the assumed homogeneity of the interstellar 
medium at all times. The inhomogeneous model of \cite{argast04} drops this assumption, which might
 increase the plausibility of the predictions, especially at early times in the galactic history. 
With the granularity of the nucleosynthesis events duly considered, one might hope to better 
account for the large scatter of the observed r-nuclide abundances at very low metallicities. 
In addition, the model of \cite{argast04} takes into account the r-process contribution from NS 
mergers (Sect.~\ref{compact_general}) on top of the one from supernovae in selected mass ranges, 
a classically adopted procedure in the field (e.g. \cite{wanajo06}). The many other simplifications 
generally made in other chemical evolution models are also adopted by \cite{argast04}. This 
concerns in particular the r-process yields from supernovae, as well as from NS mergers, that are
 just taken to be solar-like.

\begin{figure}
\center{\includegraphics[scale=0.74]{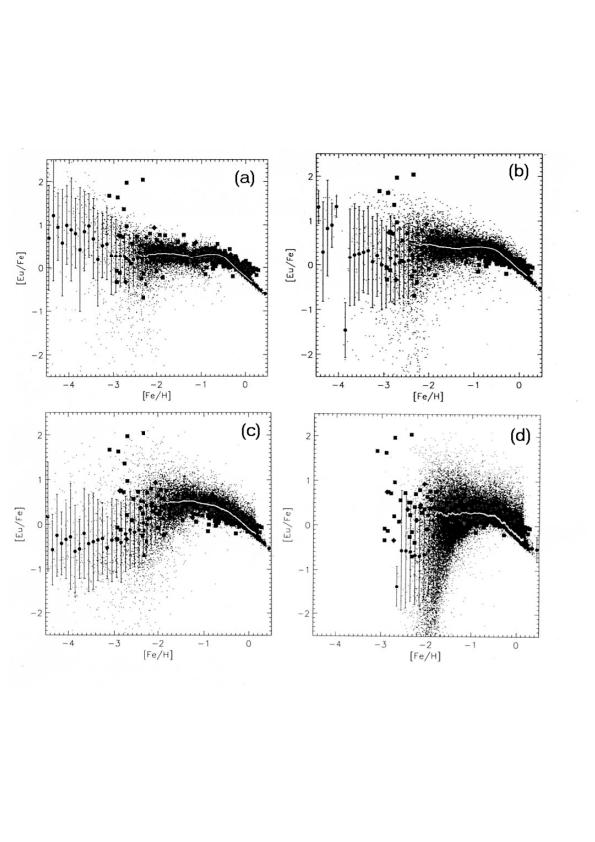}}
\vskip-4.5cm
\caption{ Same as Fig.~\ref{fig_chemevol1}, but for the inhomogeneous model of \cite{argast04}.
 Panels (a) to (c) correspond to the contribution from Type II supernovae in the 8 -- 10, 20 -- 25
 and 20 -- 50 mass ranges (in $M_\odot$). Panel (d) is obtained under the assumption that NS mergers
 are the main r-nuclide sources. These events are given a rate of $2 \times 10^{-4}$ per year, a
 coalescence timescale of $10^6$ y, and $10^{-3} M_\odot$ mass of ejected r-nuclides per event
 (from \cite{argast04})}
\label{fig_chemevol2}
\end{figure}

Figure~\ref{fig_chemevol2} displays some of the [Eu/Fe] ratios predicted by the inhomogeneous model
 of \cite{argast04}. It is concluded that the scenario assuming the predominance of SNII events in 
the 20 to 50 $M_\odot$ range allows the best fit to the observations, this result being obtained
 for total masses of r-nuclides per supernova varying from about $10^{-4} M_\odot$ down to about
 $10^{-7} M_\odot$ when going from 20 to 50 $M_\odot$ stars.  Again, this statement has to be taken 
with great care, as acknowledged by \cite{argast04}, in view of the many uncertainties and
 approximations involved in the chemical evolution model. Within the same model, it is also 
concluded that NS mergers are ruled out as the major source of r-nuclides in the Galaxy. This 
conclusion relies  on very-approximate and highly-uncertain time-dependent frequency of the events,
 coalescence timescales and amount of r-nuclides ejected per merger.
In \cite{argast04}, this amount is allowed 
to vary from about 0.1 to $10^{-4} M_\odot$ depending upon other parameters of the NS merging model,
 and in order to cope at best with observational constraints. An additional uncertainty comes from 
the disregard of various events that could possibly eject initially cold decompressed NS matter 
into the interstellar medium. In particular, magnetars of explosions of NSs below their minimum 
mass~\cite{sumiyoshi98} (in binary neutron-star systems, the mass-losing star may explode if the mass 
drops below a minimum critical mass, see \cite{sumiyoshi98} for more details) or during the spin-down 
phase of very rapidly rotating supra-massive NSs, which could lead to the equatorial shedding of 
material with high angular momentum. Also note that NS--black hole mergers have been estimated to
 be about ten times more frequent  than their NS-NS counterparts. It has to be acknowledged, 
however, that all these possibilities of mass ejection remain highly speculative and uncertain. 

All in all, 
one may conclude that the galactic chemical evolution models devised up to now 
remain highly schematic and uncertain, such that they cannot yet provide us with a trustworthy
 tool to account for the observed evolution of 
the r-nuclide content of the Galaxy, or for constraining the possible sites of the r-process.  

\subsection{Solar-system isotopic anomalies from the r-process}
\label{anomalies_obs_th}

As reviewed in Sect.~\ref{anomalies}, some 
 meteoritic anomalies are characterised by an excess with respect to the SoS mix of 
r-isotopes of certain heavy elements, this enhanced abundance
 being sometimes correlated with the one of the corresponding p-isotopes. Among these anomalies, 
the excesses of heavy isotopes of Xe and of Te making up the so-called Xe-H and Te-H are without 
doubt the most striking ones (Sect.~\ref{anomaly_Xe}), and much effort has been devoted to their 
interpretation. 

The possibility for some r-nuclide anomalies, specially those of the FUN type
 (Sect.~\ref{anomalies}), to emerge from explosive He-burning (Sect.~\ref{coco_HeC}) has been
 investigated for the first time by \cite{thielemann79}. Subsequently, a fully ad-hoc neutron 
burst model that could mimic the situation encountered in an explosive He burning 
layer\footnote{This stellar layer is of special relevance because
 its carbon  richness may be required for the formation of the presolar diamonds that carry Xe-H}
  has been  specifically tailored in order to account for the Xe-H composition of the 
\chem{129,131,132,134,136}{Xe} isotopes by calling for a SoS admixture of \chem{129,131,132}{Xe}
 isotopes to neutron-burst produced \chem{134,136}{Xe} isotopes \cite{howard92}. 

The predicted isotopic composition of Te that goes along with the resulting Xe in the diamonds
 contradicts the observed composition of Te-H. This leads \cite{richter98} to exclude the scenario 
in the form proposed by \cite{howard92} and to favour the following multi-step scenario \cite{ott96},
 referred to as the `early loss' model: (1) a `standard' r-process nucleosynthesis is followed by 
the rapid trapping of the produced nuclides within tiny (possibly pre-existing) 
grains; and (2) this
 trapping is followed by subsequent loss from these grains by recoil resulting from the radioactive 
decay of the unstable precursors that were sufficiently long-lived to decay within the grains. This 
scenario is far from being free of difficulties, however. One concerns the early-enough availability
 of trapping grains in the supernova environment, the timescale requirements being quite stringent.
 Another one is identified by \cite{richter98} as relating to the fact that a strictly SoS 
r-nuclide composition in the Te-Xe region cannot account for the observed Te-H composition within
 the early-loss model. 
Is this problem insuperable? Maybe not.  Indeed,
 the existence of a standard (or universal) r-nuclide pattern in the Te-Xe region is in no way
 demonstrated (Sect.~\ref{galaxy_universality}). Even a single proper site for the r-process 
remains to be identified, as discussed in Sect.~\ref{r_dccsn}. It is also speculated by 
\cite{richter98} that the Te-H pattern could be reproduced by a neutron burst solution constrained 
to reproduce the Xe-H \chem{134}{Xe}/\chem{136}{Xe} ratio.

The model of \cite{howard92} has also been used by \cite{meyer00} to suggest that the \chem{88}{Sr},
 \chem{95}{Mo}, \chem{97}{Mo}, \chem{96}{Zr} and \chem{138}{Ba} excesses with respect to SoS 
observed in certain type X SiC grains (Sect.~\ref{anomaly_Mo}) could be attributed to the He-shell 
neutron burst. This conclusion remains to be scrutinised further with the use of detailed stellar
 models, with an examination of the isotopic composition of the other elements accompanying Sr, Mo, 
Zr and Ba in the X-grains, and with the build-up of a plausible scenario allowing the ashes of an 
explosively burning He-shell to find their way into X-grains at a level such that isotopic 
anomalies are identifiable. 

The pattern of meteoritic isotopic anomalies of the FUN type (Sect.~\ref{anomalies}) that could 
result from the capture of neutrons produced during explosive C-burning (Sect.~\ref{coco_HeC}) 
by \reac{22}{Ne}{\alpha}{n}{25}{Mg} or by \reac{12}{C}{^{12}C}{n}{23}{Mg} has been the subject 
of a specific study \cite{lee79}, who present this scenario as an alternative to the explosive 
He-burning neutron burst or to a fully developed r-process. They acknowledge, however, that many
 uncertainties affect the predictions.

In summary, it is fair to say that the understanding of the very origin of meteoritic anomalies
 characterised by an excess of r-isotopes of a variety of heavy elements still eludes us, or 
needs to be confirmed at best,  just as the r-nuclide content of the various galactic locations
 where it is observed remains to be explained.

\section{The r-process chronometry}                           
\label{chronometry_general}

As recalled in Sects.~\ref{actinides} and \ref{GCR}, actinides enter astrophysics in different 
ways, and especially in attempts (i) to estimate the age of the Galaxy through their present SoS 
content as measured from meteoritic analyses, or through  their abundances determined
 spectroscopically at the surface of old very metal-poor stars, or (ii) to evaluate the time 
elapsed between the synthesis of the GCR actinides and their acceleration to GCR energies, 
thus helping to determine  whether GCRs were accelerated out of fresh supernova ejecta, 
supper-bubble material, or old well-mixed galactic material.

In all the fields referred to above, a necessary condition to interpret the observational
data is to have at disposal r-process predictions for the production ratios
at the sources of the actinides with half-lives typically in excess of about $10^6$ y,
as well as ratios of these actinides to lower $Z$-element abundances. Most importantly,  fair
estimates of the uncertainties in these predicted abundances have also to be evaluated.  These 
challenges have been tackled in substantial details by \cite{goriely99a} and  \cite{goriely01a}
 on grounds of the MER model (Sect.~\ref{MER}) in view of the absence of precise identifications 
of the site(s) where the r-process can develop, as made clear in Sects.~\ref{r_dccsn} and
 \ref{compact_general}. 

In this framework, \cite{goriely99a} and  \cite{goriely01a} demonstrate that the predicted 
synthesis of the actinides is sensitive to the many, and still largely uncertain, astrophysics
 and nuclear physics aspects of the r-process modelling. In fact, the problem is particularly 
acute because of the absence of stable elements heavier than $^{209}{\rm Bi}$, so that the 
actinides production can only be constrained by the stable elements lying some 30 mass units below.
 The situation is worsened further by the especially large uncertainties in the SoS r-process 
contributions to the Pb and Bi abundances, as demonstrated in Table~\ref{tab_r}.

\subsection{MER and the production of the actinides}
\label{MER_actinides}

\begin{figure}
\centerline{\epsfig{figure= 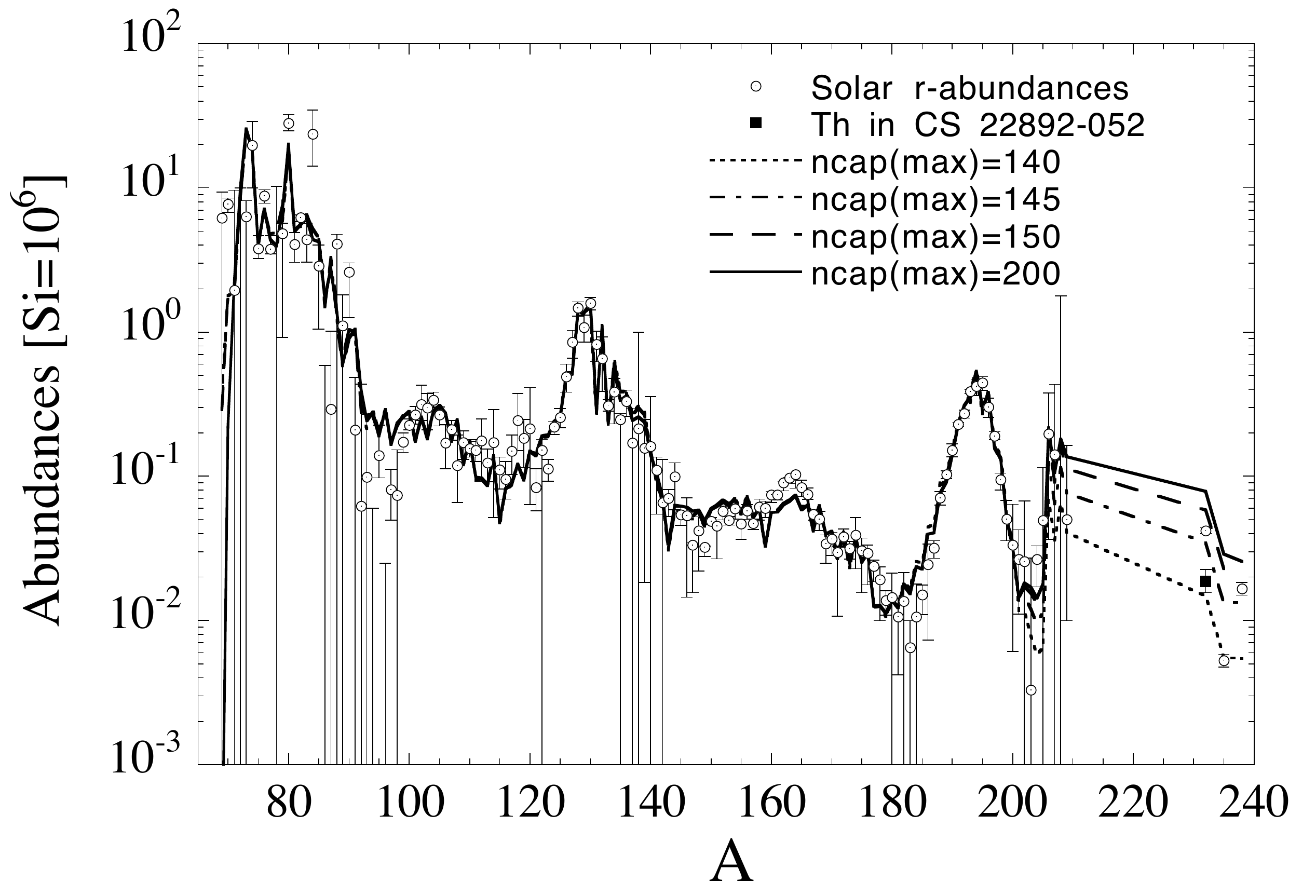,height=8cm,width=14.0cm}} 
\caption{Comparison between the SoS r-nuclides (see Fig.~\ref{fig_rsol_isot}) and the distribution
obtained by MER with a superposition of CEVs with 
$1.3  \leq T_9 \leq  1.7$ and $10 \leq n_{\rm cap} \leq n_{\rm cap}(\rm max)$ with $140 
\leq n_{\rm cap}(\rm max) \leq 200$. The Pb and actinides abundances are affected by the $n_{\rm cap}
 \geq 140$ only. The black square represents the Th abundance derived from the CS 22892-052 
spectrum. The predicted nuclear masses are from the ETFSI model (Sect.~\ref{nuc_static}),
 the $\beta$-decays (and $\beta$-delayed neutron emissions) from the GT2 version of the Gross 
Theory (Sect.~\ref{beta}). The rates of $\alpha$-decays and spontaneous, $\beta$-delayed and
 neutron-induced fissions (before and after the irradiation time $t_{\rm irr}$) are derived following 
an adaptation of the prescription of \cite{kodama75} to the ETFSI fission barriers 
(Sect.~\ref{fission}) (from \cite{goriely99a})}
\label{fig_actinides1}
\end{figure}

\begin{figure}
\centerline{\epsfig{figure= 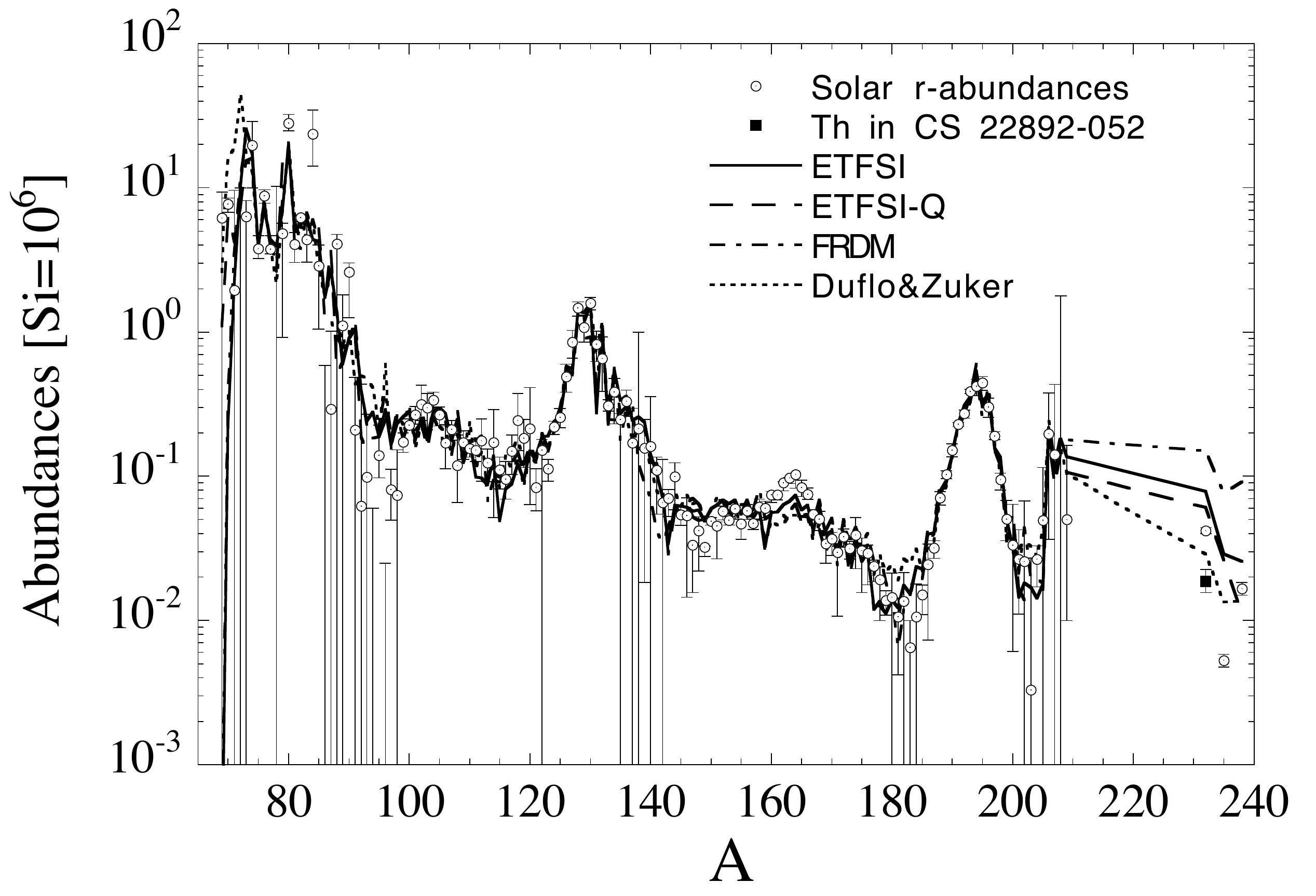,height=8cm,width=14.0cm}} 
\caption{Same as Fig.~\ref{fig_actinides1}, but with 
 different sets of mass predictions
 (from \cite{goriely99a})}
\label{fig_actinides2}
\end{figure}

Figure~\ref{fig_actinides1} provides a sample of MER fits to SoS r-nuclide abundance curve 
obtained by the superposition of CEVs with temperatures varying from 1.3 to $1.7 \times 10^9$ K 
and neutrons captures per seed \chem{56}{Fe}  $n_{\rm cap}$ (Eq.~\ref{eq_ncap}) ranging from 10 to a
 maximum value $n_{\rm cap}$(max) up to 200. The nuclear input is the same for all cases, and 
is identified in the figure caption.   It is seen that the fit to the SoS abundances of the 
 $A \lsimeq 204$ nuclides is excellent, largely independent of the adopted values of the 
maximum number of captured neutrons $n_{\rm cap}$(max), 
whereas the abundances of the Pb peak and of the actinides increase 
with increasing $n_{\rm cap}(\rm max)$. This results directly from the fact that only the CEVs with
 $n_{\rm cap} > 140$ are significant producers of Pb and and of the heavier species only, and 
consequently do not affect the calculated abundances of the lighter r-nuclides. The large 
uncertainties in the Pb peak SoS abundances 
make any reliable selection of a preferred $n_{\rm cap}$(max) value difficult.

Figure~\ref{fig_actinides2} complements Fig.~\ref{fig_actinides1} by illustrating the impact of
 changes in nuclear mass predictions on the calculated abundances
for a given set of astrophysical conditions.
 While the  fits to the SoS r-nuclide distribution are again quite
satisfactory up to the Pb region, significant differences are observed for the actinides. Given
 the lack of realistic r-process models, there is obviously no reason to favour one or another 
nuclear input, and in particular specific mass predictions, even though
 the actinides abundances are sensitive to the adopted masses. 

\begin{figure}
\centerline{\epsfig{figure= 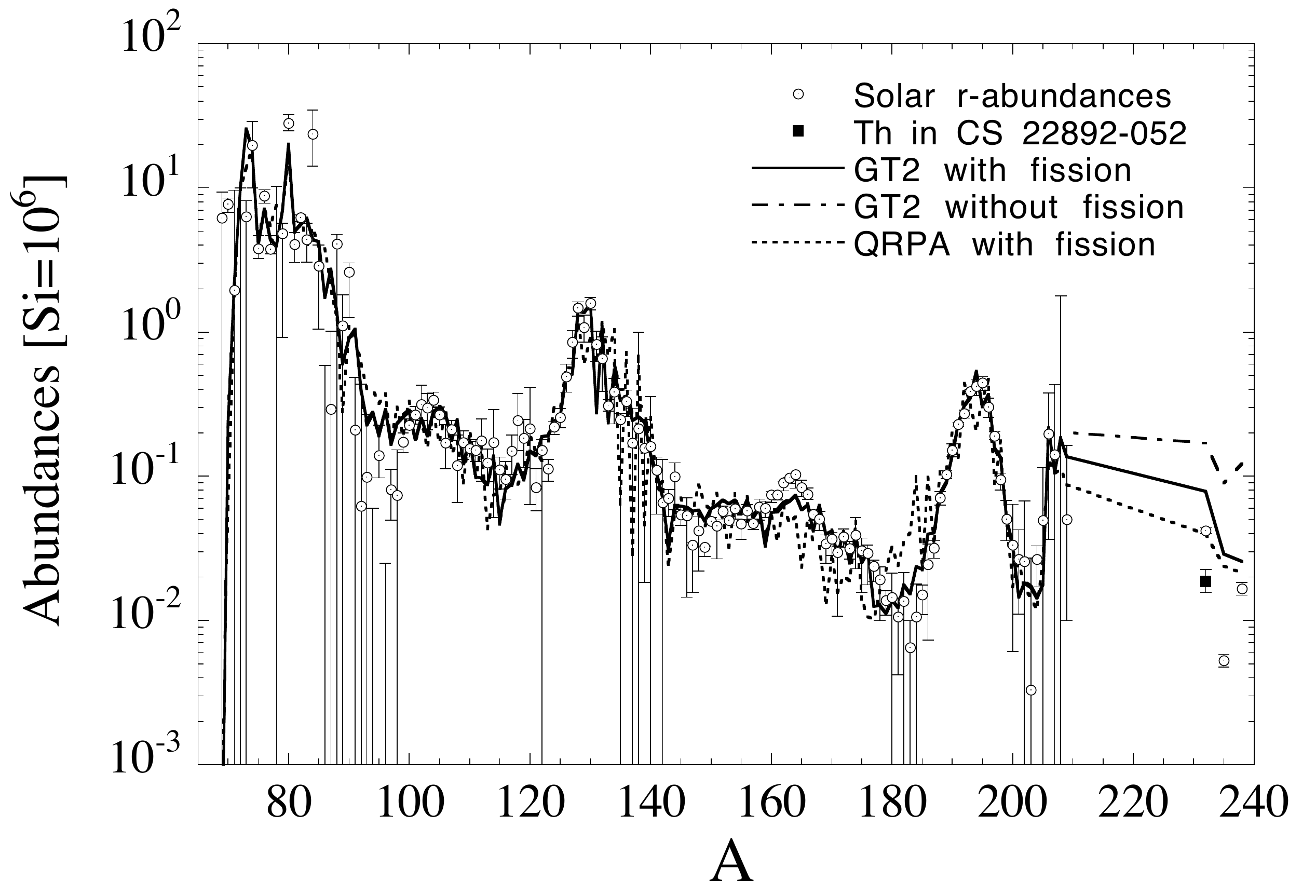,height=8cm,width=14.0cm}} 
\caption{Same as Fig.~\ref{fig_actinides1}, but for CEVs with  $n_{\rm cap}({\rm max}) = 200$ and with the different fission channels included or turned off  (from \cite{goriely99a})}
\label{fig_actinides4}
\end{figure}

Another especially important nuclear ingredient entering the predictions of the actinides yields 
concerns the spontaneous, $\beta$-delayed and neutron-induced fission, which must be described 
in the most careful way in order to take into account
the competing processes responsible for the final 
actinide abundances correctly. Figure~\ref{fig_actinides4} provides a clear illustration 
that due consideration of fission reduces the production of actinides.
  Because of the strong ETFSI shell-effect on 
the fission barriers around  $N = 184$, no fission recycling is found during neutron irradiation
 in the calculations leading to the fits of Figs.~\ref{fig_actinides1} and \ref{fig_actinides2}, 
at least before the crossing of this $N$ value. Also note that, for the CEVs selected in the
 construction of these fits, the fission fragments do not feed significantly the nuclides lighter 
than the Pb peak. A word of caution is in order here. Our inability to provide accurate and
 reliable fission rates close to the valley of nuclear stability (Sect.~\ref{fission}) makes
it close to impossible to have
good quality predictions for the fission modes of exotic neutron-rich nuclei,
 and consequently for the production of actinides by the r-process, and for the recycling efficiency.
 
\begin{table*}
\caption{Minimum and maximum abundances (normalised to Si = $10^6$) of
 Pb and of the actinides with half-lives $t_{1/2} > 10^6$ y as derived from the MER model for 32 
different combinations of CEVs mingling different astrophysics conditions and different 
nuclear-physics input in order to fit two different sets of SoS r-nuclide abundance distributions
(from \cite{goriely01a})} 
\vskip 0.3cm
\begin{center}
\begin{tabular}{ccccccccc}

\hline

Case	&	Pb	&	$^{232}$Th	&	$^{235}$U	&	$^{236}$U	&	$^{238}$U	&
$^{237}$Np &
$^{244}$Pu 	&	$^{247}$Cm	\\
\hline	

Min	&		4.11E-01	&	8.62E-03	&	7.95E-03	&	6.86E-03	&	9.42E-03	&	5.82E-03	&	3.19E-03	&	1.01E-03	\\
Max	&		8.69E-01	&	9.05E-02	&	1.68E-01	&	1.41E-01	&	2.45E-01	&	1.37E-01	&	2.28E-01	&	5.72E-02	\\
 
\hline																	 \end{tabular}
\end{center}
\label{tab_actinides}
\end{table*}

 
The production of the actinides in the MER framework has been revisited and substantially extended by  \cite{goriely01a}. A variety of 32 different superpositions of CEVs are considered, 
corresponding to different sets of astrophysical conditions and nuclear physics inputs that are not
 reviewed in detail here, as well as to different SoS r-nuclide distributions compatible with the
 uncertainties reported in Table~\ref{tab_r}.  As in Figs.~\ref{fig_actinides1}, \ref{fig_actinides2}
 and \ref{fig_actinides4}, all these superpositions give equally good fits to the SoS r-abundances, 
but deviate in their predicted actinides production. Table~\ref{tab_actinides} provides the minimum 
and maximum abundances of Pb and of the actinides with half-lives $t_{1/2} > 10^6$ y
(see Table 1 of \cite{goriely01a} for more details). The corresponding abundances
\chem{232}{Th_{\mathrm f}} and \chem{235}{U_{\mathrm f}} of \chem{232}{Th} and of \chem{235}{U} after decay of their shorter-lived progenitors are given by \chem{232}{Th_{\mathrm f}} = \chem{232}{Th} + \chem{236}{U} + \chem{244}{Pu} and by \chem{235}{U_{\mathrm f}} = \chem{235}{U} + \chem{247}{Cm}.

\subsection{The solar-system nucleo-cosmochronology}
\label{chronometry_solar}

The long-lived \chem{232}{Th}-\chem{238}{U} and \chem{235}{U}-\chem{238}{U}  
pairs have been classically used to estimate the age of the r-nuclides (assumed to be roughly
equal to the age of the Galaxy) from the present meteoritic content of these nuclides.(e.g.
 \cite{cowan91a} for a review).
The opinion has been expressed at several occasions that these pairs have just limited
chronometric virtues (e.g. \cite{yokoi83,arnould01}). This opinion does
not relate only to the uncertainties in the actinide production exemplified in
Table~\ref{tab_actinides}, which could still increase if the oversimplification coming
 from the considered CEVs was removed. An additional source of worry comes from the still large
 uncertainties affecting the meteoritic Th and U abundances, which amount to at
least 25\% and 8\%, respectively \cite{grevesse96}. Last but not least, further
problems arise because any SoS-based nucleo-cosmochronology requires the introduction of 
 chemical evolution models of the Galaxy. These models have to
satisfy in the best possible way as many astronomical observables as possible. In addition, their
internal consistency has to be checked by comparing the deduced actinides abundance ratios at the
 time of formation of the SoS with those adopted at the nucleo-synthetic source. In fact,
this consistency requirement is far from being trivial to fulfill. 

\begin{figure}
\centerline{\epsfig{figure= 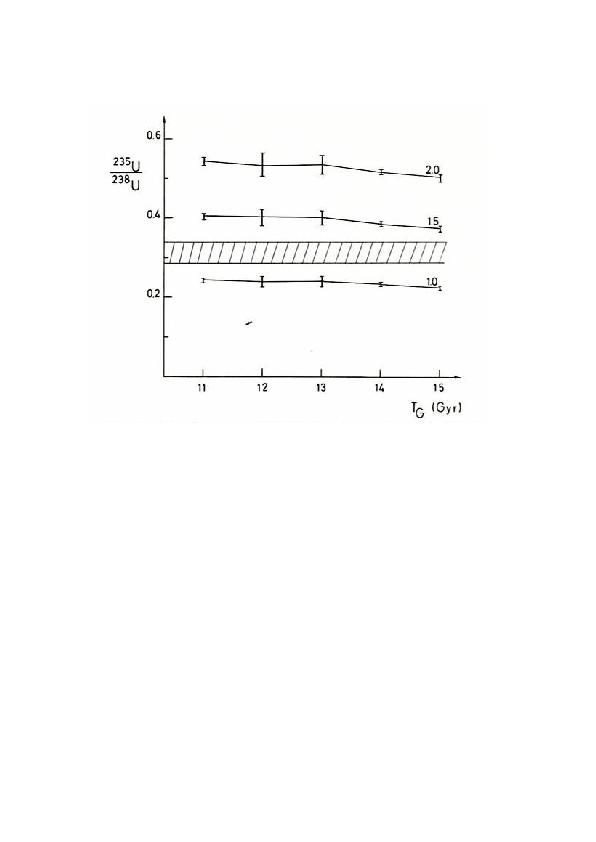,height=16cm,width=14.0cm}} 
\vskip-8.0truecm
\caption{Abundance ratio (\chem{235}{U}/\chem{238}{U})$_0$ versus  $T_G$ for different r-process
 production ratios $R_{235,238}$, as derived from a model for a chemical evolution of the Galaxy
 satisfying a variety of observational constraints. The shaded area corresponds to the measured 
meteoritic ratio with its uncertainty (from \cite{yokoi83} )}
\label{fig_ chrono_sol1}
\end{figure}

\begin{figure}
\centerline{\epsfig{figure= 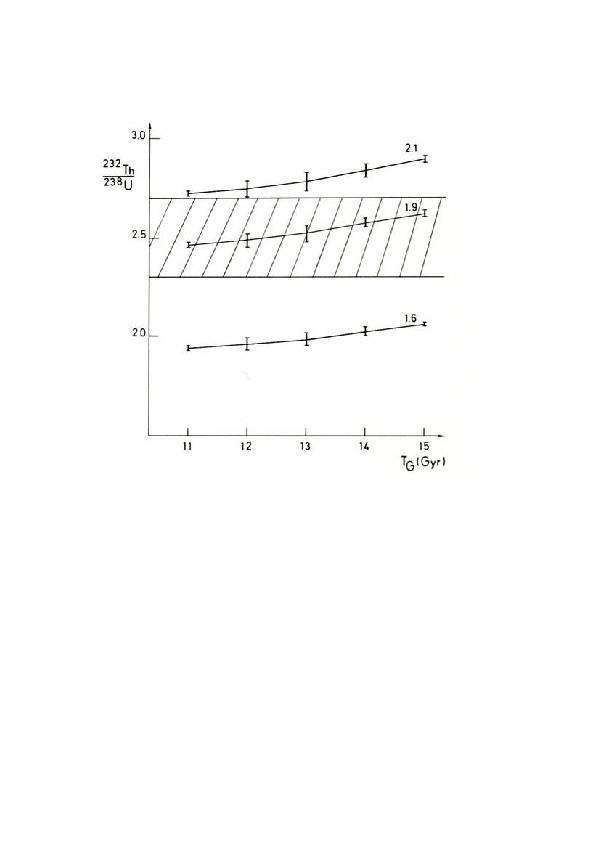,height=16cm,width=14.0cm}} 
\vskip-7truecm
\caption{ Same as Fig.~\ref{fig_ chrono_sol1}  for the (\chem{232}{Th}/\chem{238}{U})$_0$ ratio
 and different production ratios  $R_{232,238}$}
\label{fig_ chrono_sol2}
\end{figure}

 From the construction of a
galactic evolution model generalised in order to include chronometric pairs, \cite{yokoi83}
conclude first that the predicted (\chem{235}{U}/\chem{238}{U})$_0$ and
(\chem{232}{Th}/\chem{238}{U})$_0$ ratios at the time $T_\odot$ of isolation of the SoS
material from the galactic matter about 4.6 Gy ago may be
 only very weakly dependent on the galactic age
 $T_G$, at least in the explored range from about 11 to 15 Gy. 
This result is illustrated in Figs.~\ref{fig_ chrono_sol1} and \ref{fig_ chrono_sol2}. It is owing
 to the fact that the stellar birthrate (i.e. the cumulative galactic mass going into stars forming 
at a given time) is expected to be rather weakly time dependent (except possibly at early galactic
 epochs, but a reliable information on these times is largely erased by the subsequent long period 
of chemical evolution). In this situation, the \chem{232}{Th}-\chem{238}{U} and
\chem{235}{U}-\chem{238}{U} pairs are unable to provide chronometric information which cannot 
be revealed by other methods. At best, they can provide results in agreement with
conclusions derived from other techniques. As shown by \cite{yokoi83}, this is true at least if the
r-process production ratio $R_{235,238}$ of \chem{235}{U_{\mathrm f}} and of \chem{238}{U} and the ratio $R_{232,238}$ of \chem{232}{Th_{\mathrm f}} and of \chem{238}{U} lie in the approximate ranges $1 <  R_{235,238} < 1.5$ and
 $1.6 <  R_{232,238} < 2$ (\chem{232}{Th_{\mathrm f}} and \chem{235}{U_{\mathrm f}} are defined in Sect.~\ref{MER_actinides}). If this is
not the case, the adopted galactic evolution model simply does not provide any chronometric
solution in the explored 11 to 15 Gy age range. As shown by \cite{goriely01a}, these two constraints 
cannot be satisfied simultaneously by any of their considered 32 different superpositions of 
CEVs. Some suitable solutions could be obtained, however, if the above constraints on 
 $R_{232,238}$	and	$R_{235,238}$ were stretched by a value of 0.1. Considering that the 
chronological results of \cite{yokoi83} are derived from a simple approximation of the
 highly-intricate galactic chemical evolution, this small extension
certainly does not hurt unsupportably their results.  On the other hand, this does not deny the
 necessity of further improvements of the concerned nuclear models, as well as of r-process
 scenarios that produce those chronometric nuclides in a natural way.

\subsection{The chronometry of very metal-poor stars}
\label{chronometry_lowz}

One might also confront the predicted range of actinides productions reported in Table~\ref{tab_actinides} with the observations of r-nuclides in old metal-poor stars (Sect.~\ref{actinides}). Compared with the SoS case (Sect.~\ref{chronometry_solar}), this chronometry has the advantage of allowing the economy of a galactic evolution model. Even so, the difficulties are more substantial than it might appear at a first glance.
 One of the main problems lies in the necessity to make
the assumption that the r-process is `universal'. In other words, the observed patterns of
r-nuclide abundances in metal-poor stars have to be considered as exactly solar.  This is indeed
the only way to take the largest possible advantage of the observed metal-poor star content of Th and  U by bringing them to the status of chronometers.

Under the assumption of universality, the age of CS 31082-001, in which both the Th and U abundances have been derived from observation \cite{hill02}, has been estimated by \cite{goriely01a} on grounds of the Eu and actinide yields predicted in the MER framework from the 32 different CEV superpositions referred to in Sect.~\ref{MER_actinides}. The ages $T^*_{\rm U,Eu}$ and $T^*_{\rm U,Th}$ measured from today and predicted from the comparison between the observed CS 31082-001 Eu, Th and U abundances and the range of  their predicted abundances are displayed in Fig.~\ref{fig_age} only if the two ages differ by at most 5 Gy. It is seen that both ages are so uncertain, with $3.5 \lsimeq T^*_{\rm U,Eu}{\rm [Gy]} \lsimeq 15$ and $4.5 \lsimeq T^*_{\rm U,Th}{\rm [Gy]} \lsimeq 16$, that no really useful chronometry can be derived. The spread in ages would be even larger if the condition that the cases for which the two ages differ by more that 5 Gy was relaxed. Even negative ages can in fact be obtained!

\begin{figure}
 \center{\includegraphics[scale=0.4,angle=0]{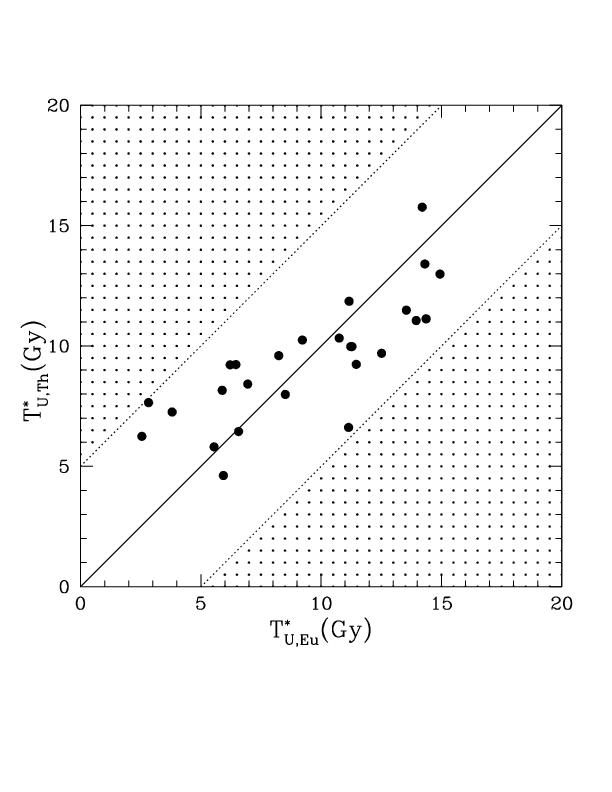}}
 \vskip-1.4cm
 \caption{Ages measured from today of the metal-poor star CS 31082-001 derived from the U-Eu ($T^*_{\rm U,Eu}$) and U-Th ($T^*_{\rm U,Th}$) chronometries, assuming the universality of the r-process. The observed abundances are from \cite{hill02}, and the range of yield predictions are based within the MER model on 32 superpositions of CEVs that fit equally well the solar r-nuclide distribution in the $A \leq 204$ mass range \cite{goriely01a}. Only the cases for which the two age predictions differ by at most 5 Gy are displayed. These cases lie inside the two dotted boundaries. The diagonal solid line is the location of equal $T^*_{\rm U,Eu}$ and $T^*_{\rm U,Th}$ ages. Note that the ages reported  here are somewhat different from those published by \cite{goriely01a} before a revision of the observed abundances proposed by \cite{hill02}}
\label{fig_age}
\end{figure}
 
Figure~\ref{fig_age} calls for additional comments. In particular, the question of the `universality' of the r-process  has already been addressed in
 Sect.~\ref{galaxy_universality}.
It may be worth reiterating the strong reservations expressed there on the 
validity of this hypothesis. In fact, the universality assumption leads to some odd chronometric 
conclusions, as stressed by \cite{arnould01}. As an example, under the universality assumption, 
the approximately 2.8 times larger Th/Eu ratio in CS 31082-001 than in CS 22892-052 would lead to the 
conclusion that CS 22892-052 (with [Fe/H]=-3.1) largely predates CS 31082-001 (with [Fe/H]=-2.9) by 21 Gy.
 Another example is provided by the  Pb/Th ratios observed in 
CS 22892-052  (log$\epsilon$(Pb/Th)=1.80~$\pm$~0.40) and in
CS 31082-001 (log$\epsilon$(Pb/Th)$=0.43~\pm 0.2$),
which may shatter the hope for the universality hypothesis as well.
 A correlation indeed exists between the r-process production of Pb and Th, the 
abundances of these two elements increasing or decreasing concomitantly (see Fig.~1 of 
\cite{goriely99a}).  In these conditions, and if the
universality of the Pb/Th ratio is assumed, the observed Pb/Th values turn out to be discrepant
by a factor of about 25, at least if the two stars have roughly the same age. If this is indeed
the case (which is not a far-fetched assumption in view of their similar [Fe/H] ratio),
either the universality assumption is invalid, and a specific actinides-producing
r-process has to be called for, or the Pb in CS 22892-052 is largely of s-process
origin (see Sect.~\ref{actinides}).

\subsection{The actinides content of GCRs}
\label{GCR_actinides}

As stressed in Sect.~\ref{GCR}, the measurement of the actinides content of GCRs is within the reach
 of present technologies. It could provide first-hand information on their origin and age, as well
 as quality constraints on the actinides production by the r-process. 

Figure~\ref{fig_actinides5} provides some abundances of interest for the quantitative 
interpretation of future GCR actinides measurements derived from the recommended, minimum and 
maximum values displayed in Table~\ref{tab_actinides}.  Predictions of this kind 
provide a necessary tool to help confirming that GCRs are not fresh supernova ejecta. They 
also provide us with 
a way of discriminating between two competing models for their acceleration: the isolated
 supernova remnant exploding in ordinary, old, ISM, or the super-bubble scenario.

\begin{figure}
 \center{\includegraphics[scale=0.5,angle=0]{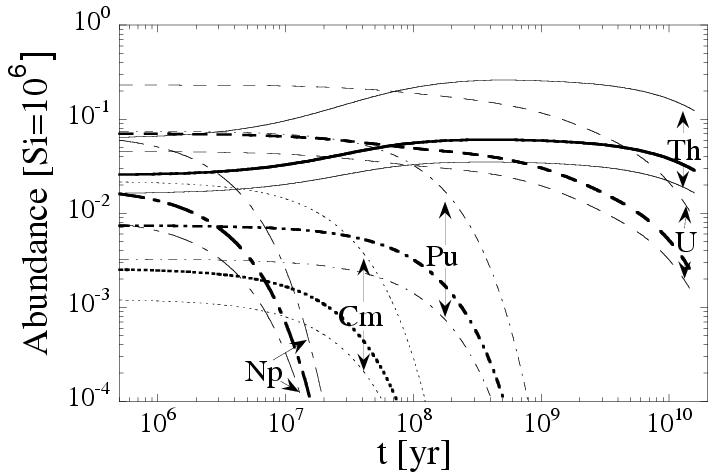}}
\caption{Time variations of the elemental abundances of the actinides of interest for GCR
studies. A single r-process production is assumed at time zero. The thick and thin lines correspond
 to the recommended values and their lower or upper limits given in Table~\ref{tab_actinides}
 (from  \cite{goriely01a}) }  
\label{fig_actinides5}
\end{figure}

\section{Summary and prospects}
\label{summary}

We briefly summarise here a selection of the main aspects of the r-process problem that have been 
reviewed above:

 (1) {\it The r-nuclide content of the solar system}.  Since \cite{BBFH57}, the bulk 
solar-system (SoS) content of r-nuclides has been a key source of information. Uncertainties 
unfortunately still affect a variety of SoS r-nuclide abundances, which weakens to some
 extent their constraining virtues on r-process models. These uncertainties stem from various 
origins, including meteoritic data or, more importantly, models for the SoS s-process content, 
from which r-abundances are traditionally derived. They need to be kept in mind in particular 
when interpreting spectroscopic data [see (2) below].  It is difficult to see how this situation 
can be improved in a decisive way. Isotopic anomalies involving some r-nuclides have been 
identified in meteorites. They do not bring much useful information on the identification of the 
site and characteristics of the r-process;

 (2) {\it The evolution of the r-nuclide content of the Galaxy}. A substantial observational
 work has been conducted in recent years in order to identify the level of r-process contamination
 of the Galaxy with time. This effort largely relies on the abundance evolution of Eu, classified
 as an r-process element on grounds of SoS abundance analyses. The main conclusions derived from
 this observational work may be summarised as follows:\\
(i) The Eu data are classically used to support the idea that the r-process has contributed very
 early to the heavy-element content of the Galaxy. However, the 
scatter of the observed Eu abundances
introduces some confusion when one tries to establish a clear trend of the Eu enrichment with
 metallicity. It is also difficult to identify the value of [Fe/H] at which the signature of the 
s-process becomes identifiable. This conclusion relies in particular on La abundances derived 
from observation, La being classically considered as an s-element in the SoS (even if a
 non-negligible r-process contribution cannot be excluded). A most useful information on the 
relative evolution of the s- and r-process efficiencies in the Galaxy would be provided by the
 knowledge of the isotopic composition of the neutron-capture elements. Such data are 
unfortunately  very scarce
 the Ba isotopic composition in a limited sample of stars. They are still 
under some debate, but suggest that the relative r-process contribution to Ba has decreased during 
the galactic evolution;\\
(ii) Much excitement has been raised by the observation that the patterns of abundances of heavy 
neutron-capture elements between Ba and Pb in r-process-rich metal-poor stars are remarkably 
similar to the SoS one. This claimed `convergence' or `universality' has to be taken with some
 care, however, as it largely relies on the assumption that the decomposition between s- and 
r-process contributions in metal-poor stars is identical to 
that in the SoS, a point that has yet to be demonstrated. 
A case is made that the convergence does not tell much about the astrophysics 
of the r-process, and is largely a signature of nuclear properties in the
 $58 \lsimeq Z \lsimeq 76$ range. This claim is in marked contrast with a statement that is
often found in the literature. On the other hand, no `universality' appears to hold for
$Z \gsimeq 76$. This concerns in particular the Pb-peak elements and the actinides. This situation 
has far-reaching consequences. In particular, it invalidates the many attempts that have been made
 to build detailed galactic chronologies based on the actinides content of very metal-poor stars. 
At best, some lower limits to the age can be be derived. For $Z \lsimeq 58$ elements, the existence
 of a universality appears to be limited to stars with a large Eu overabundance only;\\
(iii) the different behaviours of the abundance patterns of the elements below and above Ba have
 been the ground for the hypothesised existence of two so-called r-process `components'. The 
question of the number of such components brings us more than forty years back, when it was 
discussed in the framework of the newly-constructed canonical r-process model. 
At a time when one is desperately trying to identify
a suitable astrophysical site for the development of an (even limited) r-process [see (4) below], 
numbering the `components' may not deserve a very high priority;

(3) {\it The nuclear physics for the r-process}. Much effort has been put recently in the
 development of microscopic nuclear models aiming at reliable predictions of nuclear ground-state
 properties of thousands of nuclides located between the valley of nuclear stability and the
 neutron-drip line that may be involved in the r-process. Their decay characteristics ($\beta$-decay,
 $\alpha$-decay, spontaneous or various kinds of delayed $\beta$-decays, spontaneous or induced 
fission) and reactivity (nucleon or $\alpha$-particle captures, photo-reactions, or neutrino
 captures) have been scrutinised as well in the framework of models that are microscopic to the
 largest possible extent. Much emphasis has also been put on the coherence of the models 
constructed to predict different nuclear data, in particular through the use of the same basic 
nuclear inputs, like the effective nuclear forces. It has to be noted here that the relevant 
nuclear information entering the modelling of an r-process depend on the astrophysical conditions
 under which the nucleosynthesis is considered to take place. These conditions are still very
 poorly identified, as summarised below.

Beside some remarkable achievements in the field, much obviously remains to be done. The present 
situation may be summarised as follows:\\
(i) the large-scale microscopic-type calculations of $\beta$-decay rates, as needed for the 
r-process simulations,  have not reached a satisfactory level of accuracy and predictive power.
At this  stage, macroscopic models of the `Gross Theory'-type still have to be preferred in the 
nucleosynthesis calculations. The predictability of the probabilities of $\beta$-delayed processes
 is even poorer. Further developments in the microscopic description of the $\beta$-decay
 or delayed $\beta$-decay processes of exotic nuclei are eagerly awaited;\\
(ii) much progress has also to be made in the description of the neutrino-matter interaction.
 This question is important not only for the r-process itself, but also for supernova simulations;\\
(iii) in spite of much recent effort, fission properties remain very-poorly predictable. This is
 even more so for spontaneous fission probabilities. The difficulties culminate for $\beta$-delayed
 fission or neutron-induced fission,  and even more so for neutrino-induced fission,
which is suggested to play an important role in the r-process,
although at a highly speculative level so far.
 There is still a long way to go before being able to acquire
 reliable estimates of transformations involving fission;\\
(iv) the predictability of the rates of the reactions relevant to the r-process has improved
 greatly. The statistical (Hauser-Feshbach) model calculations have benefited largely from 
advances in the evaluation of nuclear level densities, optical potentials, or $\gamma$-ray strength
 functions. Existing sophisticated codes (like Talys or Empire) are currently extended in order
 to meet the specific astrophysics requirements (e.g. the contribution from target excited states),
 and will lead to improved global Hauser-Feshbach rate predictions. The situation concerning 
neutron-induced fission remains very unsatisfactory in view of the problems raised by the
 description of fission [see (iii) above]. Direct captures (as opposed to the formation of a
 compound nucleus, as described by statistical models) are also expected to play an important 
role in the r-process, as exotic neutron-rich nuclei with low neutron separation-energies are 
likely to be involved. A simple model for the direct captures has been devised, but progress has 
clearly to be made in this matter;

(4) {\it The astrophysics aspects of the r-process}. This is clearly and by far the most
 unsatisfactorily understood facet of the r-process modelling, and the one that calls most 
desperately for progress. After some fifty years of research on this subject, the identification
 of a fully convincing r-process astrophysical site remains an elusive dream.
 The attempts conducted thus far  may be briefly summarised as follows:\\
(i) some simplified site-free r-process models have been devised. They have the virtue of shedding
 some light on the conditions that are required in  order for an r-process to develop. Broadly 
 speaking, they can be divided into high-temperature scenarios
 and high-density scenarios. The former ones
 rely on the assumption that high neutron concentrations and high temperatures are both mandatory. 
They include, in order of increasing complexity (or decreasing number of simplifying assumptions),
 (a) the canonical waiting-point approximation model, (b) the multi-event model, and (c) dynamical
approaches. The high-density models rely on the possibility of extensive neutronisation as a result
 of endothermic free-electron captures at high enough densities, 
 followed by the decompression of this highly-neutronised material. 
 A simple steady-flow model is constructed, and shows that the location and width of the
 SoS r-process peaks can be roughly accounted for;\\
(ii) inspired by simulations that attempt to account for successful supernova explosions, 
(semi-)analytic models have been constructed in order to estimate the properties of the material
 ablated from a proto-neutron-star following the deposition of energy and momentum by the neutrinos
 streaming out of it. These models help identifying the physical quantities that have a significant
 impact on the nucleosynthesis in the ablated material, and the suitable ranges of their values for
 the development of a successful r-process. It appears that entropy (depending on temperature and
 density), expansion timescales  (depending on the energy of the ablated material and on its 
mass-loss rate), and electron fraction are decisive quantities for the r-process nucleosynthesis, as 
illustrated by some network integrations. Recall also that the r-process efficiency
 tends to be reduced for increasing neutrino luminosities (which also affect the electron
 fraction);\\
(iii) the conclusions derived from the analytic models referred to above largely ruin the hope of
 having a successful r-process developing in the ablated material from a proto-neutron star, at
 least if one relies on realistic simulations of prompt or delayed core-collapse supernovae. The 
identified necessary conditions for the r-process are indeed not met in these models. True, one may 
get some hope that this conclusion will be invalidated when successful explosion simulations will
 at last be available!\\
(iv) some hope might also come from a high-density r-process that could develop following the
 decompression of crust material ejected as a result of neutron-star coalescence. This is made 
plausible by some yield calculations based on numerical neutron-star merger simulations. Additional
 possibilities could be offered by the outflow of material from the discs forming around the
 coalesced neutron-stars. Considering the present status of the simulations of core-collapse 
supernovae and of neutron-star coalescence, one may conclude that the latter scenario offers 
better potentialities for the r-process. Of course, uncertainties of different natures affect 
this scenario. They concern some modelling details and the calculated r-process yields. The limited
 efficiency of the scenario at early times in the galactic history based on the assumed low 
coalescence frequency in the young Galaxy is often considered as incompatible with the observed 
r-nuclide abundance at very low metallicity.  This may well be the case. One has to acknowledge,
 however, that the frequency of the coalescence or magnetar events along the galactic history, 
as well as the amount of ejected matter per event, remain very uncertain;\\
(v) attempts to interpret the observed trends in the galactic r-nuclide abundances have been
 conducted on grounds of models for the chemical evolution of the Galaxy. This is a task that
is seemingly impossible to complete in the present state of affairs, 
given the absence of a reliable identification of an r-process site, and
 the schematic  nature of  the available galactic chemical evolution models, 
not to mention various intricacies of  observational nature;\\
(vi) the actinides produced by the r-process enter in particular attempts to estimate the age of 
the Galaxy through their present SoS content, or through their abundances evaluated at the surface 
of very metal-poor stars. These attempts face some severe problems related to the nuclear-
 and astrophysics uncertainties that affect the predictions of the actinides production. The
 situation is worsened further by the especially large uncertainties in the contribution of the
 r-process to the solar-system Pb and Bi content. Concerning  the \chem{232}{Th} - \chem{238}{U}
 and \chem{235}{U} -  \chem{238}{U} pairs classically used to date the Galaxy from their present 
meteoritic abundances, 
it may be worth reiterating an opinion first expressed more than two decades
 ago that they have just limited chronometric virtues. 
 The  chronometric predictions based on the observations of Th and U in very metal-poor 
stars have to  be considered with great care as well.
 In order for them to be reliable, it is not only required that the
 production of the actinides by the r-process is well known, but, and very decisively, that the
 production of Th with respect to U and to the Pt peak is `universal.'
 Observation seemingly demonstrates now that this is not the case.

\vskip0.5truecm
The bottom line of this review is that unanswered questions are by far more numerous than solved 
problems when one is dealing with the r-process. This is a very pleasing situation, as hope for 
many exciting discoveries is still ahead of us.
 
\vskip0.5truecm
{\bf Acknowledgements}

In writing this review, the authors have benefited from communications with A. Burrows, I. Borzov,
 H.-Th. Janka and T. Tachibana. They are grateful to the anonymous referee for a careful reading of 
the manuscript and his/her comments. Thanks are also due to L.-S.~The and B.S.~Meyer 
for having kindly provided Fig. 81. They acknowledge the partial financial support of the Konan 
University - Universit\'e Libre de Bruxelles Convention, and of the Belgian Interuniversity 
Attraction Pole PAI 5/07. S.G. is FNRS senior research associate.



\begin{thebibliography}{999}

\def\ApJ#1#2{{\it Astrophys. J.}, {\bf #1}, #2}
\def\ApJS#1#2{{\it Astrophys. J. Suppl.}, {\bf #1}, #2}
\def\A&A#1#2{{\it Astron. Astrophys.}, {\bf #1}, #2}
\def\AJ#1#2{{\it Astron. J.}, {\bf #1}, #2}
\def\MNRAS#1#2{{\it Mon. Not. Roy. Astron. Soc.}, {\bf #1}, #2}
\def\PR#1#2{{\it Phys. Rev.}, {\bf #1}, #2}
\def\PRC#1#2{{\it Phys. Rev. C}, {\bf #1}, #2}
\def\PRD#1#2{{\it Phys. Rev. D}, {\bf #1}, #2}
\def\PL#1#2{{\it Phys. Lett.}, {\bf #1}, #2}
\def\PLB#1#2{{\it Phys. Lett. B}, {\bf #1}, #2}
\def\PRL#1#2{{\it Phys. Rev. Lett.}, {\bf #1}, #2}
\def\Prep#1#2{{\it Phys. Rep.}, {\bf #1}, #2}
\def\RMP#1#2{{\it Rev. Mod. Phys.}, {\bf #1}, #2}
\def\ARAA#1#2{{\it Ann. Rev. Astron. Astrophys.}, {\bf #1}, #2}
\def\ARNS#1#2{{\it Ann. Rev. Nucl. Sci.}, {\bf #1}, #2}
\def\ARNPS#1#2{{\it Ann. Rev. Nucl. Part. Sci.}, {\bf #1}, #2}
\def\PASP#1#2{{\it Pub. Astron. Soc. Pac.}, {\bf #1}, #2}
\def\PTPS#1#2{{\it Prog. Theor. Phys. Suppl.}, {\bf #1}, #2}
\def\RPP#1#2{{\it Rep. Prog. Phys.}, {\bf #1}, #2}
\def\PTP#1#2{{\it Prog. Theor. Phys.}, {\bf #1}, #2}
\def\NP#1#2{{\it Nucl. Phys.}, {\bf #1}, #2}
\def\NPA#1#2{{\it Nucl. Phys. A}, {\bf #1}, #2}
\def\ADNDT#1#2{{\it At. Data Nucl. Data Tables}, {\bf #1}, #2}
\def\Nat#1#2{{\it Nature}, {\bf #1}, #2}
\def\ZP#1#2{{\it Z. Phys.}, {\bf #1}, #2}
\def\ZPA#1#2{{\it Z. Phys. A}, {\bf #1}, #2}
\def\EPJ#1#2{{\it Eur. Phys. J.}, {\bf #1}, #2}
\def\EPJA#1#2{{\it Eur. Phys. J. A}, {\bf #1}, #2}
\def\EPJG#1#2{{\it Eur. Phys. J. G}, {\bf #1}, #2}


\bibitem{BBFH57} Burbidge  E.M., Burbidge G.R., Fowler W.A., Hoyle F. 1957,  \RMP{29}{547}

\bibitem{arnould03} Arnould M., Goriely S. 2003, \Prep{384}{1}

\bibitem{busso99} Busso M., Gallino R., Wasserburg G.J. 1999, \ARAA{37}{239}

\bibitem{meyer94} Meyer B.S. 1994, \ARAA{32}{153}

\bibitem{AK99} Arnould M., Takahashi K. 1999, \RPP{62}{395}

\bibitem{lodders03} Lodders K. 2003, \ApJ{591}{1220}

\bibitem{asplund05} Asplund M, Grevesse N., Sauval A.J. 2005, 
 in {\it Cosmic Abundances as Records of Stellar Evolution and Nucleosynthesis},
 eds. T.G. Barnes \& F.N. Bash,  ASP Conf. Series, vol. 336, p.~25

\bibitem{simmerer05} Simmerer J., Sneden C., Cowan J.J., et al. 2004,  \ApJ{617}{1091}

\bibitem{goriely99} Goriely S. 1999, \A&A{342}{881}

\bibitem{clayton61} Clayton D.D., Fowler W.A., Hull T.E., Zimmerman B.A. 1961, {\it Ann. Phys.}, {\bf 12}, 331

\bibitem{kappeler89} K\"appeler F., Beer, H., Wisshak K 1989, \RPP{52}{945}

\bibitem{arnould97} Arnould M., Paulus G., Meynet G. 1997, \A&A{321}{452}

\bibitem{rayet00} Rayet M, Hashimoto M. 2000, \A&A{354}{740}

\bibitem{goriely05} Goriely S., Siess L. 2005,
 in {\it From Lithium to Uranium: Elemental tracers of early cosmic evolution} 
  eds: V. Hill et al.,  Proc. of IAU symposium Nr 228, (Cambridge: Cambridge University Press), p. 451

\bibitem{arlandini99} Arlandini C., K\"appeler F., Wisshak K., et al. 1999, \ApJ{525}{886}
 
\bibitem{goriely97} Goriely S. 1997, \A&A{327}{845}

\bibitem{palme93} Palme H., Beer H. 1993,
 in {\it Landolt B\"ornstein}, New Series, Group VI, Astron. \& Astrophys.,
 Vol. 3, Subvol. a, (Berlin: Springer), p.~196

\bibitem{anders89} Anders E, Grevesse N. 1989,
 {\it Geochim. Cosmochim. Acta}, {\bf 53}, 197
 
\bibitem{zinner03} Zinner E.K. 2003,
 in {\it Treatrise on Geochemistry}, eds. H.D.Holland \& K.K. Turekian,
 (Amdterdam: Elsevier), {\bf 1}, p.~17

\bibitem{begemann93} Begemann F. 1993,
 in {\it Origin and Evolution of the Elements}, 
 eds. N. Prantzos et al.,  (Cambridge: Cambridge Univ. Press), p. 517

\bibitem{meyer05} Meyer B.S., Zinner E.K. 2006, 
in {\it Meteorites and the Early Solar System II}, 
 eds. D.S. Lauretta \& H.Y. McSween Jr., (Tucson: U. of Arizona Press), p. 69

\bibitem{dauphas02} Dauphas N., Marty B., Reisberg L. 2002, \ApJ{565}{640}

\bibitem{yin02} Yin Q.Z., Jacobsen S.B., Yamashita K. 2002, \Nat{415}{881}

\bibitem{chen04} Chen J.H., Papanastassiou D.A., Wasserburg G.J., Ngo H.H. 2004,
 {\it Lunar Planet Sci.} {\bf XXXV}, Abstract no. 1431
  
\bibitem{pellin99} Pellin M.J., Davis A.M., Lewis R.S., et al. 1999,
 {\it Lunar Planet Sci.} {\bf XXX}, Abstract no. 1969

\bibitem{pellin00} Pellin M.J., Calaway W.F., Davis A.M., et al. 2000,
 {\it Lunar Planet. Sci.}, {\bf XXXI}, Abstract no. 1917 

\bibitem{savina03} Savina M.R., Tripa C.E., Pellin M.J., et al. 2003,
 {\it Lunar Planet. Sci.} {\bf XXXIV}, Abstract no. 2079 

\bibitem{huss95} Huss G.R., Lewis R.S. 1995, {\it Geochim. Cosmochim. Acta}, {\bf 59}, 115

\bibitem{pepin95} Pepin R.O., Becker R.H., Rider P.E. 1995, {\it Geochim. Cosmochim. Acta}, {\bf 59}, 4997

\bibitem{richter98} Richter S., Ott U, Begemann F. 1998, \Nat{391}{261}

\bibitem{hidaka03} Hidaka H., Ohta Y., Yoneda S. 2003, {\it Earth Planet. Sci. Lett.}, {\bf 214}, 455
 
\bibitem{burris00} Burris D.L., Pilachowski C.A., Armandroff T.E., et al. 2000, \ApJ{544}{302}

\bibitem{christlieb04} Christlieb N., Beers T.C., Barklem P.S., et al. 2004, \A&A{428}{1027}

\bibitem{barklem05} Barklem P.S., Christlieb N, Beers T.C., et al. 2005, \A&A{439}{129} 

\bibitem{ishimaru04} Ishimaru Y., Wanajo S., Aoki W., Ryan S.G. 2004, \ApJ{600}{L47}

\bibitem{jonsell06} Jonsell K., Barklem P.S., Gustafsson B., et al.  2006, \A&A{451}{651} 

\bibitem{cescutti06} Cescutti G., Fran\c{c}ois P., Matteucci F., et al. 2006, \A&A{448}{557}

\bibitem{bensby05} Bensby T., Feltzing S., Lundstr\"om I., Ilyin I. 2005, \A&A{433}{185} 

\bibitem{truran02} Truran J.W., Cowan J.J., Pilachowski A., Sneden C. 2002, \PASP{114}{1293}

\bibitem{otsuki06} Otsuki K., Honda S., Aoki W., Kajino T. 2006,  \ApJ{641}{L117}
 
\bibitem{mashonkina06} Mashonkina L., Zhao G. 2006, \A&A{456}{313}

\bibitem{sneden02} Sneden C, Cowan J.J., Lawler J.E., et al. 2002, \ApJ{566}{L25}

\bibitem{aoki03} Aoki W., Honda S., Beers T.C., Sneden C. 2003, \ApJ{586}{506}

\bibitem{magain95} Magain P. 1995, \A&A{297}{686}

\bibitem{lambert02} Lambert D.L., Allende Prieto C. 2002, \MNRAS{335}{325}

\bibitem{mashonkina99} Mashonkina L., Gehren T., Bikmaev I. 1999, \A&A{343}{519}
 
\bibitem{goriely97a} Goriely S., Arnould M. 1997, \A&A{322}{L29}

\bibitem{goriely99a} Goriely S., Clerbaux B. 1999, \A&A{346}{798}

\bibitem{goriely01a} Goriely S., Arnould M. 2001, \A&A{379}{1113}

\bibitem{plez04} Plez B., Hill V., Cayrel R., et al. 2004, \A&A{428}{L9}

\bibitem{yushchenko05} Yushchenko A., Gopka V., Goriely S., et al. 2005, \A&A{430}{255}
 
\bibitem{fowler60} Fowler W.A., Hoyle F, 1960, {\it Ann. Phys.}, {\bf 10}, 280

\bibitem{butcher87} Butcher H.R. 1987, \Nat{328}{127}

\bibitem{francois93} Fran\c{c}ois P., Spite M, Spite F. 1993, \A&A{274}{821}

\bibitem{honda04} Honda S., Aoki W., Kajino T., et al. 2004, \ApJ{607}{474}

\bibitem{vaneck03} Van Eck S., Goriely S., Jorissen A., Plez B. 2003, \A&A{404}{291}

\bibitem{paul01} Paul M., Valenta A., Ahmad I., et al. 2001, \ApJ{558}{L133}

\bibitem{browlee96} Brownlee D.E., Burnett D., Clark B., et al. 1996,
 in {\it  Physics, Chemistry, and Dynamics of Interplanetary Dust}, 
  ed. B.A.S. Gustafson \& M.S. Hanner,  ASP Conf. Series, vol. 104, p.~223

\bibitem{meyer97} Meyer J.-P., Drury L.O'C., Ellison D.C. 1997, \ApJ{487}{182}

\bibitem{westphal98} Westphal A.J., Price P.B., Weaver B.A., Afanasiev V.G. 1998, \Nat{396}{50}
 
\bibitem{westphal00} Westphal A.J., Weaver B.A., Tarl\'e G., et al. 2001, {\it Adv. Space Res.}, {\bf 27}, 797
  
\bibitem{meyer99} Meyer  J.-P., Ellison, D.C. 1999, 
 in {\it  LiBeB, Cosmic Rays, and Related X- and Gamma-Rays},  
 ed. R. Ramaty et al., ASP Conf. Series, vol. 171, p.~187

\bibitem{ellison99} Ellison  D.C., Meyer J.-P. 1999,
 in  {\it  LiBeB, Cosmic Rays, and Related X- and Gamma-Rays},
 ed. R. Ramaty et al.,  ASP Conf. Series, vol.171, p.~207
 
\bibitem{spitzer90} Spitzer L.Jr. 1990, \ARAA{28}{71}
 
\bibitem{higdon98} Higdon J.C., Lingenfelter R.E., Ramaty R 1998, \ApJ{509}{L33}
 
\bibitem{parizot00} Parizot E. 2000, \A&A{362}{786}
 
\bibitem{parizot01} Parizot E. 2001, {\it Space Sci. Rev.}, {\bf 99}, 61
 
\bibitem{bykov01} Bykov A.M. 2001, {\it Space Sci. Rev.}, {\bf 99}, 317
 
\bibitem{binns05} Binns W.R., Wiedenbeck M.E., Arnould M., et al 2005, \ApJ{634}{351}
 
\bibitem{weaver05} Weaver B.A., Westphal A.J. 2005, {\it Adv. Space Res.}, {\bf 35}, 167

\bibitem{lunney03} Lunney D., Pearson J.M., Thibault C. 2003, \RMP{75}{1021}

\bibitem{sch01} Schwarz S., Ames F., Audi G., et al. 2001, \NPA{693}{533}

\bibitem{nov02} Novikov Yu.N., Attallah F., Bosch F., et al. 2002, \NPA{697}{92}

\bibitem{awt03} Audi G., Wapstra A.H.,  Thibault C. 2003, \NPA{729}{337}

\bibitem{aw95} Audi G., Wapstra A.H. 1995,  \NPA{595}{409}

\bibitem{we35} von Weizs\"acker C.F. 1935,  \ZP{96}{431}

\bibitem{frdm} M\"{o}ller P., Nix J.R., Myers W.D., Swiatecki W.J. 1995,\ADNDT{59}{185}

\bibitem{abo95} Aboussir Y., Pearson J.M., Dutta A.K., Tondeur F. 1995, \ADNDT{61}{127}

\bibitem{sg00} Goriely S., Tondeur F., Pearson J.M. 2001, \ADNDT{77}{311}

\bibitem{Vautherin72} Vautherin D., Brink D.M. 1972, \PRC{5}{626} 

\bibitem{sam01} Samyn M., Goriely S., Heenen P.-H., et al. 2002, \NPA{700}{142}

\bibitem{aw01} Audi G., Wapstra A.H. 2001, private communication

\bibitem{sg02} Goriely S., Samyn M., Heenen P.-H., et al.  2002, \PRC{66}{024326}

\bibitem{sam03} Samyn M., Goriely S., Pearson J.M. 2003, \NPA{725}{69} 

\bibitem{gor03} Goriely S.,  Samyn M.,  Bender M.,  Pearson J.M. 2003, \PRC{68}{054325}
 
\bibitem{sam04} Samyn M., Goriely S.,  Bender M.,  Pearson J.M. 2004, \PRC{70}{044309}

\bibitem{gor04} Goriely S.,  Samyn M.,  Pearson J.M., Onsi M. 2005, \NPA{750}{425}

\bibitem{gar99} Garrido E., Sarriguren P., Moya de Guerra  E., Schuck P. 1999, \PRC{60}{064312}

\bibitem{zuo99} Zuo W., Bombaci I., Lombardo U. 1999, \PRC{60}{024605}

\bibitem{rayet82} Rayet M., Arnould M., Paulus G., Tondeur F. 1982, \A&A{116}{183}

\bibitem{fp81} Friedman B., Pandharipande  V.R. 1981,  \NPA{361}{502}

\bibitem{ang04} Angeli I. 2004, \ADNDT{87}{185}
 
\bibitem{khan04} Goriely S., Khan E., Samyn M. 2004, \NPA{739}{331}

\bibitem{bender05} Bender M., Bertsch G.F., Heenen P.-H. 2005, \PRL{94}{102503}

\bibitem{bu99} Bulgac A., Shaginyan V.R. 1999, \PLB{469}{1}

\bibitem{borzov05} Borzov I. 2006, \NPA{777}{645}

\bibitem{langanke03} Langanke K., Mart\'inez-Pinedo G. 2003, \RMP{75}{819}

\bibitem{grotz90} Grotz K., Klapdor H.V. 1990, 
 {\it The Weak Interaction in  Nuclear, Particle and Astrophysics}, (Bristol: Adam Hilger)

\bibitem{audi03a} Audi G., Bersillon O., Blachot J., Wapstra A.H. 2003, \NPA{729}{3} 

\bibitem{ikeda63} Ikeda K., Fujii S., Fujita J.-I. 1963, \PL{3}{271}
 
\bibitem{konopinski66} Konopinski E.J. 1966, 
 {\it The Theory of Beta Rradioactivity}, (London: Oxford University Press)

\bibitem{konopinski66a} Konopinski E.J., Rose M.E. 1965, 
 in {\it Alpha-, beta- and gamma-ray spectroscopy}, 
 ed. K. Siegbahn, vol.2, (Amsterdam: North-Holland), p.~1327

\bibitem{blatt52} Blatt J.M., Weisskopf V.F. 1952, 
 {\it Theoretical Nuclear  Physics}, (New York: John Wiley) 

\bibitem{gove71} Gove N.B., Martin M.J. 1971,  {\it Nucl. Data Tables}, {\bf 10}, 205

\bibitem{yamada65} Yamada M. 1965,  {\it Bull. Sci. Eng. Res. Lab. Waseda Uni.}, No.31/32, 146

\bibitem{takahashi69} Takahashi K., Yamada M. 1969, \PTP{41}{1470}

\bibitem{alford61} Alford W.P., French J.B. 1961, \PRL{6}{119}

\bibitem{bainum80} Bainum D.E., Rapaport J., Goodman C.D., et al. 1980, \PRL{44}{1751}

\bibitem{goodman80} Goodman C.D., Goulding C.A., Greenfield M.B., et al. 1980, \PRL{44}{1755}  

\bibitem{doering75} Doering R.R., Galonsky A., Patterson D.M., Bertsch G.F. 1975, \PRL{35}{1691}

\bibitem{gaarde81} Gaarde C., Rapaport J., Taddeucci T.~N., et al 1981,  \NPA{369}{258}

\bibitem{borge91} Borge M.J.G., Hansen P.G., Johansen L., et al. 1991, \ZPA{340}{255} 

\bibitem{hamamoto94} Hamamoto I. 1994, \NPA{577}{19c}

\bibitem{gaarde80} Gaarde C., Larsen J.S., Harakeh M.N., et al. 1980, \NPA{334}{248}

\bibitem{gaarde83} Gaarde C. 1983, \NPA{396}{127c}

\bibitem{wakasa97} Wakasa T., Sakai H., Okamura H., et al. 1997, \PRC{55}{2909}

\bibitem{yako05} Yako K., Sakai H., Greenfield M.B., et al. 2005, \PLB{615}{193}

\bibitem{bertch82} Bersch G.F., Hamamoto I. 1982, \PRC{26}{1323}

\bibitem{koyama70} Koyama S.I., Takahashi K., Yamada M. 1970, \PTP{44}{663}

\bibitem{takahashi71} Takahashi K. 1971, \PTP{45}{1466}

\bibitem{takahashi73} Takahashi K., Yamada M., Kondoh T. 1973, \ADNDT{12}{101}

\bibitem{tachibana90} Tachibana T., Yamada M., Yoshida Y. 1990, \PTP{84}{641}

\bibitem{nakata97} Nakata H., Tachibana T., Yamada M. 1997, \NPA{625}{521}

\bibitem{takahashi88} Takahashi K. 1987,
 in {\it Origin and Distribution of the Elements},  ed. G.J.Mathews,  (Singapore: World Scientific), p.~542

\bibitem{myers66} Myers W.D., Swiatecki W.J. 1966, \NP{81}{1}

\bibitem{lederer67} Lederer C.M., Hollander J.M., Perlman I. 1967 
 {\it Table of Isotopes}, 6th ed., (New York: John Wiley \& Sons)

\bibitem{kodama75} Kodama T., Takahashi K. 1975, \NPA{239}{489}

\bibitem{eisenberg72} Eisenberg J.M., Greiner W. 1972,
 {\it Nuclear Theory}, Vol.~III, (Amsterdam: North-Holland)

\bibitem{fujita65} Fujita J.-I., Ikeda K. 1965,  \NP{67}{145}

\bibitem{morita71} Morita M., Yamada M., Fujitamyers
 J.-I., et al. 1970, \PTPS{48}{41}

\bibitem{ejiri71} Ejiri H. 1971, \NPA{166}{594}

\bibitem{klapdor84} Klapdor H.V., Metzinger J., Oda T. 1984, \ADNDT{31}{81}

\bibitem{moeller97} M\"oller P., Nix J.R., Kratz K.-L. 1997, \ADNDT{66}{131}

\bibitem{staudt90} Staudt A., Bender E., Muto T., Klapdor-Kleingrothaus H.V. 1990, \ADNDT{44}{79}

\bibitem{terasaki05} Terasaki J., Engel J., Bender M., et al. 2005, \PRC{71}{034310}        

\bibitem{borzov03} Borzov I.N. 2003, \PRC{67}{025802}

\bibitem{migdal65} Migdal A.B. 1968,
 {\it Theory of Finite Fermi Systems and Aplications to Atomic Nuclei},
(New York: Interscience) [{\it Teoriya Konechnykh Fermi-Sistem i Atomnykh Yader}, (Moscow: Nauka, 1965)]

\bibitem{krumlinde84} Krumlinde J. 1984, \NPA{413}{223}

\bibitem{borzov00} Borzov I.N., Goriely S. 2000, \PRC{62}{035501}

\bibitem{engel99} Engel J., Bender M., Dobaczewski J., et al. 1999, \PRC{60}{014302}

\bibitem{niksic05} Nik$\breve{\rm s}$i\'c\ T., Marketin T., Vretenar D., et al. 2005, \PRC{71}{014308}

\bibitem{caurier05} Caurier E., Matr\'inez-Pinedo G., Nowacki F., et al. 2005, \RMP{77}{427}

\bibitem{whitehead80} Whitehead R.R. 1980, 
 in {\it Moment Method in Many Fermions Systems}, 
 eds. B.J.Dalton et al., (New York: Plenum), p.~235

\bibitem{hausman76}  Hausman R.F.Jr. 1976, 
 Ph.D. thesis, Univ. California Radiation Laboratory Report  UCRL-52178 (unpublished)

\bibitem{caurier99} Caurier E., Nowacki F. 1999,  {\it Acta. Phys. Pol. B} {\bf 30}, 705   

\bibitem{mathews83} Mathews G.J., Bloom S.D. 1983, \PRC{28}{1367}

\bibitem{takahashi86} Takahashi K., Mathews G.J., Bloom S.D., 1986,
 in {\it Nuclei off the line of stability} 
 eds. R.A. Meyer \& D.S. Brenner, ACS Symp. Series, vol. 324, p.~145

\bibitem{borzov05a}  Borzov I.N. 2005, \PRC{71}{065801}

\bibitem{tachibana05}  Tachibana T. 2005,  private communication

\bibitem{moeller03} M\"oller P., Pfeiffer B., Kratz K.-L. 2003, \PRC{67}{055802}

\bibitem{martinez99}  Mart\'inez-Pinedo G., Langanke K. 1999, \PRL{83}{4502}

\bibitem{hilf76} Hilf E.R., von Groote H., Takahashi K. 1976, 
 in {\it Nuclei Far from Stability}, CERN-report 73-13, p.~142 
        
\bibitem{duflo95} Duflo J., Zuker A.P. 1995, \PRC{52}{R23}

\bibitem{tachibana95} Tachibana T., Arnould M. 1995, \NPA{588}{333c}

\bibitem{takahashi87} Takahashi K., Yokoi K. 1987, \ADNDT{36}{375}

\bibitem{sato74} Sato K. 1974,  \PTP{51}{726}

\bibitem{hillebrandt76} Hillebrandt W., Takahashi K., Kodama T. 1976,  \A&A{52}{63}

\bibitem{hm80} Howard W. M. , M\"oller  P. 1980, \ADNDT{25}{219}

\bibitem{msi04} M\"{o}ller,P., Sierk A.J., Iwamoto A. 2004, \PRL{92}{072501}

\bibitem{mprt98} Mamdouh A., Pearson J.M., Rayet  M., Tondeur F.  1998, \NPA{644}{389}

\bibitem{mprt01} Mamdouh A., Pearson J.M., Rayet  M., Tondeur F.  2001, \NPA{679}{337}

\bibitem{ms99} Myers W.D., Swiatecki W.J. 1999, \PRC{60}{014606}

\bibitem{sam05} Samyn M., Goriely S.,  Pearson J.M 2005, \PRC{72}{044316}

\bibitem{sam05b} Samyn M., Goriely S. 2005, 
 in {\it Nuclear Data for Science and Technology}  
 eds. R.C. Haight et al.,  AIP Conf. Proc., vol. 769, p.~1382

\bibitem{bd72} Brack M., Damgaard J.,  Jensen A.S., et al. 1972,  \RMP{44}{320}

\bibitem{dutta00} Dutta A., Pearson J.M., Tondeur F. 2000, \PRC{61}{054303}

\bibitem{bqs04} Bonneau L.,  Quentin P., Samsoen D. 2004,  \EPJA{21}{391}
 
\bibitem {dem05} Demetriou P.,  Samyn M.,  Goriely S. 2005,
 in {\it Nuclear Data for Science and Technology},  
 eds. R.C. Haight et al.,  AIP Conf. Proc., vol. 769, p.~1319

\bibitem{fires96} Firestone R.B., Shirley V.S., Baglin C.M., et al. 1996,
 {\it Table of Isotopes}, 8th edition,  (New York: Wiley-Interscience)

\bibitem{hau52} Hauser W. and Feshbach H. 1952, \PR{87}{366}

\bibitem{tepel74} Tepel J.W., Hofmann H.M., Weidenm\"uller H.A. 1974, \PLB{49}{1}

\bibitem{hofman80} Hofmann H.M., Mertelmeier T., Herman M., Tepel J.W. 1980, \ZPA{297}{153}

\bibitem{cox68} Cox J.P., Giuli R.T.  1968, 
 {\it Principles of Stellar Structure}, (New York: Gordon and Breach)

\bibitem{hol76} Holmes J.A., Woosley S.E., Fowler W.A., Zimmerman B.A 1976, \ADNDT{18}{305}

\bibitem{go97} Goriely S. 1997, \A&A{325}{414}

\bibitem{go98}  Goriely S. 1998, \PLB{436}{10}

\bibitem{ripl} {\it Handbook for Calculations of Nuclear Reaction Data, RIPL-2} 2006,
 IAEA-Tecdoc-1506

\bibitem{bethe37} Bethe H.A. 1936, \PR{50}{332}

\bibitem{dem00} Demetriou P., Goriely S. 2001, \NPA{695}{95}

\bibitem{hi05} Hilaire, S., Goriely S. 2006,  \NPA{779}{63}


\bibitem{ndst02} Goriely S. 2002, 
 in {\it Nuclear Data for Science and Technology}, ed. K. Shibata, 
 {\it J. Nucl. Science and Technology}, Suppl. 2, vol. 1, p.~536

\bibitem{hi01} Hilaire S., Delaroche J.P., Girod M. 2001,  \EPJA{12}{169}

\bibitem{koning02} Koning A.J., Delaroche J.P. 2003, \NPA{713}{231}

\bibitem{jlm77} Jeukenne J.-P., Lejeune A., Mahaux C. 1977, \PRC{16}{80}

\bibitem{bdg01} Bauge E., Delaroche J.P., Girod M. 2001, \PRC{63}{024607}

\bibitem{goriely02} Goriely S. 2003, \NPA{718}{287c}

\bibitem{fadden} McFadden L., Satchler G.R. 1966, \NP{84}{177}

\bibitem{nolt}Nolte M., Machner H., Bojowald J., et al. 1987, \PRC{36}{1312}

\bibitem{avr94} Avrigeanu V., Hodgson P.E., Avrigeanu M. 1994, \PRC{49}{2136}

\bibitem{gr98} Grama C., Goriely S. 1998, 
 in {\it Nuclei in the Cosmos V},  eds. N.Prantzos \& S. Harissopulos, 
 (Gif-sur-Yvette: Editions Fronti\`eres), p.~463

\bibitem{mohr00} Mohr P. 2000, \PRC{61}{045802}

\bibitem{Dem02} Demetriou P., Grama C., Goriely S. 2002, \NPA{707}{253}

\bibitem{mc81} McCullagh C.M., Stelts M.L., Chrien R.E. 1981, \PRC{23}{1394}

\bibitem{ko90} Kopecky J., Uhl M. 1990, \PRC{41}{1941}

\bibitem{my77} Myers W.D., Swiatecki  W.J.,  Kodoma T., et al. 1977,  \PRC{15}{2032}

\bibitem{gov98} Govaert K., Bauwens F., Bryssinck J., et al. 1998, \PRC{57}{2229}

\bibitem{zil02} Zilges A., Volz S., Babilon M., et al. 2002, \PLB{542}{43}

\bibitem{va92} Van Isacker P., Nagarajan M.A., Warner D.D. 1992, \PRC{45}{R13}

\bibitem{ca97} Catara F., Lanza E.G., Nagarajan M.A, Vitturi A. 1997, \NPA{624}{449}

\bibitem{kh01} Goriely S., Khan E. 2002, \NPA{706}{217}

\bibitem{kh04} Goriely S., Khan E., Samyn M.  2004, \NPA{739}{331} 

\bibitem{wag91} Wagemans C. 1991, {\it The Nuclear Fission Process}, (Boca Raton: CRC Press)

\bibitem{Lyn80} Lynn J.E., Back B.B. 1974, {\it J.Phys.} {\bf A7}, 395

\bibitem{sin06} Sin M., Capote R., Ventura A, et al. 2006, \PRC{74}{014608}

\bibitem{bruslib1} Arnould M., Goriely S. 2006, \NPA{777} {157}

\bibitem{bruslib2} Aikawa M., Arnould M., Goriely S., et al. 2005, \A&A{441}{1195}

\bibitem{talys} Koning A.J., Hilaire S., Duijvestijn M.C. 2005, 
 in {\it Nuclear Data for Science and Technology},
 eds. R.C. Haight et al.,  AIP Conf. Proc., Vol. 769, p.~1154


\bibitem{empire} Herman M., Capote-Noy R., Oblozinsky P., et al. 2002, in
in {\it Nuclear Data for Science and Technology}, ed. K. Shibata,
{\it J. Nucl. Science and Technology}, Suppl. 2, vol. 1,  p.~116

\bibitem{roro01} Rauscher T., Thielemann F.-K. 2001, \ADNDT{79}{47}

\bibitem{bao00} Bao Z.Y., Beer H., K\"appeler F., et al. 2000, \ADNDT{76}{70}

\bibitem{Dem03} P. Demetriou, M. Samyn, S. Goriely 2004, 
 in {\it Seminar on Fission Point d'Oye V}, eds C. Wagemans et al., (Singapore: World Scientific), p.~21


\bibitem{cameron70} Cameron A.G.W., Delano M.D., Truran J.W. 1970, CERN Rep. 70-30 vol.~2, 735  

\bibitem{kodama73} Kodama T., Takahashi K. 1973, \PLB{43}{167}

\bibitem{blake73} Blake J.B., Schramm D.N. 1973, {\it Astrophys. Lett.}, {\bf 14}, 207

\bibitem{berlovich69} Berlovich E.Ye., Novikov Yu.N. 1969,  {\it Dokl. Akad. Nauk SSSR}, {\bf 185}, 1025

\bibitem{skobelev72} Skobelev N.K. 1972, {\it Yad. Fiz.}, {\bf 15}, 444

\bibitem{wene75} Wene C.-O 1975, \A&A{44}{233}

\bibitem{thielemann83} Thielemann F.-K., Mezinger J., Klapdor H.V. 1983, \A&A{123}{162}

\bibitem{takahashi72} Takahashi K. 1972, \PTP{47}{1500}

\bibitem{meyer89} Meyer B.S., Howard W.M., Mathews G.J. , et al. 1989, \PRC{39}{1876}

\bibitem{cowan91} Cowan J.J., Thielemann F.-K., Truran J.W. 1991, \Prep{208}{267}

\bibitem{panov05} Panov I.V., Kolbe E., Pfeiffer B., et al. 2005, \NPA{747}{633}

\bibitem{meyer95} Meyer B.S. 1995, \ApJ{449}{L55}

\bibitem{hoffman92} Hofman A.D., Woosley S.E. 1992, unpublished [http://ie.lbl.gov/astro.html]

 \bibitem{qian03} Qian Y.-Z. 2003, {\it Prog. Part. Nucl. Phys.}, {\bf 50}, 153

\bibitem{kolbe04} Kolbe E., Langanke K., Fuller G.M. 2004, \PRL{92}{111101}

\bibitem{seeger65} Seeger P.A., Fowler W.A., Clayton D.D. 1965, \ApJS{11}{121}

\bibitem{goriely96} Goriely S., Arnould M., 1996, \A&A{312}{327}

\bibitem{bouquelle96} Bouquelle V., Cerf N., Arnould M., et al. 1995, \A&A{305}{1005}

\bibitem{sneden96} Sneden C., McWilliam A., Preston G.W., et al. 1996, \ApJ{467}{819}

\bibitem{delano71} Delano M.D., Cameron A.G.W. 1971, {\it Astophys. Space Sci.}, {\bf 10}, 203

\bibitem{schramm73} Schramm D.N. 1973, \ApJ{185}{293}

\bibitem{clayton68} Clayton D.D. 1968,
{\it Principles of Stellar Evolution and Nucleosynthesis}, (Chicago: Univ. Chicago Press)

\bibitem{ruffert97} Ruffert M., Janka H.-Th.,  Takahashi K., Sch\"afer 1997, \A&A{319}{122}

\bibitem{witti94a} Witti J. 1994, Diplom thesis, Max-Planck-Institut f\"ur Astrophysik, Garching, unpublished

\bibitem{witti94} Witti J., Janka H.-Th., Takahashi K. 1994, \A&A{286}{841}

\bibitem{hoffman97} Hoffman R.D., Woosley S.E., Qian Y.-Z. 1997, \ApJ{482}{951}

\bibitem{meyer98a} Meyer B.S., Krishnan T.D., Clayton D.D. 1998, \ApJ{498}{808}

\bibitem{woosley92} Woosley S.E., Hoffman R.D. 1992, \ApJ{395}{202}

\bibitem{tsuruta65} Tsuruta S., Cameron A.G.W. 1965, {\it Can. J. Phys.}, {\bf 43}, 2056

\bibitem{baym71} Baym G., Bethe H.A., Pethick C.J., 1971, \NPA{175}{225}

\bibitem{lattimer77} Lattimer J.M., Mackie F., Ravenhall D.G., Schramm D.N. 1977, \ApJ{213}{225}
 
\bibitem{meakin06} Meakin C.A., Arnett D. 2006, \ApJ{637}{L53}

\bibitem{sumiyoshi05} Sumiyoshi K., Yamada S, Suzuki H., et al. 2005, \ApJ{629}{922}

\bibitem{liebendorfer05} Liebend\"orfer M., Rampp M., Janka H.-Th, Mezzacappa A. 2005, \ApJ{620}{840}

\bibitem{siess06} Siess L. 2006,
 {\it Stars and Nuclei: A Tribute to Manuel Forestini},  eds. T. Montmerle \& C. Kahane,
 EAS Pub. Series, (Les Ulis: EDP Sciences), vol. 19, p.~103

\bibitem{hillebrandt84} Hillebrandt W., Wolff R.G., Nomoto K. 1984, \A&A{133}{175}

\bibitem{nomoto87} Nomoto K. 1987, \ApJ{322}{206}

\bibitem{kitaura06} Kitaura F.S., Janka H.-Th, Hillebrandt W. 2006, \A&A{450}{345}

\bibitem{mayle88} Mayle R., Wilson J.R. 1988, \ApJ{334}{909}

\bibitem{maeder96} Maeder A. 1996,
 in {\it Wolf-Rayet Stars in the Framework of Stellar Evolution}, 
 eds. J.-M. Vreux et al., (Li\`ege: Universit\'e de Li\`ege), p.~280

\bibitem{mezzacappa05} Mezzacappa A. 2005,
 in {\it Supernovae as Cosmological Lighthouses}, eds. M. Turetto et al., 
ASP Conf. Series, vol. 342, p.~175

\bibitem{burrows06} Burrows A., Livne E., Dessart L., et al. 2006, \ApJ{640}{878}

\bibitem{blondin03} Blondin J.M., Mezzacappa A., DeMarino C. 2003, \ApJ{584}{971}

\bibitem{scheck06} Scheck L., Kifonidis K., Janka H.-Th., M\"uller E. 2006, \A&A{457}{963}

\bibitem{wang02} Wang L., Wheeler J.C., H\"oflich P., et al. 2002, \ApJ{579}{671}

\bibitem{warren04} Fryer C., Warren M.S. 2004, \ApJ{601}{391}

\bibitem{sawai05} Sawai H., Kotake K., Yamada S. 2005, \ApJ{631}{446}

\bibitem{akiyama03} Akiyama S., Wheeler, J.C., Meier D.L., Lichtenstadt I. 2003, \ApJ{584}{954}

\bibitem{moiseenko06} Moiseenko S.G., Bisnovatyi-Kogan G.S., Ardeljan N.V. 2006, \MNRAS{370}{501}

\bibitem{vietri98} Vietri M, Stella L. 1998, \ApJ{507}{L45}

\bibitem{vietri99} Vietri M., Stella L. 1999, \ApJ{527}{L43}

\bibitem{nomoto06} Nomoto K., Tominaga N., Umeda H., et al. 2006, \NPA{777}{424}

\bibitem{piran05} Piran T. 2004, \RMP{76}{1143}

\bibitem{cardall03} Cardall C.Y., Mezzacappa A. 2003, \PRD{ 68}{023006}

\bibitem{strack05} Strack P., Burrows A. 2005, \PRD{ 71}{093004}

\bibitem{buras05} Buras R., Rampp M., Janka H.-Th., Kifonidis K. 2006, \A&A{447}{1049}

\bibitem{buras05a}  Buras R., Rampp M., Janka H.-Th., Kifonidis K. 2006, \A&A{457}{281}

\bibitem{buras03} Buras R., Rampp M., Janka H.-Th., Kifonidis K. 2003, \PRL{90}{241101}

\bibitem{pruet05} Pruet J., Woosley S.E., Buras R., et al. 2005, \ApJ{623}{325}

\bibitem{dessart06} Dessart L., Burrows A., Ott C.D., et al. 2006, \ApJ{644}{1063} 

\bibitem{fryer99} Fryer C., Benz W., Herant M., Colgate S.A. 1999, \ApJ{516}{892}

\bibitem{woosley94} Woosley S.E., Wilson J.R., Mathews G.J., et al. 1994, \ApJ{433}{229}

\bibitem{takahashi94} Takahashi K., Witti J., Janka H.-Th. 1994, \A&A{286}{857}

\bibitem{thompson01} Thompson T.A., Burrows A., Meyer B.S. 2001, \ApJ{562}{887}

\bibitem{takahashi97} Takahashi K., Janka H.-Th., 1997, 
 in {\sl Origin of Matter and Evolution of Galaxies}, 
 eds. T.~Kajino et al., (Singapore: World Scientific), p.~213

\bibitem{janka93} Janka H.-Th. 1993, 
 {\it Frontier Objects in Astrophysics and Particle Physics},
 eds. F. Givanelli \& G. Mannocchi, (Bologna: SIdF), p.~345

\bibitem{qian96} Qian Y.-Z., Woosley S.E. 1996, \ApJ{471}{331}

\bibitem{meyerbs97} Meyer B.S., Brown J.S. 1997, \ApJS{112}{199}

\bibitem{otsuki00} Otsuki K., Tagoshi H., Kajino T., Wanajo S. 2000, \ApJ{533}{424}

\bibitem{wang06} Wang Z., Kaplan D.L., Chakrabarty D. 2007, \ApJ{655}{261}

\bibitem{janka95} Janka H.-Th., M\"uller E. 1995, \ApJ{448}{L109}

\bibitem{sumiyoshi00} Sumiyoshi K., Suzuki H., Otsuki K., et al. 2000, 
{\it Pub. Astron. Soc. Japan}, {\bf 52}, 601

\bibitem{terasawa02} Terasawa M., Sumiyoshi K., Yamada S., et al. 2002, \ApJ{578}{L137}

\bibitem{fuller95} Fuller G.M., Meyer B.S. 1995, \ApJ{453}{792}

\bibitem{mclaughlin96} McLaughlin G.C., Fuller G.M., Wilson J.R. 1996, \ApJ{472}{440}

\bibitem{meyer98} Meyer B.S., McLaughlin G.C., Fuller G.M. 1998, \PRC{58}{3696}

\bibitem{woosley90} Woosley S.E., Hartmann D.H., Hoffman R.D., Haxton W.C. 1990, \ApJ{356}{272}

\bibitem{wanajo03} Wanajo S., Tamamura M., Itoh, N., et al. 2003, \ApJ{593}{968}

\bibitem{wheeler98} Wheeler, J.C., Cowan J.J., Hillebrandt W. 1998, \ApJ{493}{L101}

\bibitem{frohlich06} Fr\"ohlich C., Mart\'{\i}nez-Pinedo G., Liebend\"orfer M., et al. 2006, \PRL{96}{142502}

\bibitem{suzuki05} Suzuki T.K., Nagataki S. 2005, \ApJ{628}{914}

\bibitem{thompson03} Thompson T.A. 2003, \ApJ{585}{L33}

\bibitem{ito05} Ito, H., Yamada S., Sumiyoshi K., Nagataki S. 2005, \PTP{114}{995}

\bibitem{pruet03} Pruet J., Woosley S.E., Hoffman R.D. 2003, \ApJ{586}{1254}

\bibitem{cameron03} Cameron A.G.W. 2003, \ApJ{587}{327}

\bibitem{nishimura06} Nishimura S., Kotake K., Hashimoto M., et al. 2006, \ApJ{642}{410}

\bibitem{falle02} Falle S.A.E.G. 2002, \ApJ{577}{L123}

\bibitem{macfadyen99} MacFadyen A.I., Woosley S.E. 1999, \ApJ{524}{262}

\bibitem{fujimoto04} Fujimoto S., Hashimoto M., Arai K., Matsuba R. 2004, \ApJ{614}{847}

\bibitem{surman06} Surman R., McLaughlin G.C., Hix W.R. 2006, \ApJ{643}{1057}

\bibitem{daigne02} Daigne F., Mochkovitch R. 2002, \A&A{388}{189}

\bibitem{fryer06} Fryer C., Herwig F., Hungerford A., Timmes, F.X. 2006, \ApJ{646}{L131}

\bibitem{epstein88} Epstein R.I., Colgate S.A., Haxton W.C. 1988, \PRL{61}{2038}

\bibitem{rauscher02} Rauscher T., Heger A., Hoffman R.D., Woosley S.E. 2002, \ApJ{576}{323}

\bibitem{meyer04} Meyer B.S., The L.-S., Clayton D.D., El Eid M.F. 2004, 
{\it Lunar Planet. Sci.}, {\bf XXXV}, Abstract no. 1908

\bibitem{wefel81} Wefel J.P., Schramm D.N., Blake J.B., Pridmore-Brown D. 1981, \ApJS{45}{565}

\bibitem{blake81} Blake J.B., Woosley S.E., Weaver, T.A., Schramm D.N. 1981, \ApJ{248}{315}

\bibitem{oechslin06} Oechslin R., Janka H.-Th. 2006, \MNRAS{368}{1489}

\bibitem{rosswog04} Rosswog S., Speith R., Wynn G.A. 2004, \MNRAS{351}{1121}

\bibitem{rosswog05} Rosswog S. 2005, \ApJ{634}{1202}

\bibitem{sumiyoshi98} Sumiyoshi K., Yamada S., Suzuki H., Hillebrandt W. 1998, \A&A{334}{159}

\bibitem{woods06} Woods P.M., Thompson C. 2006, 
 in {\it Compact Stellar X-ray Sources}, 
 eds. W. Lewin \& M. van der Klis, (Cambridge: Cambridge Univ. Press)
 Cambridge Astrophysics Series, Vol. 39, p.~547

\bibitem{harding06} Harding A.K., Lai D. 2006, \RPP{69}{2631}

\bibitem{gelfand05} Gelfand J.D., Lyubarsky Y.E., Eichler D., et al. 2005, \ApJ{634}{L89}

\bibitem{symbalisty82} Symbalisty E., Scramm D.N. 1982,  \ApJ{22}{L143}

\bibitem{meyer89a} Meyer B.S. 1989, \ApJ{343}{254}

\bibitem{frei99} Freiburghaus C., Rosswog S., Thielemann F.-K. 1999, \ApJ{525}{L121}

\bibitem{go05} Goriely S., Demetriou P., Janka H.-Th., et al.  2005, \NPA{758}{587c}

\bibitem{onsi97} Onsi M., Przysiezniak H., Pearson J.M. 1997, \PRC{55}{3139}

\bibitem{tim99} Timmes F.X.,  Arnett D. 1999, \ApJS{125}{277}

\bibitem{argast04} Argast D., Samland M., Thielemann F.-K., Qian Y.-Z. 2004, \A&A{416}{997}

\bibitem{wanajo06} Wanajo S., Ishimaru Y. 2006,  \NPA{777}{676}

\bibitem{thielemann79} Thieleman F.-K., Arnould M., Hillebrandt W. 1979,  \A&A{74}{175}

\bibitem{howard92} Howard W.M., Meyer B.S., Clayton D.D. 1992, {\it Meteoritics}, {\bf 27}, 404

\bibitem{ott96} Ott U. 1996,  \ApJ{463}{344}

\bibitem{meyer00} Meyer B.S., Clayton D.D., The L.-S. 2000, \ApJ{540}{L49}

\bibitem{lee79} Lee T., Schramm D.N., Wefel J.P., Blake J.B. 1979,  \ApJ{232}{854}

\bibitem{cowan91a} Cowan J.J., Thielemann F.-K., Truran J.W., 1991,  \ARAA{29}{447}

\bibitem{yokoi83} Yokoi K., Takahashi K., Arnould M. 1983,  \A&A{117}{65}

\bibitem{arnould01} Arnould M., Goriely S. 2001,
 in {\it  Astrophysical Ages and Time Scales}, eds. T. von Hippel et al., 
  ASP Conference Series, Vol. 245, p.~252

\bibitem{grevesse96} Grevesse N., Noels A.,  Sauval, A.J. 1996, 
 in  {\it Cosmic Abundances},  eds. S.S. Holt, \& G. Sonneborn, 
 ASP Conf. Series, vol. 99, p.~117
 
 \bibitem{hill02} Hill V., Plez B., Cayrel R., et al. 2002, \A&A{387}{560}
 
 
\end{thebibliography}
\end{document}